%% file: SUS-13-011_temp.tex
\pdfoutput=1

\documentclass[11pt,twoside,a4paper,cmspaper,final,collab]{cms-tdr}
\def\svnVersion{222095}\def\cmsCernNo{2013-148}\def\cmsCernDate{\today}\def\cmsMessage{Published in the European Physical Journal C as \href{http://dx.doi.org/10.1140/epjc/s10052-013-2677-2}{\doi{10.1140/epjc/s10052-013-2677-2.}}}

\begin{document}\cmsNoteHeader{SUS-13-011}

\hyphenation{had-ron-i-za-tion}
\hyphenation{cal-or-i-me-ter}
\hyphenation{de-vices}

\RCS$Revision: 212076 $
\RCS$HeadURL: svn+ssh://svn.cern.ch/reps/tdr2/papers/SUS-13-011/trunk/SUS-13-011.tex $
\RCS$Id: SUS-13-011.tex 212076 2013-10-17 03:36:20Z vimartin $
\newlength\cmsFigWidth
\ifthenelse{\boolean{cms@external}}{\setlength\cmsFigWidth{0.95\columnwidth}}{\setlength\cmsFigWidth{0.4\textwidth}}
\ifthenelse{\boolean{cms@external}}{\providecommand{\cmsLeft}{top}}{\providecommand{\cmsLeft}{left}}
\ifthenelse{\boolean{cms@external}}{\providecommand{\cmsRight}{bottom}}{\providecommand{\cmsRight}{right}}
\ifthenelse{\boolean{cms@external}}{\providecommand{\breakhere}{\linebreak[4]}}{\providecommand{\breakhere}{\relax}}
\newlength\cmsFigWidthTwo
\ifthenelse{\boolean{cms@external}}{\setlength\cmsFigWidthTwo{0.50\linewidth}}{\setlength\cmsFigWidthTwo{0.4\textwidth}}
\newcommand{\lumin}{19.5\fbinv}
\newcommand{\Ttt}{\ensuremath{\PSQt \to \cPqt \PSGczDo}\xspace}
\newcommand{\TbW}{\ensuremath{\PSQt \to \cPqb \PSGcp}\xspace}
\newcommand{\ttll}{\ensuremath{\ttbar\to\ell\ell}\xspace}
\newcommand{\ttlj}{\ensuremath{\ttbar\to\ell+\text{jets}}\xspace}
\newcommand{\ttdl}{\ensuremath{\ttbar\to\ell\ell}}
\newcommand{\ttbarW}{\ensuremath{\ttbar\PW}\xspace} %
\newcommand{\ttbarZ}{\ensuremath{\ttbar\cPZ}\xspace} %
\newcommand{\ttbarg}{\ensuremath{\ttbar\PGg}\xspace} %
\newcommand{\wjets}{\ensuremath{\PW+\text{jets}}\xspace}
\newcommand{\zjets}{\ensuremath{\cPZ+\text{jets}}\xspace}
\newcommand{\mtw}{\ensuremath{M_{\cmsSymbolFace{T}2}^{\PW}}\xspace}
\newcommand{\lsp}{\PSGczDo\xspace}
\newcommand{\chipo}{\ensuremath{\chipm_1}}
\newcommand{\Lep}{\ensuremath{\ell}\xspace}
\newcommand{\htratio}{\ensuremath{H_{\cmsSymbolFace{T}}^{\text{ratio}}}\xspace}
\newcommand{\mindphi}{\ensuremath{\min \Delta \phi}}
\providecommand{\MT}{\ensuremath{M_{\cmsSymbolFace{T}}}\xspace}
\providecommand{\FASTJET}{\textsc{fastjet}\xspace}
\cmsNoteHeader{SUS-13-011} 
\title{Search for top-squark pair production in the single-lepton final state in pp collisions at \texorpdfstring{$\sqrt{s}=8$\TeV}{sqrt(s) = 8 TeV}}

\date{\today}

\abstract{
This paper presents a search for the pair production of top squarks in events with a single isolated electron or muon, jets,
large missing transverse momentum, and large transverse mass.
The data sample corresponds to an integrated luminosity of
19.5\fbinv of pp collisions collected in 2012 by the CMS experiment at the LHC
at a center-of-mass energy of $\sqrt{s}=8$\TeV.
No significant excess in data is observed above the expectation from standard model processes.
The results are interpreted in the context of supersymmetric models with pair production of top squarks that decay
either to a top quark and a neutralino or to a bottom quark and a chargino.
For small mass values of the lightest supersymmetric particle,
top-squark mass values up to around 650\GeV are excluded.
}

\hypersetup{%
pdfauthor={CMS Collaboration},%
pdftitle={Search for top-squark pair production in the single-lepton final state in pp collisions at sqrt(s) = 8 TeV},%
pdfsubject={CMS},%
pdfkeywords={CMS, physics, SUSY}}

\maketitle 

\section{Introduction}
\label{sec:intro}
The standard model (SM) has been extremely successful at describing particle physics phenomena.
However, it suffers from such shortcomings as the hierarchy problem, where fine-tuned cancellations of large quantum corrections
are required in order for the Higgs boson to have a mass at the electroweak
symmetry breaking scale of order 100\GeV~\cite{SUSY1,SUSY2,SUSY3,SUSY4,SUSY5,SUSY5,SUSY6}.
Supersymmetry (SUSY) is a popular extension of the SM that postulates
the existence of a superpartner for every SM particle, with the same quantum numbers but differing by
one half-unit of spin.
SUSY potentially provides a ``natural'', \ie, not fine-tuned, solution to the hierarchy problem through the
cancellations of the quadratic divergences
of the top-quark and
top-squark loops. In addition, it provides a connection to cosmology, with the lightest
supersymmetric particle (LSP), if neutral and stable, serving as a dark matter candidate in R-parity conserving SUSY models.

This paper describes a search for the pair production of top squarks using the full dataset collected at $\sqrt{s}=8$\TeV
by the Compact Muon Solenoid (CMS) experiment~\cite{JINST} at the Large Hadron Collider
(LHC) during 2012, corresponding to an integrated luminosity of \lumin.
This search is motivated by the consideration
that relatively light top squarks, with masses below around 1\TeV,
are necessary if SUSY is to be the natural solution to the
hierarchy
problem~\cite{Barbieri:1987fn,deCarlos1993320,Dimopoulos1995573,Barbieri199676,Papucci:2011wy}.
These constraints are especially relevant given the recent discovery of a particle that closely resembles
a Higgs boson, with a mass of $\sim$125\GeV~\cite{ATLAS_Higgs,CMS_Higgs,CMS_HiggsLong}.
Searches for top-squark pair production have also been performed by
the ATLAS Collaboration at the LHC in several final states~\cite{ATLAS1,ATLAS2,ATLAS3,ATLAS4,ATLAS5},
and by the CDF~\cite{CDFstop} and \DZERO~\cite{D0stop} Collaborations at the Tevatron.

\begin{figure}[hbt]
  \begin{center}
        \includegraphics[width=\cmsFigWidthTwo]{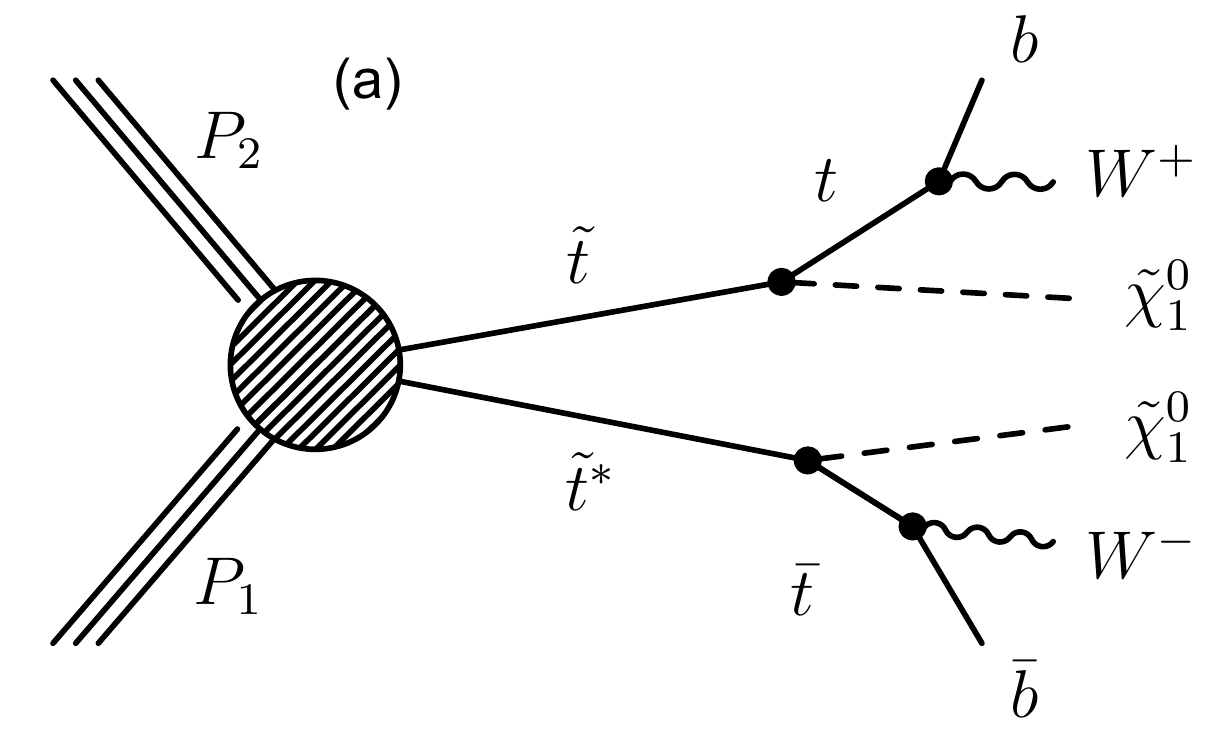}%
        \includegraphics[width=\cmsFigWidthTwo]{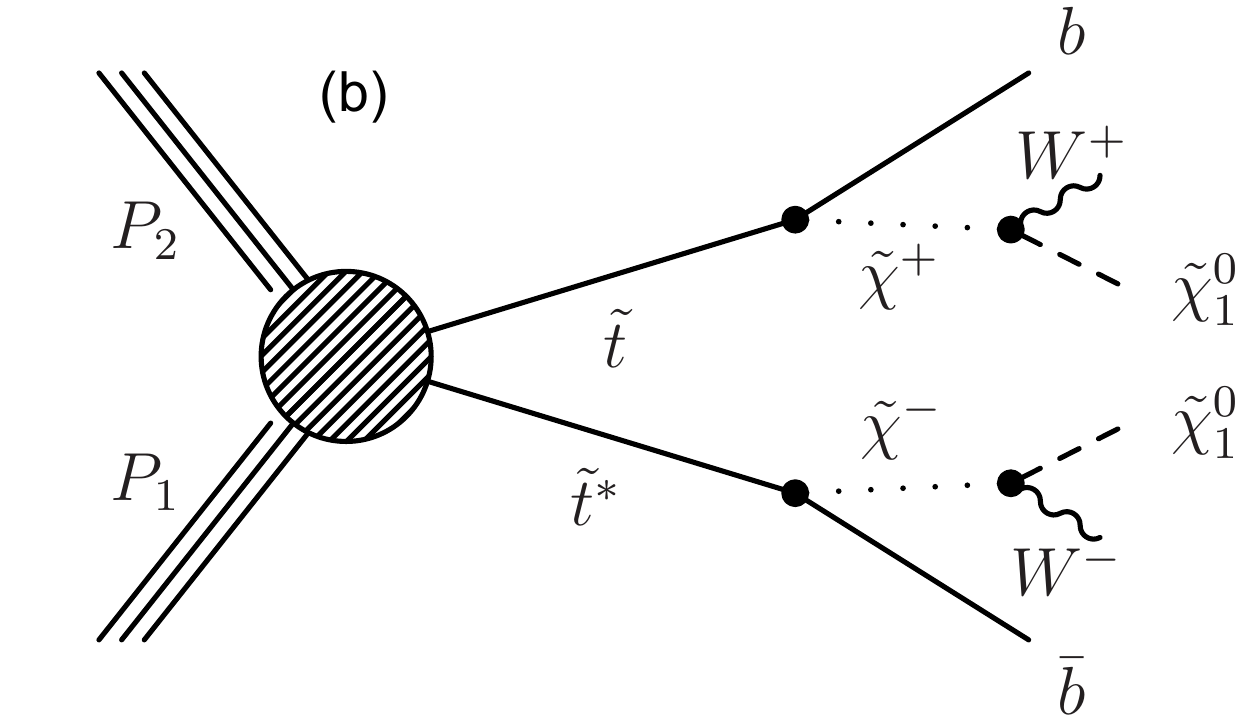}%
	\caption{Diagram for top-squark pair production for (a) the
          $\Ttt \to \cPqb \PW \chiz_1$ decay mode and (b) the $\TbW \to \cPqb \PW \chiz_1$ decay mode.
	\label{fig:SigDiagram}}
      \end{center}
\end{figure}

The search presented here focuses on two decay modes of the
top squark (\PSQt):
\Ttt\ and \TbW.
These modes are expected to have large branching fractions if
kinematically allowed. Here $\cPqt$ and $\cPqb$ are the top and bottom quarks, and
the neutralinos (\chiz) and charginos (\chipm) are the mass eigenstates
formed by the linear combination of the gauginos and higgsinos, which
are the fermionic
superpartners of the gauge and Higgs bosons, respectively.
The charginos are unstable and may subsequently decay into neutralinos and \PW\ bosons, leading to the following processes
of interest: $\Pp\Pp\to\PSQt\PSQt^{*}\to \ttbar\chiz_1\chiz_1 \to \bbbar\PW^+\PW^-\chiz_1\chiz_1$
and $\Pp\Pp\to\PSQt\PSQt^{*}\to \bbbar\chip_1\chim_1 \to \bbbar\PW^+\PW^-\chiz_1\chiz_1$, as displayed in Fig.~\ref{fig:SigDiagram}.
The lightest neutralino $\chiz_1$ is considered to be the stable LSP,
which escapes without detection.

The analysis is based on events where one of the \PW\ bosons decays leptonically and the other hadronically. This
results in one isolated lepton and four jets, two of which originate
from b quarks. The two neutralinos and the neutrino from the \PW\
decay can result in large missing transverse momentum (\MET).

The largest backgrounds in this search arise
from events with a top-antitop (\ttbar) quark pair where one top quark
decays hadronically and the other leptonically, and from events with a
$\PW$ boson produced in association with jets (\wjets).
These backgrounds, like the signal, contain a single leptonically
decaying \PW\ boson.  The transverse mass, defined as
$\MT \equiv \sqrt{\smash[b]{2 \MET \pt^{\ell} (1-\cos(\Delta\phi))}}$, where $\pt^{\ell}$ is the transverse momentum of the
lepton and $\Delta\phi$ is the difference in azimuthal
angles between the lepton and \MET directions,
has a kinematic endpoint $\MT\ <M_\PW$ for these backgrounds, where $M_\PW$ is the \PW\ boson mass.
For signal events, the presence of LSPs in the final state allows \MT\ to exceed $M_\PW$. Hence we search for an excess
of events with large \MT. The dominant background with large \MT\
arises from the ``dilepton \ttbar" channel,
\ie, \ttbar\ events where both \PW\ bosons decay leptonically but
with one of the leptons not identified.
In these events the presence of two neutrinos can lead to large values
of \MET\ and \MT.

\section{The CMS detector}

The central feature of the CMS apparatus is a superconducting
solenoid, 13\unit{m} in length and 6\unit{m} in diameter, which provides
an axial magnetic field of 3.8\unit{T}. Within the field volume are
several particle detection systems.
Charged-particle
trajectories are measured with silicon pixel and strip trackers,
covering $0 \leq \phi < 2\pi$ in azimuth and $\abs{\eta} < 2.5$ in pseudorapidity,
where $\eta \equiv -\ln [\tan (\theta/2)]$ and $\theta$ is the
polar angle of the trajectory of the particle with respect to
the counterclockwise proton beam direction. A lead-tungstate crystal electromagnetic calorimeter
and a brass/scintillator hadron calorimeter surround the
tracking volume, providing energy  measurements of electrons, photons, and
hadronic jets. Muons are identified and measured in gas-ionization detectors embedded in
the steel flux return yoke of the solenoid. The CMS detector is nearly
hermetic, allowing momentum balance measurements in the plane
transverse to the beam direction. A two-tier trigger system selects pp collision events
of interest for use in physics analyses.
A more detailed description of the CMS detector can be found
elsewhere~\cite{JINST}.

\section{Signal and background Monte Carlo simulation}
\label{sec:mc}

Simulated samples of SM processes are generated using the
\POWHEG \cite{POWHEG},
\MCATNLO \cite{MCatNLO1,MCatNLO2}, and
\MADGRAPH \breakhere 5.1.3.30 \cite{madgraph5} 
Monte Carlo (MC) event
generator programs with the
CT10 \cite{ct10} (\POWHEG),
CTEQ6M \cite{cteq6lm} (\MCATNLO), and
CTEQ6L1 \cite{cteq6lm} (\MADGRAPH)
parton distribution functions.
The reference sample for \ttbar\ events is generated with \POWHEG, while
the \MADGRAPH and \MCATNLO generators are used for crosschecks
and validations.
All SM processes are normalized to cross section calculations
valid to next-to-next-to-leading order (NNLO)~\cite{xsec_WZ} or approximate
NNLO~\cite{xsec_ttbar} when available, and otherwise to
next-to-leading order (NLO)~\cite{MCatNLO1,MCatNLO2,xsec_ttbarW,xsec_ttbarZ,xsec_MCFM,MCatNLO3}.

For the signal events, the production of top-squark pairs is generated with \MADGRAPH, including up to two additional
partons at the matrix element level. The
decays of the top squarks are generated with \PYTHIA~\cite{Pythia} assuming
100\% branching fraction for either \Ttt\ or \TbW.
A grid of signal events is generated as a function of the top-squark and neutralino masses in 25\GeV steps.
We also consider scenarios with off-shell top quarks (for \Ttt) and
off-shell \PW\ bosons (for \TbW\ followed by $\chip_1 \to \PWp \chiz_1$).
For the $\TbW$ decay mode, the chargino mass is specified by a third parameter $x$ defined as
$m_{\chipm_1} = x \cdot m_{\PSQt} + (1-x) \cdot m_{\chiz_1}$.
We consider three mass spectra, namely $x=0.25$, 0.50, and 0.75.
The lowest top squark mass that we consider is $m_{\PSQt}$ = 100\GeV
for \Ttt, and $m_{\PSQt}$ = 200 (225, 150)\GeV for
\TbW\ with $x=0.25$ (0.50, 0.75).

The polarizations of the final- and intermediate-state particles (top
quarks in the $\Ttt$ scenario, and
charginos and \PW\ bosons in the $\TbW$ case) are model dependent and are
non-trivial functions of the
top-squark, chargino, and neutralino mixing matrices~\cite{polarization1,polarization2}.
The SUSY MC events are generated without polarization. The effect of this choice
on the final result is discussed in Section~\ref{sec:interpretation}.
Expected signal event rates are normalized to cross sections
calculated at NLO in the strong coupling constant, including the resummation of soft gluon emission at
next-to-leading-logarithmic accuracy (NLO+NLL)~\cite{bib-nlo-nll-01,bib-nlo-nll-02,bib-nlo-nll-03,bib-nlo-nll-04,bib-nlo-nll-05,ref:xsec}.

For both signal and background events,
multiple proton-proton interactions in the same or nearby bunch
crossings (pileup)
are simulated using \PYTHIA and superimposed on the hard
collision.
The simulation of the detector response to
SUSY signal events is performed using the CMS fast simulation
package~\cite{Abdullin:2011zz}, whereas almost all
SM samples are simulated  using a
\GEANTfour-based~\cite{Geant} model of the CMS detector.
The
exceptions are the \MADGRAPH \ttbar\ samples used to
study the sensitivity of estimated backgrounds to the details of the generator settings;
these samples are processed with the fast simulation.
The two simulation methods provide consistent results for the
acceptances of processes of interest to this analysis.
The simulated
events are
reconstructed and analyzed with the same software used to process the data.

\section{Event selection}
\label{sec:eventSel}

\subsection{Object definition and event preselection}
\label{sec:presel}

The data used for this search were collected using high transverse
momentum (\pt),
isolated, single-electron and single-muon triggers with \pt\ thresholds of 27 and
24\GeV, respectively.
The electron (muon) trigger efficiency, as measured with a sample of
$\Z \to \ell \ell$ events, varies between 85\% and 97\%
(80\% and 95\%), depending on the $\eta$ and \pt\ of the leptons.
Data
collected with high-\pt\
double-lepton triggers (ee, e$\mu$, or $\mu\mu$, with \pt\ thresholds of
17\GeV for one lepton
and 8\GeV for the other) are used for studies of
dilepton control regions.

Events are required to have an electron (muon)
with
$\pt > 30\,(25)\GeV$.
Electrons are required to
lie
in the barrel region of the
electromagnetic calorimeter ($\abs{\eta} < 1.4442$),
while muons are considered up to $\abs{\eta} = 2.1$.
Electron candidates are reconstructed starting from a cluster of energy deposits in the
electromagnetic calorimeter. The cluster is then matched to
a reconstructed track.
The electron selection is based on the shower shape,
track-cluster matching, and consistency between the cluster energy and
the track momentum~\cite{EGMPAS}.
Muon candidates are reconstructed by performing a global fit that
requires
consistent hit patterns
in the tracker and the muon system~\cite{MUOART}.

The particle flow (PF) method~\cite{CMS-PAS-PFT-10-002} is used to
reconstruct final-state particles.
Leptons
are required to be isolated from other activity in the event.
A measure of lepton isolation is the scalar sum ($\pt^\text{sum}$)
of the \pt\ of all
PF particles,
excluding the lepton itself, within
a cone of radius $\Delta R \equiv\sqrt{\smash[b]{(\Delta\eta)^2+(\Delta\phi)^2}} = 0.3$,
where $\Delta \eta$ ($\Delta \phi$) is the difference in $\eta$ ($\phi$)
between the lepton and the PF particle at the primary interaction vertex.
The average contribution of particles from pileup interactions is estimated and
subtracted from the $\pt^\text{sum}$ quantity.
The isolation requirement is
$\pt^\text{sum} <  \min(5\GeV,\, 0.15 \cdot \pt^{\ell})$.
Typical lepton identification and isolation efficiencies, measured in
samples of $\Z \to \ell \ell$ events, are 84\% for electrons and 91\% for muons, with variations at the level of a few percent depending
on \pt\ and $\eta$.

To reduce the background from \ttbar events
in which both W bosons decay leptonically
(denoted as \ttll\ in the following), events are rejected
if they contain evidence for an additional
lepton.  Specifically, we reject events with a second
isolated lepton of $\pt > 5$\GeV, identified with requirements
that are considerably looser than for the primary lepton.
We also reject events with an isolated track of $\pt > 10$\GeV
with opposite sign with respect to the primary lepton, as well
as events with a jet of $\pt> 20$\GeV
consistent with the hadronic decay of a $\tau$ lepton~\cite{tau}.
The isolation algorithm used at this stage differs slightly from the one
used  in the selection of primary leptons: the cone has radius $\Delta R =
0.4$, the $\pt^\text{sum}$ variable is constructed using charged PF particles only,
and the isolation requirement is $\pt^\text{sum} < \alpha \cdot \pt$, where
$\pt$ is the transverse momentum of the track (lepton)
and $\alpha = 0.1\,(0.2)$,
for tracks (leptons).

The PF particles are clustered to form jets using the anti-\kt clustering algorithm~\cite{antikt}
with a distance parameter of 0.5, as implemented in the {\FASTJET} package~\cite{Cacciari:2005hq,FastJet}.
The contribution to the jet energy from pileup is estimated on an event-by-event basis using the
jet area method described in Ref.~\cite{cacciari-2008-659}, and is subtracted from the overall jet \pt.
Jets from pileup interactions are suppressed using a multivariate
discriminant based on the multiplicity of objects clustered in the jet,
the topology of the jet shape, and the impact parameters of the charged tracks
with respect to the primary interaction vertex. The jets must be separated from
the lepton by $\Delta  R  >  0.4$ in order to resolve overlaps.

Selected events are required to contain at least four jets with $\pt > 30\GeV$ and  $\abs{\eta} < 2.4$.
At least one of these jets must be consistent with containing
the decay of a heavy-flavor hadron, as identified using the medium
operating point of the combined secondary vertex bottom-quark
(b-quark) tagging algorithm (CSVM)~\cite{ref:btag}. We refer to such jets as ``b-tagged jets''.
The efficiency of this algorithm for
bottom-quark jets in the $\pt$ range 30--400\GeV varies between
approximately 60 and 75\% for $\abs{\eta} < 2.4$.
The nominal misidentification rate for light-quark or gluon jets is 1\%~\cite{ref:btag}.

The missing transverse momentum
is defined as the magnitude of
the vector sum of the transverse
momenta of all PF particles over the full calorimeter coverage
($\abs{\eta} < 5$).
The \MET vector is the negative of that
same vector sum.
The calibrations that are applied to the energy measurements of jets are
propagated consistently as a correction to the \MET obtained from PF
particles.
We require $\MET > 100\GeV$.

To summarize, events are required to contain one isolated
lepton (e or $\mu$),
no additional isolated track or hadronic $\tau$-lepton candidate, at least four jets with at least one
b-tagged jet, and $\MET > 100\GeV$; this is referred to below as the ``event preselection''.
Signal regions are defined
demanding $\MT>120\GeV$.
This requirement
provides large suppression of the SM backgrounds while retaining high signal efficiency.
Requirements on several kinematic quantities
or on the output of boosted decision tree (BDT) multivariate discriminants
are also used to define the signal regions, as described below.

\subsection{Kinematic quantities}
\label{sec:sigsel}

To reduce the dominant \ttll\ background,
we make use of the quantity \mtw,
defined as the minimum ``mother'' particle mass compatible with all
the transverse momenta and mass-shell constraints~\cite{mt2w}.
This variable is a variant of the $M_{\mathrm{T}2}$ observable~\cite{mt2-1,mt2-2,Burns:2008va},
and is designed specifically to suppress
the \ttll\ background with one undetected lepton in the top squark search.
By construction, for the dilepton \ttbar\ background without mismeasurement effects, \mtw\ has an endpoint
at the top-quark mass.
The \mtw\ calculation relies on the correct
identification of the \cPqb-quark jets and the correct
pairing of the \cPqb-quark jets with the leptons.
The \mtw\ value in the event is defined
as the minimum of the \mtw\ values calculated from all possible combinations
of \cPqb-quark jets and the lepton.  For events with only one \cPqb-tagged jet,
the combinations are performed using each of the three remaining highest
$\pt$ jets as the possible second \cPqb-quark jet.

In the \Ttt\ search, the dilepton \ttbar\ background is suppressed
by requiring that three of the jets in the event be consistent with
the $\cPqt \to \cPqb \PW \to \cPqb \cPq \bar{\cPq}$ decay chain.
For each triplet of jets in the event we construct a hadronic top $\chi^2$ as:
\begin{equation}
\chi^2 = \frac{(M_{j_1 j_2 j_3}-M_{\text{top}})^2}{\sigma_{j_1 j_2 j_3}^2} + \frac{(M_{j_1 j_2}-M_\PW)^2}{\sigma_{j_1 j_2}^2}.
\label{eqn:chi2var}
\end{equation}

Here $M_{j_1 j_2 j_3}$ is the mass of the three-jet system,
$M_{j_1 j_2}$ is the mass of two of the jets posited to originate from
\PW\ boson decay,
and $\sigma_{j_1 j_2 j_3}$ and $\sigma_{j_1 j_2}$ are the uncertainties on
these masses calculated from the jet energy resolutions~\cite{JER}.
The three-jet mass
$M_{j_1 j_2 j_3}$ is computed after requiring $M_{j_1 j_2}=M_\PW$
using a constrained kinematic fit, while $M_{j_1 j_2}$ in
Eqn.~\ref{eqn:chi2var} is the two-jet mass before the fit.
Finally,
$M_{\text{top}}=173.5$\GeV ($M_\PW=80.4$\GeV) is the mass of the top quark (\PW\ boson)~\cite{Beringer:1900zz}.
The three jets are required to have $\pt > 30$\GeV and $\abs{\eta} < 2.4$
and to be among the six leading selected jets. The jet assignments are made consistently with the
\cPqb-tagging information, \ie, $j_3$ must be \cPqb-tagged if there are
at least two \cPqb-tagged jets and $j_1$ and $j_2$ cannot be \cPqb-tagged
unless there are at least three \cPqb-tagged jets in the event.
The minimum hadronic top $\chi^2$ amongst all
possible jet combinations is used as a discriminant on an event-by-event basis.

Two topological variables are used in the selection of signal candidates.
The first is the minimum $\Delta\phi$ value between the
\MET\ vector and either of the two highest \pt\ jets, referred to below as ``\mindphi''.
Background $\ttbar$ events tend to have high-\pt\
top quarks, and thus objects in these events tend to
be collinear in the transverse plane, resulting in smaller values
of $\Delta \phi$ than is typical for signal events.
The second variable is
\htratio, defined as the fraction of the total scalar sum of the jet transverse energies ($H_\mathrm{T}$)
with $\pt > 30\GeV$ and $\abs{\eta} < 2.4$
that lies in the same hemisphere as the \MET\ vector.
This quantity tends to be smaller for signal than for background events, because
in signal events the visible particles
recoil against the LSPs, resulting on average in events with
more energy in the opposite hemisphere to the \MET.

In the $\TbW$ decay mode, the bottom quarks arise from the decay of the top squark, while in
background events they originate from the decay of the top quark.
As a result,
in most of the signal parameter space the $\pt$ spectrum of the bottom quarks is harder for signal than
for background events. Conversely, in the \Ttt\ decay mode, if the top quark is off-shell,
the $\pt$ spectrum of the bottom quarks is softer for signal than for
the background. The $\pt$ value of the highest-$\pt$ \cPqb-tagged jet is therefore a
useful discriminant. An additional, related, discriminating variable is
the $\Delta R$ separation between this jet and the lepton.
Finally, the $\pt$ spectrum of the lepton can be used to discriminate
between on-shell and off-shell leptonic $\PW$ decays, which occur in
the \TbW\ mode when the mass splitting between the chargino and the
LSP
is smaller than the $\PW$ boson mass.

\begin{figure*}[htbp]
\centering
\includegraphics[width=0.30\textwidth]{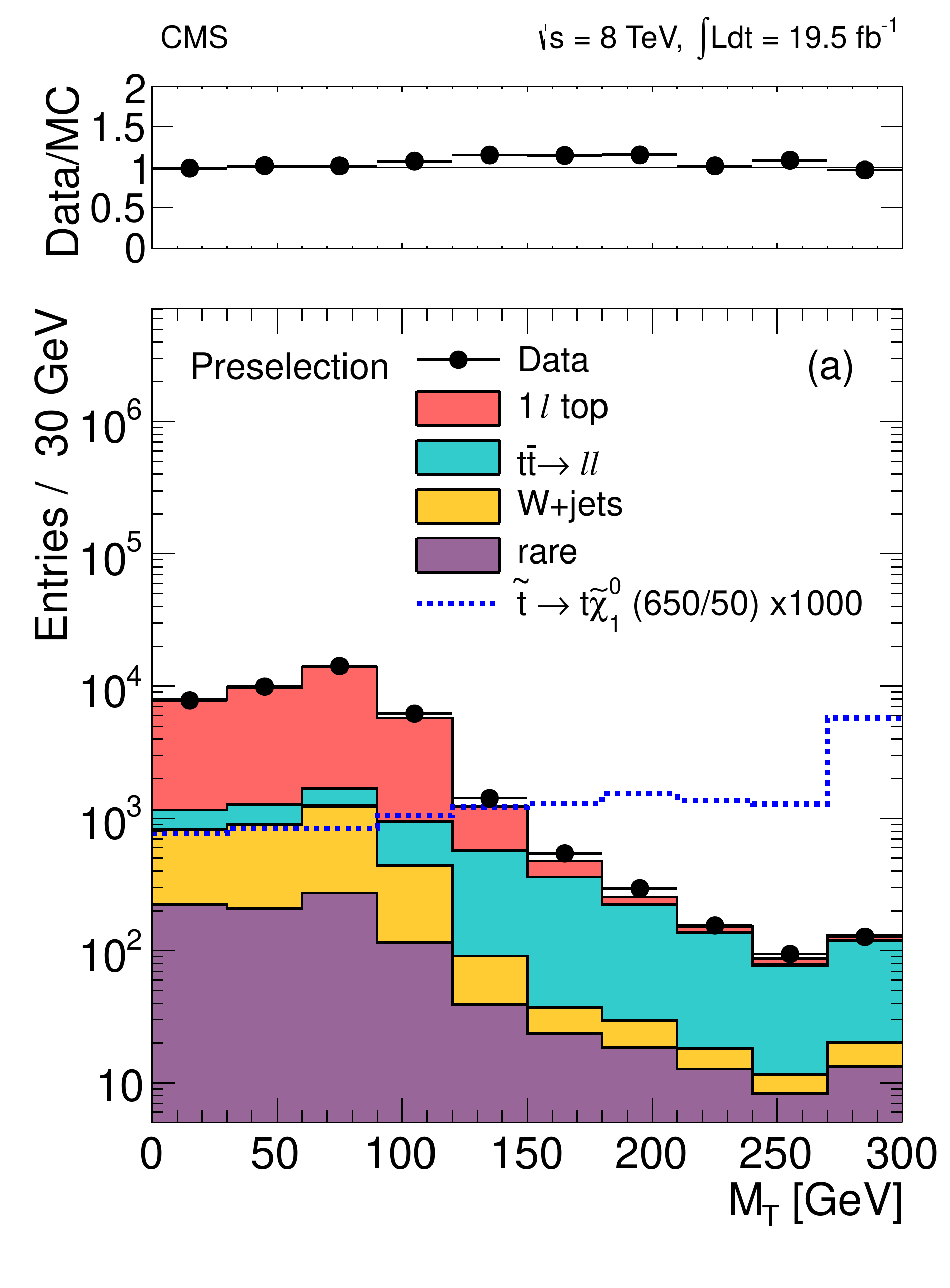}
\includegraphics[width=0.30\textwidth]{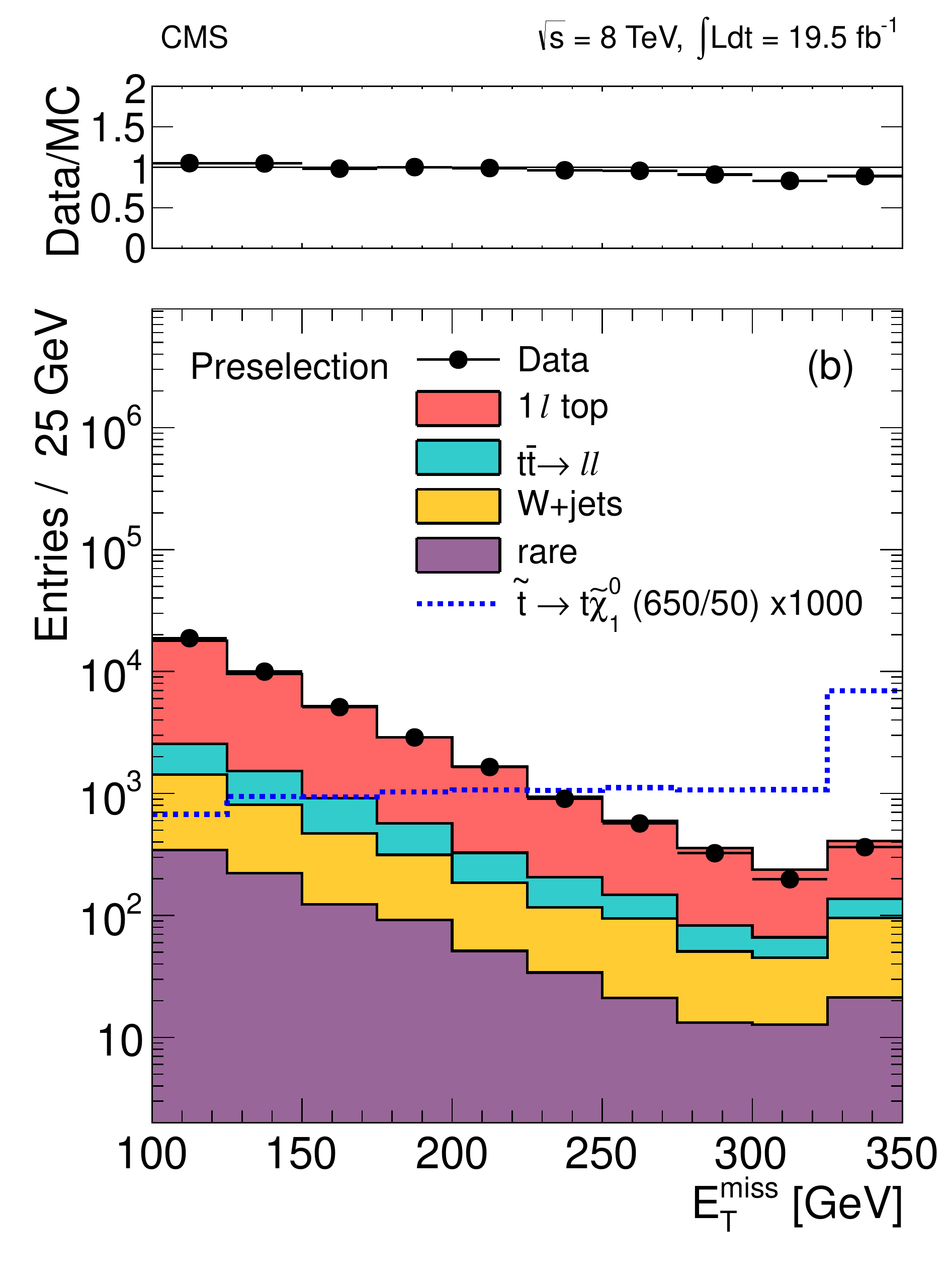}
\includegraphics[width=0.30\textwidth]{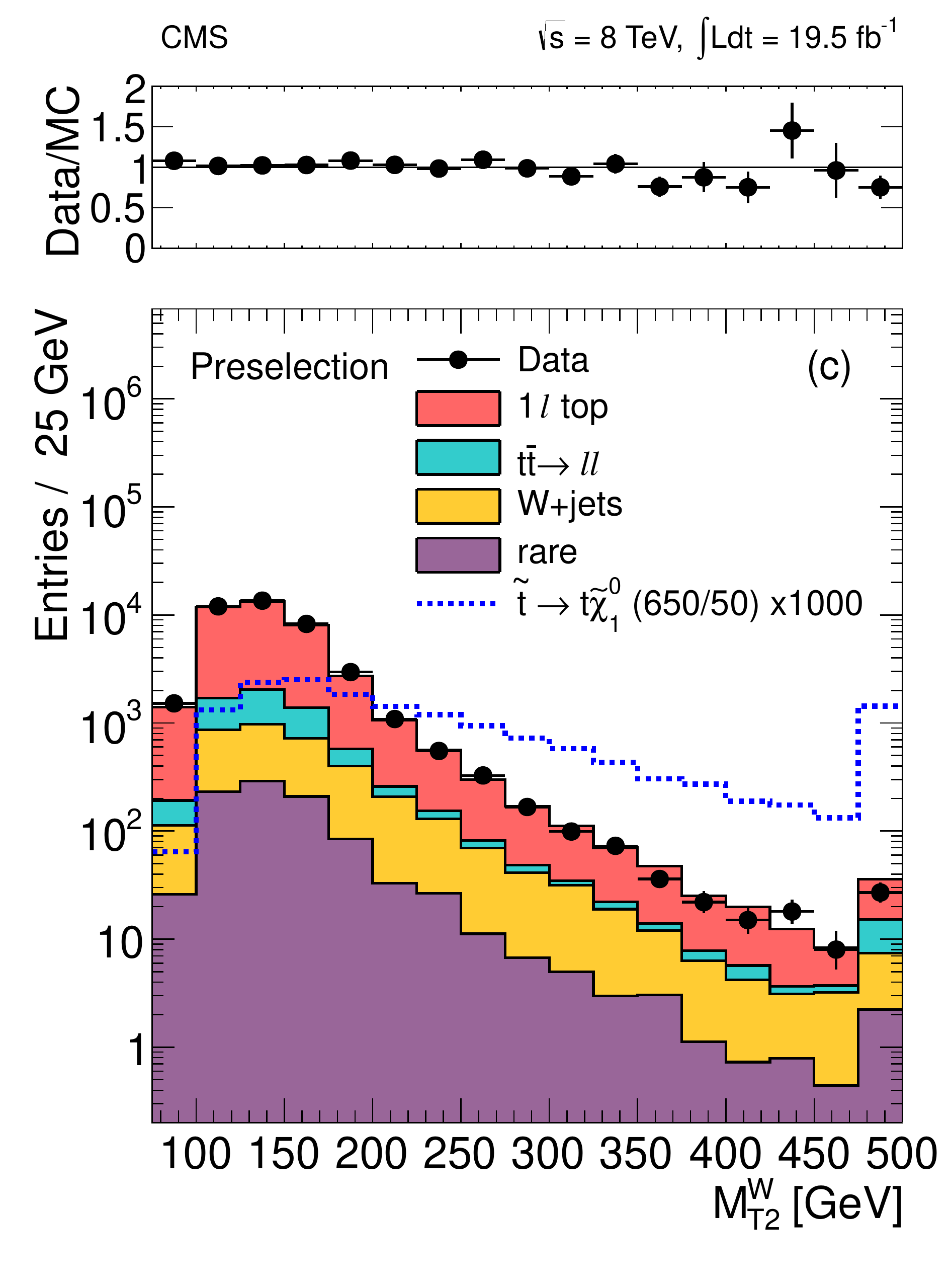}
\includegraphics[width=0.30\textwidth]{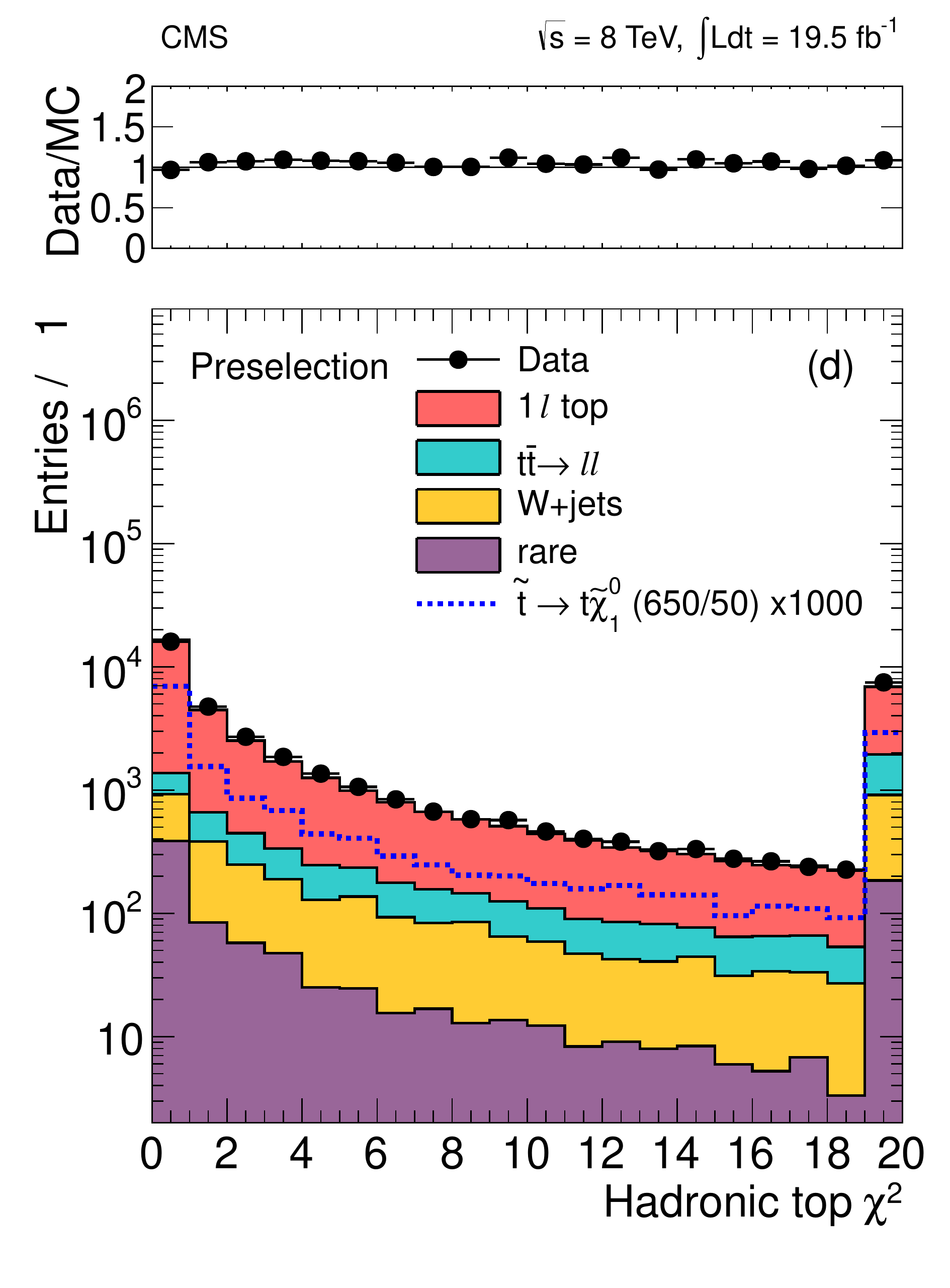}
\includegraphics[width=0.30\textwidth]{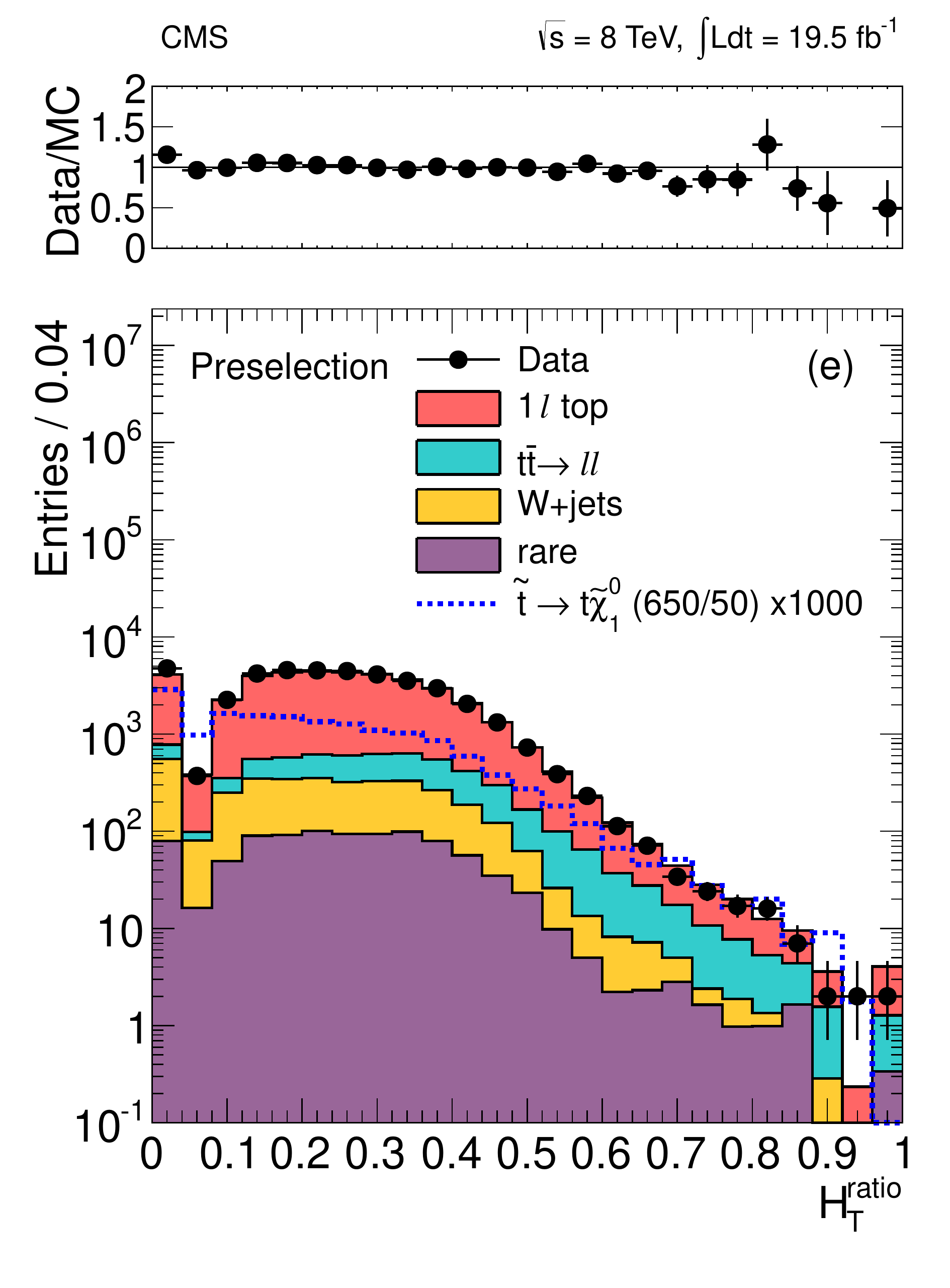}
\includegraphics[width=0.30\textwidth]{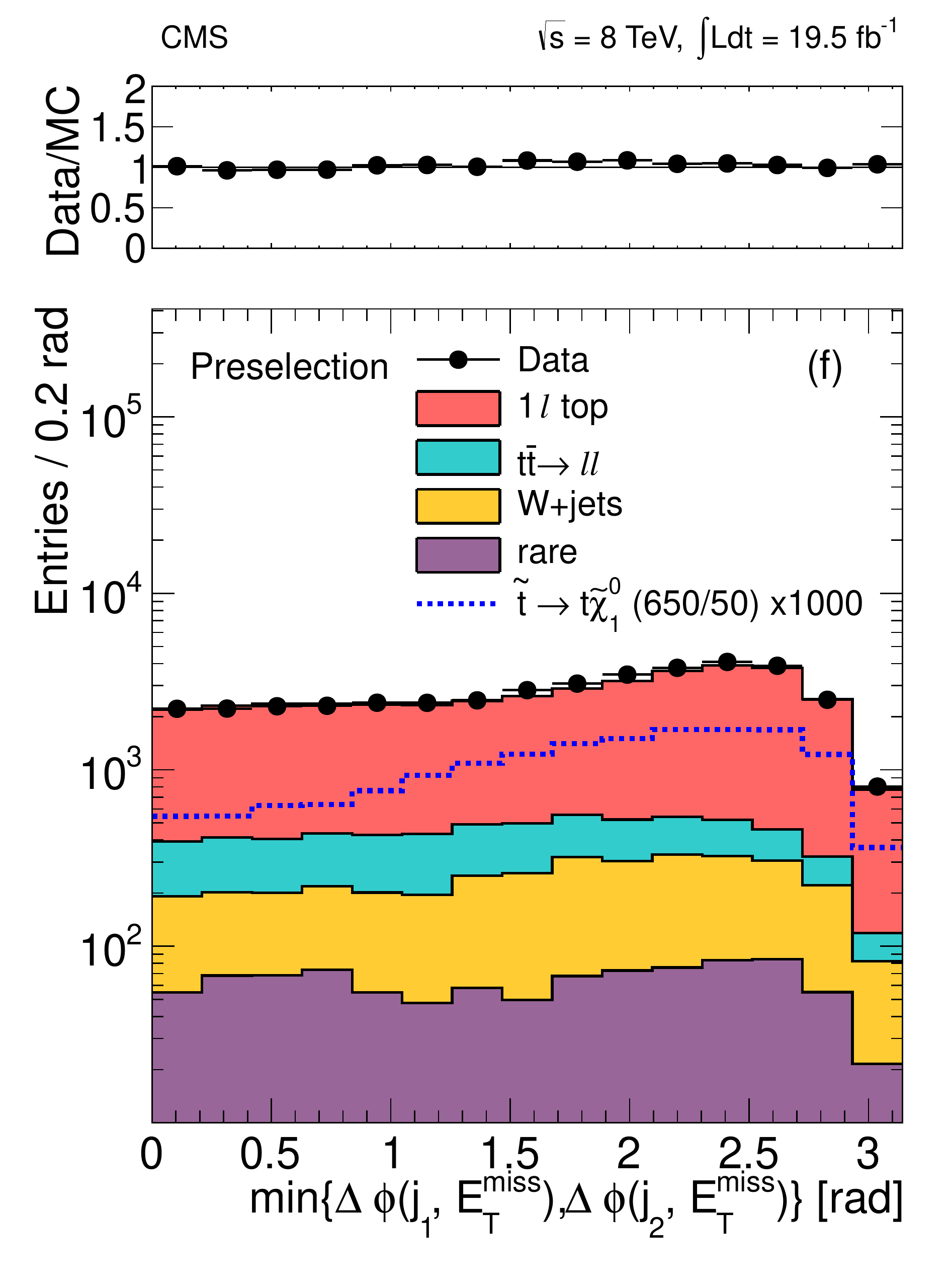}
\includegraphics[width=0.30\textwidth]{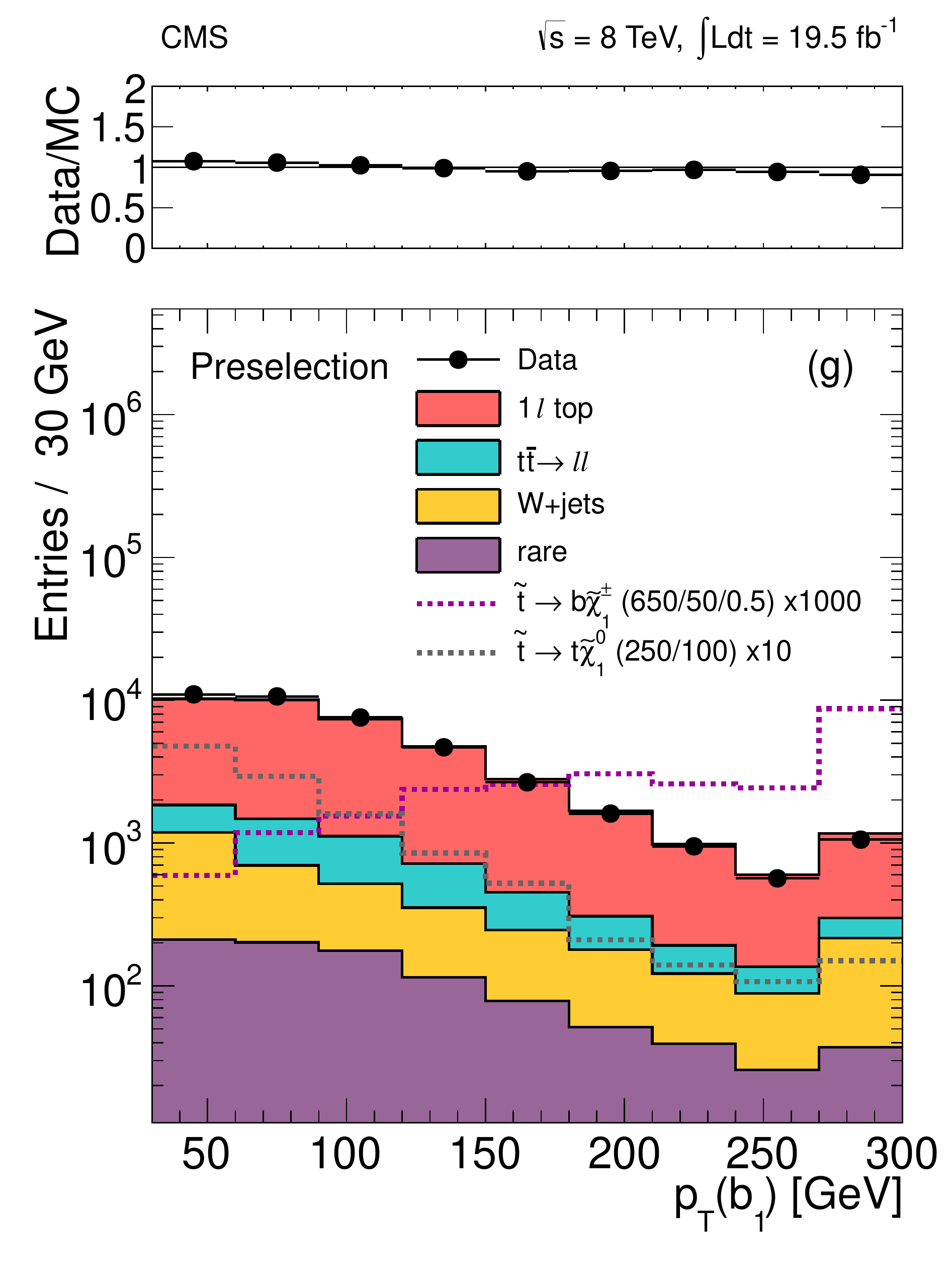}
\includegraphics[width=0.30\textwidth]{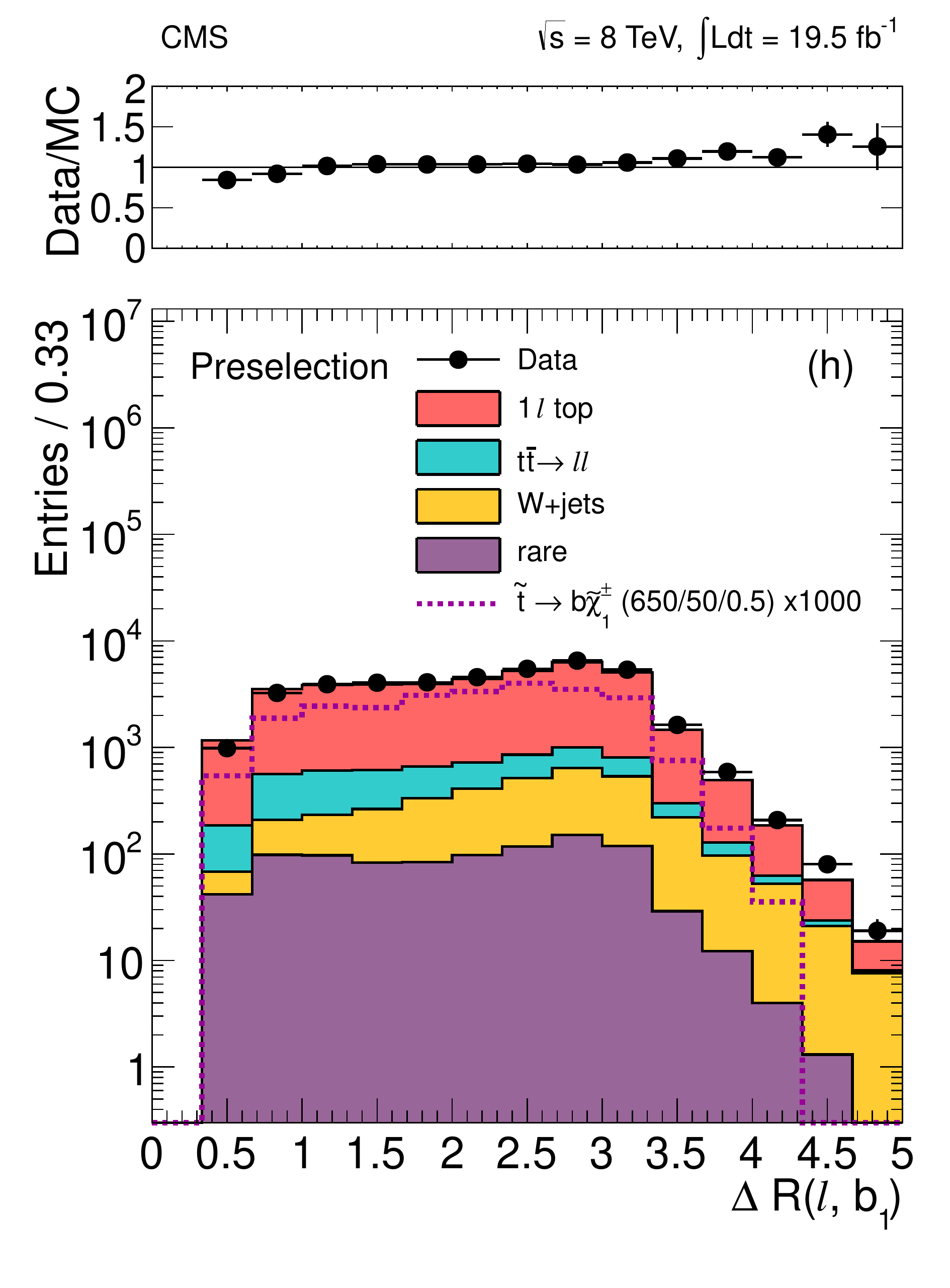}
\includegraphics[width=0.30\textwidth]{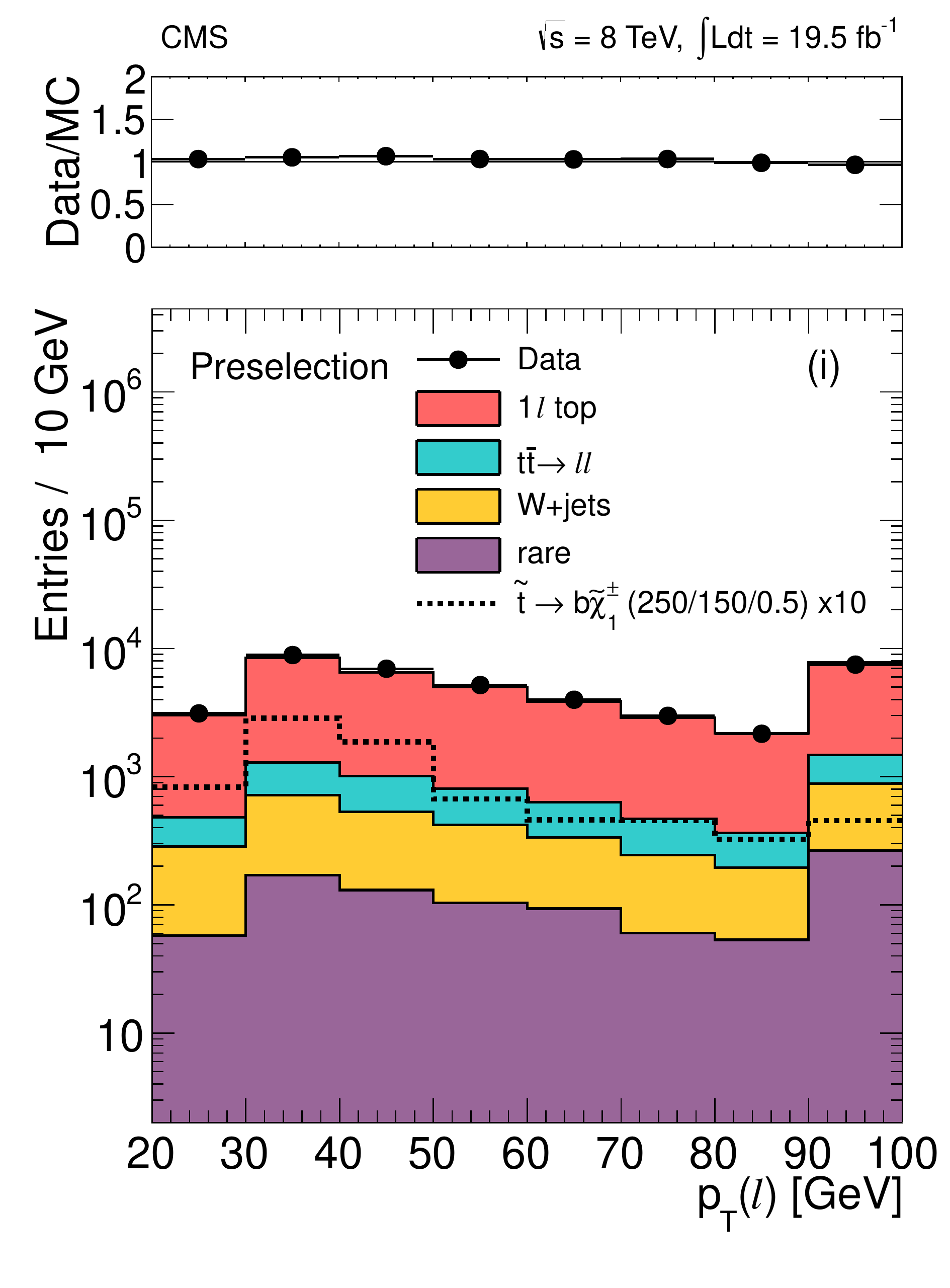}
\caption{\label{fig:quantitiesNew}
Comparison of data with MC simulation for
the distributions of (a) \MT, (b) \MET, (c) \mtw, (d) hadronic top $\chi^2$,
(e) \htratio, (f) minimum $\Delta\phi$
between the \MET\ vector and the two leading jets, (g) \pt\ of the leading b-tagged jet,
(h) $\Delta R$ between the leading b-tagged jet and the lepton,
and (i) lepton \pt, after the preselection.
For the plots (a)-(f),
distributions for the \Ttt\ model with $m_{\PSQt}=650$\GeV and $m_{\lsp}=50$\GeV,
scaled by a factor of 1000, are overlayed.
We also show distributions
of \Ttt\ with
$m_{\PSQt}=250$\GeV and $m_{\lsp}=100$\GeV for (g), scaled by 10,
and of \TbW\ with
$m_{\PSQt}=650$\GeV, $m_{\lsp}=50$\GeV, and $x=0.5$ for (h) and (i), scaled by 1000,
as well as of
$m_{\PSQt}=250$\GeV, $m_{\lsp}=150$\GeV, and $x=0.5$ for (i), scaled by 10.
In all distributions the last bin contains the overflow.
}
\end{figure*}

The distributions after the preselection of \MET, \MT, and the
kinematic quantities described above, are shown in Fig.~\ref{fig:quantitiesNew}.
These quantities are seen to be in agreement with the simulation
of the SM background processes that will be discussed in
more
detail in Section~\ref{sec:backgroundmodeling}.

\subsection{Signal region definition}
\label{sec:SR}

Two approaches are pursued to define the signal regions (SRs): a
``cut-based'' approach based on sequential selections on individual
variables, and a BDT multivariate approach implemented via the {\sc
  TMVA} package~\cite{TMVA}.
In both methods, we apply the preselection requirements of Section~\ref{sec:presel}.
The cut-based signal regions are defined by adding requirements
on
individual kinematic variables.
In contrast, the BDT combines the kinematic variables into
a single discriminant, and the BDT SRs are defined by
requirements on this discriminant.
The BDT approach improves the expected sensitivity of the search
by up to 40\% with respect to the cut-based approach, at the cost of additional complexity.
The primary result of our search is obtained with the BDT,
while the cut-based analysis serves as a crosscheck.
Table~\ref{tab:SRsummary}
lists the variables used in the training of the BDTs (Section~\ref{sec:bdtSR})
and summarizes the requirements for the cut-based SRs
(Section~\ref{sec:cncT2tt}).

\begin{table*}[htb]
\centering
\topcaption{
\label{tab:SRsummary}
Summary of the variables used as inputs for the BDTs and of the kinematic requirements
in the cut-based analysis. All signal regions include the requirement $\MT>120$\GeV.
For the \Ttt\ BDT trained in the region where the top quark is off-shell, the hadronic top $\chi^2$
is not included and the leading b-tagged jet \pt\ is included.  The lepton
\pt is used only in the training of the \TbW\ BDT in the case where
the $\PW$ boson is off-shell.
}
{\ifthenelse{\boolean{cms@external}}{\footnotesize}{\scriptsize}
\begin{tabular}{l|c | c c | c | c c }
\cline{2-7}
&  \multicolumn{3}{c|}{\Ttt} & \multicolumn{3}{c}{\TbW}\\\cline{2-7}
 & &  \multicolumn{2}{c|}{Cut-based}  & & \multicolumn{2}{c}{Cut-based} \\
 Selection     & BDT & Low $\Delta M$ & High $\Delta M$  & BDT & Low $\Delta M$ & High $\Delta M$ \\ \hline
\multirow{2}{*}{\MET (\GeVns{})}        & yes & $>$~150,~200, & $>$~150,~200, & yes & $>$~100,~150, & $>$~100,~150, \\
            & & \phantom{>}~250,~300      & \phantom{>}~250,~300      &     & \phantom{>}~200,~250     &            \phantom{>}~200,~250 \\
$\mtw$ (\GeVns{})                 & yes &      & $>$200             & yes
&                        & $>$200             \\
\mindphi & yes  &  $>$0.8 &   $>$0.8 & yes & $>$0.8 &   $>$0.8   \\
\htratio    &  yes & & & yes & & \\
Hadronic top $\chi^2$    & (on-shell top) & $<$5  & $<$5 & & & \\
Leading b-tagged jet \pt (\GeVns{}) &  (off-shell top) & & & yes & & $>$100 \\
$\Delta R$($\ell$,leading \cPqb-tagged jet)    & & & & yes & & \\
Lepton \pt (\GeVns{})                            & & & & (off shell $\PW$) & & \\
\hline
\end{tabular}
}
\end{table*}

\begin{figure*}[htb]
\begin{center}
\includegraphics[width=0.47\linewidth]{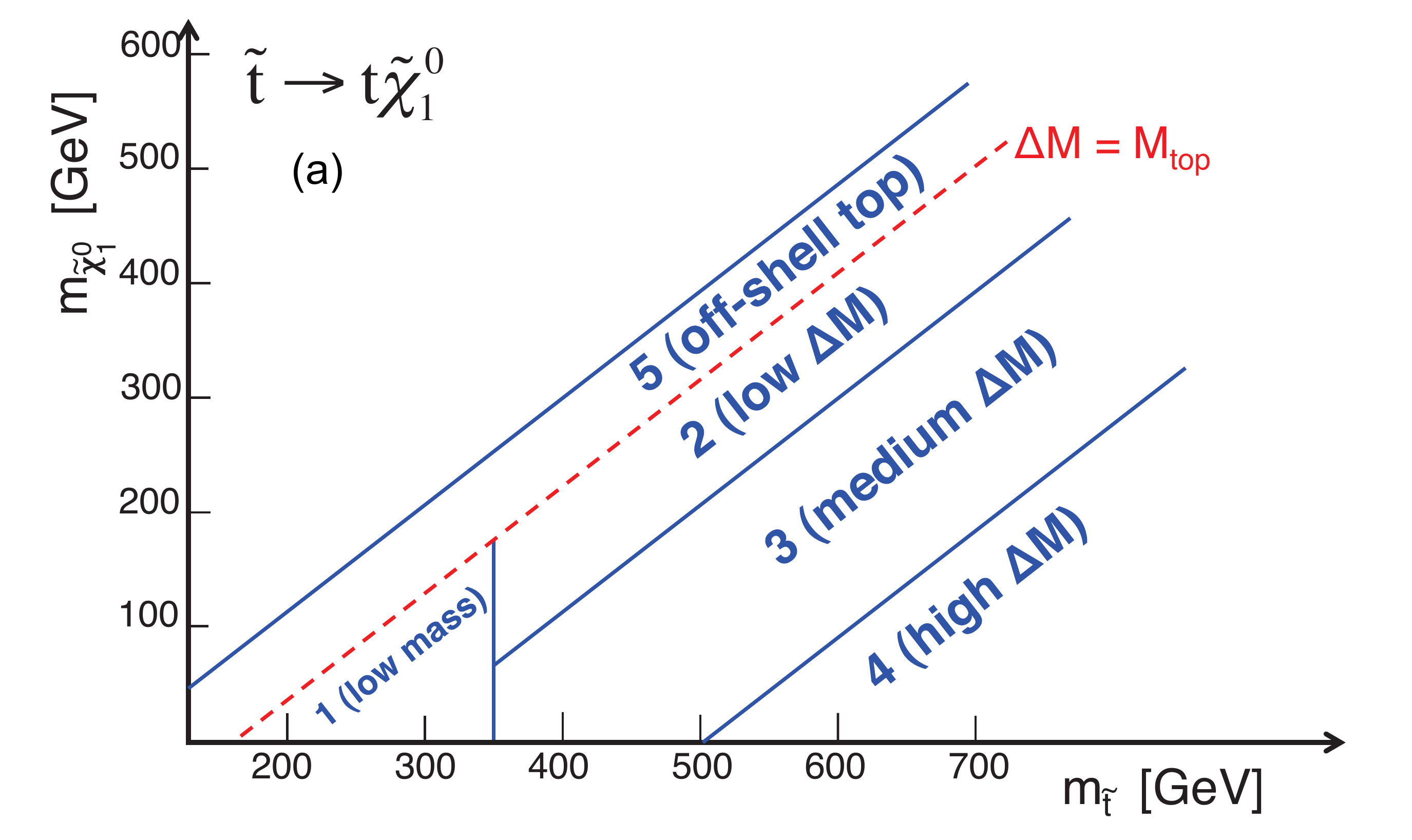}
\includegraphics[width=0.47\linewidth]{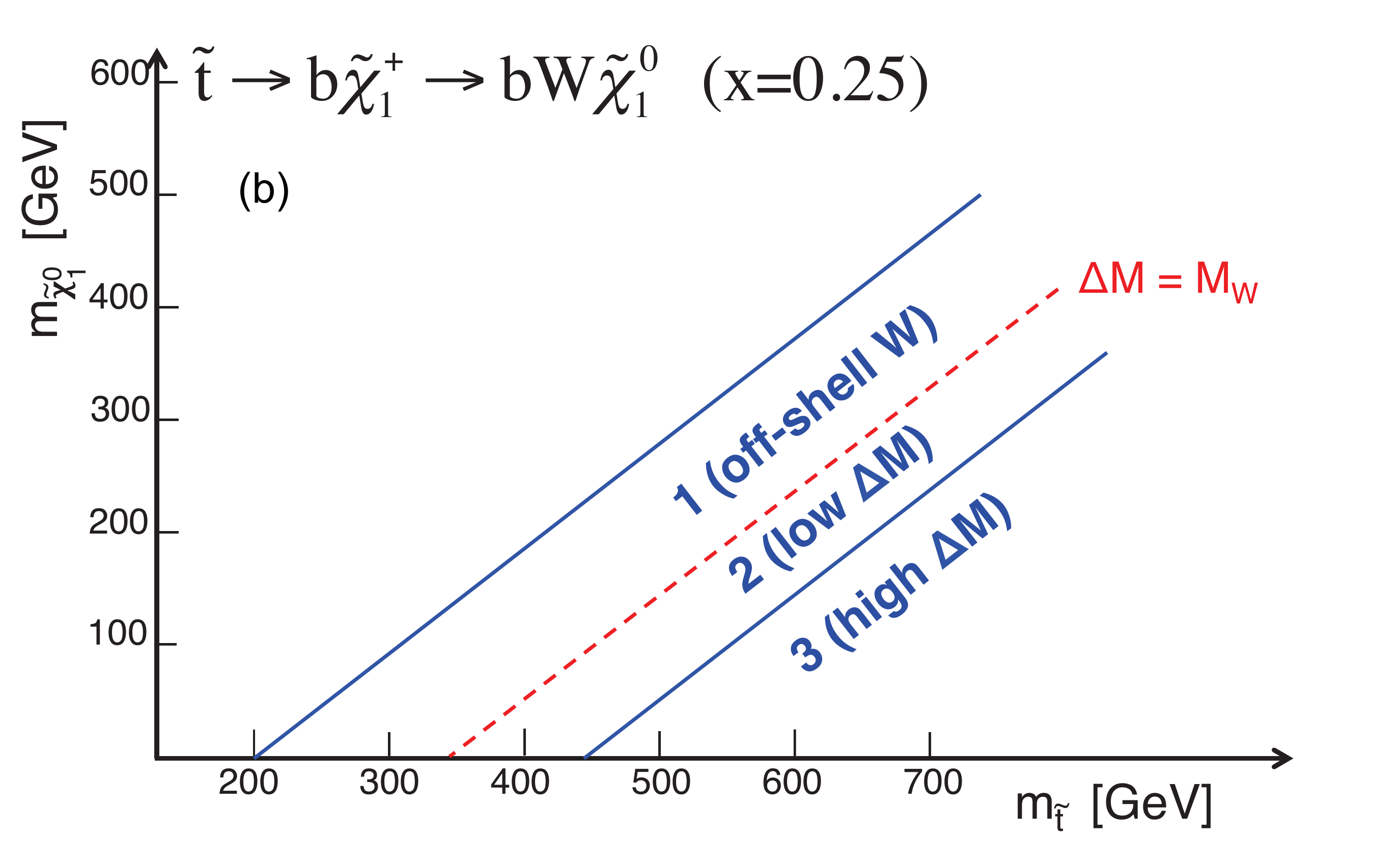}
\includegraphics[width=0.47\linewidth]{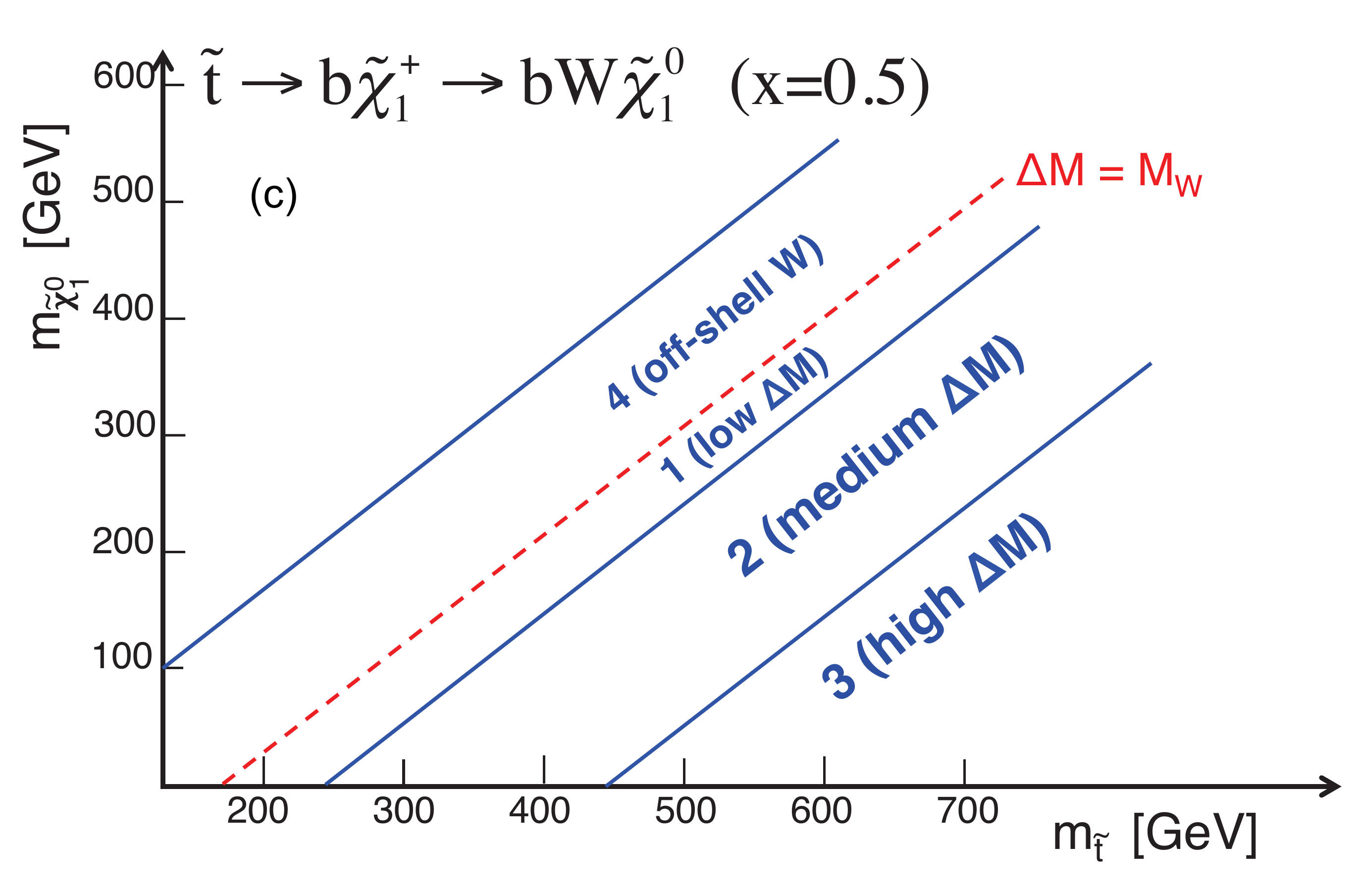}
\includegraphics[width=0.47\linewidth]{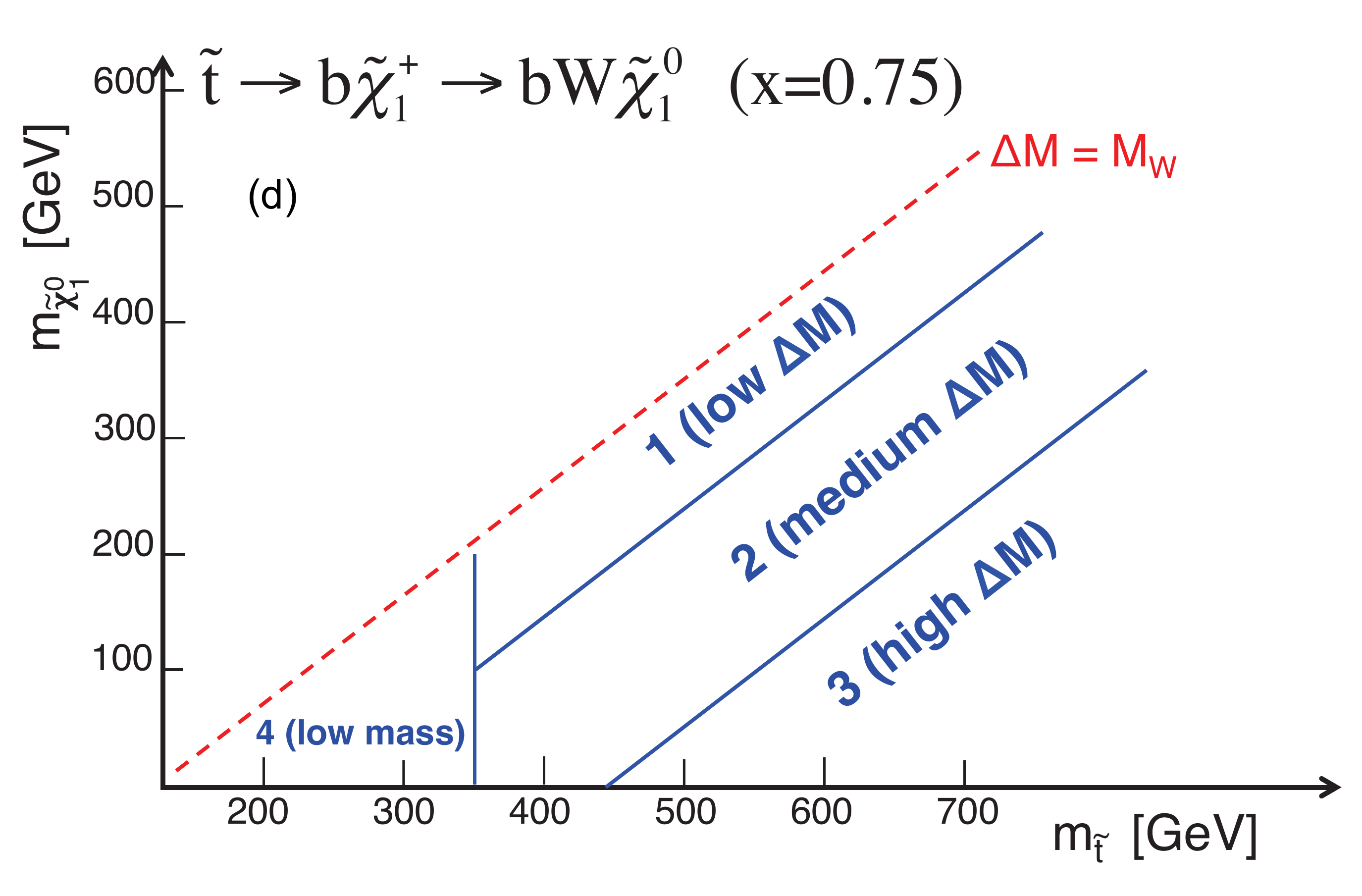}
\caption{\label{fig:BDT_regions}
The regions used to train the BDTs, in the $m_{\lsp}$ vs. $m_{\PSQt}$
parameter space, for (a) the $\Ttt$ scenario, and for (b) the $\TbW$
$x=0.25$, (c) $0.5$, and (d) $0.75$ scenarios.
 The dashed lines correspond to
$\Delta M \equiv m_{\PSQt} - m_{\chiz_1} = M_{\text{top}}$
for $\Ttt$, and
$\Delta M \equiv m_{\chi_1^+} - m_{\chiz_1} = M_\PW$
for $\TbW$.
}
\end{center}
\end{figure*}

\subsubsection{BDT signal regions}
\label{sec:bdtSR}

The BDTs are trained on samples of MC signal and
background events satisfying the preselection requirements and with
$\MT > 120$\GeV.
The BDTs are trained with \MADGRAPH samples
for  $\Ttt$ and a mixture of \MADGRAPH and \PYTHIA samples for
$\TbW$. The choice of generators has little impact on the final result.
The background MC sample contains all the
expected SM processes.

Separate BDTs are trained for the $\Ttt$ and $\TbW$ decay modes and for
different regions of parameter space.
In what follows we refer to the different BDTs as BDT$n$, where
$n$ is the region number defined in Fig.~\ref{fig:BDT_regions}.
In general, for a given BDT,
the optimal requirement
does not depend strongly on the point in parameter space within each region.
Thus, for almost all regions
a single BDT requirement is sufficient, and each such
requirement defines a BDT signal region.
The exceptions are BDT1 for the $\Ttt$ signal model and BDT2
for the $\TbW$ signal model with parameter $x=0.5$; in these regions
we choose two BDT operating
points, referred to as ``tight''  and ``loose''.

\begin{figure*}[thb]
\begin{center}
\includegraphics[width=0.35\linewidth]{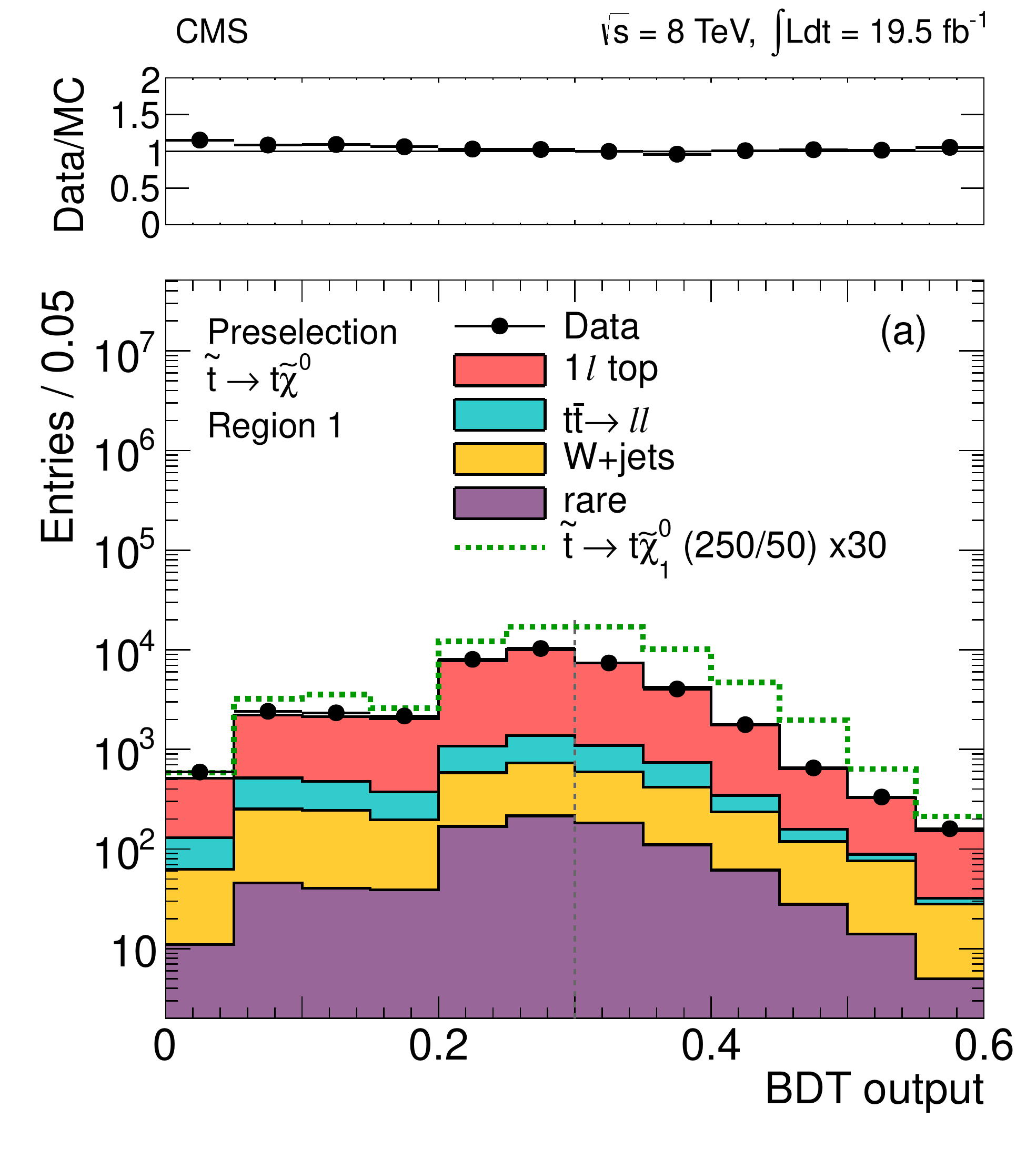}
\includegraphics[width=0.35\linewidth]{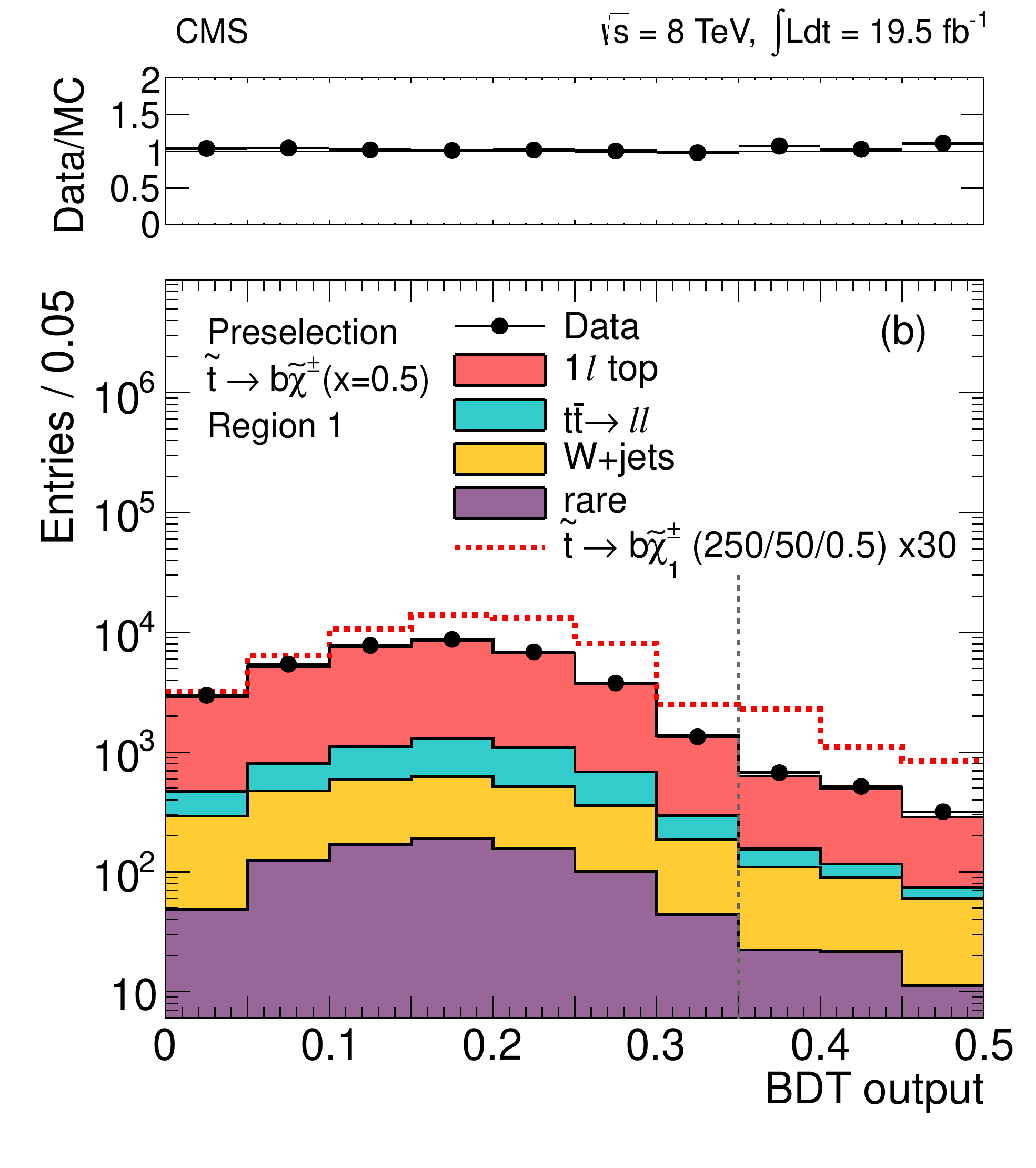}
\includegraphics[width=0.35\linewidth]{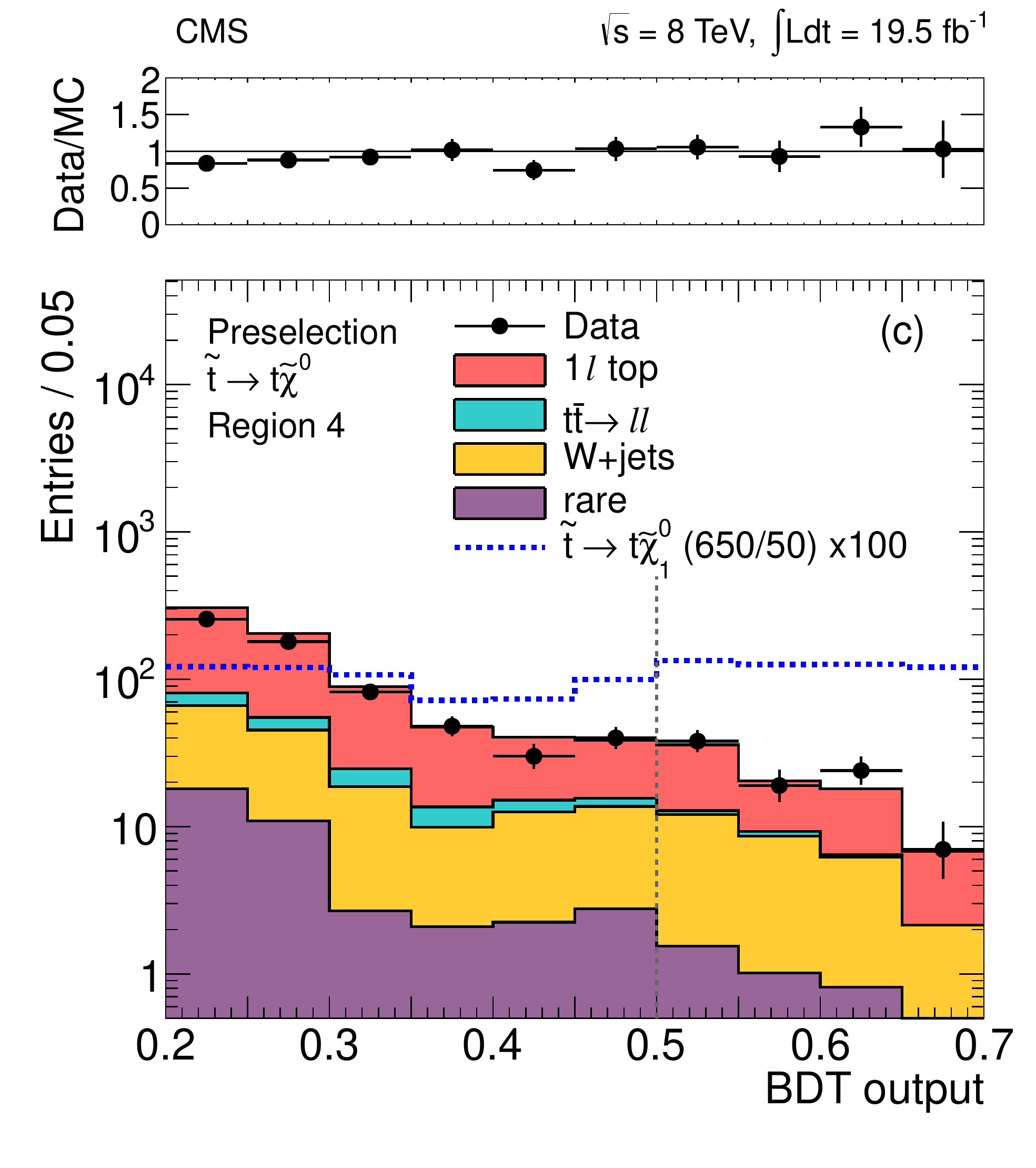}
\includegraphics[width=0.35\linewidth]{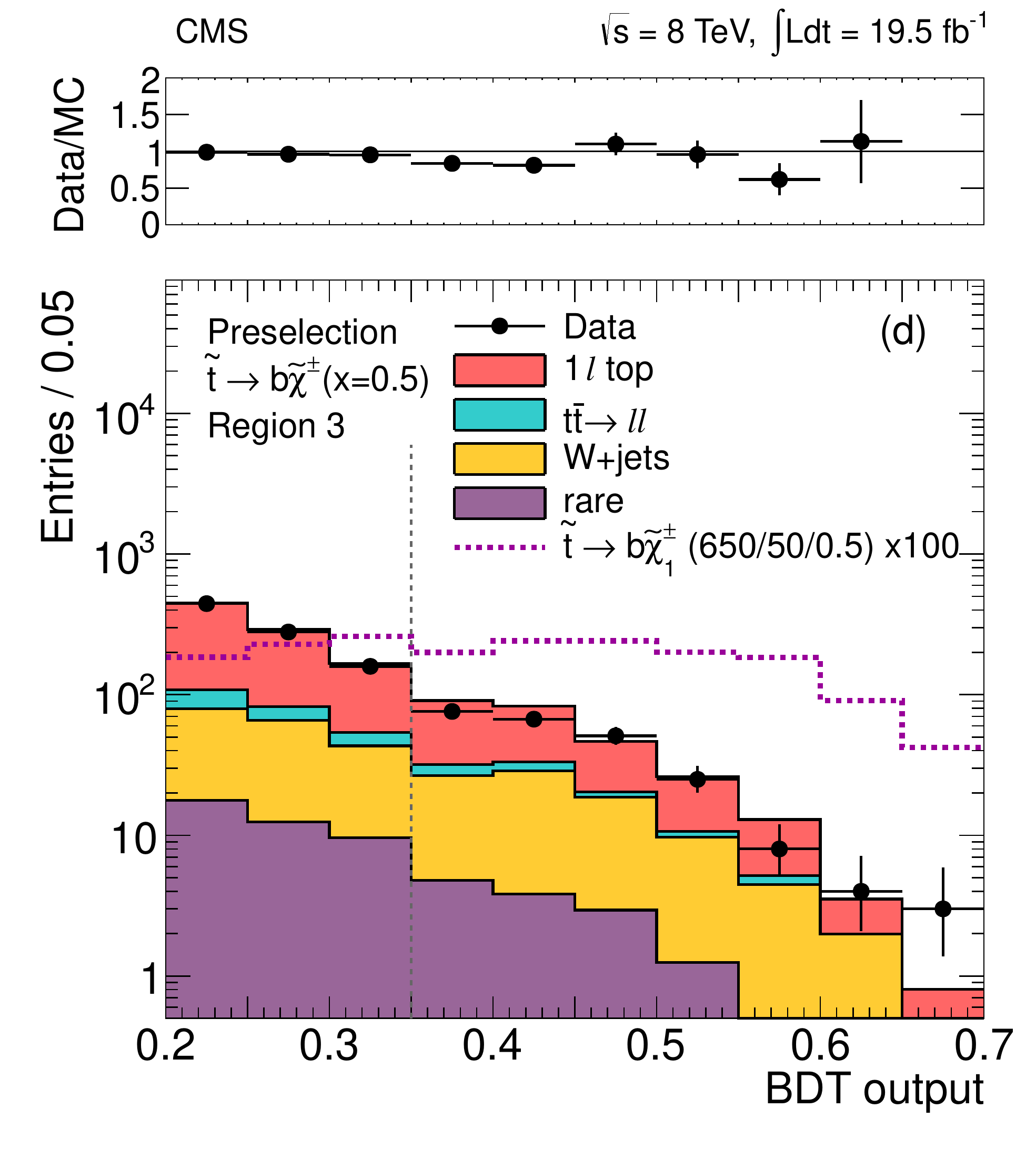}
\caption{\label{fig:quantities2}
Comparison of data and MC simulation for sample BDT outputs.
(a) \Ttt\ scenario in training region 1; 
(b) \TbW scenario with $x=0.5$ in training region 1;
(c) \Ttt\ scenario in training region 4; 
(d) \TbW scenario with $x=0.5$ in training region 3.
Only the event preselection is applied, and in all cases the last bin contains the overflow.
Events in the signal regions are further selected by requiring $\MT > 120$\GeV and
by applying BDT requirements as indicated by the
vertical dashed lines.
We also overlay expectations for possible signals with 
$m_{\PSQt}=250$\GeV and $m_{\lsp}=50$\GeV (panels (a) and (b)) 
and $m_{\PSQt}=650$\GeV and $m_{\lsp}=50$\GeV (panels (c) and (d)).
For display purposes, these are scaled up by factors of 30 and 100 
respectively.} 
\end{center}
\end{figure*}

BDT distributions after the preselection are shown in
Fig.~\ref{fig:quantities2} for four of the
16 BDTs (two tight and two loose BDTs).
The data are in agreement with the MC simulation of SM processes. 

\subsubsection{Cut-based signal regions}
\label{sec:cncT2tt}

For the \Ttt\ model, two types of signal regions are distinguished: those targeting ``small $\Delta
M$'' and those targeting ``large $\Delta M$'', where $\Delta M \equiv m_{\PSQt}-m_{\chiz}$.
Both categories include the requirement that the azimuthal angular difference between the two
leading jets and the \MET\ vector exceed $0.8$~radians, in addition to the requirement
that the value of the hadronic top $\chi^2$ be less than 5.
The $\mtw > 200$\GeV requirement
is applied only for the large $\Delta M$ signal regions.
Within each set, the SRs are distinguished by four successively
tighter \MET\ requirements: $\MET>150,$ 200, 250, and 300\GeV.

For the \TbW\ model, the same approach is followed as for $\Ttt$
by defining two sets of signal regions, one for
small $\Delta M$ and one for high $\Delta M$, where $\Delta M$
here is the mass difference between the chargino and the LSP.
Just as in the $\Ttt$ case, SRs are distinguished by
increasingly tighter
requirements on \MET.  Since in the case of \TbW\ the signal has no top
quark in its decay products, the requirement on the hadronic
top $\chi^2$ is not used.
The large $\Delta M$ selection includes the \mtw\ requirement, as well
as the requirement that the leading b-tagged jet have \pt\ larger than
100\GeV.

\subsubsection{Signal regions summary}
\label{sec:srsummary}

To summarize, this search uses two complementary approaches:
one a cut-based approach and the other a BDT multivariate
method. Correspondingly, there are two distinct sets of signal
regions. In the BDT case, the SRs are defined by requirements on the BDT outputs.
The BDT SRs provide the primary result, since the BDT method has
better expected sensitivity.
There are a total of 16 cut-based SRs (eight each for the $\Ttt$ and
$\TbW$ cases) and 18 BDT SRs (six for the $\Ttt$ mode and 12 for
the $\TbW$ mode).
The expected  number
of background events in the SRs varies between approximately
4 and 1600 (see Section~\ref{sec:results}).

\section{Background estimation methodology}
\label{sec:backgroundmodeling}

The SM background is divided into four categories that are evaluated separately.
The largest background contribution after full selection
is \ttbar production in which both W bosons decay leptonically (\ttll),
but one of the leptons is not identified.
The second largest background consists of \ttbar production in which one W boson decays leptonically
and the other hadronically (\ttlj), as well as single-top-quark production
in the s- and t-channels: These are collectively referred to as
``single-lepton-top-quark''
processes. The third largest background consists of a variety of SM processes
with small cross sections, including \ttbar events produced in association with a vector boson (\ttbarW, \ttbarZ, \ttbarg),
processes with two ($\PW\PW$, $\PW\Z$, $\Z\Z$) and three ($\PW\PW\PW$, $\PW\PW\Z$, $\PW\Z\Z$, $\Z\Z\Z$) electroweak vector bosons,
and single-top-quark production in the tW-channel. These processes are collectively referred to as the ``rare'' processes.
The fourth and final background contribution is from the production of
\PW\ bosons
with jets
(\wjets).
The multijet contribution to the background is negligible in the signal regions
due to the requirement of a high-\pt\ isolated lepton, large \MT, large \MET, and a b-tagged jet.
Here, ``multijet" refers to events composed entirely of jets, without a
lepton, \PW\ or Z boson, or top quark.

Backgrounds are estimated from MC simulations, with small corrections
(see below).
The simulation is validated in control regions (CRs)
designed to enrich the data sample in specific sources of
background while maintaining kinematic properties that
are similar to those in the signal regions (see Section~\ref{sec:controlRegions}).
In the CRs the kinematic variables used
in the cut-based and BDT selections are examined to verify that they are
properly modeled.  A key distribution in each CR is that
of \MT\ after the cut-based or BDT selection requirements, since 
$\MT> 120$\GeV is the final criterion that defines each 
signal region. The data/MC comparison
of the number of events with $\MT> 120$\GeV is then a direct
test of the ability of the method to correctly predict the SM background in the signal
regions.

The CR studies are designed to extract data/MC scale factors
to be applied to the MC predictions for the background in the signal regions.
We find that the only scale factor required is related to an underestimation of the \MT\ tail
for single-lepton-top-quark and \wjets\ events,
as discussed in more detail in Section~\ref{sec:controlRegions}.

 \begin{figure}[htbp]
  \begin{center}
        \includegraphics[width=\cmsFigWidth]{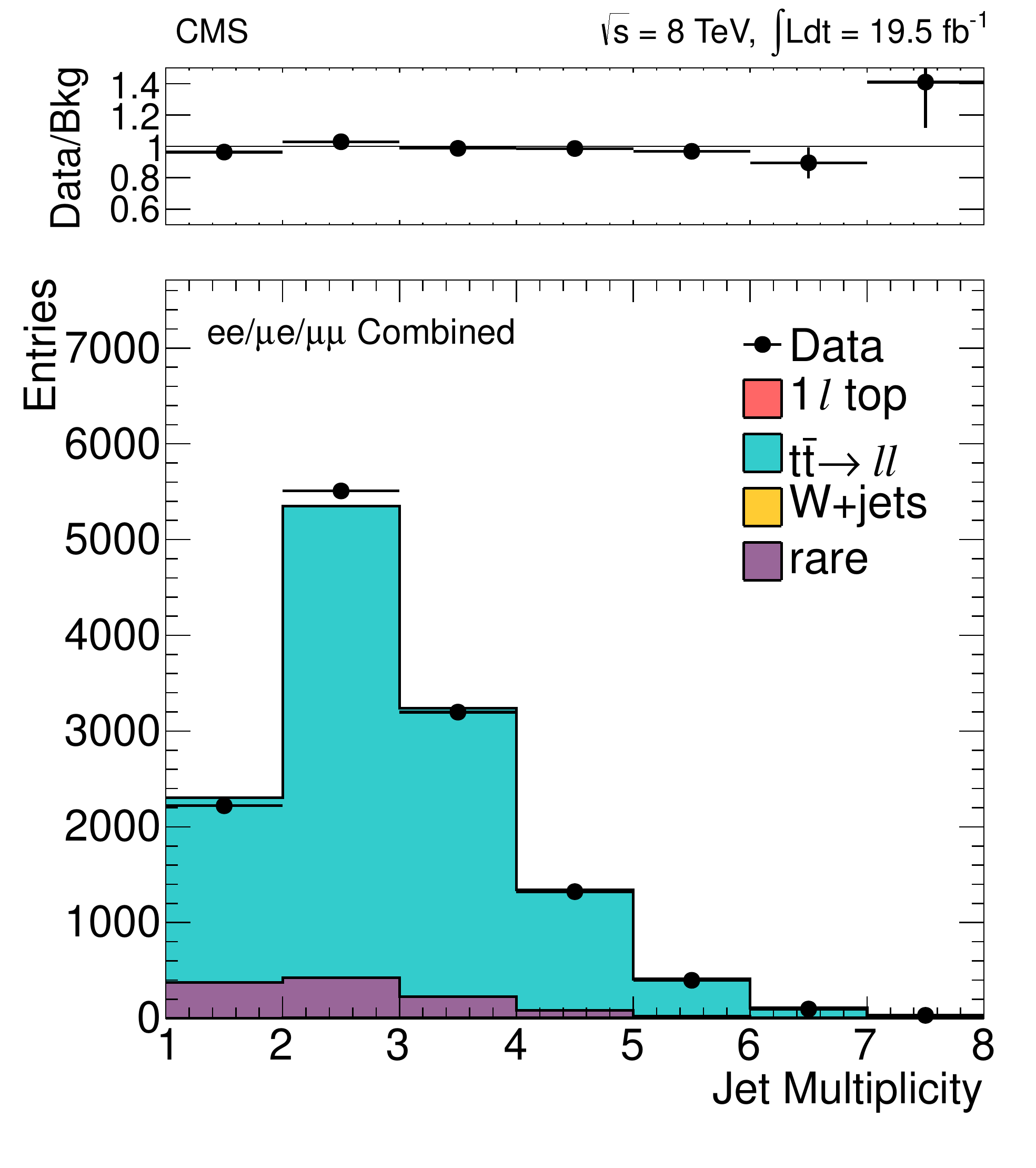}
    \caption{
      Comparison of the jet multiplicity distributions in data and MC simulation in the
      sample dominated by \ttll\ events.
\label{fig:ttllnjets}
}
 \end{center}
\end{figure}

The selection of signal events requires at least four hadronic jets.
As mentioned above, the dominant background consists of
\ttll\ events with one unidentified lepton.  These events satisfy the signal region selection
only if there are two additional jets from initial- or final-state
radiation (ISR/FSR) or if there is one such jet in conjunction with a second lepton identified as
a jet (\eg, in the case of
hadronic $\tau$-lepton decays).
To validate the modeling of ISR/FSR, a data control sample of
\ttll\ events is defined by requiring the presence of exactly two opposite-sign leptons
(electrons or muons) in events satisfying dilepton triggers. To
suppress the \zjets\ background that is present in this control sample, same-flavor (ee or $\mu\mu$)
events with an invariant mass in the range $76<m_{\ell\ell}<106$\GeV are rejected,
the presence of at least one b-tagged jet is required, and minimum requirements are imposed on \MET.
We then compare the
distribution of the number of jets
in data and MC simulation, as displayed
in Fig.~\ref{fig:ttllnjets}.
The fraction of \ttll\ events with three or more jets is found to be
in agreement
with the expectation from the MC simulation within a 3\% statistical
uncertainty.

To minimize systematic uncertainties associated with the \ttbar\ production cross section,
integrated luminosity, lepton efficiency, and jet energy scale, the
\ttbar\ MC backgrounds at high \MT\ are always normalized to the number of events in data in the
transverse-mass peak region, defined as $50 < \MT < 80$\GeV,
after subtracting the contribution from rare backgrounds. 
We refer to this normalization factor as the ``tail-to-peak ratio''.
Background contributions from rare processes are taken directly from
the simulated samples. Their rates are normalized
using the corresponding NLO cross sections.

\section{Control region studies}
\label{sec:controlRegions}
Three CRs are used in this analysis.
A sample dominated by \ttll\ events is obtained by requiring the
presence of two leptons (CR-2$\ell$).
A sample dominated by a mixture of $\ttbar\to\ell+\text{jets}$ and \ttll\ events is obtained by requiring the presence of
a lepton and one isolated track or hadronic $\tau$-lepton candidate  (CR-$\ell$t).
A sample dominated by \wjets\ events is obtained by
vetoing events with b-tagged jets (CR-0b).

In all CRs, we apply the various SR selections and compare data and MC
yields with $\MT > 120$\GeV after normalizing the \MT\ distribution
to the transverse-mass peak region
as described in Section~\ref{sec:backgroundmodeling}.
In the case of CR-2$\ell$, the definition of \MT\ is ambiguous
because there are two identified leptons;
we take the \MT\ value constructed from the leading lepton and the
\MET\ vector.

\begin{figure*}[htbp]
\begin{center}
\includegraphics[width=0.35\textwidth]{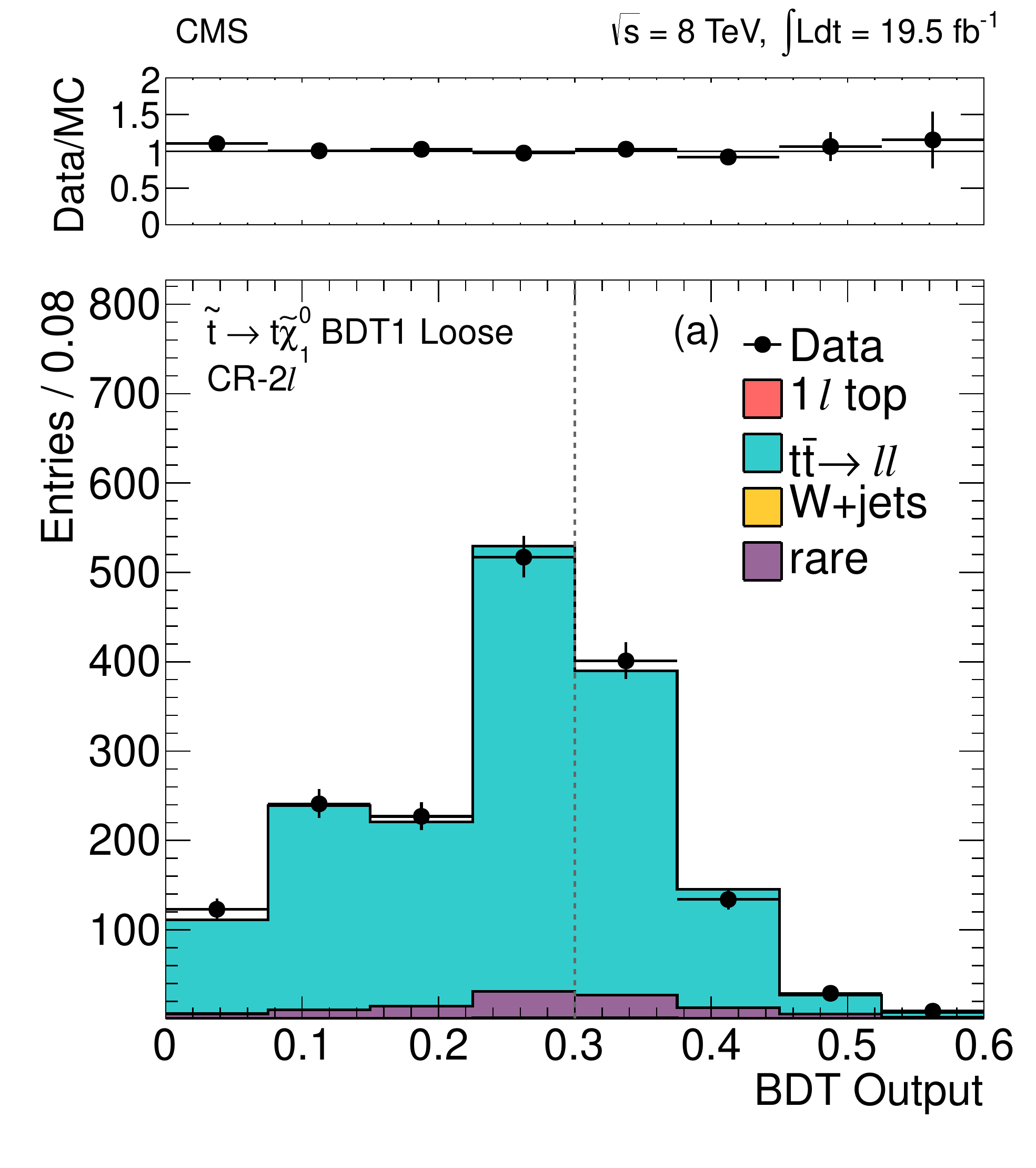}
\includegraphics[width=0.35\textwidth]{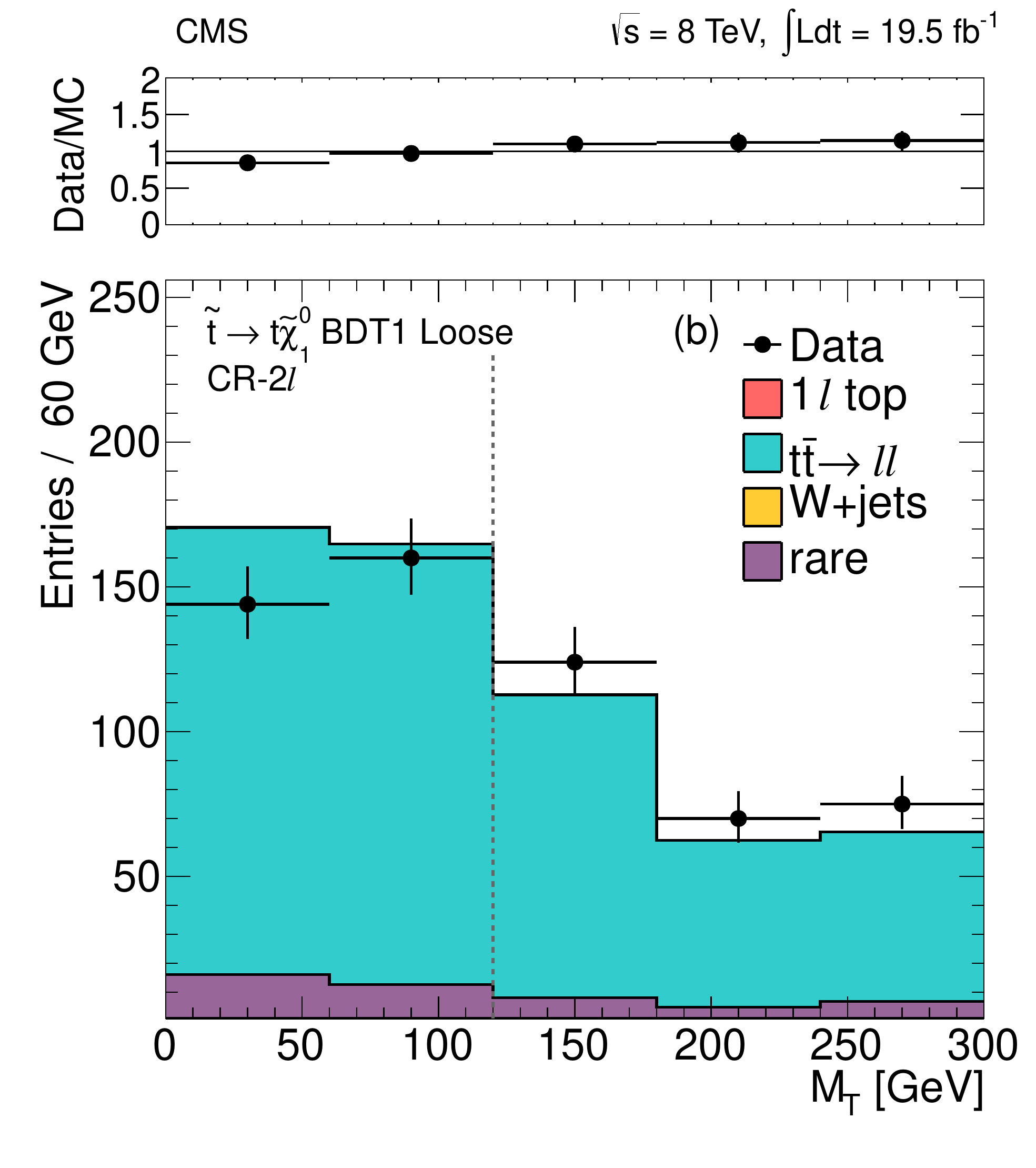}
\includegraphics[width=0.35\textwidth]{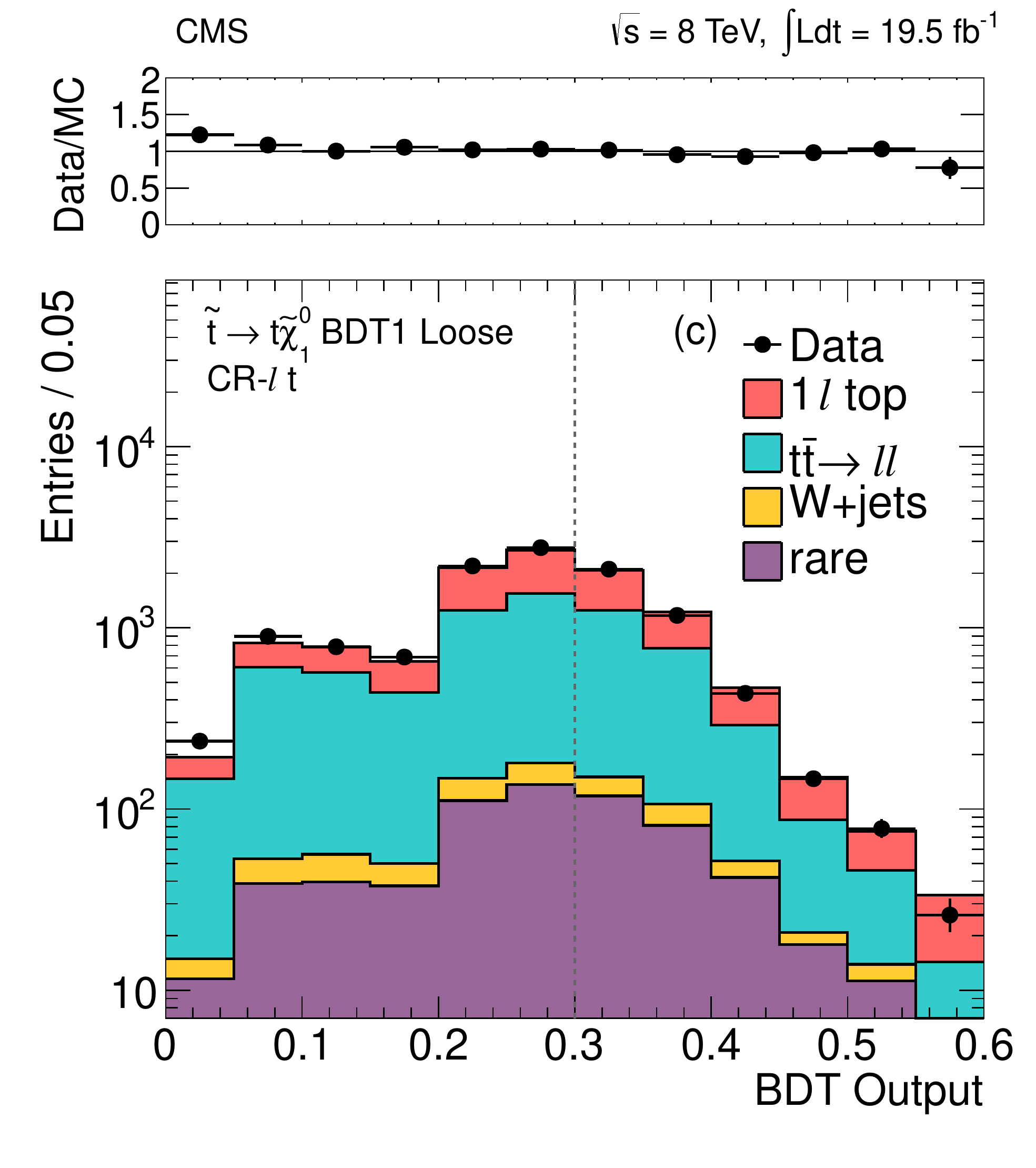}
\includegraphics[width=0.35\textwidth]{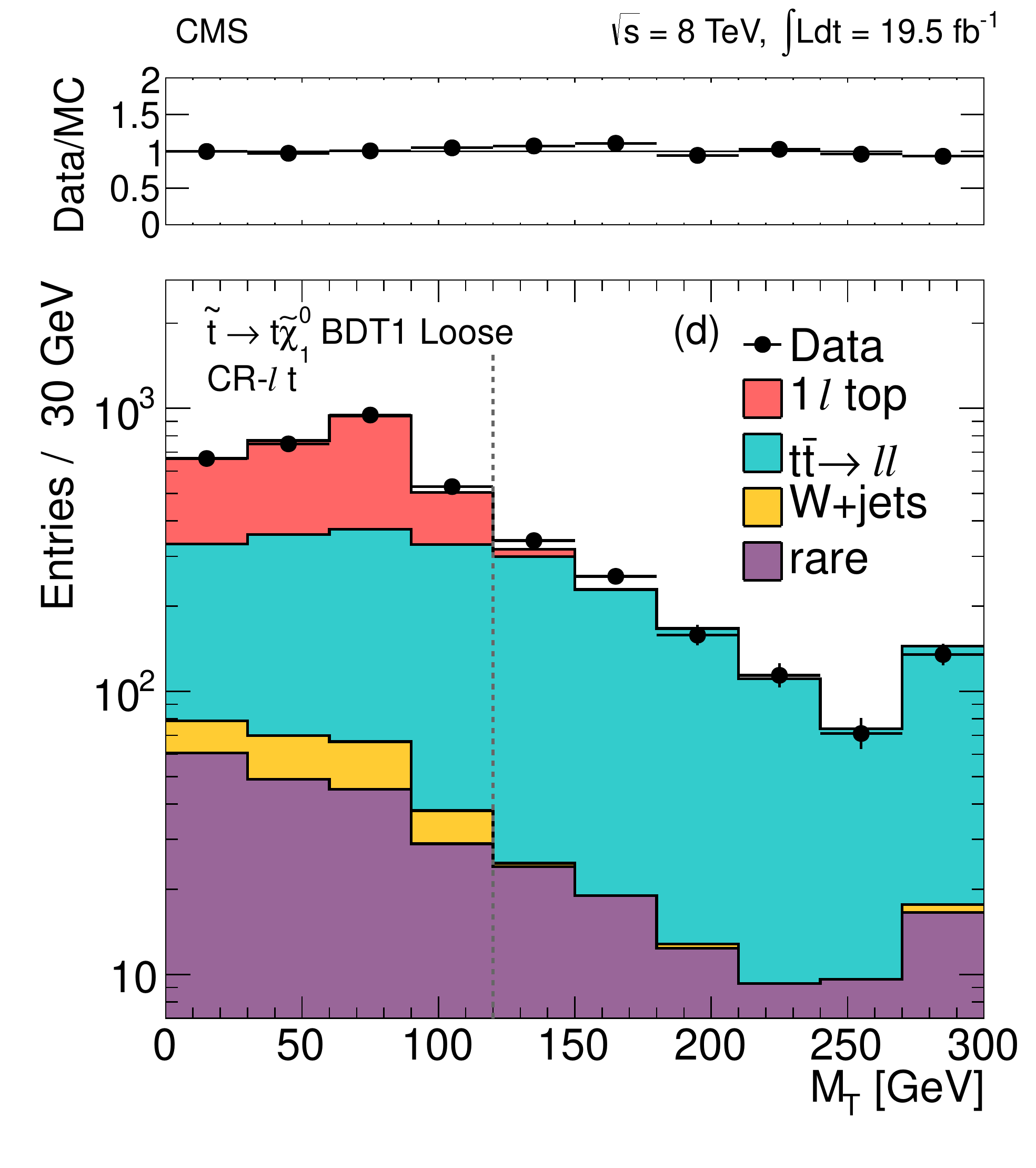}
\includegraphics[width=0.35\textwidth]{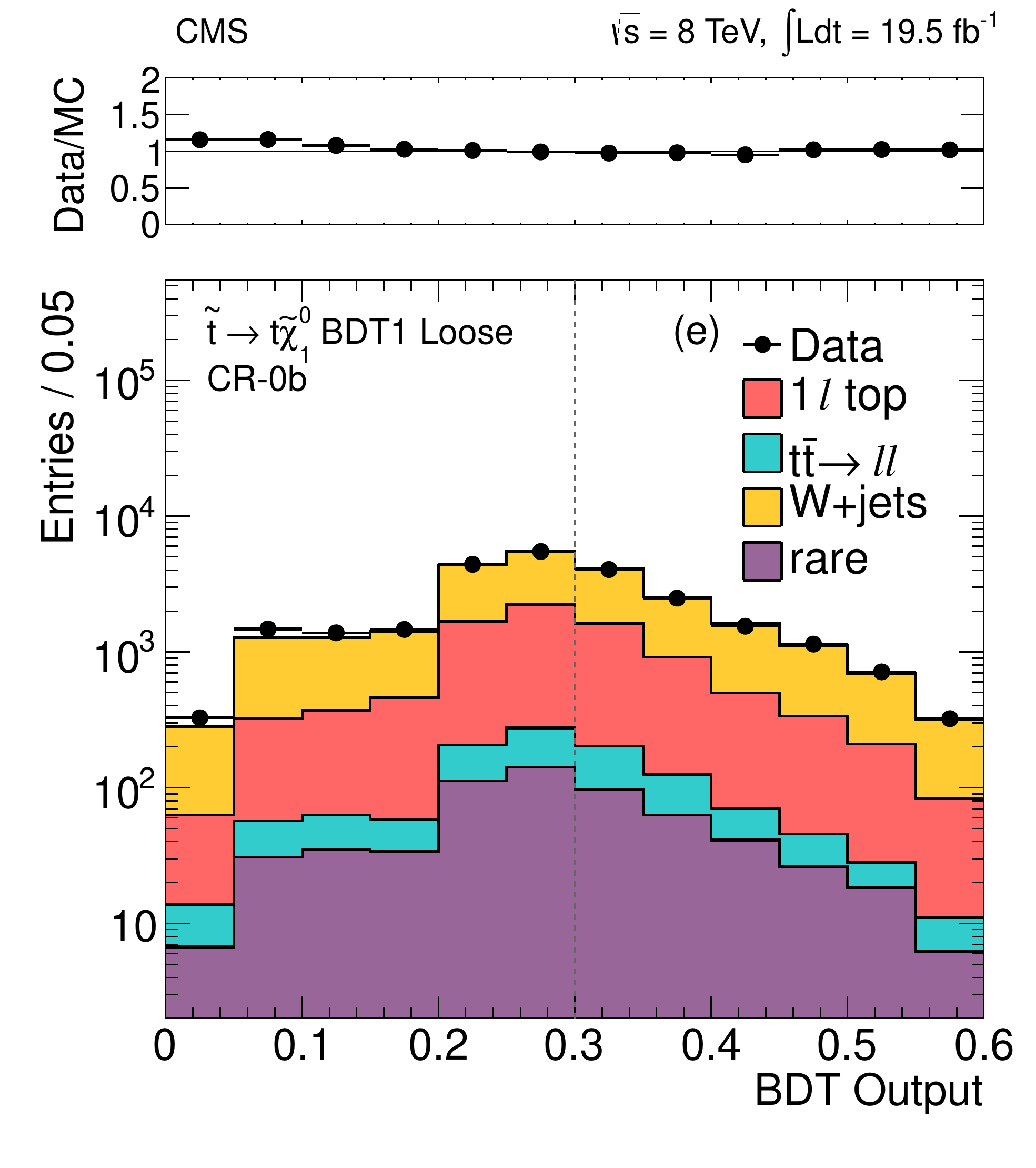}
\includegraphics[width=0.35\textwidth]{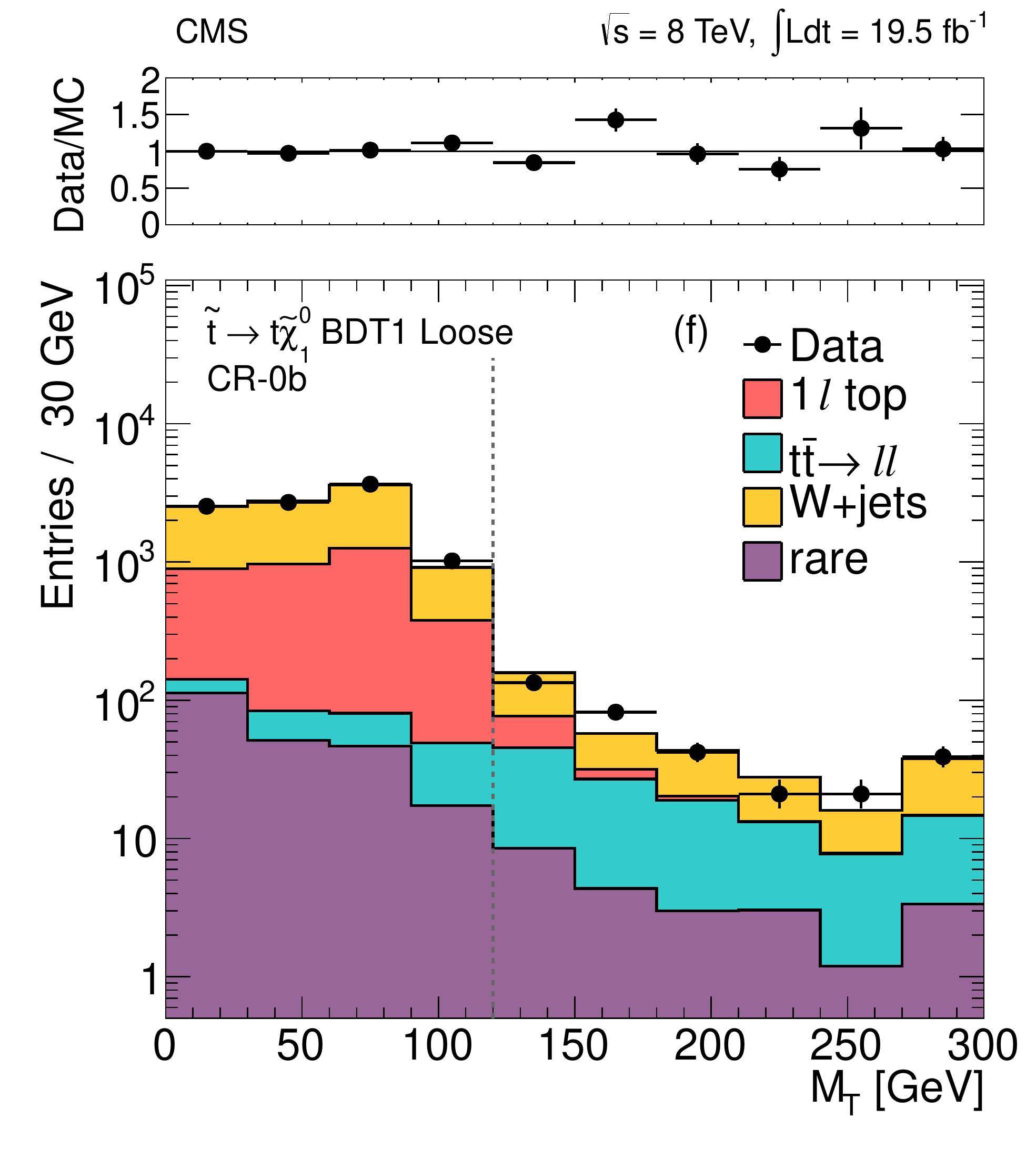}
\caption{\label{fig:CRplots}
Comparison of data and MC simulation for the distributions of \MT\ and BDT output for
the control regions associated with the BDT trained in region 1 for the \Ttt\ scenario.
The \MT\ distributions are shown after the ``BDT1 loose'' requirement indicated
by vertical dashed lines on the BDT output plots.  
(a)-(b): CR-2$\ell$; (c)-(d): CR-$\ell$t; (e)-(f): CR-0b. 
The vertical dashed lines in the \MT\ plots correspond to the $\MT > 120$\GeV
selection requirement.
For CR-0b, the scale factors are applied to the MC distribution in the
$\MT$ tail. The last bin in all distributions contains the overflow.}
\end{center}
\end{figure*}

The BDT output distribution trained in \Ttt\ region 1 (BDT1) is shown
in Fig.~\ref{fig:CRplots} for the three control regions. The \MT\
distribution after the BDT signal region requirement is also
displayed (in the case of CR-0b this is corrected using the
scale factor discussed below).
Similar levels of agreement between data and MC simulation are found for the other
SR-like selections.

For CR-2$\ell$ and CR-$\ell$t,
the number of data events with $\MT > 120$\GeV is consistent with the MC
prediction.
The level of agreement is used to assess
a systematic uncertainty for the \ttll\ background prediction.
The uncertainty ranges from 5\% for the loosest signal regions
to 70\% for the tightest signal regions, reflecting the limited statistical
precision of the control samples after applying the \MT\ and BDT requirements.
The fraction of events in CR-2$\ell$ and CR-$\ell$t with $\MT>120$\GeV that could
be from stop pair production varies between approximately 1\% and
20\%, depending on the CR and the masses of the top squark and the
LSP. This contribution is always much smaller than the statistical 
uncertainty on the data event counts.

\begin{figure}[htbp]
\begin{center}
\includegraphics[width=\cmsFigWidth]{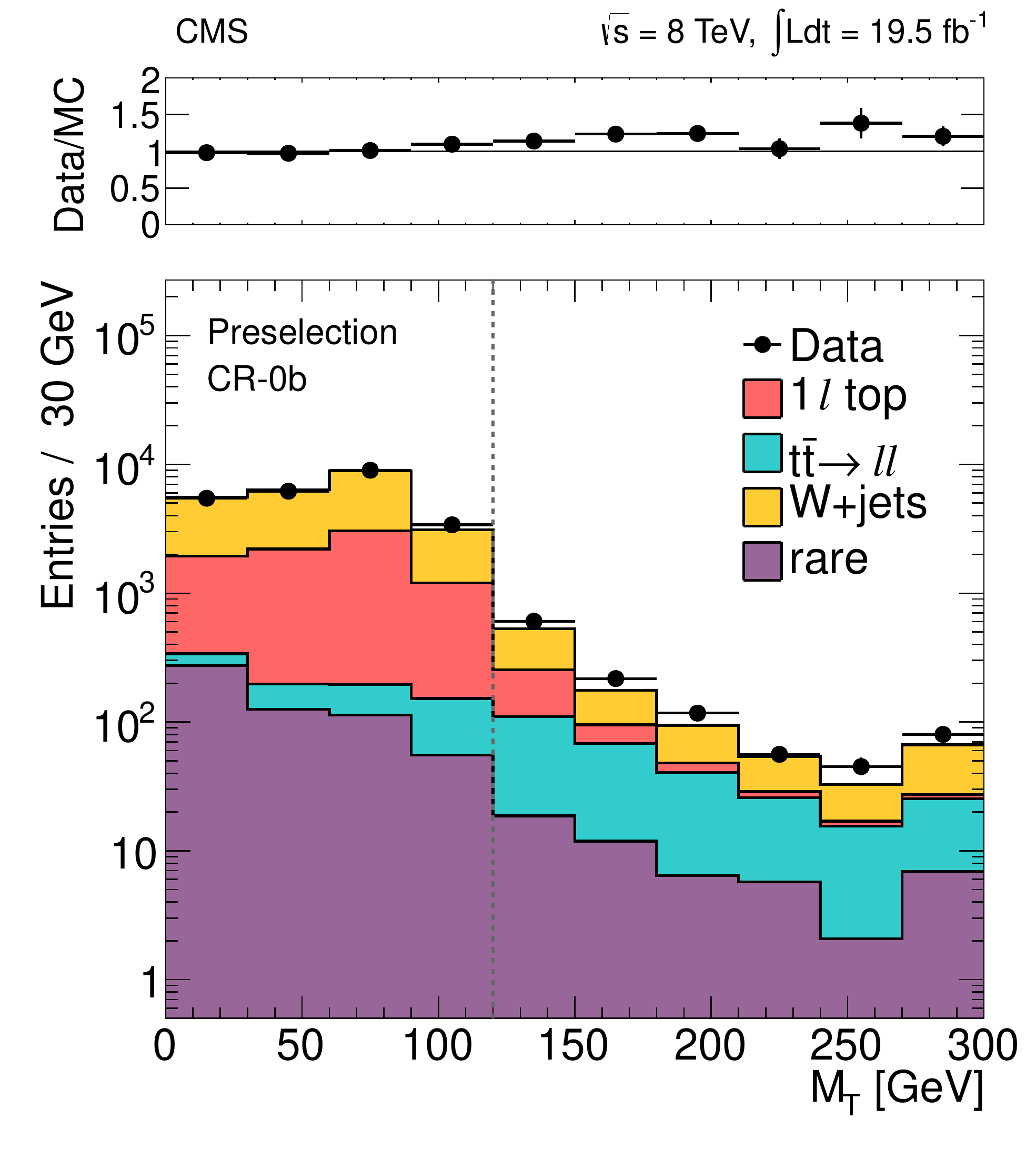}
\caption{\label{fig:CR1unscaledplot}
Comparison of data and MC simulation for the \MT\ distribution in the
CR-0b control region, after the preselection. The \MT\ tail is
underestimated by the simulation. A scale factor derived from this
control region is used to correct the predictions of the \wjets\ and
single-lepton-top-quark backgrounds. The last bin of the distribution includes the overflow.}
\end{center}
\end{figure}

In the case of CR-0b, the transverse-mass distribution of events exhibits a small excess
at high \MT\ with respect to the MC prediction.
This discrepancy, illustrated in Fig. ~\ref{fig:CR1unscaledplot} using
the high-statistics samples of the preselection level, is attributed
to imperfect modeling of the tails of the \MET\ resolution in
\wjets\ events. The data/MC agreement in the CR-0b \MT\ tail can be restored
by rescaling the \wjets\ contribution by a factor of $1.2 \pm 0.3$,
as seen for example in Fig.~\ref{fig:CRplots}, bottom right.
We find that this factor is
insensitive to the details
of the selection of the kinematic variables in Table~\ref{tab:SRsummary}
for the CR-0b event sample.

The observation
that the simulation underestimates the \MT\ tail in the \wjets\
sample suggests that a similar effect should exist in the
single-lepton-top-quark background.
However, the \MT\ tail is more populated for the  \wjets\ background
than for the single-lepton-top-quark background, due to a significant
contribution from very off-shell \PW\ bosons.
This contribution is much less pronounced for the
single-lepton-top-quark background because,
ignoring the top-quark width,
the lepton-neutrino mass $M_{\ell \nu}$
cannot exceed the difference between the top- and bottom-quark masses,
$M_{\ell \nu} < M_{\text{top}}-M_\cPqb$.  This bound can be violated only if
both the top quark and $\PW$ boson from top quark decays are off-shell.
For this reason the scale factor of $1.2 \pm 0.3$ measured in \wjets\
events cannot be simply applied
to the single-lepton-top-quark simulated sample.
The scale factor is larger in the single-lepton-top-quark sample
because the fraction of events that have $\MT > 120$\GeV
due to $\MET$ mismeasurement is larger than in the \wjets\ sample.

Following the arguments given above, a lower bound on the data
tail-to-peak ratio for the single-lepton-top-quark sample ($R_{\text{top}}$)
can be obtained by scaling the MC value of $R_{\text{top}}$ by the
\wjets\ scale factor ($1.2 \pm 0.3$).  Conversely, an upper bound for
$R_{\text{top}}$ is $R_{\text{top}} = R_{{\wjets}}$, where $R_{{\wjets}}$ is the tail-to-peak
ratio for \wjets\ in the data, i.e., its MC value scaled up by $1.2 \pm
0.3$.  This is an overestimate of the true value of $R_{\text{top}}$ because, as
mentioned above, the \MT\ tail is more populated for the \wjets\ sample
than for the one-lepton-top sample. 
Since the true value of $R_{\text{top}}$ lies between these
two extremes, we take the average of the upper and lower bounds.
The resulting scale factor for $R_{\text{top}}$
with respect to its uncorrected MC
result lies between 1.5 and 2, depending on the signal region.
The associated uncertainty includes the statistical uncertainty
in the data/MC scale factor from CR-0b, and half the
difference between these upper and lower bounds.

\section{Systematic uncertainties of the background prediction}
\label{sec:syst}

All backgrounds except for the rare contribution are normalized to data in the \MT-peak region, so the statistical
uncertainties of the data and MC yields in the \MT-peak region
contribute to the uncertainty of the background predictions in the
high-\MT\ signal regions. This normalization is repeated after varying the
\wjets\ background yield in the \MT-peak region by $\pm$50\% to estimate the associated systematic uncertainty.

For the \ttll\ background, the dominant uncertainty is assessed by
comparing the data and MC yields in the high-\MT\ regions of the CR-2$\ell$
and CR-$\ell$t samples after applying the kinematic requirements
for the corresponding signal region.
This uncertainty varies between 5\% and 70\%.
The uncertainty for the modeling of
additional jets from radiation in \ttll\ events
results in a 3\% uncertainty on the
dilepton background. The uncertainty from the limited number of events in the
\ttll\ MC sample also contributes, particularly in the tight signal
regions.

An additional uncertainty is associated with
the efficiency  to identify
a second lepton (e, $\mu$, or one-prong hadronic $\tau$-lepton decay)
as an isolated track.
We verify that the simulation reproduces the efficiency
of the isolated track requirement
through
studies of Z$\to\ell\ell$ events in data, and we assign a systematic uncertainty of 6\%.
An uncertainty of
7\%, based on studies of the efficiency for $\tau$-lepton identification in
data and simulation, is applied to events with a hadronic
$\tau$-lepton in the hadronic
$\tau$-lepton veto acceptance. We also verify the stability of the \ttll\ MC background prediction
by comparing the results of the nominal \POWHEG sample with those
obtained using \MADGRAPH and \MCATNLO,  by varying the \MADGRAPH
scale parameters for
renormalization and factorization, as well as the scale
for the matrix element and parton shower matching,
up and down by a factor of two, and by
varying the top-quark mass
in the range
178.5 to 166.5\GeV. Since the resulting background predictions are consistent within the systematic
uncertainties discussed above, we do not assess an additional
uncertainty from
the \ttbar\ MC stability tests.

The uncertainty of the
\wjets\ background prediction is dominated by the uncertainty from the tail-to-peak ratio, as determined
from data/MC comparisons in the CR-0b control region.
The main uncertainty for the single-lepton-top-quark background arises from
the difference in the
tail-to-peak ratios for \wjets\ and single-lepton-top-quark events.

\begin{table*}[htb]
\topcaption{The bottom row of this table shows
the relative uncertainty (in percent) of the total background
  predictions for the \Ttt\ BDT signal regions.
The breakdown of this total uncertainty in terms of its individual
components is also shown.}
\label{tab:relativeuncertaintycomponents_t2tt_bdt}
\centering
{\footnotesize
\begin{tabular}{l|cccccc}
\multicolumn{7}{c}{$\Ttt$} \\
\hline
 Sample              & BDT1--Loose & BDT1--Tight & BDT2 & BDT3 & BDT4 & BDT5\\
\hline
\MT-peak data and MC (stat) 	 	 & $1.0$ 	 & $2.1$ 	 & $2.7$ 	 & $5.3$ 	 & $8.7$ 	 & $3.0$  \\
\ttll\ N$_{\text{jets}}$ modeling 	 	 & $1.7$ 	 & $1.6$ 	 & $1.6$ 	 & $1.1$ 	 & $0.4$ 	 & $1.7$  \\
\ttll\ (CR-$\ell$t and CR-2$\ell$ tests) 	 	 & $4.0$ 	 & $8.2$ 	 & $11.0$ 	 & $12.5$ 	 & $7.2$ 	 & $13.8$  \\
2nd lepton veto 	 	 & $1.5$ 	 & $1.4$ 	 & $1.4$ 	 & $0.9$ 	 & $0.3$ 	 & $1.4$  \\
\ttll\ (stat.) 	 	 & $1.1$ 	 & $2.8$ 	 & $3.4$ 	 & $7.0$ 	 & $7.4$ 	 & $3.3$  \\
W+jets cross section 	 	 & $1.6$ 	 & $2.2$ 	 & $2.8$ 	 & $1.7$ 	 & $2.7$ 	 & $2.2$  \\
W+jets (stat.) 	 	 & $1.1$ 	 & $1.9$ 	 & $2.0$ 	 & $4.6$ 	 & $10.8$ 	 & $5.2$  \\
W+jets SF uncertainty 	 	 & $8.3$ 	 & $7.7$ 	 & $6.8$ 	 & $8.1$ 	 & $9.7$ 	 & $8.6$  \\
$1-\ell$  top (stat.) 	 	 & $0.4$ 	 & $0.8$ 	 & $0.8$ 	 & $1.4$ 	 & $4.4$ 	 & $1.2$  \\
$1-\ell$  top tail-to-peak ratio 	 	 & $9.0$ 	 & $11.4$ 	 & $12.4$ 	 & $19.6$ 	 & $28.5$ 	 & $9.1$  \\
Rare processes cross section 	 	 & $1.8$ 	 & $3.0$ 	 & $4.0$ 	 & $8.1$ 	 & $15.7$ 	 & $0.7$  \\
\hline
Total 	 	 & $13.4$ 	 & $17.1$ 	 & $19.3$ 	 & $27.8$ 	 & $38.4$ 	 & $20.2$  \\
\hline
\end{tabular}}
\end{table*}

The main contributors to the rare SM backgrounds are
$\Pp\Pp\to\ttbarZ$ and $\Pp\Pp\to\ttbarW$; these processes
have not yet been measured accurately. As mentioned in Section~\ref{sec:mc},
we normalize their rates to the respective NLO cross-section
calculations~\cite{xsec_ttbarW,xsec_ttbarZ}. We assign an overall
conservative uncertainty of 50\% to account for missing higher order terms, as well as
possible mismodeling of their kinematical properties (see for example
the discussion of Ref.~\cite{xsec_ttbarW}).

The systematic uncertainties for the \Ttt\ BDT analysis are summarized in
Table~\ref{tab:relativeuncertaintycomponents_t2tt_bdt}. The
uncertainties for all other signal regions are presented in Appendix~\ref{app:othersyst}.

\section{Results}
\label{sec:results}

\begin{figure*}[htbp]
  \begin{center}
    \includegraphics[width=0.49\linewidth]{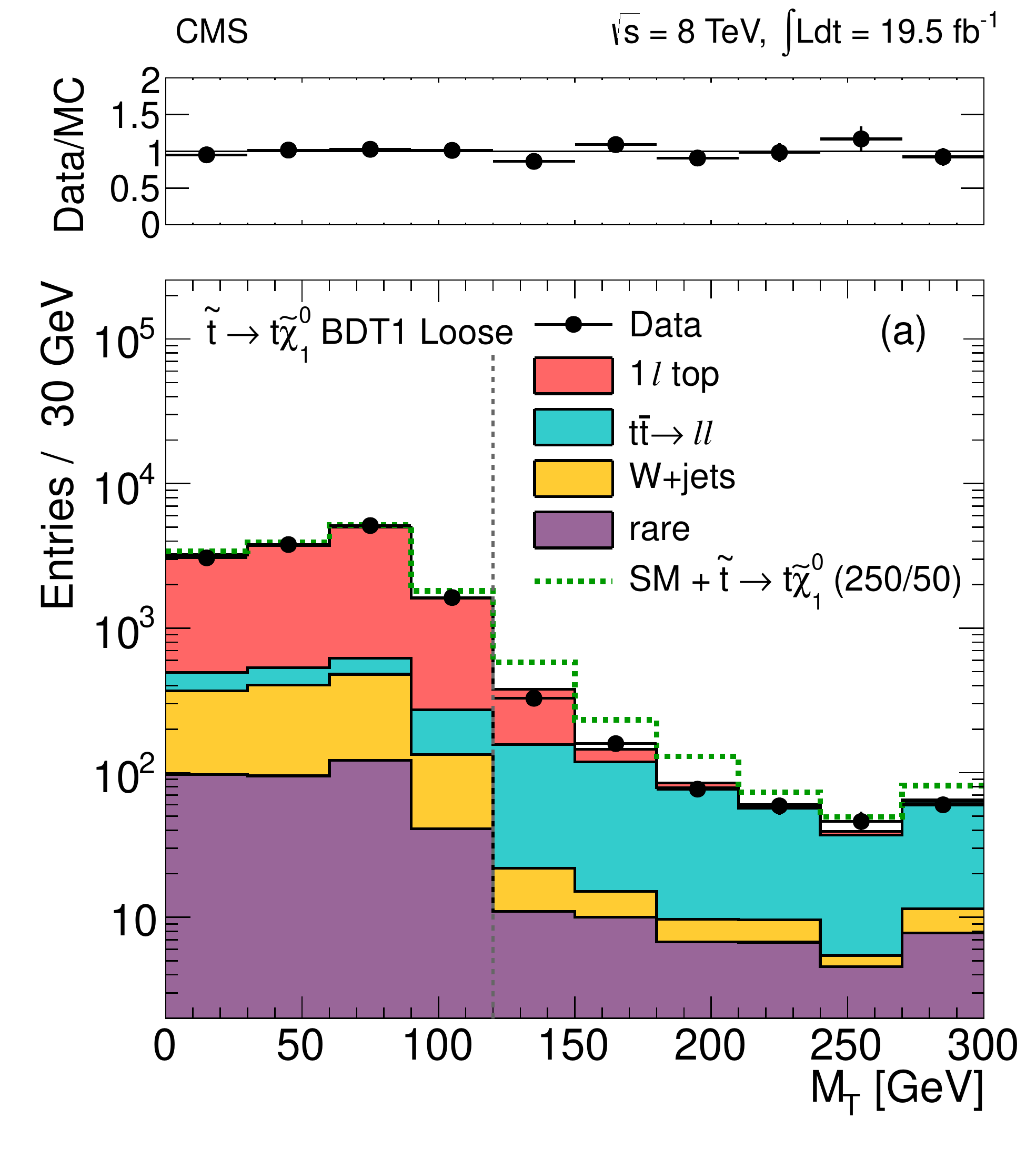}%
    \includegraphics[width=0.49\linewidth]{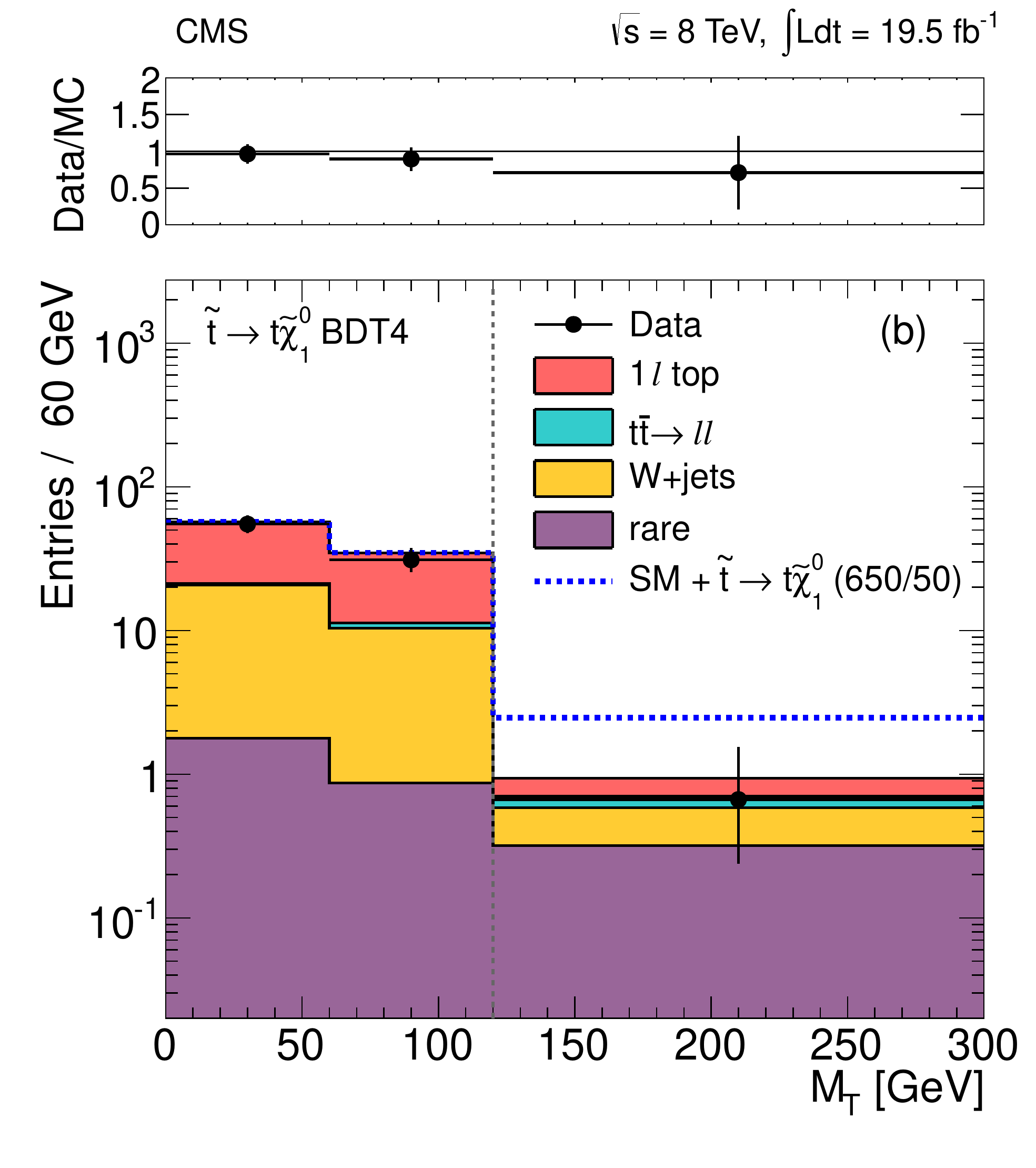} \\
    \includegraphics[width=0.49\linewidth]{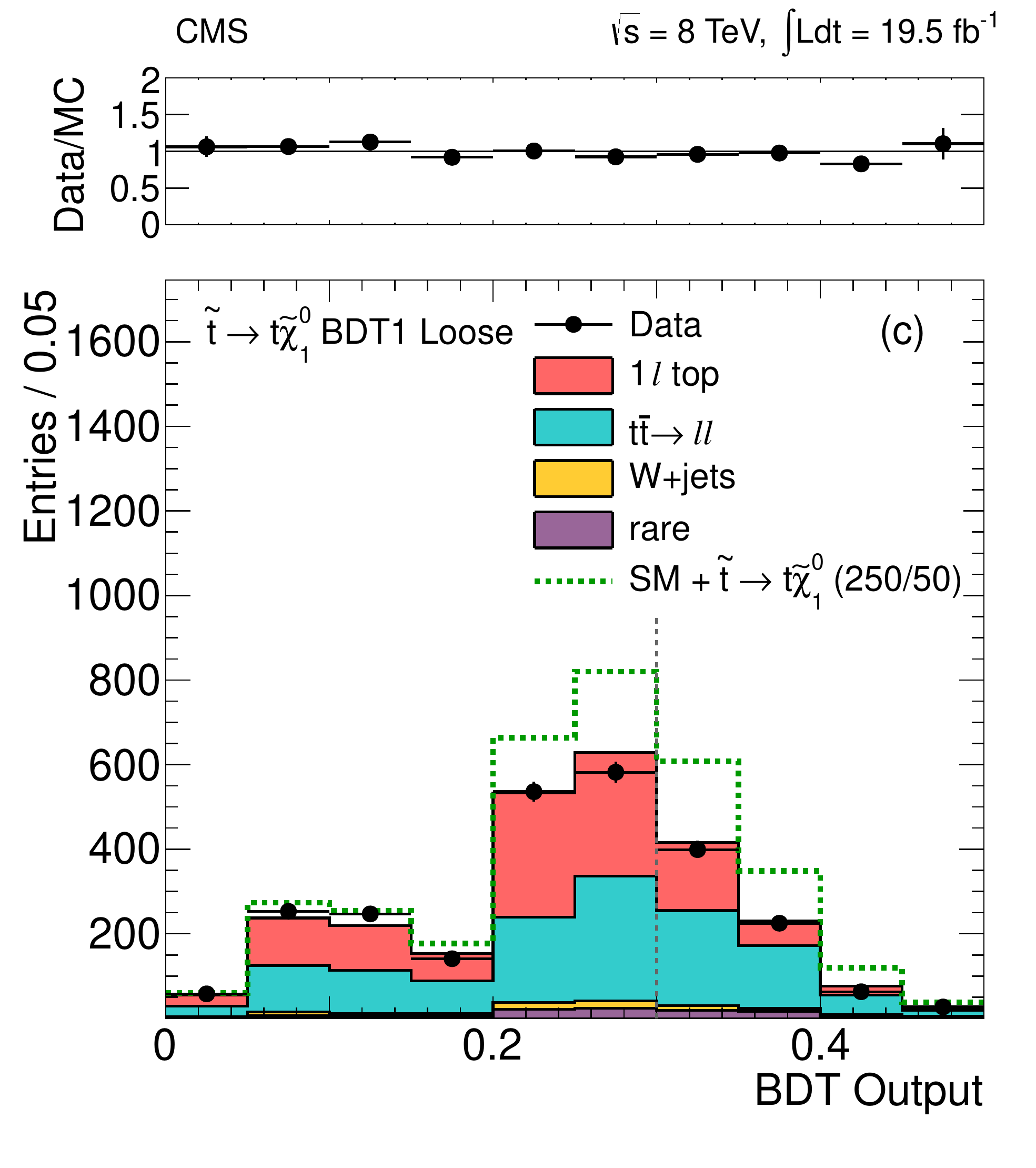}%
    \includegraphics[width=0.49\linewidth]{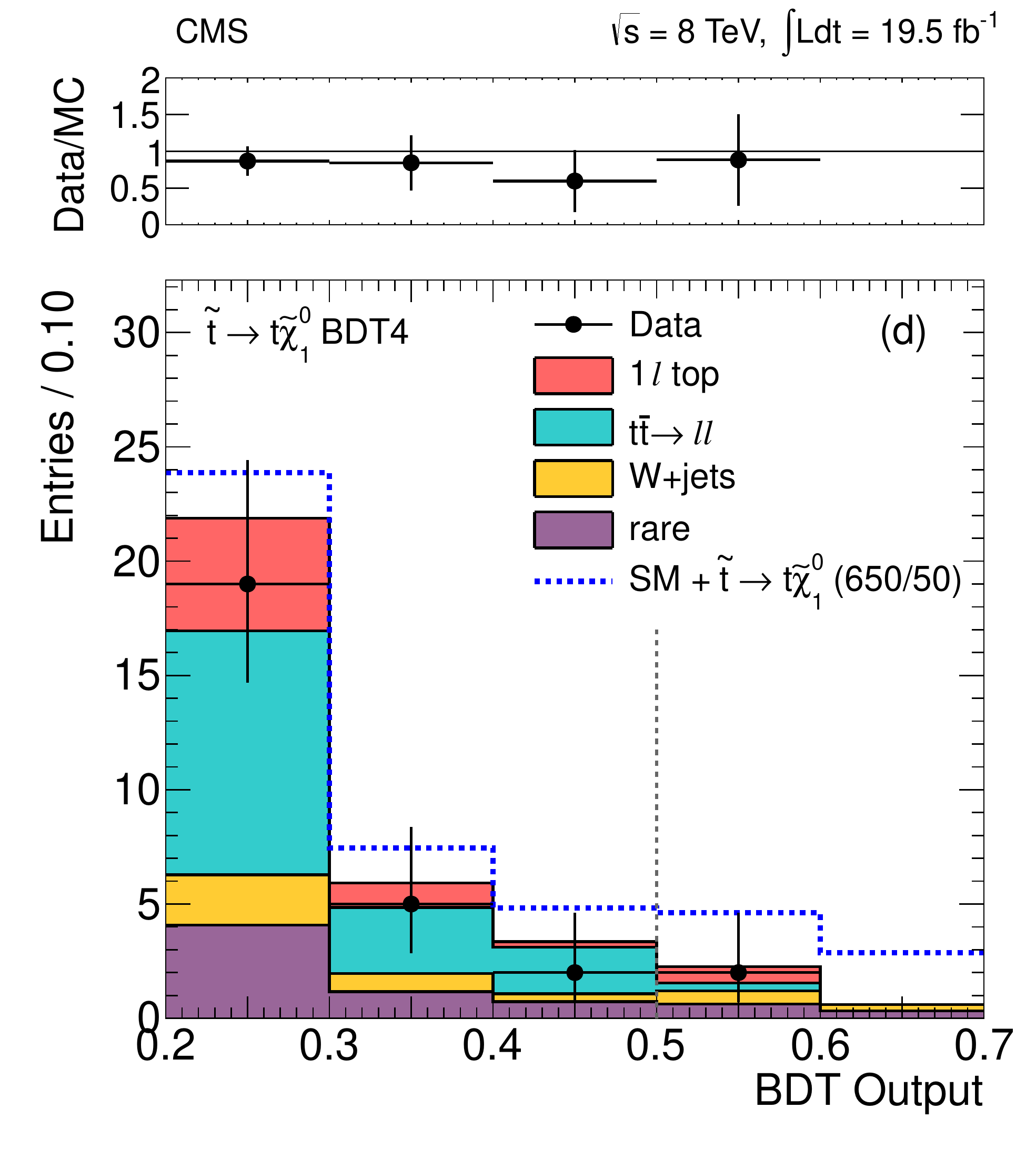}%
    \caption{
      Comparison of data and MC simulation for the distributions of BDT output and \MT\ corresponding
      to the tightest and loosest signal region selections in the \Ttt\ scenario.  The \MT\ distributions
      are shown after the requirement on the BDT output, and the BDT output distributions are shown after the
      $\MT> 120\GeV$ requirement (these requirements are also indicated by vertical dashed lines
      on the respective distributions).
      (a) \MT\ after the loose cut on the BDT1 output; 
      (b) \MT\ after the cut on the BDT4 output;
      (c) BDT1 output after the \MT\ cut;
      (d) BDT4 output after the \MT\ cut.
      Expected signal distributions for $m_{\lsp}=50$\GeV and $m_{\PSQt}=250$\GeV or 650 GeV
      are also overlayed, as indicated in the figures. 
      In plot (b), the bin to the right of the vertical line
      contains all events with $\MT > 120$\GeV, and has been scaled
      by a factor of 1/3 to indicate the number of events per 60\GeV.
      In all distributions the last bin contains the overflow.
      \label{fig:mt_t2tt}
    }
      \end{center}
\end{figure*}

A summary of the background expectations
and the corresponding data counts
for each signal region is shown in
Table~\ref{tab:result_t2tt_bdt} for the \Ttt\ BDT analysis,
Table~\ref{tab:result_t2tt_cnc} for the \Ttt\ cut-based analysis,
Table~\ref{tab:result_t2bw_bdt} for the \TbW\ BDT analysis,
and Table~\ref{tab:result_t2bw_cnc} for the \TbW\ cut-based analysis.
Figure~\ref{fig:mt_t2tt} presents a comparison of data with MC
simulation for the \MT\ and BDT-output distributions of events that
satisfy a loose and a tight \Ttt\ BDT signal-region requirement.
Equivalent plots for \TbW\ are shown in Fig.~\ref{fig:mt_t2bw}.
The \MT\ and BDT output distributions for the other signal regions are presented in Appendix~\ref{app:plots}.

\begin{table*}[htb]
\centering
\topcaption{The result of the $\Ttt$ BDT analysis. For each signal region the individual background contributions, total
background, and observed yields are indicated. The uncertainty
includes both the statistical and systematic components.
The expected yields for two example signal models are also indicated (statistical uncertainties only).
The first and second numbers in parentheses indicate the top-squark and neutralino
masses, respectively, in \GeVns{}.
}
\label{tab:result_t2tt_bdt}
{\footnotesize
\begin{tabular}{l|cccccc}
\multicolumn{7}{c}{$\Ttt$} \\
\hline
 Sample              & BDT1--Loose & BDT1--Tight & BDT2 & BDT3 & BDT4 & BDT5\\
\hline
\ttdl\ 	 	 & $438 \pm 37$ 	 & $68 \pm 11$ 	 & $46 \pm 10$ 	 & $5 \pm 2$ 	 & $0.3 \pm 0.3$ 	 & $48 \pm 13$ \\
$1\ell$ top 	 	 & $251 \pm 93$ 	 & $37 \pm 17$ 	 & $22 \pm 12$ 	 & $4 \pm 3$ 	 & $0.8 \pm 0.9$ 	 & $30 \pm 12$ \\
\wjets\  	 & $27 \pm 7$ 	 & $7 \pm 2$ 	 & $6 \pm 2$ 	 & $2 \pm 1$ 	 & $0.8 \pm 0.3$ 	 & $5 \pm 2$ \\
Rare 	 	 & $47 \pm 23$ 	 & $11 \pm 6$ 	 & $10 \pm 5$ 	 & $3 \pm 1$ 	 & $1.0 \pm 0.5$ 	 & $4 \pm 2$ \\
\hline
Total 	 	 & $763 \pm 102$ 	 & $124 \pm 21$ 	 & $85 \pm 16$ 	 & $13 \pm 4$ 	 & $2.9 \pm 1.1$ 	 & $87 \pm 18$ \\
\hline
Data 		 & $728$& $104$& $56$& $8$& $2$& $76$ \\
\hline
\Ttt\ (250/50)    & $285 \pm 8.5$    &   $50 \pm 3.5$   &   $28 \pm 2.6$   &   $4.4 \pm 1.0$   &   $0.3 \pm 0.3$   &   $34 \pm 2.9$   \\
\Ttt\ (650/50)    & $12 \pm 0.2$   &   $7.2 \pm 0.2$   &   $9.8 \pm 0.2$   &   $6.5 \pm 0.2$   &   $4.3 \pm 0.1$   &   $2.9 \pm 0.1$   \\
\hline
\end{tabular}}
\end{table*}

The observed and predicted yields agree in all signal regions within
about 1.0--1.5 standard deviations.
Therefore, we observe no evidence for top-squark pair production.
We note that there is a tendency for the background predictions to lie somewhat above the observed yields;
however,
the yields and background predictions in different signal regions are
correlated, both for the BDT and cut-based analysis.
The interpretation of the results in the context of models of top-squark pair production
is presented in Section~\ref{sec:interpretation}.

\begin{table*}[htb]
\centering
\topcaption{The result of the $\Ttt$ cut-based analysis. For each signal region the individual background contributions, total
background, and observed yields are indicated. The uncertainty
includes both the statistical and systematic components.
The expected yields for two example signal models are also indicated (statistical uncertainties only).
The first and second numbers in parentheses indicate the top-squark and neutralino
masses, respectively, in \GeVns.
}
{\footnotesize
\begin{tabular}{l|cccc}
 \multicolumn{1}{c}{Sample} 	 	 & $\MET>150$\GeV	 & $\MET>200$\GeV	 & $\MET>250$\GeV & $\MET>300$\GeV\\
\hline \multicolumn{5}{c}{}\\[-0.5ex]
\multicolumn{5}{c}{Low $\Delta M$ Selection} \\
\hline
\ttdl\ 	 	 & $131 \pm 15$ 	 & $42 \pm 7$ 	 & $17 \pm 5$ 	 & $5.6 \pm 2.5$  \\
$1\ell$ top 	 	 & $94 \pm 47$ 	 & $30 \pm 19$ 	 & $9 \pm 6$ 	 & $3.1 \pm 2.4$  \\
\wjets\  	 & $10 \pm 3$ 	 & $5 \pm 1$ 	 & $2 \pm 1$ 	 & $1.0 \pm 0.4$  \\
Rare 	 	 & $16 \pm 8$ 	 & $7 \pm 4$ 	 & $4 \pm 2$ 	 & $1.8 \pm 0.9$  \\
\hline
Total 	 	 & $251 \pm 50$ 	 & $83 \pm 21$ 	 & $31 \pm 8$ 	 & $11.5 \pm 3.6$  \\
\hline
Data 		 & $227$& $69$& $21$& $9$ \\
\hline
\Ttt\ (250/50)    & $108 \pm 3.7$   &   $32 \pm 2.0$   &   $12 \pm 1.2$   &   $5.2 \pm 0.8$  \\
\Ttt\ (650/50)    & $8.0 \pm 0.1$   &   $7.2 \pm 0.1$   &   $6.2 \pm 0.1$   &   $4.9 \pm 0.1$ \\
\hline \multicolumn{5}{c}{}\\[-0.5ex]
\multicolumn{5}{c}{High $\Delta M$ Selection} \\
\hline
\ttdl\ 	 	 & $8 \pm 2$ 	 & $5 \pm 2$ 	 & $3.2 \pm 1.4$ 	 & $1.4 \pm 0.9$ \\
$1\ell$ top 	 	 & $13 \pm 6$ 	 & $6 \pm 4$ 	 & $3.0 \pm 2.2$ 	 & $1.4 \pm 1.0$ \\
\wjets\  	 & $4 \pm 1$ 	 & $2 \pm 1$ 	 & $1.5 \pm 0.5$ 	 & $0.9 \pm 0.3$ \\
Rare 	 	 & $4 \pm 2$ 	 & $3 \pm 1$ 	 & $1.8 \pm 0.9$ 	 & $1.0 \pm 0.5$ \\
\hline
Total 	 	 & $29 \pm 7$ 	 & $17 \pm 5$ 	 & $9.5 \pm 2.8$ 	 & $4.7 \pm 1.4$ \\
\hline
Data 		 & $23$& $11$& $3$& $2$ \\
\hline
\Ttt\ (250/50)    &  $10 \pm 1.1$   &   $4.6 \pm 0.8$   &   $2.3 \pm 0.5$   &   $1.4 \pm 0.4$  \\
\Ttt\ (650/50)    & $4.9 \pm 0.1$   &   $4.7 \pm 0.1$   &   $4.3 \pm 0.1$   &   $3.7 \pm 0.1$  \\
\hline

\end{tabular}}
\label{tab:result_t2tt_cnc}
\end{table*}

\begin{table*}[htb]
\centering
\topcaption{The result of the $\TbW$ BDT analysis. For each signal region the individual background contributions, total
background, and observed yields are indicated. The uncertainty
includes both the statistical and systematic components.
The expected yields for several example signal models are also
indicated (statistical uncertainties only). The first number in
parentheses indicates the top-squark mass, the second the
gluino mass, and the third the chargino mass parameter $x$.
The units of the two mass values are \GeVns.
}
\label{tab:result_t2bw_bdt}
{\footnotesize
\begin{tabular}{l| cccccccccc cccccccccc cccccccccc cccccccccc cccccccccc cccccccccc}
\multicolumn{61}{c}{$\TbW$ $x=0.25$} \\
\hline
Sample              & \multicolumn{20}{c}{\phantom{0}\phantom{0}\phantom{0}\phantom{0}\phantom{0}\phantom{0}BDT1\phantom{0}\phantom{0}\phantom{0}\phantom{0}\phantom{0}\phantom{0}} & \multicolumn{20}{c}{\phantom{0}\phantom{0}\phantom{0}\phantom{0}\phantom{0}\phantom{0}BDT2\phantom{0}\phantom{0}\phantom{0}\phantom{0}\phantom{0}\phantom{0}} & \multicolumn{20}{c}{\phantom{0}\phantom{0}\phantom{0}\phantom{0}\phantom{0}BDT3\phantom{0}\phantom{0}\phantom{0}\phantom{0}\phantom{0}} \\
\hline
\ttdl\ 	 	 & \multicolumn{20}{c}{$18 \pm 4$}  	 & \multicolumn{20}{c}{$2.2 \pm 1.3$}  	 & \multicolumn{20}{c}{$1.2 \pm 1.0$} \\
$1\ell$ top 	 & \multicolumn{20}{c}{$10 \pm 5$}  	 & \multicolumn{20}{c}{$4.0 \pm 1.8$}  	 & \multicolumn{20}{c}{$1.5 \pm 0.8$}  \\
\wjets\  	         & \multicolumn{20}{c}{$3 \pm 1$}           & \multicolumn{20}{c}{$2.0 \pm 0.7$}  	 & \multicolumn{20}{c}{$0.7 \pm 0.3$}  \\
Rare 	 	 & \multicolumn{20}{c}{$4 \pm 2$}  	 & \multicolumn{20}{c}{$1.6 \pm 0.8$}  	 & \multicolumn{20}{c}{$1.0 \pm 0.5$}  \\
\hline
Total 	 	 & \multicolumn{20}{c}{$35 \pm 6$}  	 & \multicolumn{20}{c}{$9.8 \pm 2.4$}  	 & \multicolumn{20}{c}{$4.4 \pm 1.4$}  \\
\hline
Data 		 & \multicolumn{20}{c}{$29$}  	 & \multicolumn{20}{c}{$7$}  	 & \multicolumn{20}{c}{$2$}  \\
\hline
\TbW\ (450/50/0.25)   & \multicolumn{20}{c}{$19 \pm 2.9$}  	 & \multicolumn{20}{c}{$11 \pm 2.2$}  	 & \multicolumn{20}{c}{$5.2 \pm 1.5$}  \\
\TbW\ (600/100/0.25) & \multicolumn{20}{c}{$8.8 \pm 0.8$}  	 & \multicolumn{20}{c}{$7.5 \pm 0.8$}  	 & \multicolumn{20}{c}{$5.6 \pm 0.7$}  \\
\hline \multicolumn{61}{c}{}\\[-0.5ex]
\multicolumn{61}{c}{$\TbW$ $x=0.5$} \\
\hline
Sample &\multicolumn{12}{c}{BDT1}  & \multicolumn{12}{c}{BDT2--Loose}& \multicolumn{12}{c}{BDT2--Tight} & \multicolumn{12}{c}{BDT3} & \multicolumn{12}{c}{BDT4} \\
\hline
\ttdl\ 	 	         & \multicolumn{12}{c}{$40 \pm 5$} 	 & \multicolumn{12}{c}{$21 \pm 4$} 	 & \multicolumn{12}{c}{$4 \pm 2$} 	 & \multicolumn{12}{c}{$6 \pm 2$} & \multicolumn{12}{c}{$100 \pm 16$ } \\
$1\ell$ top 	 	 & \multicolumn{12}{c}{$24 \pm 10$} 	 & \multicolumn{12}{c}{$15 \pm 7$} 	 & \multicolumn{12}{c}{$4 \pm 3$} 	 & \multicolumn{12}{c}{$4 \pm 2$} & \multicolumn{12}{c}{$33 \pm 12$} \\
\wjets\  	                 & \multicolumn{12}{c}{$5 \pm 1$} 	 & \multicolumn{12}{c}{$5 \pm 1$} 	 & \multicolumn{12}{c}{$2 \pm 1$} 	 & \multicolumn{12}{c}{$3 \pm 1$} & \multicolumn{12}{c}{$5 \pm 1$} \\
Rare 	 	         & \multicolumn{12}{c}{$8 \pm 4$} 	 & \multicolumn{12}{c}{$8 \pm 4$} 	 & \multicolumn{12}{c}{$3 \pm 1$} 	 & \multicolumn{12}{c}{$4 \pm 2$} & \multicolumn{12}{c}{$8 \pm 4$} \\
\hline
Total 	 	         & \multicolumn{12}{c}{$77 \pm 12$} 	 & \multicolumn{12}{c}{$50 \pm 9$} 	 & \multicolumn{12}{c}{$13 \pm 4$} 	 & \multicolumn{12}{c}{$17 \pm 4$} & \multicolumn{12}{c}{$146 \pm 21$} \\
\hline
Data 		         & \multicolumn{12}{c}{$67$} & \multicolumn{12}{c}{$35$}& \multicolumn{12}{c}{$12$}& \multicolumn{12}{c}{$13$} & \multicolumn{12}{c}{$143$} \\
\hline
\TbW\ (250/50/0.5) & \multicolumn{12}{c}{$45 \pm 7.6$}       &   \multicolumn{12}{c}{$24 \pm 5.2$}   &   \multicolumn{12}{c}{$5.7 \pm 2.4$}   &   \multicolumn{12}{c}{$5.2 \pm 2.6$} &   \multicolumn{12}{c}{$55 \pm 8.1$}\\
\TbW\ (650/50/0.5) &  \multicolumn{12}{c}{$3.5 \pm 0.4$}    &   \multicolumn{12}{c}{$9.5 \pm 0.7$}   &   \multicolumn{12}{c}{$5.6 \pm 0.5$}   &   \multicolumn{12}{c}{$8.3 \pm 0.6$} &   \multicolumn{12}{c}{$3.2 \pm 0.4$}  \\
\hline \multicolumn{61}{c}{}\\[-0.5ex]
\multicolumn{61}{c}{$\TbW$ $x=0.75$} \\
\hline
Sample &\multicolumn{15}{c}{\phantom{0}\phantom{0}\phantom{0}BDT1\phantom{0}\phantom{0}\phantom{0}}  & \multicolumn{15}{c}{\phantom{0}\phantom{0}\phantom{0}BDT2\phantom{0}\phantom{0}\phantom{0}} & \multicolumn{15}{c}{\phantom{0}\phantom{0}\phantom{0}BDT3\phantom{0}\phantom{0}\phantom{0}} & \multicolumn{15}{c}{\phantom{0}\phantom{0}BDT4\phantom{0}\phantom{0}} \\
\hline
\ttdl\ 	 	  & \multicolumn{15}{c}{$37 \pm 5$} 	 & \multicolumn{15}{c}{$9 \pm 2$} 	 & \multicolumn{15}{c}{$3.1 \pm 1.3$} 	 & \multicolumn{15}{c}{$248 \pm 22$} \\
$1\ell$ top       & \multicolumn{15}{c}{$17 \pm 9$} 	 & \multicolumn{15}{c}{$6 \pm 5$} 	 & \multicolumn{15}{c}{$1.6 \pm 1.6$} 	 & \multicolumn{15}{c}{$188 \pm 70$} \\
\wjets\  	         & \multicolumn{15}{c}{$4 \pm 1$}          & \multicolumn{15}{c}{$4 \pm 1$} 	 & \multicolumn{15}{c}{$1.6 \pm 0.6$} 	 & \multicolumn{15}{c}{$22 \pm 6$} \\
Rare 	 	 & \multicolumn{15}{c}{$4 \pm 2$} 	 & \multicolumn{15}{c}{$4 \pm 2$} 	 & \multicolumn{15}{c}{$1.8 \pm 0.9$} 	 & \multicolumn{15}{c}{$20 \pm 10$} \\
\hline
Total 	 	 & \multicolumn{15}{c}{$61 \pm 10$} 	 & \multicolumn{15}{c}{$22 \pm 6$} 	 & \multicolumn{15}{c}{$8.1 \pm 2.3$} 	 & \multicolumn{15}{c}{$478 \pm 74$} \\
\hline
Data 		 & \multicolumn{15}{c}{$50$} &\multicolumn{15}{c}{$13$} & \multicolumn{15}{c}{$5$} & \multicolumn{15}{c}{$440$} \\
\hline
\TbW\ (250/50/0.75) & \multicolumn{15}{c}{$115 \pm 13$}     &   \multicolumn{15}{c}{$21 \pm 5.6$}          &       \multicolumn{15}{c}{$8.0 \pm 3.7$}      &   \multicolumn{15}{c}{$518 \pm 28$}  \\
\TbW\ (650/50/0.75) & \multicolumn{15}{c}{$3.9 \pm 0.4$}   &   \multicolumn{15}{c}{$8.4 \pm 0.6$}   &   \multicolumn{15}{c}{$6.8 \pm 0.6$}   &   \multicolumn{15}{c}{$5.5 \pm 0.5$} \\
\hline
\end{tabular}}
\end{table*}

\begin{table*}[htb]
\centering
\topcaption{The result of the $\TbW$ cut-based analysis. For each signal region the individual background contributions, total
background, and observed yields are indicated. The uncertainty
includes both the statistical and systematic components.
The expected yields for several sample signal models are also
indicated (statistical uncertainties only). The first number in
parentheses indicates the top-squark mass, the second the
gluino mass, and the third the chargino mass parameter $x$.
The units of the two mass values are \GeVns.
}
\label{tab:result_t2bw_cnc}
{\footnotesize
\begin{tabular}{l|cccc}
\multicolumn{1}{c}{Sample} 	 	 & $\MET>100$\GeV	 & $\MET>150$\GeV	 & $\MET>200$\GeV & $\MET>250$\GeV\\
\hline\multicolumn{5}{c}{}\\[-0.5ex]
\multicolumn{5}{c}{Low $\Delta M$ Selection} \\
\hline
\ttdl\ 	 	 & $875 \pm 57$ 	 & $339 \pm 23$ 	 & $116 \pm 14$ 	 & $40 \pm 9$ \\
$1\ell$ top 	 	 & $658 \pm 192$ 	 & $145 \pm 70$ 	 & $41 \pm 24$ 	 & $14 \pm 9$ \\
\wjets\  	 & $59 \pm 15$ 	 & $21 \pm 5$ 	 & $8 \pm 2$ 	 & $4 \pm 1$ \\
Rare 	 	 & $70 \pm 35$ 	 & $33 \pm 17$ 	 & $16 \pm 8$ 	 & $8 \pm 4$ \\
\hline
Total 	 	 & $1662 \pm 203$ 	 & $537 \pm 75$ 	 & $180 \pm 28$ 	 & $66 \pm 13$ \\
\hline
Data 		 & $1624$& $487$& $151$& $52$ \\
\hline

\TbW\ (450/50/0.25)  & $47 \pm 3.3$   &    $33 \pm 2.7$   &   $19 \pm 2.0$   &   $8.7 \pm 1.4$  \\
\TbW\ (600/100/0.25) & $15 \pm 0.7$   &    $13 \pm 0.7$   &   $11 \pm 0.6$   &   $7.9 \pm 0.5$  \\

\TbW\ (250/50/0.5)   & $419 \pm 17$   &   $157 \pm 9.9$   &   $52 \pm 5.4$   &   $21 \pm 3.4$   \\
\TbW\ (650/50/0.5)   &  $14 \pm 0.6$  &    $13 \pm 0.5$   &   $11 \pm 0.5$   &   $8.4 \pm 0.4$  \\

\TbW\ (250/50/0.75)  & $854 \pm 26$   &   $399 \pm 18$    &   $144 \pm 10$   &   $56 \pm 6.4$   \\
\TbW\ (650/50/0.75)  &  $17 \pm 0.7$  &    $16 \pm 0.6$   &   $13 \pm 0.6$   &   $11 \pm 0.5$   \\

\hline\multicolumn{5}{c}{}\\[-0.5ex]
\multicolumn{5}{c}{High $\Delta M$ Selection} \\
\hline
\ttdl\ 	 	 & $25 \pm 5$ 	 & $12 \pm 3$ 	 & $7 \pm 2$ 	 & $2.9 \pm 1.5$ \\
$1\ell$ top 	 	 & $35 \pm 10$ 	 & $15 \pm 6$ 	 & $6 \pm 3$ 	 & $2.7 \pm 1.8$ \\
\wjets\  	 & $9 \pm 2$ 	 & $5 \pm 1$ 	 & $2 \pm 1$ 	 & $1.8 \pm 0.6$ \\
Rare 	 	 & $9 \pm 5$ 	 & $7 \pm 3$ 	 & $4 \pm 2$ 	 & $2.4 \pm 1.2$ \\
\hline
Total 	 	 & $79 \pm 12$ 	 & $38 \pm 7$ 	 & $19 \pm 5$ 	 & $9.9 \pm 2.7$ \\
\hline
Data 		 & $90$& $39$& $18$& $5$ \\
\hline

\TbW\ (450/50/0.25)   & $30 \pm 2.7$   &    $23 \pm 2.3$   &    $15 \pm 1.8$   &   $7.3 \pm 1.3$ \\
\TbW\ (600/100/0.25)  & $11 \pm 0.6$   &   $9.7 \pm 0.6$   &   $8.4 \pm 0.6$   &   $6.1 \pm 0.5$ \\

\TbW\ (250/50/0.5) & $37 \pm 4.8$    &   $23 \pm 3.8$    &    $11 \pm 2.6$   &   $5.0 \pm 1.7$ \\
\TbW\ (650/50/0.5) & $11 \pm 0.5$    &   $9.8 \pm 0.5$   &   $8.6 \pm 0.4$   &   $6.7 \pm 0.4$ \\

\TbW\ (250/50/0.75) &  $32 \pm 5.2$   &   $23 \pm 4.4$   &   $11 \pm 2.9$    &   $3.6 \pm 1.4$ \\
\TbW\ (650/50/0.75) & $9.2 \pm 0.5$   &  $8.4 \pm 0.5$   &   $7.5 \pm 0.4$   &   $6.3 \pm 0.4$ \\

\hline
\end{tabular}}
\end{table*}

\begin{figure*}[htbp]
  \begin{center}
    \includegraphics[width=0.49\linewidth]{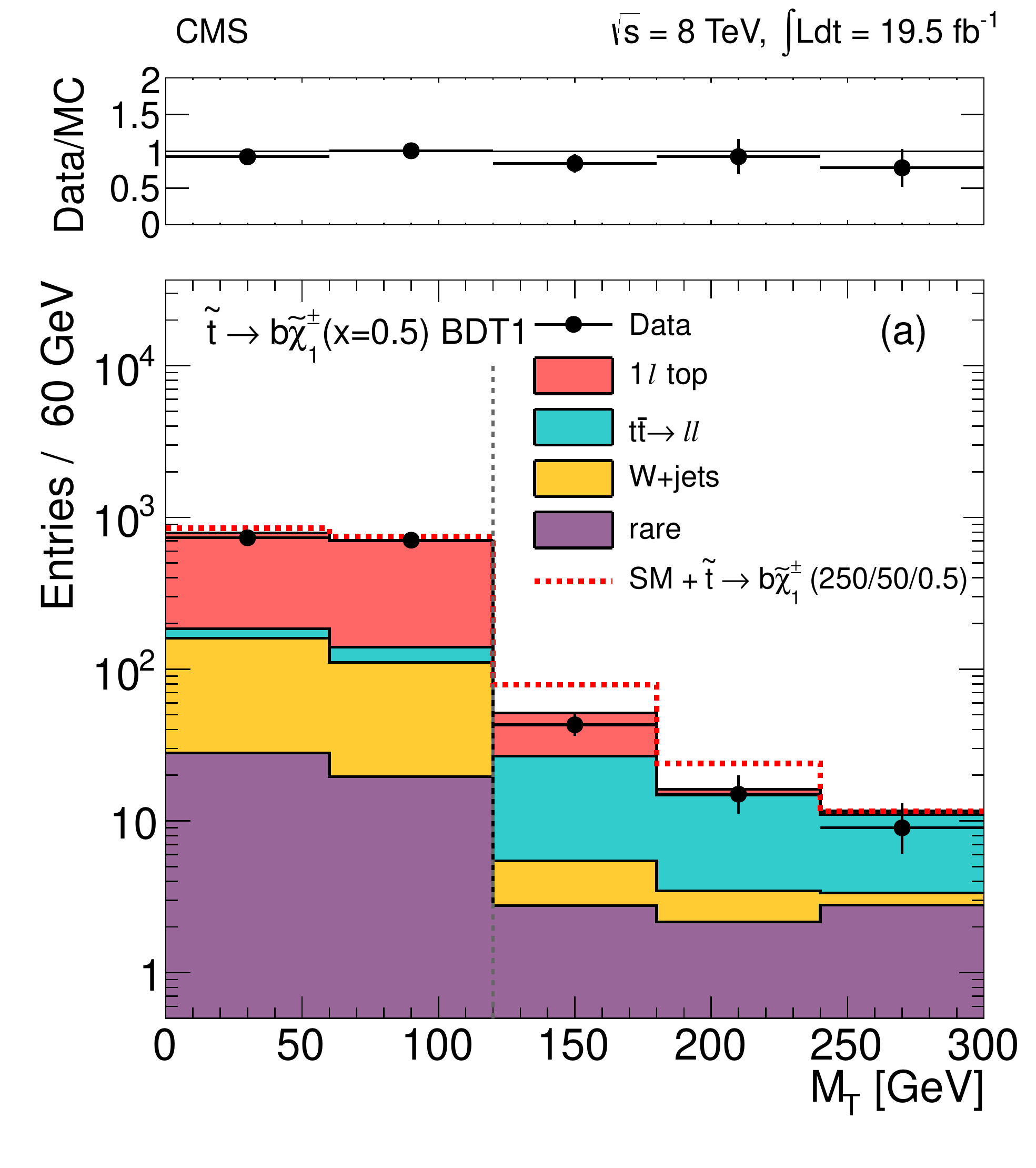}%
    \includegraphics[width=0.49\linewidth]{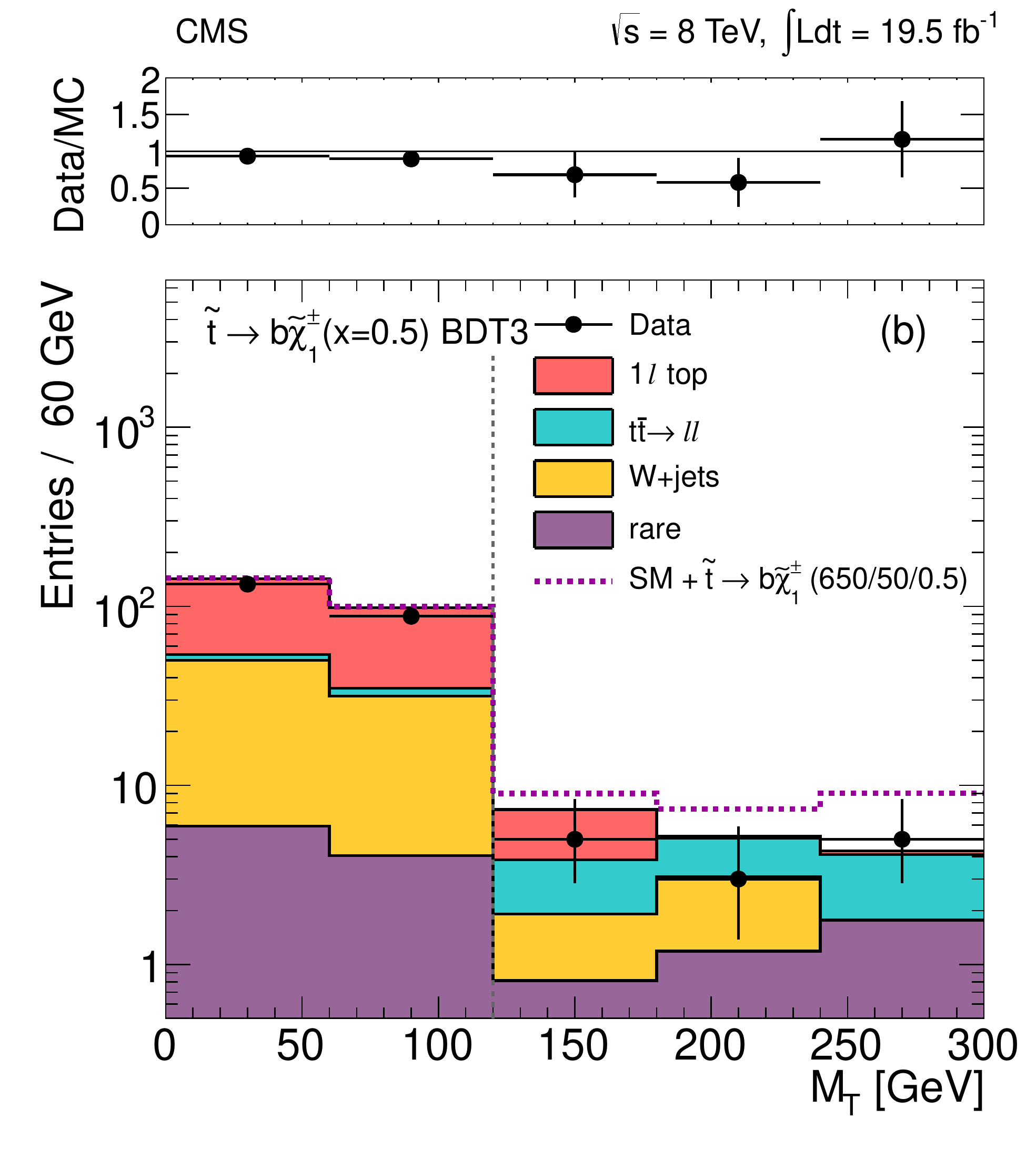} \\
    \includegraphics[width=0.49\linewidth]{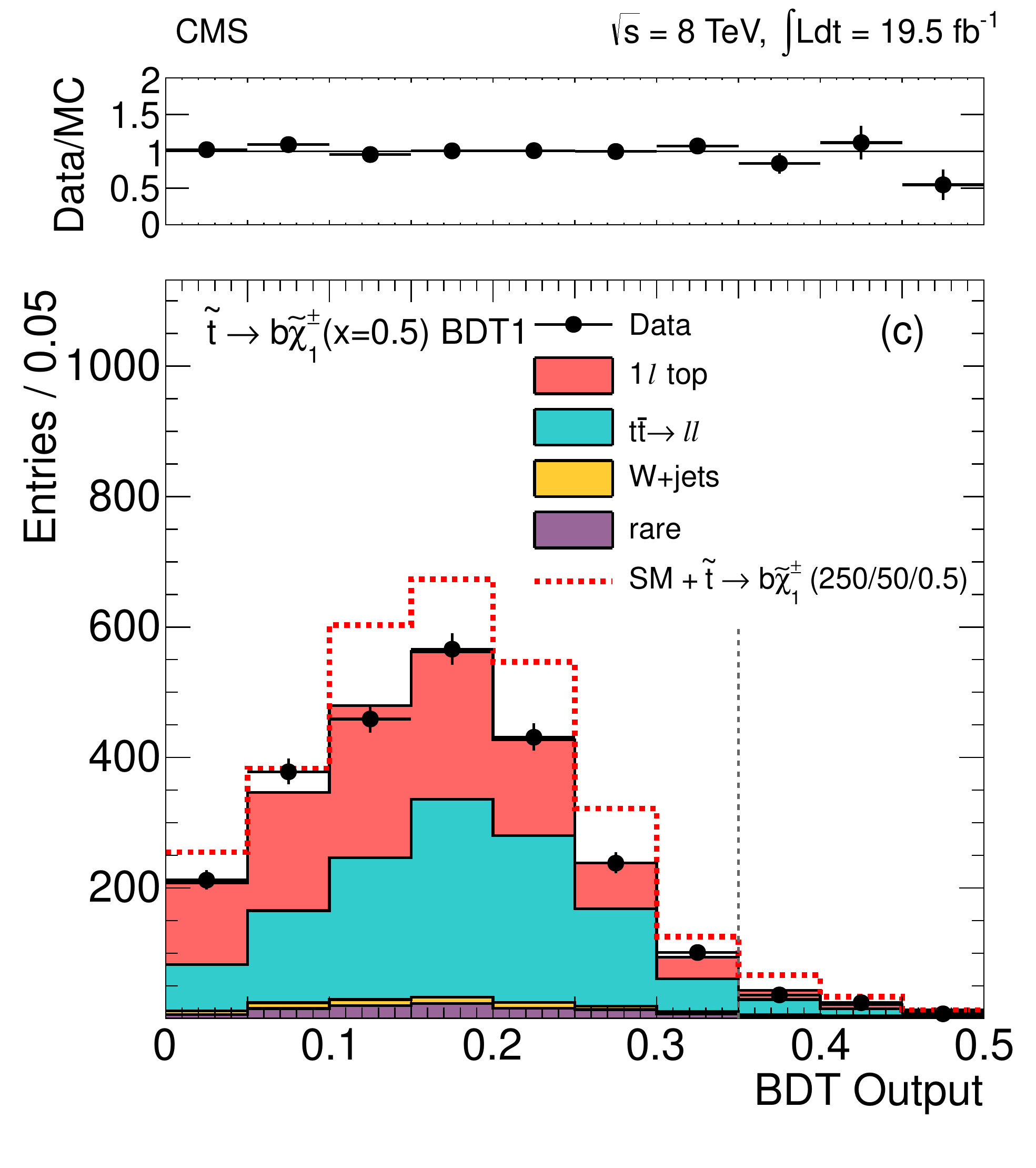}%
    \includegraphics[width=0.49\linewidth]{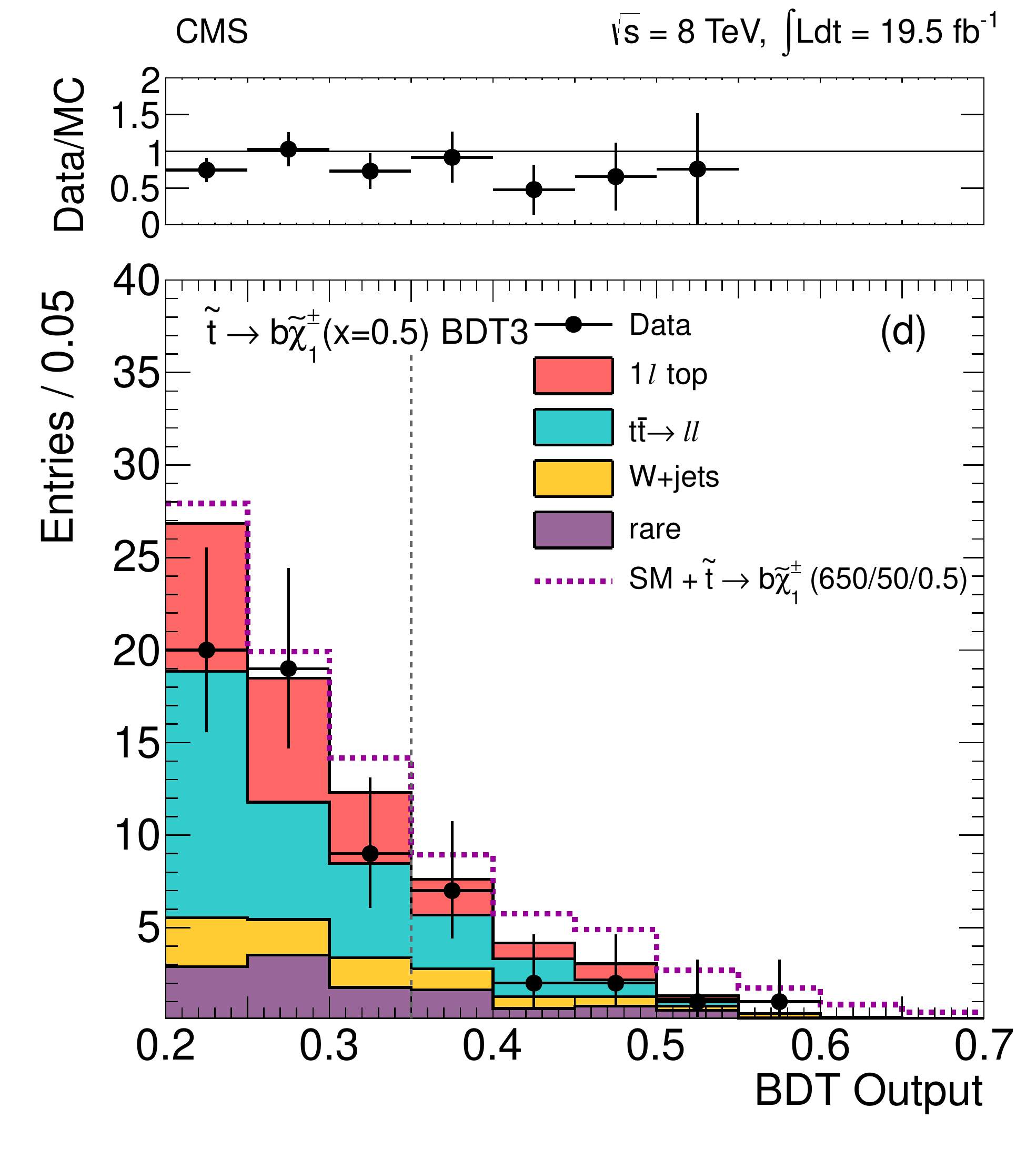}%
    \caption{
      Comparison of data and MC simulation for the distributions of BDT output and \MT\ corresponding
      to the tightest and loosest signal region selections in the $x=0.5$ \TbW\ scenario with an on-shell 
      $\PW$ boson.  The \MT\ distributions
      are shown after the requirement on the BDT output, and the BDT output distributions are shown after the
      $\MT> 120\GeV$ requirement (these requirements are also indicated by vertical dashed lines
      on the respective distributions).
      (a) \MT\ after the cut on the BDT1 output; 
      (b) \MT\ after the cut on the BDT3 output;
      (c) BDT1 output after the \MT\ cut;
      (d) BDT3 output after the \MT\ cut.
      Expected signal distributions for $x=0.5$ with $m_{\lsp}=50$\GeV and $m_{\PSQt}=250$\GeV or 650 GeV
      are also overlayed, as indicated in the figures. In all distributions the last bin contains the overflow. 
      \label{fig:mt_t2bw}
    }
      \end{center}
\end{figure*}

\section{Interpretation}
\label{sec:interpretation}

The results of the search are interpreted in the context of models of top-squark pair
production.
As discussed in Section~\ref{sec:mc}, we separately consider two possible decay
modes
of the top squark, \Ttt\ and $\TbW \to \cPqb \PW
\chiz_{1}$, each with $100\%$ branching fraction.
Using the results of Section~\ref{sec:results}, we compute 95\% confidence level (CL) cross section upper limits for top-squark
pair production in the $m_{\chiz_{1}}$ vs.\ $m_{\PSQt}$ parameter space.  Then,
based on the expected $\Pp\Pp \to \PSQt\PSQt^*$ production rate, these
cross section limits are used to exclude regions of SUSY parameter space.
For the \TbW\ scenario, the mass of the intermediate $\chipm_{1}$ is
specified by the parameter $x$ defined in Section~\ref{sec:mc}.

In setting limits,
we account for the following sources of systematic uncertainty
associated with the signal event acceptance and efficiency. The
uncertainty of the integrated luminosity determination is 4.4\%~\cite{LUMIPAS}.
Samples of $\Z \to \ell\ell$ events are used to measure the lepton
efficiencies, and the corresponding uncertainties are propagated
to the signal event acceptance and efficiency.  These uncertainties
are 3\% for the trigger efficiency
and a combined 5\% for the lepton identification and isolation efficiency,
where we also account for additional uncertainties in the modeling
of the lepton isolation due to the differences in the hadronic activity
in $\Z \to \ell\ell$ and SUSY events.
The uncertainty of the efficiency to tag bottom-quark jets results in
an uncertainty for the acceptance that depends on model details but is typically less than 1\%.
The energy scale of hadronic jets is known to 1--4\%, depending on
$\eta$ and \pt, yielding an uncertainty of 3--15\% for the signal event
selection efficiency.
The larger uncertainties correspond to models for which the difference
between the
masses of the top squark and LSP is small.

The experimental acceptance for signal events depends on the level
of ISR activity, especially in the small $\Delta$M region
where an initial-state boost may be required for an event to satisfy the selection requirements, including those on \MET, \MT, and the number of reconstructed jets.
The modeling of ISR in \MADGRAPH is investigated by comparing the
predicted and measured \pt\ spectra of the system recoiling against
the ISR jets in \zjets, \ttbar, and $\PW\Z$ events.  Good agreement is observed at lower \pt, while the simulation is found to over predict the data by about 10\%
at a \pt\ value of 150\GeV, rising to 20\% for $\pt> 250$\GeV.  The predictions from the MC signal samples are weighted to account for this difference,
by a factor of 0.8--1.0, depending on the \pt\ of the system recoiling against the ISR jets, and the deviation of this weight from 1 is taken as a
systematic uncertainty. Further details are given in Appendix~\ref{app:isr}.

\begin{figure*}[tbhp]
\centering
\includegraphics[width=0.45\textwidth]{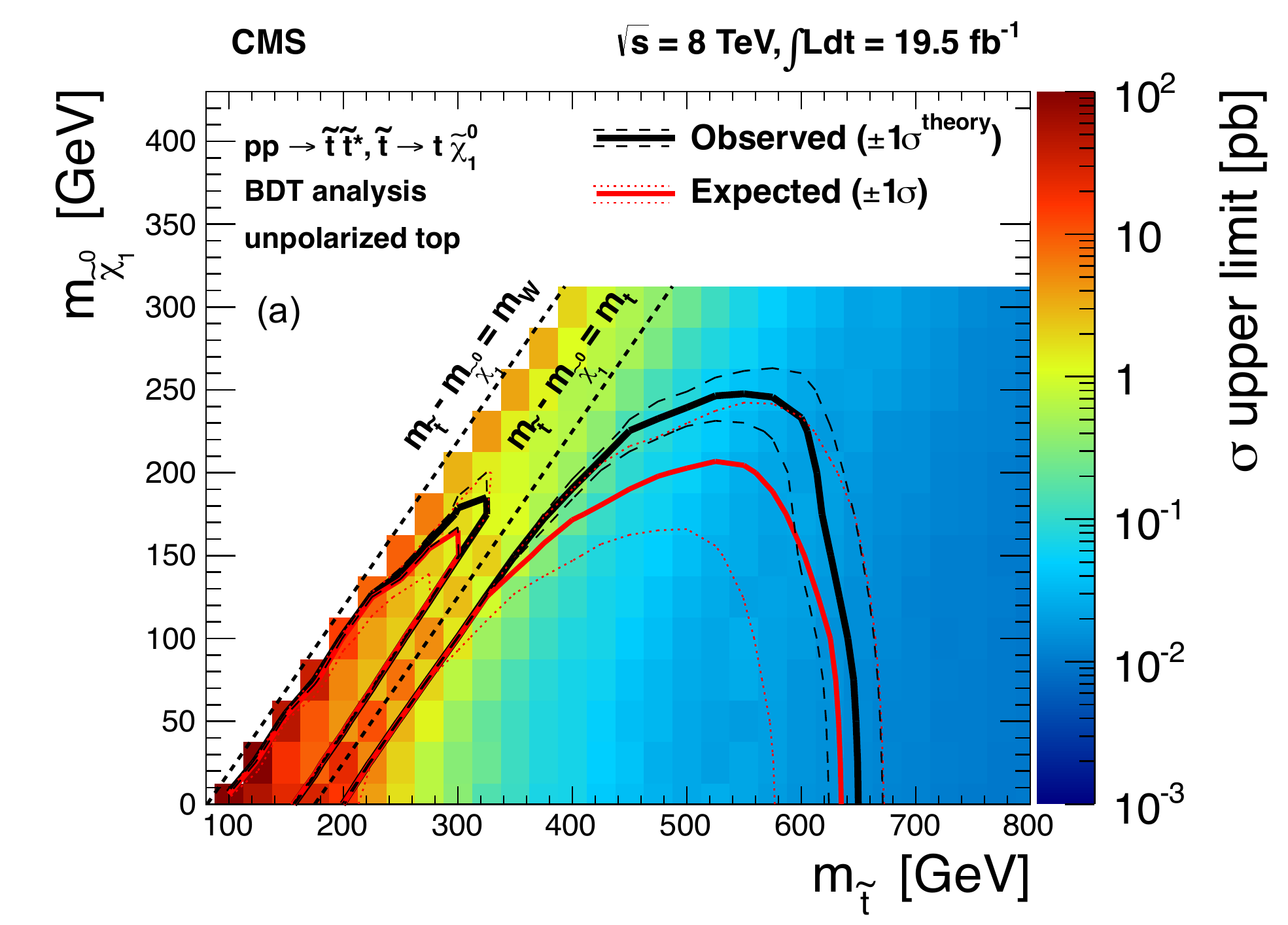} %
\includegraphics[width=0.45\textwidth]{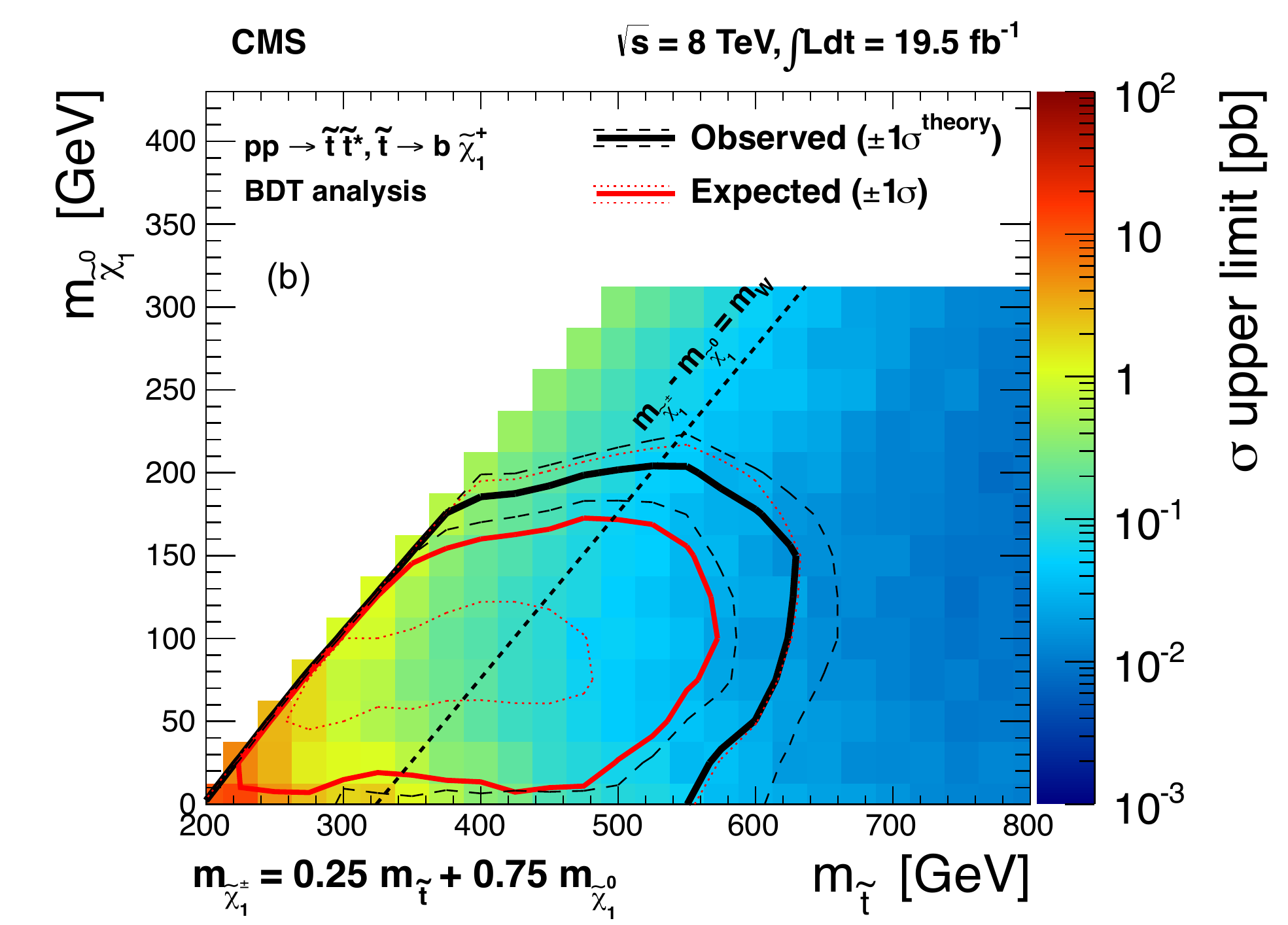}
\includegraphics[width=0.45\textwidth]{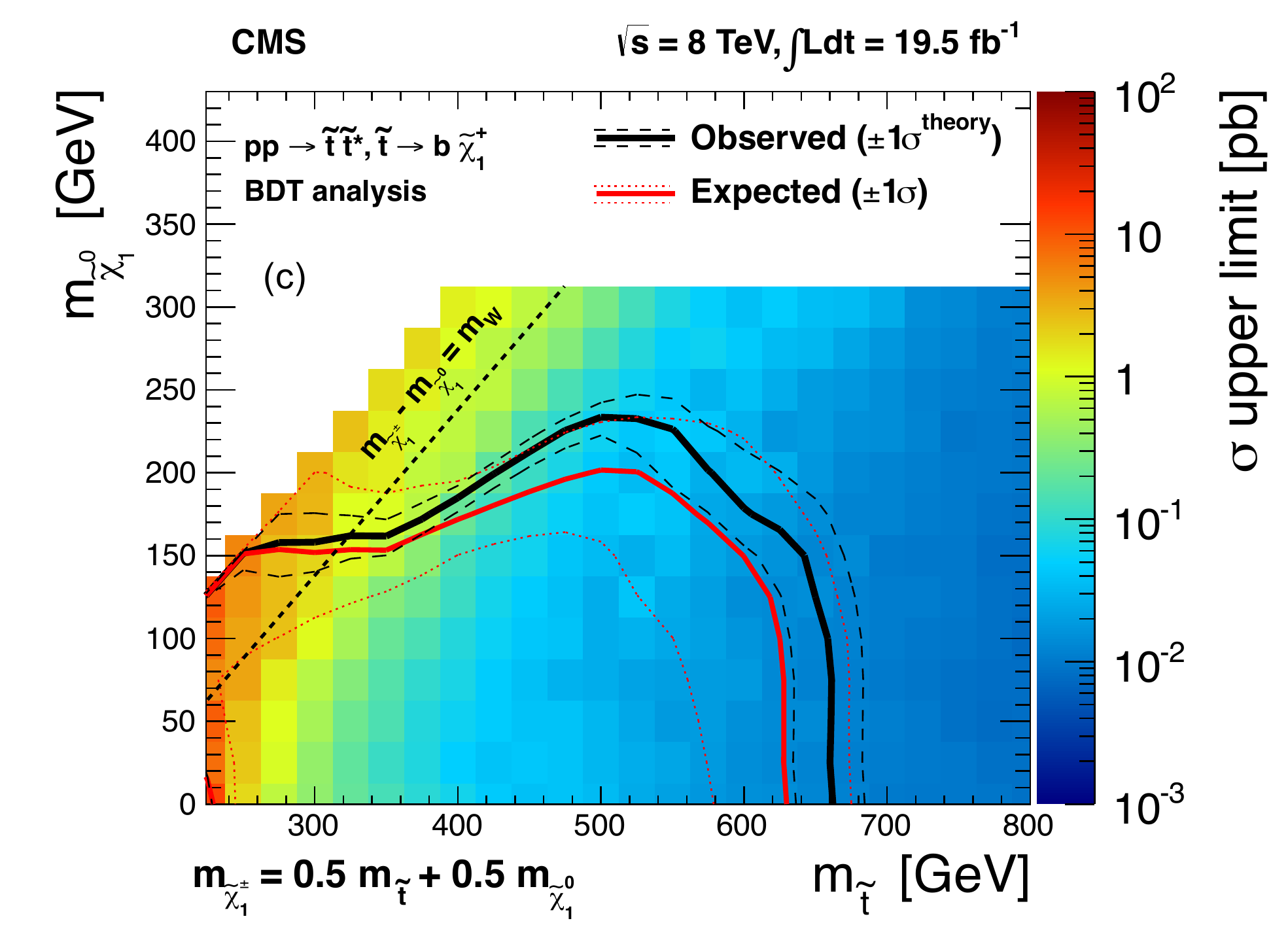} %
\includegraphics[width=0.45\textwidth]{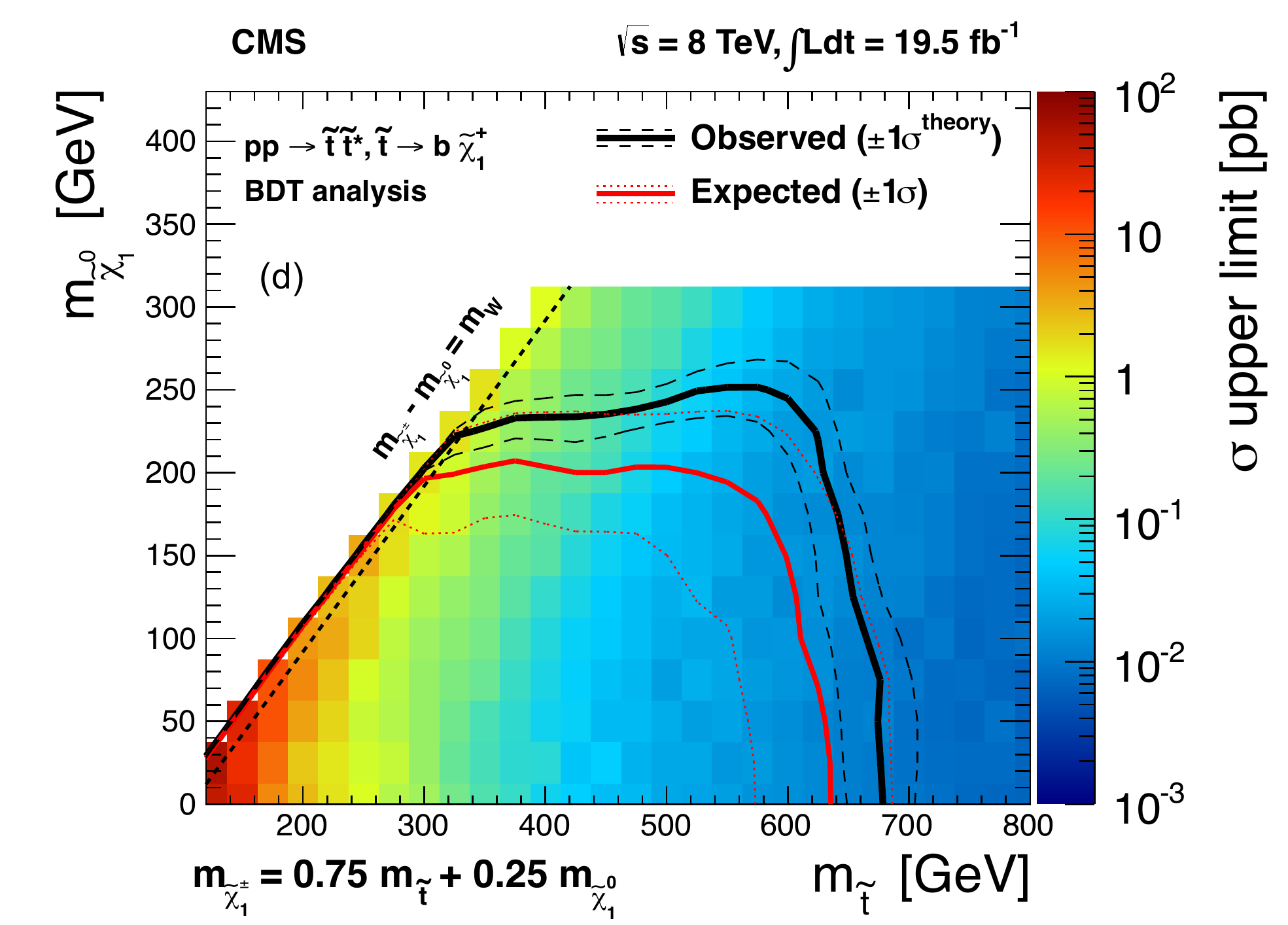}
\caption{
Interpretations using the primary results from the BDT method.
(a) \Ttt\ model; 
(b) \TbW\ model with $x=0.25$;
(c) \TbW\ model with $x=0.50$;
(d) \TbW\ model with $x=0.75$;
The color scale indicates the observed cross section upper limit.
The observed, median expected, and $\pm1$ standard deviation ($\sigma$)
expected 95\% CL exclusion contours are indicated.
The variations in the excluded region due to ${\pm}1\sigma$ uncertainty
of the theoretical prediction of the cross section for top-squark pair
production are also indicated.
\label{fig:bdt_interpretations}
}
\end{figure*}

Upper limits on the cross section for top-squark pair production are calculated separately for each SR, incorporating the
uncertainties of the acceptance and efficiency discussed above, using
the LHC-style CL$_\mathrm{s}$
criterion~\cite{Read:2002hq,Junk:1999kv,LHC-HCG}. For each point in the signal model parameter space, the observed limit is taken
from the signal region with the best expected limit. The results from the BDT analysis are displayed in
Fig.~\ref{fig:bdt_interpretations}.
The corresponding results from the cut-based analysis, and maps of
the most sensitive signal regions for each of the top-squark decay modes,
are presented in Appendix~\ref{app:models}.
The cross section limits from the BDT analysis improve upon those from the cut-based analysis by up to approximately 40\%,
depending on the model parameters.

Our results probe top squarks with masses between approximately 150 and 650\GeV,
for neutralinos with masses up to approximately 250\GeV, depending on the details of the model.
For the \Ttt\ search, the results are not sensitive to the model points with $m_{\PSQt}-m_{\lsp}=M_{\text{top}}$ because
the \lsp\ is produced at rest in the top-quark
rest frame. However the results are sensitive to scenarios with
$m_{\PSQt}-m_{\lsp}<M_{\text{top}}$
in which the top quark in the decay \Ttt\ is off-shell, including
regions of parameter space
with the top squark lighter than the top quark.

\begin{figure*}[htb]
\centering
\includegraphics[width=0.48\textwidth]{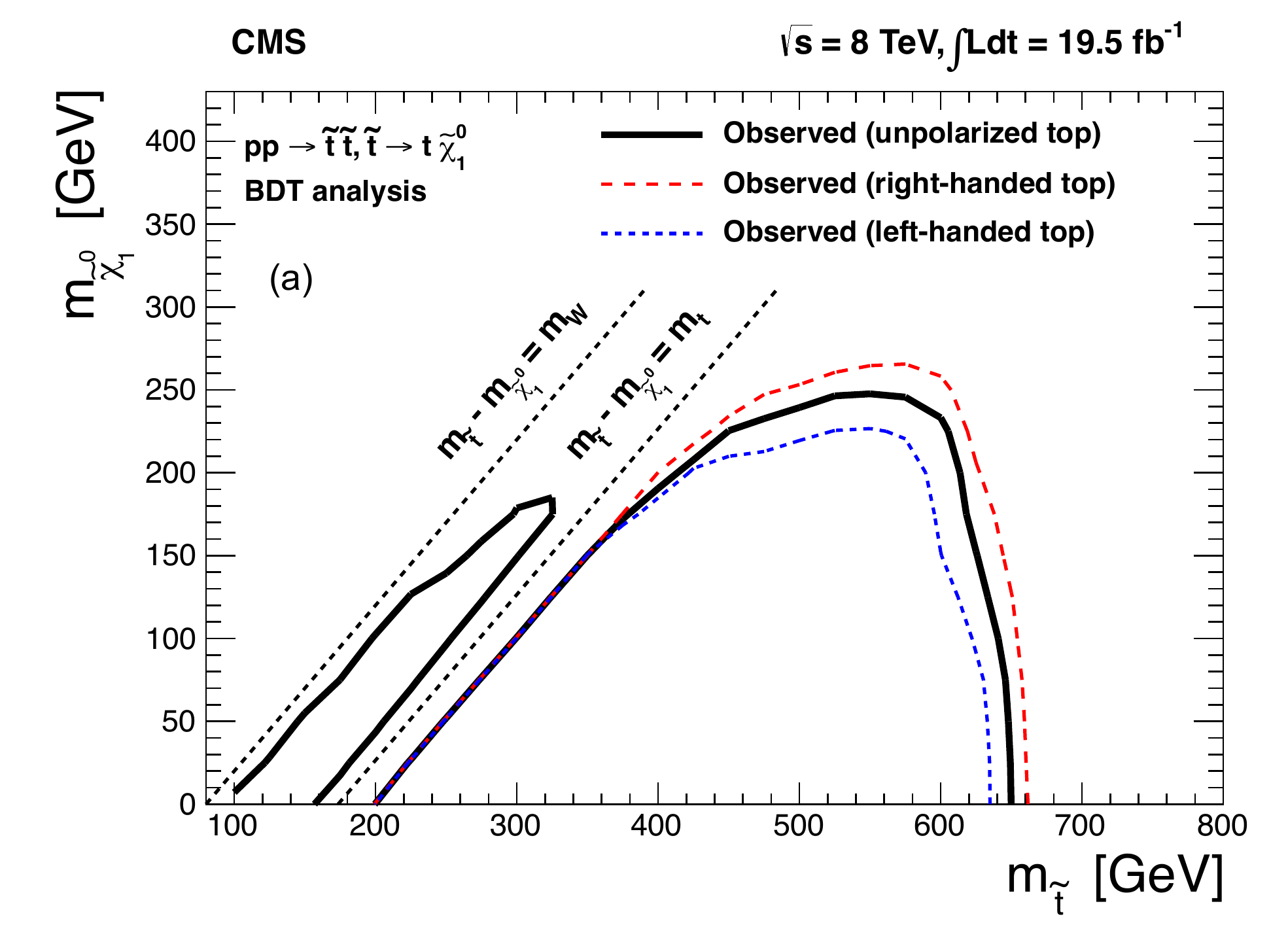}
\includegraphics[width=0.48\textwidth]{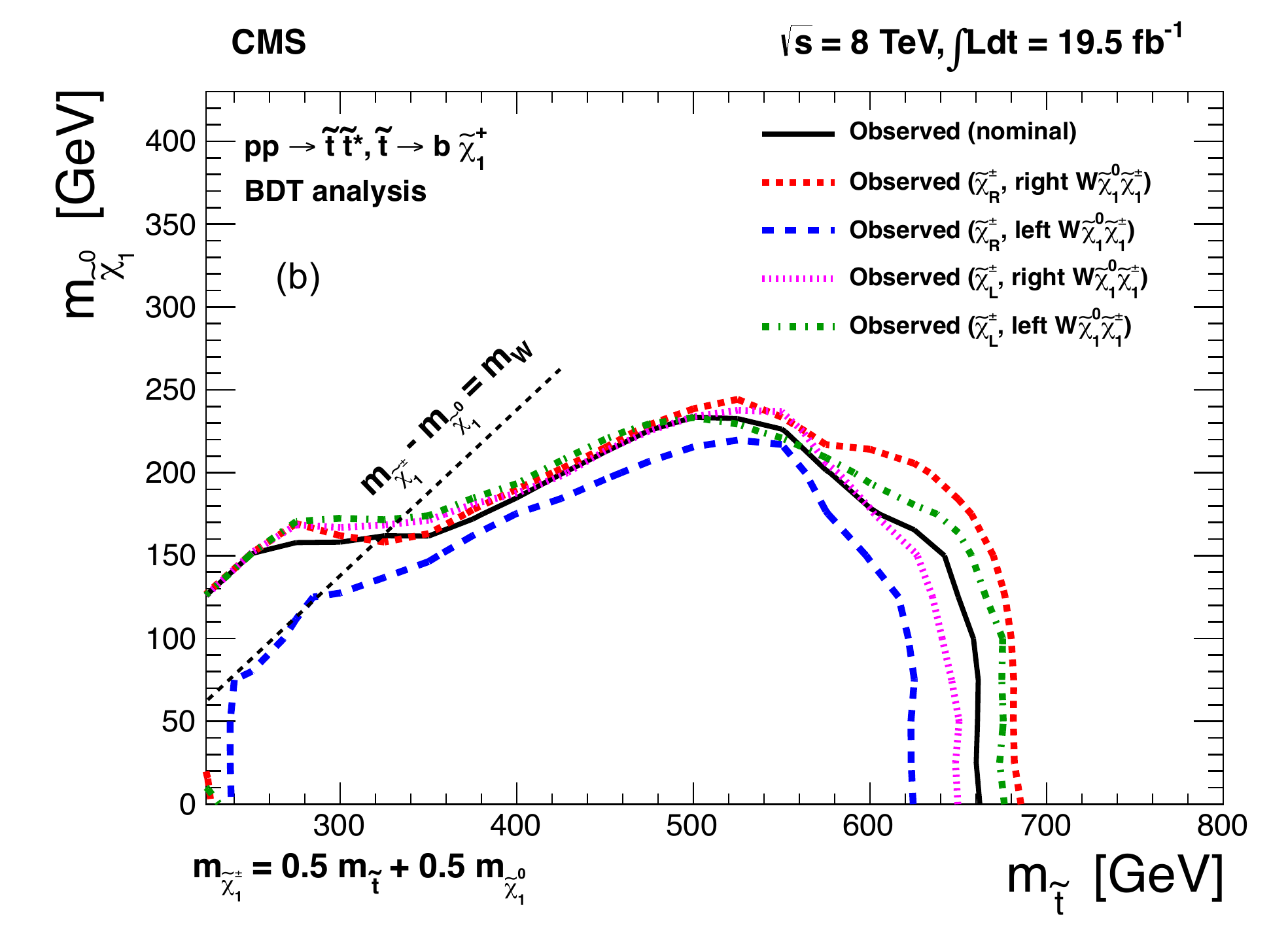}
\caption{
(a) the observed 95\% CL excluded regions for the \Ttt\ model for
the case of unpolarized, right-handed, and left-handed top quarks.
(b) the observed 95\% CL excluded regions for the \TbW\ model with $x=0.5$ for the nominal
scenario, right- vs. left-handed charginos ($\tilde{\chi}_{R}^{\pm}$ and $\tilde{\chi}_{L}^{\pm}$, respectively),
and right- vs. left-handed \PW\lsp\chipo\ couplings.
\label{fig:T2tt_polarization}
}
\end{figure*}

The acceptance depends on the polarization of
the top quarks in the \Ttt\ scenario, and on the polarization of
the charginos and \PW\ bosons in the \TbW\ scenario. These
polarizations depend on the left/right mixing of
the top squarks and on the mixing matrices of the neutralino and chargino~\cite{polarization1,polarization2}.
The exclusion regions obtained in the nominal \Ttt\ scenario with unpolarized top quarks are compared to those
obtained with pure left-handed and pure right-handed top quarks in Fig.~\ref{fig:T2tt_polarization} (left).
The limits on the top-squark and \lsp\ masses
vary by $\pm$10--20\GeV depending on the top-quark polarization.

In the \TbW\ scenario, the acceptance depends on the polarization of the chargino, and on whether the \PW\lsp\chipo\ coupling is
left-handed or right-handed.
In the nominal interpretations for the \TbW\ models presented in Fig.~\ref{fig:bdt_interpretations}, the signal events
are generated with an unpolarized chargino and a left/right-symmetric \PW\lsp\chipo\ coupling.
We have studied the dependence of our results on these assumptions.
We find that the scenarios in which the limits deviate the most from
the nominal result correspond to
right-handed charginos with either a right-handed \PW\lsp\chipo\ coupling (maximum sensitivity) or
a left-handed \PW\lsp\chipo\ coupling (minimum sensitivity).
This is shown for the \TbW\ $x=0.5$ model
in Fig.~\ref{fig:T2tt_polarization} (right).
The corresponding results for the $x=0.25$ and $0.75$ scenarios can be found in
Appendix~\ref{app:models}.

\begin{figure}[htb]
\centering
\includegraphics[width=\cmsFigWidth]{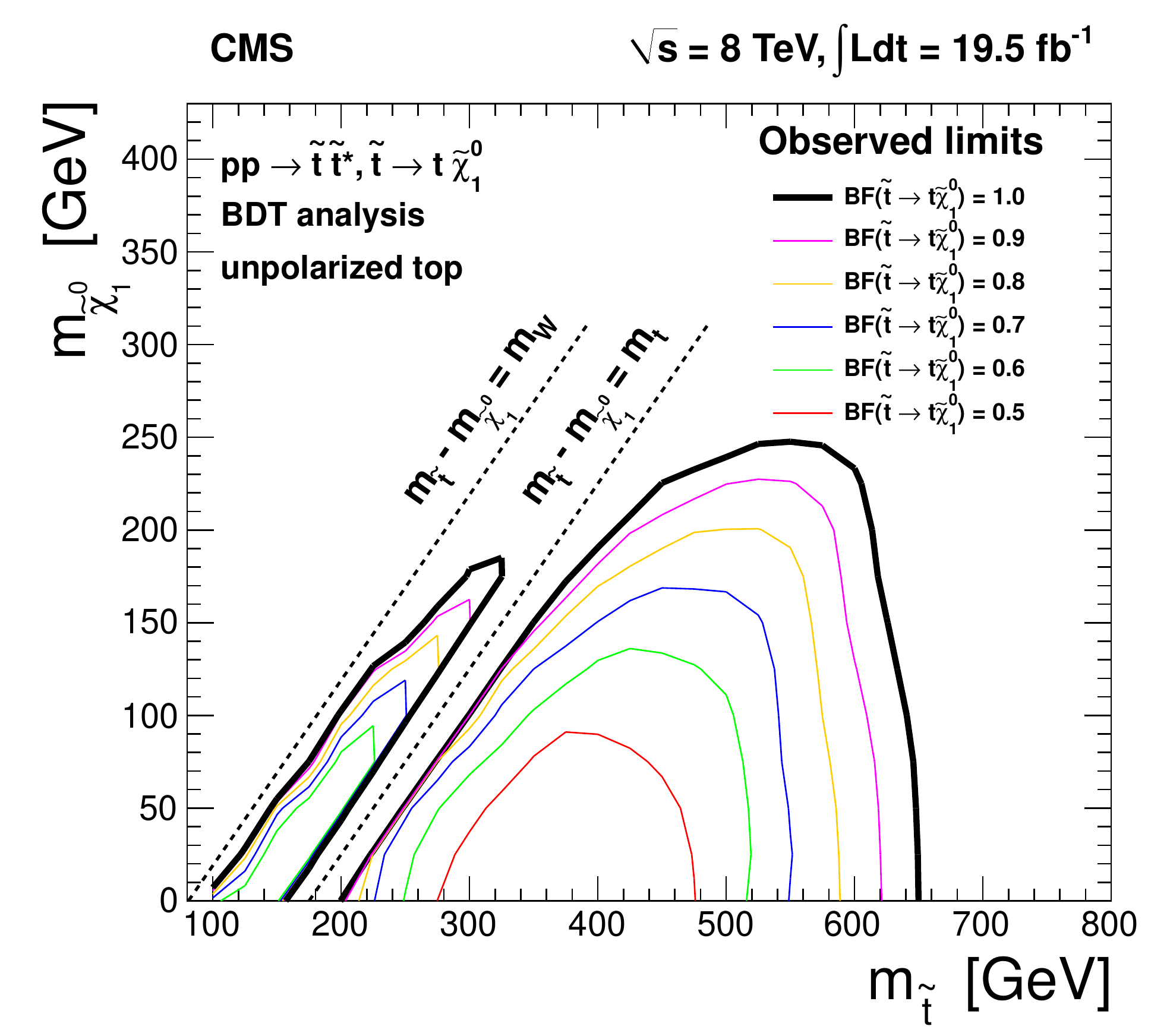}
\caption{
The observed 95\% CL excluded regions as a function
of the assumed branching fraction for the \Ttt\ decay mode.
The results are based on the assumption that the search has no
acceptance for top-squark pair events if one of the top squarks decays in a different mode.
See text for details.
\label{fig:T2tt_BF}
}
\end{figure}

Mixed-decay scenarios, \ie, scenarios with non-zero
top-squark decay branching fractions into both
\Ttt\ and \TbW, have not been considered
here.
However, our results can be used to draw useful conclusions
about these possibilities.  We must distinguish between
two typical SUSY spectra: one in which the chargino and LSP are nearly
mass-degenerate, and the other in which the chargino is considerably
heavier. In the degenerate case, corresponding to $x \approx 0$, the acceptance
is small for top-squark pairs with one or more \TbW\ decays.
This is because the visible decay products in the
$\chip_1 \to \chiz_1 +$X process are
soft and likely to escape detection.  Thus, to a good
approximation, in these scenarios the top-squark pair
cross section limit
can be extracted by scaling the corresponding limit in the
100\% \Ttt\ model by $\mathcal{B}^2$, where $\mathcal{B}$
is the branching fraction
for \Ttt.  Exclusion regions for a few choices of
$\mathcal{B}$ are shown in Fig.~\ref{fig:T2tt_BF}.
In the mixed case with a chargino much heavier than the LSP, a
conservative approximate cross section limit can be obtained as
$\sigma(\Pp\Pp\to\PSQt\PSQt^{*}) < \min({\sigma_0 / \mathcal{B}}^2, \sigma_{+}/(1 - \mathcal{B})^2)$,
where
$\sigma_{0}$ and $\sigma_{+}$ are the cross section limits for the
100\% \Ttt\ and 100\% \TbW\
scenarios, respectively,
and $\mathcal{B}$ is the branching fraction defined above.  (The limits
$\sigma_0$  and $\sigma_{+}$ shown in
Fig.~\ref{fig:bdt_interpretations} are available
electronically~\cite{website}\ifthenelse{\boolean{cms@external}}{ and as supplementary material to this paper}{}.)
This approach is conservative as it
uses only one out of the three possible
decay modes of the top-squark pair.
It should also be noted that in the heavier-chargino scenario it is
possible for one additional neutralino ($\chiz_2$) to be nearly degenerate with the chargino.  The decay
$\PSQt \to \cPqt \chiz_2$
followed by, for example, $\chiz_2 \to \Z \chiz_1$
or $\PH \chiz_1$
would then also be possible.  This would further
complicate the interpretation of the experimental results.

\section{Summary}

We have performed a search for the direct pair production of top
squarks in a final state consisting of a single isolated lepton, jets,
large missing transverse momentum, and large transverse mass.
Signal regions are defined both with requirements on the output of a BDT multivariate discriminator, and with requirements
on several kinematic discriminants.
The observed yields in the signal regions agree with the predicted backgrounds within the assessed uncertainties.
The results are interpreted in the context of models of top-squark
pair production and decay.  The analysis probes top squarks with masses up to
about 650\GeV and significantly restricts the allowed parameter space
of natural SUSY scenarios.

\section*{Acknowledgements}
{\tolerance=500
We thank Ian Low for assistance with polarization issues in top squark decays and Jiayin Gu for help in implementing the code for the variables of
Ref.~\cite{mt2w}.

We congratulate our colleagues in the CERN accelerator departments for the excellent performance of the LHC and thank the technical and administrative staffs at CERN and at other CMS institutes for their contributions to the success of the CMS effort. In addition, we gratefully acknowledge the computing centres and personnel of the Worldwide LHC Computing Grid for delivering so effectively the computing infrastructure essential to our analyses. Finally, we acknowledge the enduring support for the construction and operation of the LHC and the CMS detector provided by the following funding agencies: BMWF and FWF (Austria); FNRS and FWO (Belgium); CNPq, CAPES, FAPERJ, and FAPESP (Brazil); MEYS (Bulgaria); CERN; CAS, MoST, and NSFC (China); COLCIENCIAS (Colombia); MSES (Croatia); RPF (Cyprus); MoER, SF0690030s09 and ERDF (Estonia); Academy of Finland, MEC, and HIP (Finland); CEA and CNRS/IN2P3 (France); BMBF, DFG, and HGF (Germany); GSRT (Greece); OTKA and NKTH (Hungary); DAE and DST (India); IPM (Iran); SFI (Ireland); INFN (Italy); NRF and WCU (Republic of Korea); LAS (Lithuania); CINVESTAV, CONACYT, SEP, and UASLP-FAI (Mexico); MSI (New Zealand); PAEC (Pakistan); MSHE and NSC (Poland); FCT (Portugal); JINR (Armenia, Belarus, Georgia, Ukraine, Uzbekistan); MON, RosAtom, RAS and RFBR (Russia); MSTD (Serbia); SEIDI and CPAN (Spain); Swiss Funding Agencies (Switzerland); NSC (Taipei); ThEPCenter, IPST and NSTDA (Thailand); TUBITAK and TAEK (Turkey); NASU (Ukraine); STFC (United Kingdom); DOE and NSF (USA).

Individuals have received support from the University of California Institute for Mexico and the United States; the Marie-Curie programme and the European Research Council and EPLANET (European Union); the Leventis Foundation; the A. P. Sloan Foundation; the Alexander von Humboldt Foundation; the Belgian Federal Science Policy Office; the Fonds pour la Formation \`a la Recherche dans l'Industrie et dans l'Agriculture (FRIA-Belgium); the Agentschap voor Innovatie door Wetenschap en Technologie (IWT-Belgium); the Ministry of Education, Youth and Sports (MEYS) of Czech Republic; the Agence Nationale de la Recherche ANR-12-JS05-002-01 (France); the Council of Science and Industrial Research, India; the Compagnia di San Paolo (Torino); the HOMING PLUS programme of Foundation for Polish Science, cofinanced by EU, Regional Development Fund; the Thalis and Aristeia programmes cofinanced by EU-ESF and the Greek NSRF.
\par}

\ifthenelse{\boolean{cms@external}}{\vspace*{25ex}}{} 
\bibliography{auto_generated}
\clearpage
\appendix

\section{Additional tables and figures}

\subsection{Further information about systematic uncertainties}
\label{app:othersyst}

The systematic uncertainties for the \Ttt\ cut-based, \TbW\ BDT,
and \TbW\ cut-based analyses are shown in
Tables~\ref{tab:relativeuncertaintycomponents_t2tt_cnc},
~\ref{tab:relativeuncertaintycomponents_t2bw_bdt},
and
~\ref{tab:relativeuncertaintycomponents_t2bw_cnc},
respectively. The corresponding information for \Ttt\ BDT analysis is
given in the body of the paper (see Table~\ref{tab:relativeuncertaintycomponents_t2tt_bdt}).

\begin{table*}[htb]
\centering
\topcaption{The bottom row of this table shows
the relative uncertainty (in percent) of the total background
  predictions for the \Ttt\ cut-based signal regions.
The breakdown of this total uncertainty in terms of its individual
components is also shown.}
\label{tab:relativeuncertaintycomponents_t2tt_cnc}
{\footnotesize
\begin{tabular}{l|cccc}
\multicolumn{1}{c}{Sample} 	 	 & $\MET>150$\GeV	 & $\MET>200$\GeV	 & $\MET>250$\GeV & $\MET>300$\GeV\\
\hline\multicolumn{5}{c}{}\\[-0.5ex]
\multicolumn{5}{c}{Low $\Delta M$ Selection} \\
\hline
\MT\ peak data and MC (stat.) 	 	 & $1.4$ 	 & $2.4$ 	 & $4.0$ 	 & $6.3$  \\
\ttll\ N$_{\text{jets}}$ modeling 	 	 & $1.6$ 	 & $1.5$ 	 & $1.6$ 	 & $1.5$  \\
\ttll\ (CR-$\ell$t and CR-2$\ell$ tests) 	 	 & $5.2$ 	 & $7.6$ 	 & $13.1$ 	 & $19.6$  \\
2nd lepton veto 	 	 & $1.3$ 	 & $1.2$ 	 & $1.3$ 	 & $1.2$  \\
\ttll\ (stat.) 	 	 & $1.9$ 	 & $3.2$ 	 & $5.2$ 	 & $8.0$  \\
W+jets cross section 	 	 & $1.1$ 	 & $1.1$ 	 & $1.8$ 	 & $2.2$  \\
W+jets (stat.) 	 	 & $2.1$ 	 & $3.2$ 	 & $4.1$ 	 & $5.6$  \\
W+jets SF uncertainty	 	 & $9.4$ 	 & $9.0$ 	 & $7.5$ 	 & $7.0$  \\
$1-\ell$  top (stat.) 	 	 & $0.6$ 	 & $0.9$ 	 & $1.1$ 	 & $1.5$  \\
$1-\ell$  top tail-to-peak ratio 	 	 & $16.0$ 	 & $20.7$ 	 & $18.3$ 	 & $18.5$  \\
Rare processes cross sections 	 	 & $2.0$ 	 & $2.6$ 	 & $3.8$ 	 & $5.9$  \\
\hline
Total 	 	 & $19.8$ 	 & $24.6$ 	 & $25.5$ 	 & $30.9$  \\
\hline\multicolumn{5}{c}{}\\[-0.5ex]
\multicolumn{5}{c}{High $\Delta M$ Selection} \\
\hline
\MT\ peak data and MC (stat.) 	 	 & $3.9$ 	 & $4.8$ 	 & $6.0$ 	 & $8.5$  \\
\ttll\ N$_{\text{jets}}$ modeling 	 	 & $0.8$ 	 & $0.9$ 	 & $1.0$ 	 & $0.9$  \\
\ttll\ (CR-$\ell$t and CR-2$\ell$ tests) 	 	 & $4.1$ 	 & $6.1$ 	 & $11.7$ 	 & $14.9$  \\
2nd lepton veto 	 	 & $0.7$ 	 & $0.7$ 	 & $0.8$ 	 & $0.7$  \\
\ttll\ (stat.) 	 	 & $4.2$ 	 & $5.9$ 	 & $8.4$ 	 & $10.2$  \\
W+jets cross section 	 	 & $0.6$ 	 & $0.5$ 	 & $1.3$ 	 & $1.8$  \\
W+jets (stat.) 	 	 & $3.8$ 	 & $4.7$ 	 & $5.7$ 	 & $7.7$  \\
W+jets SF uncertainty	 	 & $11.7$ 	 & $10.3$ 	 & $8.8$ 	 & $8.8$  \\
$1-\ell$  top (stat.) 	 	 & $1.8$ 	 & $1.9$ 	 & $2.1$ 	 & $3.4$  \\
$1-\ell$  top tail-to-peak ratio 	 	 & $17.1$ 	 & $21.3$ 	 & $20.9$ 	 & $17.3$  \\
Rare processes cross sections 	 	 & $6.1$ 	 & $6.9$ 	 & $7.8$ 	 & $9.2$  \\
\hline
Total 	 	 & $23.1$ 	 & $27.0$ 	 & $29.3$ 	 & $30.6$  \\
\hline
\end{tabular}}
\end{table*}

\begin{table*}[htb]
\centering
\topcaption{
The bottom row of this table shows
the relative uncertainty (in percent) of the total background
  predictions for the \TbW\ BDT signal regions.
The breakdown of this total uncertainty in terms of its individual
components is also shown.}
\label{tab:relativeuncertaintycomponents_t2bw_bdt}
{\footnotesize
\begin{tabular}{l| cccccccccc cccccccccc cccccccccc cccccccccc cccccccccc cccccccccc}
\multicolumn{61}{c}{$\TbW$ $x=0.75$} \\
\hline
Sample &\multicolumn{15}{c}{\phantom{0}\phantom{0}\phantom{0}BDT1\phantom{0}\phantom{0}\phantom{0}}  & \multicolumn{15}{c}{\phantom{0}\phantom{0}\phantom{0}BDT2\phantom{0}\phantom{0}\phantom{0}} & \multicolumn{15}{c}{\phantom{0}\phantom{0}\phantom{0}BDT3\phantom{0}\phantom{0}\phantom{0}} & \multicolumn{15}{c}{\phantom{0}\phantom{0}BDT4\phantom{0}\phantom{0}} \\
\hline
\MT\ peak data and MC (stat.) 	 	 & \multicolumn{15}{c}{$3.5$} 	 & \multicolumn{15}{c}{$5.3$} 	 & \multicolumn{15}{c}{$7.8$} 	 & \multicolumn{15}{c}{$1.2$}  \\
\ttll\ N$_{\text{jets}}$ modeling 	 	 & \multicolumn{15}{c}{$1.8$} 	 & \multicolumn{15}{c}{$1.2$} 	 & \multicolumn{15}{c}{$1.1$} 	 & \multicolumn{15}{c}{$1.6$}  \\
\ttll\ (CR-$\ell$t and CR-2$\ell$ tests) 	 	 & \multicolumn{15}{c}{$6.0$} 	 & \multicolumn{15}{c}{$8.2$} 	 & \multicolumn{15}{c}{$11.3$} 	 & \multicolumn{15}{c}{$3.6$}  \\
2nd lepton veto 	 	 & \multicolumn{15}{c}{$1.7$} 	 & \multicolumn{15}{c}{$1.1$} 	 & \multicolumn{15}{c}{$1.0$} 	 & \multicolumn{15}{c}{$1.4$}  \\
\ttll\ (stat.) 	 	 & \multicolumn{15}{c}{$4.3$} 	 & \multicolumn{15}{c}{$5.9$} 	 & \multicolumn{15}{c}{$9.6$} 	 & \multicolumn{15}{c}{$1.4$}  \\
W+jets cross section 	 	 & \multicolumn{15}{c}{$2.7$} 	 & \multicolumn{15}{c}{$2.3$} 	 & \multicolumn{15}{c}{$2.7$} 	 & \multicolumn{15}{c}{$1.4$}  \\
W+jets (stat.) 	 	 & \multicolumn{15}{c}{$4.5$} 	 & \multicolumn{15}{c}{$5.3$} 	 & \multicolumn{15}{c}{$6.4$} 	 & \multicolumn{15}{c}{$2.4$}  \\
W+jets SF uncertainty 	 	 & \multicolumn{15}{c}{$6.9$} 	 & \multicolumn{15}{c}{$7.7$} 	 & \multicolumn{15}{c}{$7.0$} 	 & \multicolumn{15}{c}{$9.9$}  \\
$1-\ell$  top (stat.) 	 	 & \multicolumn{15}{c}{$1.2$} 	 & \multicolumn{15}{c}{$1.2$} 	 & \multicolumn{15}{c}{$1.2$} 	 & \multicolumn{15}{c}{$0.6$}  \\
$1-\ell$  top tail-to-peak ratio 	 	 & \multicolumn{15}{c}{$11.3$} 	 & \multicolumn{15}{c}{$19.5$} 	 & \multicolumn{15}{c}{$17.6$} 	 & \multicolumn{15}{c}{$10.7$}  \\
Rare processes cross sections 	 	 & \multicolumn{15}{c}{$1.9$} 	 & \multicolumn{15}{c}{$6.2$} 	 & \multicolumn{15}{c}{$8.9$} 	 & \multicolumn{15}{c}{$1.1$}  \\
\hline
Total 	 	 & \multicolumn{15}{c}{$16.8$} 	 & \multicolumn{15}{c}{$25.4$} 	 & \multicolumn{15}{c}{$27.8$} 	 & \multicolumn{15}{c}{$15.5$}  \\
\hline\multicolumn{61}{c}{}\\[-0.5ex]
\multicolumn{61}{c}{$\TbW$ $x=0.5$} \\
\hline
Sample &\multicolumn{12}{c}{\phantom{0} BDT1 \phantom{0}}  & \multicolumn{12}{c}{BDT2--Loose}& \multicolumn{12}{c}{BDT2--Tight} & \multicolumn{12}{c}{\phantom{0} BDT3 \phantom{0}} & \multicolumn{12}{c}{\phantom{0}BDT4\phantom{0}} \\
\hline
\MT\ peak data and MC (stat.) 	 	 & \multicolumn{12}{c}{$3.0$} 	 & \multicolumn{12}{c}{$3.3$} 	 & \multicolumn{12}{c}{$6.0$} 	 & \multicolumn{12}{c}{$5.8$} 	 & \multicolumn{12}{c}{$2.4$}  \\
\ttll\ N$_{\text{jets}}$ modeling 	 	 & \multicolumn{12}{c}{$1.6$} 	 & \multicolumn{12}{c}{$1.3$} 	 & \multicolumn{12}{c}{$1.0$} 	 & \multicolumn{12}{c}{$1.1$} 	 & \multicolumn{12}{c}{$2.1$}  \\
\ttll\ (CR-$\ell$t and CR-2$\ell$ tests) 	 	 & \multicolumn{12}{c}{$5.2$} 	 & \multicolumn{12}{c}{$6.4$} 	 & \multicolumn{12}{c}{$17.2$} 	 & \multicolumn{12}{c}{$11.1$} 	 & \multicolumn{12}{c}{$10.3$}  \\
2nd lepton veto 	 	 & \multicolumn{12}{c}{$1.4$} 	 & \multicolumn{12}{c}{$1.2$} 	 & \multicolumn{12}{c}{$1.0$} 	 & \multicolumn{12}{c}{$1.0$} 	 & \multicolumn{12}{c}{$1.9$}  \\
\ttll\ (stat.) 	 	 & \multicolumn{12}{c}{$3.5$} 	 & \multicolumn{12}{c}{$4.0$} 	 & \multicolumn{12}{c}{$6.6$} 	 & \multicolumn{12}{c}{$6.2$}  	 & \multicolumn{12}{c}{$2.8$} \\
W+jets cross section 	 	 & \multicolumn{12}{c}{$2.5$} 	 & \multicolumn{12}{c}{$2.6$} 	 & \multicolumn{12}{c}{$1.4$} 	 & \multicolumn{12}{c}{$3.3$} 	 & \multicolumn{12}{c}{$2.8$}  \\
W+jets (stat.) 	 	 & \multicolumn{12}{c}{$2.3$} 	 & \multicolumn{12}{c}{$2.2$} 	 & \multicolumn{12}{c}{$4.1$} 	 & \multicolumn{12}{c}{$3.4$} 	 & \multicolumn{12}{c}{$2.3$}  \\
W+jets SF uncertainty	 	 & \multicolumn{12}{c}{$8.0$} 	 & \multicolumn{12}{c}{$8.0$} 	 & \multicolumn{12}{c}{$8.1$} 	 & \multicolumn{12}{c}{$7.3$} 	 & \multicolumn{12}{c}{$5.7$}  \\
$1-\ell$  top (stat.) 	 	 & \multicolumn{12}{c}{$1.0$} 	 & \multicolumn{12}{c}{$1.2$} 	 & \multicolumn{12}{c}{$1.5$} 	 & \multicolumn{12}{c}{$1.6$} 	 & \multicolumn{12}{c}{$0.8$}  \\
$1-\ell$  top tail-to-peak ratio 	 	 & \multicolumn{12}{c}{$10.3$} 	 & \multicolumn{12}{c}{$11.5$} 	 & \multicolumn{12}{c}{$18.4$} 	 & \multicolumn{12}{c}{$11.7$} 	 & \multicolumn{12}{c}{$5.6$}  \\
Rare processes cross sections 	 	 & \multicolumn{12}{c}{$3.3$} 	 & \multicolumn{12}{c}{$6.8$} 	 & \multicolumn{12}{c}{$8.7$} 	 & \multicolumn{12}{c}{$9.4$} 	 & \multicolumn{12}{c}{$1.3$}  \\
\hline
Total 	 	 & \multicolumn{12}{c}{$15.7$} 	 & \multicolumn{12}{c}{$18.0$} 	 & \multicolumn{12}{c}{$29.7$} 	 & \multicolumn{12}{c}{$22.3$} 	 & \multicolumn{12}{c}{$14.4$}  \\
\hline\multicolumn{61}{c}{}\\[-0.5ex]
\multicolumn{61}{c}{$\TbW$ $x=0.25$} \\
\hline
Sample              & \multicolumn{20}{c}{\phantom{0}\phantom{0}\phantom{0}\phantom{0}\phantom{0}\phantom{0}BDT1\phantom{0}\phantom{0}\phantom{0}\phantom{0}\phantom{0}\phantom{0}} & \multicolumn{20}{c}{\phantom{0}\phantom{0}\phantom{0}\phantom{0}\phantom{0}\phantom{0}BDT2\phantom{0}\phantom{0}\phantom{0}\phantom{0}\phantom{0}\phantom{0}} & \multicolumn{20}{c}{\phantom{0}\phantom{0}\phantom{0}\phantom{0}\phantom{0}BDT3\phantom{0}\phantom{0}\phantom{0}\phantom{0}\phantom{0}} \\
\hline
\MT\ peak data and MC (stat.) 	 	 & \multicolumn{20}{c}{$4.0$} 	 & \multicolumn{20}{c}{$9.0$} 	 & \multicolumn{20}{c}{$10.6$}  \\
\ttll\ N$_{\text{jets}}$ modeling 	 	 & \multicolumn{20}{c}{$1.5$} 	 & \multicolumn{20}{c}{$0.7$} 	 & \multicolumn{20}{c}{$0.8$}  \\
\ttll\ (CR-$\ell$t and CR-2$\ell$ tests) 	 	 & \multicolumn{20}{c}{$7.7$} 	 & \multicolumn{20}{c}{$11.4$} 	 & \multicolumn{20}{c}{$19.1$}  \\
2nd lepton veto 	 	 & \multicolumn{20}{c}{$1.4$} 	 & \multicolumn{20}{c}{$0.6$} 	 & \multicolumn{20}{c}{$0.8$}  \\
\ttll\ (stat.) 	 	 & \multicolumn{20}{c}{$5.0$} 	 & \multicolumn{20}{c}{$6.5$} 	 & \multicolumn{20}{c}{$11.8$}  \\
W+jets cross section 	 	 & \multicolumn{20}{c}{$3.0$} 	 & \multicolumn{20}{c}{$1.0$} 	 & \multicolumn{20}{c}{$1.5$}  \\
W+jets (stat.) 	 	 & \multicolumn{20}{c}{$2.4$} 	 & \multicolumn{20}{c}{$5.3$} 	 & \multicolumn{20}{c}{$6.7$}  \\
W+jets SF uncertainty 	 	 & \multicolumn{20}{c}{$7.2$} 	 & \multicolumn{20}{c}{$11.3$} 	 & \multicolumn{20}{c}{$9.5$}  \\
$1-\ell$  top (stat.) 	 	 & \multicolumn{20}{c}{$1.3$} 	 & \multicolumn{20}{c}{$3.2$} 	 & \multicolumn{20}{c}{$4.2$}  \\
$1-\ell$  top tail-to-peak ratio 	 	 & \multicolumn{20}{c}{$10.8$} 	 & \multicolumn{20}{c}{$12.6$} 	 & \multicolumn{20}{c}{$13.2$}  \\
Rare processes cross sections 	 	 & \multicolumn{20}{c}{$4.5$} 	 & \multicolumn{20}{c}{$6.2$} 	 & \multicolumn{20}{c}{$9.6$}  \\
\hline
Total 	 	 & \multicolumn{20}{c}{$17.7$} 	 & \multicolumn{20}{c}{$24.9$} 	 & \multicolumn{20}{c}{$32.3$}  \\
\hline
\end{tabular}}
\end{table*}

\begin{table*}[htb]
\centering
\topcaption{
The bottom row of this table shows
the relative uncertainty (in percent) of the total background
  predictions for the \TbW\ cut-based signal regions.
The breakdown of this total uncertainty in terms of its individual
components is also shown.}
\label{tab:relativeuncertaintycomponents_t2bw_cnc}
{\footnotesize
\begin{tabular}{l|cccc}
\multicolumn{1}{c}{Sample} 	 	 & $\MET>100$\GeV	 & $\MET>150$\GeV	 & $\MET>200$\GeV & $\MET>250$\GeV\\
\hline\multicolumn{5}{c}{}\\[-0.5ex]
\multicolumn{5}{c}{Low $\Delta M$ Selection} \\
\hline
\MT\ peak data and MC (stat.) 	 	 & $0.7$ 	 & $1.3$ 	 & $2.2$ 	 & $3.5$  \\
\ttll\ N$_{\text{jets}}$ modeling 	 	 & $1.6$ 	 & $1.9$ 	 & $1.9$ 	 & $1.9$  \\
\ttll\ (CR-$\ell$t and CR-2$\ell$ tests) 	 	 & $2.6$ 	 & $3.2$ 	 & $6.4$ 	 & $12.4$  \\
2nd lepton veto 	 	 & $1.3$ 	 & $1.5$ 	 & $1.5$ 	 & $1.5$  \\
\ttll\ (stat.) 	 	 & $0.7$ 	 & $1.4$ 	 & $2.4$ 	 & $3.9$  \\
W+jets cross section 	 	 & $1.5$ 	 & $2.0$ 	 & $2.5$ 	 & $3.2$  \\
W+jets (stat.) 	 	 & $0.8$ 	 & $1.1$ 	 & $1.6$ 	 & $2.2$  \\
W+jets SF uncertainty	 	 & $9.9$ 	 & $6.8$ 	 & $5.7$ 	 & $5.4$  \\
$1-\ell$  top (stat.) 	 	 & $0.3$ 	 & $0.4$ 	 & $0.5$ 	 & $0.7$  \\
$1-\ell$  top tail-to-peak ratio 	 	 & $5.9$ 	 & $11.0$ 	 & $11.7$ 	 & $12.1$  \\
Rare processes cross sections 	 	 & $1.1$ 	 & $1.7$ 	 & $2.6$ 	 & $3.7$  \\
\hline
Total 	 	 & $12.2$ 	 & $14.0$ 	 & $15.6$ 	 & $19.7$  \\
\hline\multicolumn{5}{c}{}\\[-0.5ex]
\multicolumn{5}{c}{High $\Delta M$ Selection} \\
\hline
\MT\ peak data and MC (stat.) 	 	 & $2.9$ 	 & $3.3$ 	 & $4.3$ 	 & $5.5$  \\
\ttll\ N$_{\text{jets}}$ modeling 	 	 & $1.0$ 	 & $0.9$ 	 & $1.1$ 	 & $0.9$  \\
\ttll\ (CR-$\ell$t and CR-2$\ell$ tests) 	 	 & $4.8$ 	 & $6.3$ 	 & $10.6$ 	 & $13.4$  \\
2nd lepton veto 	 	 & $0.9$ 	 & $0.8$ 	 & $0.9$ 	 & $0.8$  \\
\ttll\ (stat.) 	 	 & $2.6$ 	 & $3.9$ 	 & $5.6$ 	 & $7.1$  \\
W+jets cross section 	 	 & $2.3$ 	 & $1.5$ 	 & $1.6$ 	 & $1.5$  \\
W+jets (stat.) 	 	 & $1.8$ 	 & $2.5$ 	 & $3.2$ 	 & $4.3$  \\
W+jets SF uncertainty	 	 & $11.5$ 	 & $10.2$ 	 & $8.4$ 	 & $8.3$  \\
$1-\ell$  top (stat.) 	 	 & $1.3$ 	 & $1.5$ 	 & $1.5$ 	 & $1.6$  \\
$1-\ell$  top tail-to-peak ratio 	 	 & $5.2$ 	 & $12.5$ 	 & $15.3$ 	 & $16.6$  \\
Rare processes cross sections 	 	 & $4.1$ 	 & $7.0$ 	 & $8.7$ 	 & $10.2$  \\
\hline
Total 	 	 & $15.0$ 	 & $19.7$ 	 & $23.7$ 	 & $27.1$  \\
\hline
\end{tabular}}
\end{table*}
\ifthenelse{\boolean{cms@external}}{}{\clearpage}

\subsection{Additional \texorpdfstring{\MT}{Mtransverse} and BDT output distributions}
\label{app:plots}

\begin{figure*}[htb]
  \begin{center}
    \includegraphics[width=0.49\textwidth]{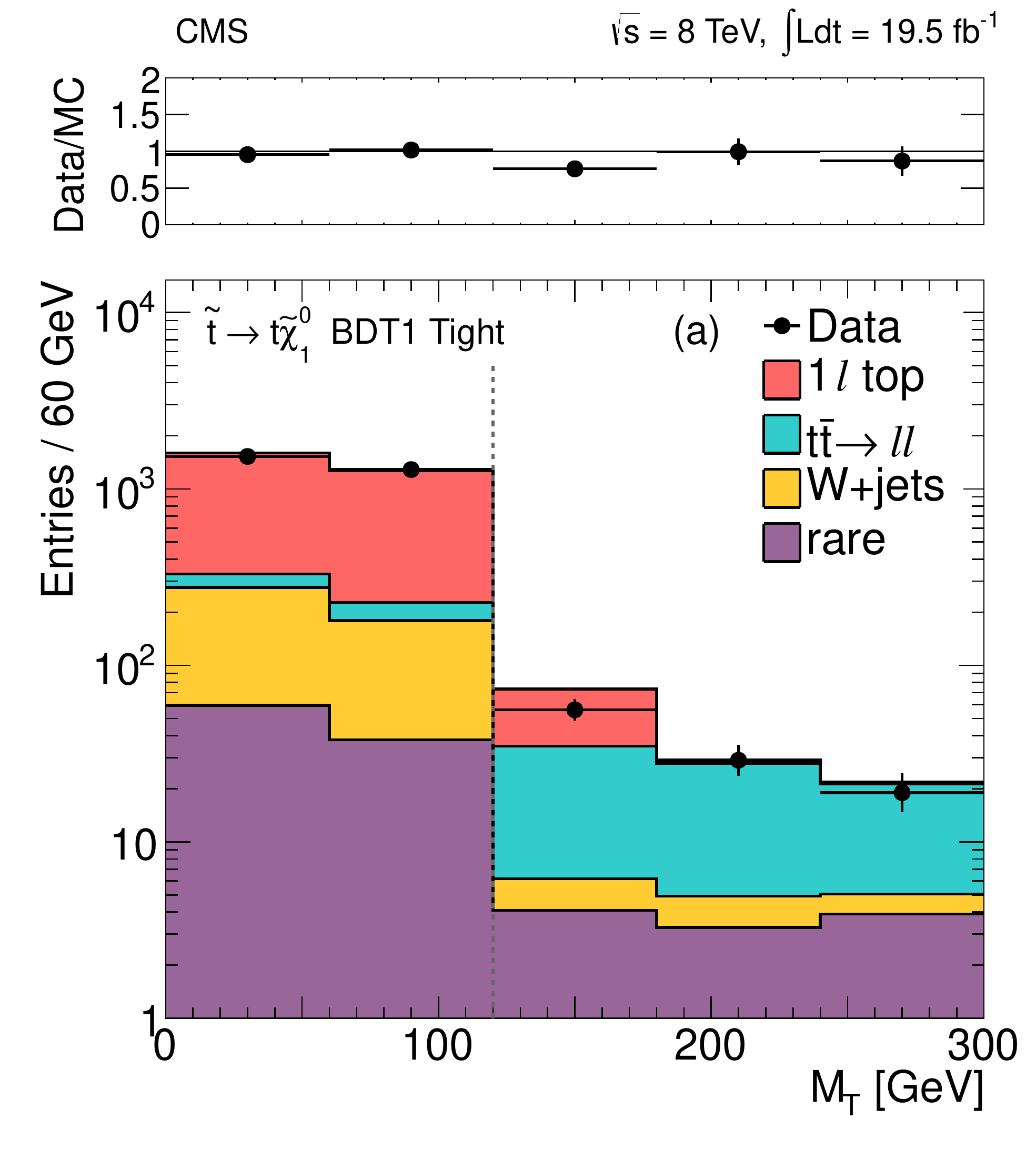}%
    \includegraphics[width=0.49\textwidth]{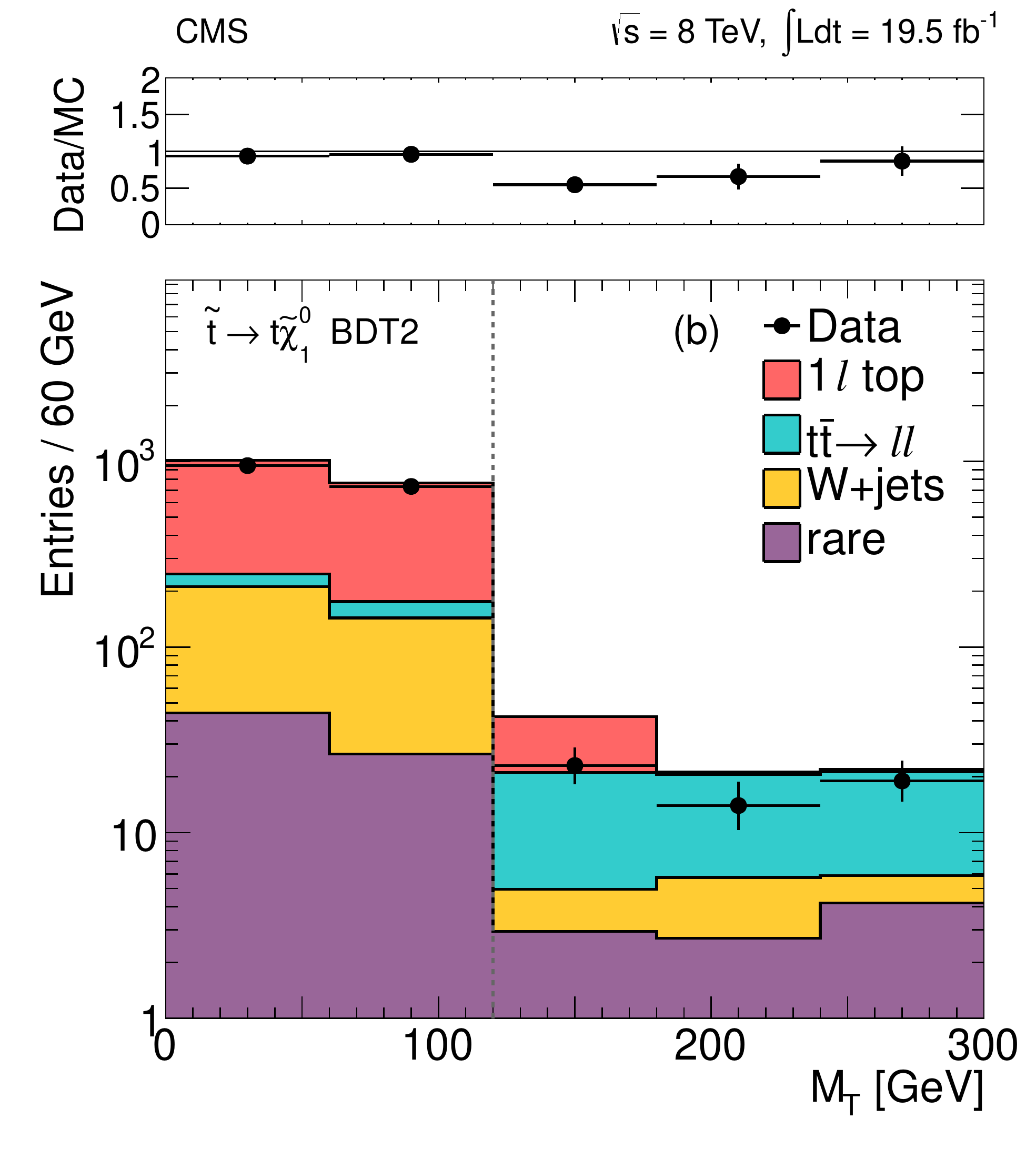} \\
    \includegraphics[width=0.49\textwidth]{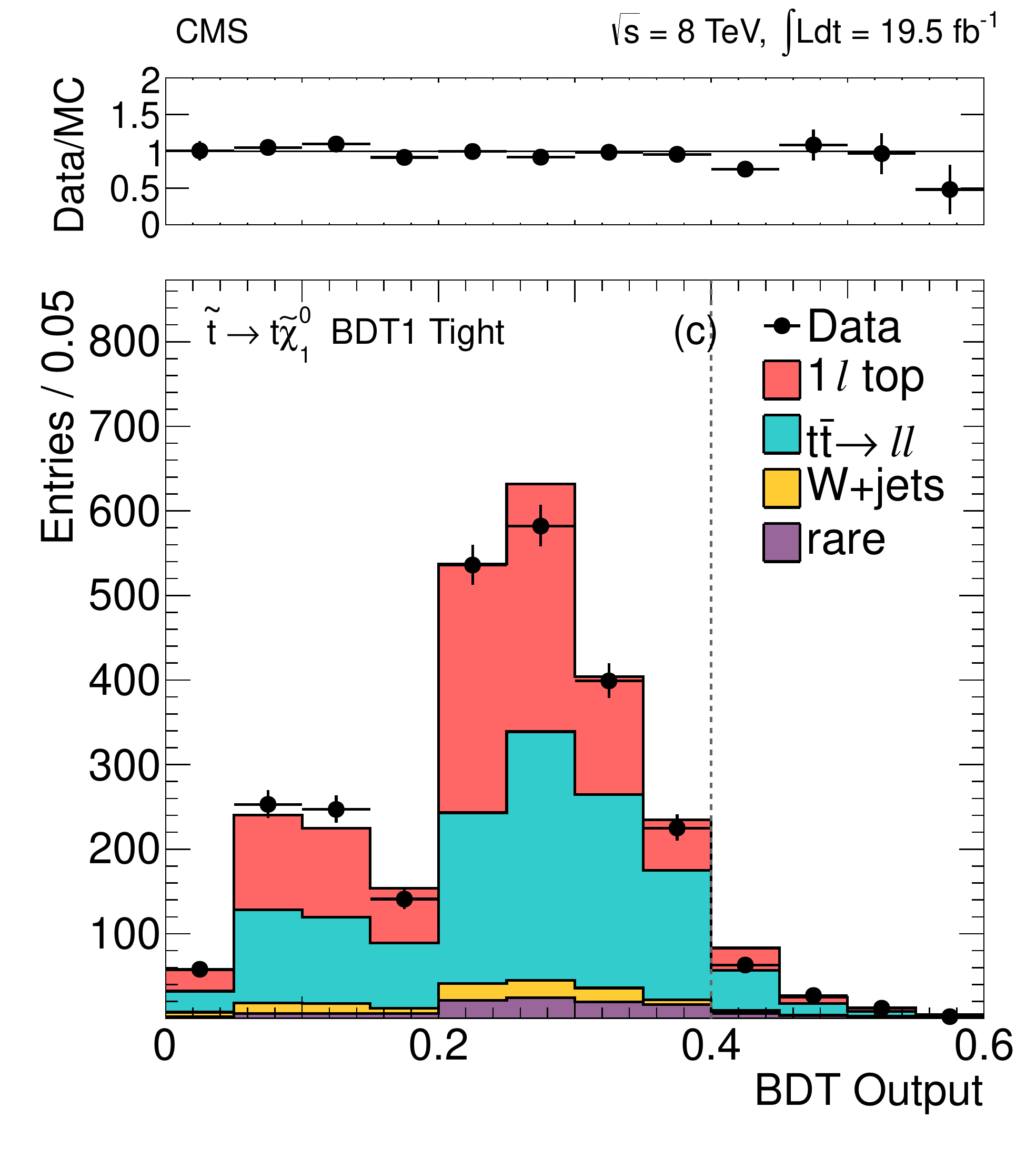}%
    \includegraphics[width=0.49\textwidth]{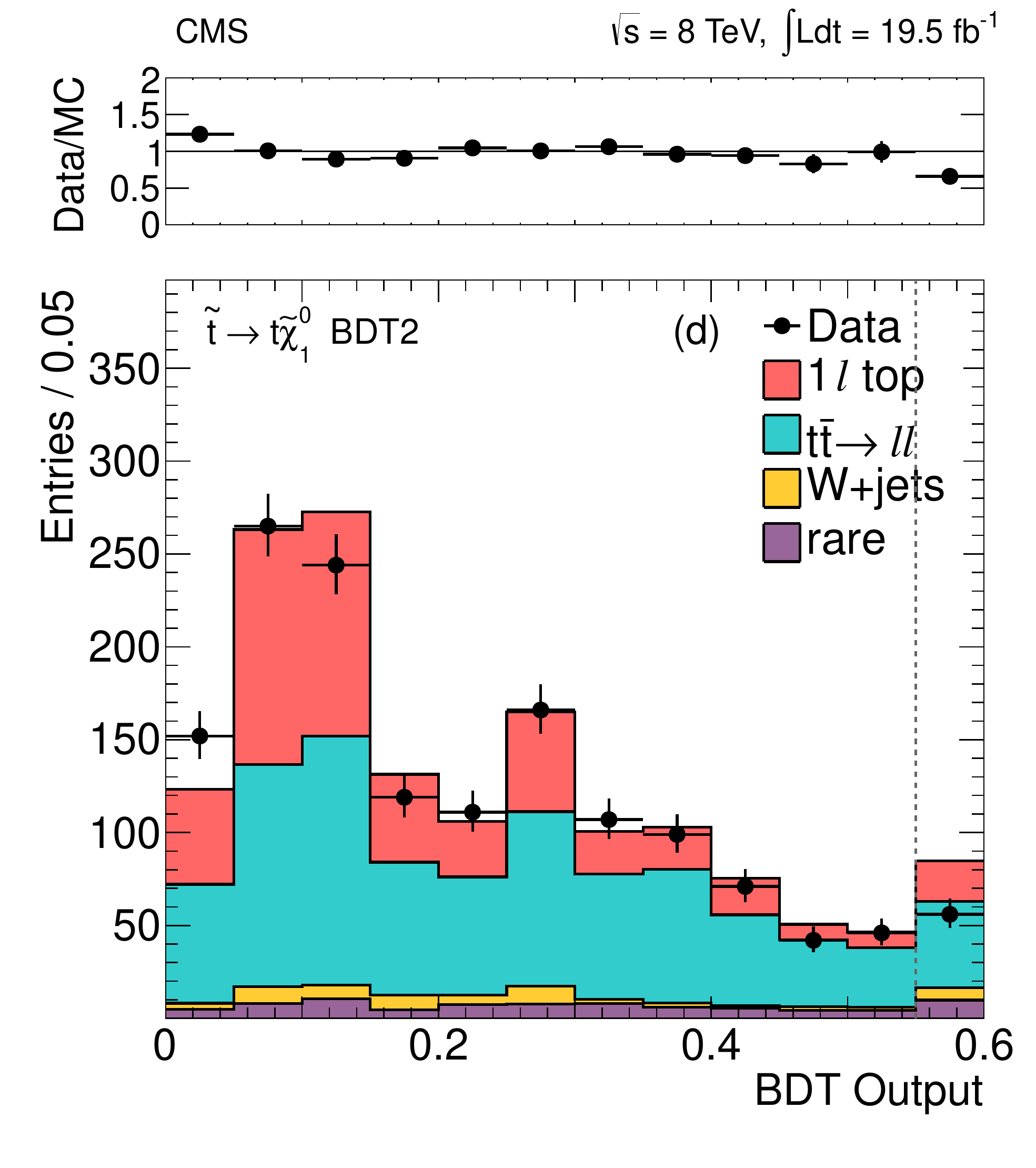}%
    \caption{
      Comparison of data and MC simulation for the distributions of BDT output and \MT\ corresponding
      to the \Ttt\ scenario in training regions 1 and 2.  The \MT\ distributions
      are shown after the requirement on the BDT output, and the BDT output distributions are shown after the
      $\MT> 120\GeV$ requirement (these requirements are also indicated by vertical dashed lines
      on the respective distributions).
      (a) \MT\ after the tight cut on the BDT1 output; 
      (b) \MT\ after the cut on the BDT2 output;
      (c) BDT1 output after the \MT\ cut;
      (d) BDT2 output after the \MT\ cut. In all distributions the last bin contains the overflow.
      \label{fig:plots1}
    }
      \end{center}
\end{figure*}

\begin{figure*}[htb]
  \begin{center}
    \includegraphics[width=0.49\textwidth]{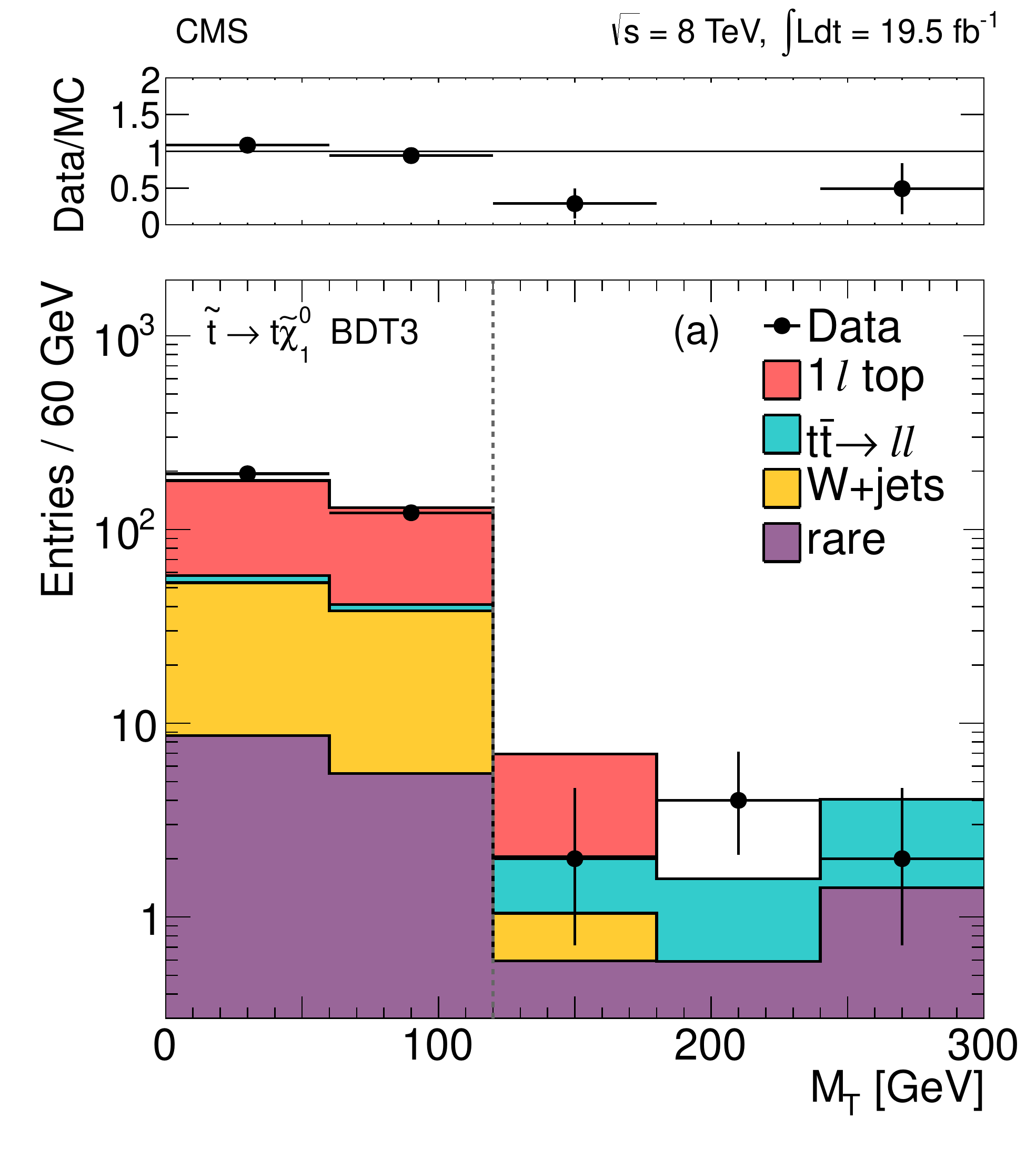}%
    \includegraphics[width=0.49\textwidth]{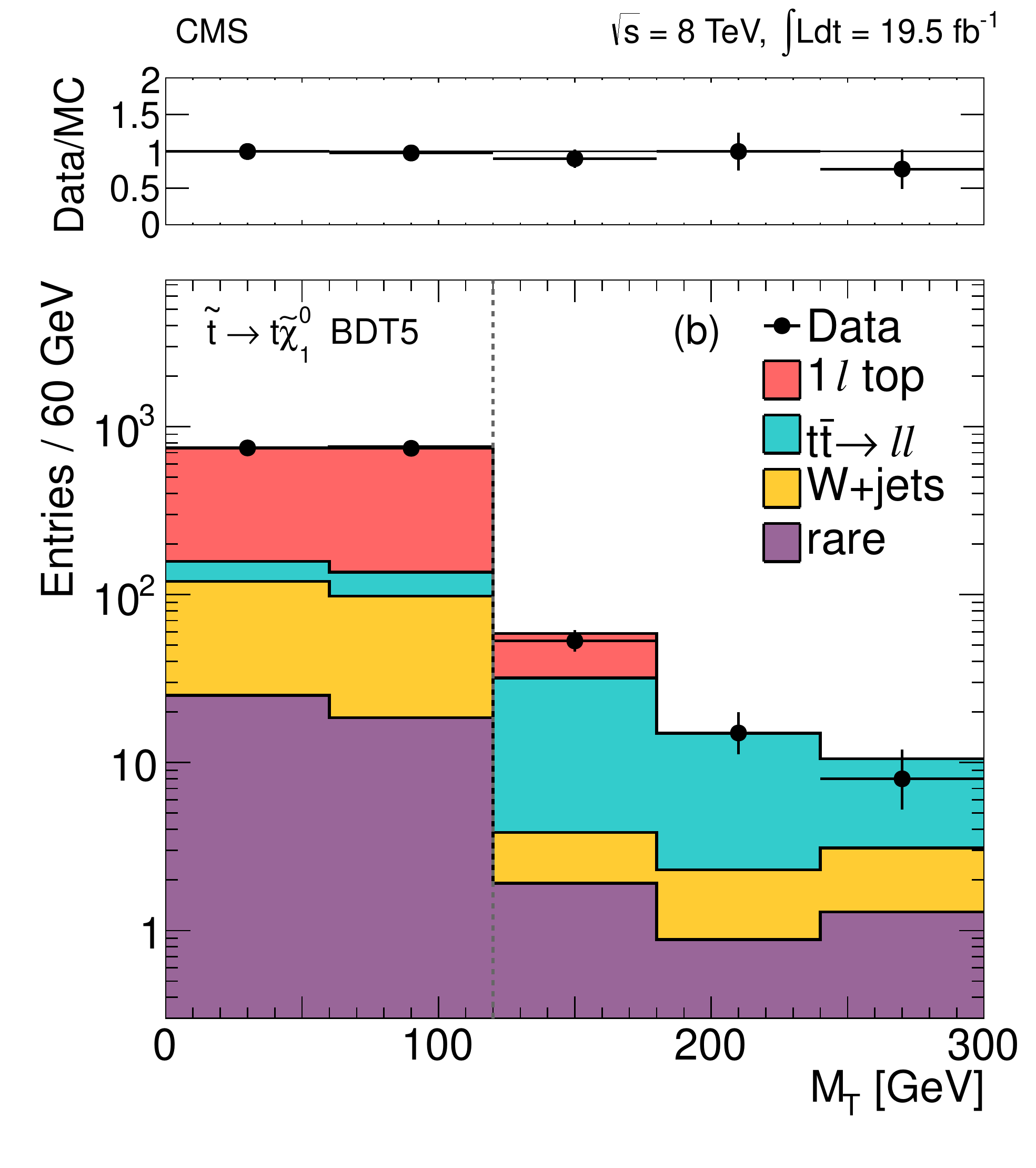} \\
    \includegraphics[width=0.49\textwidth]{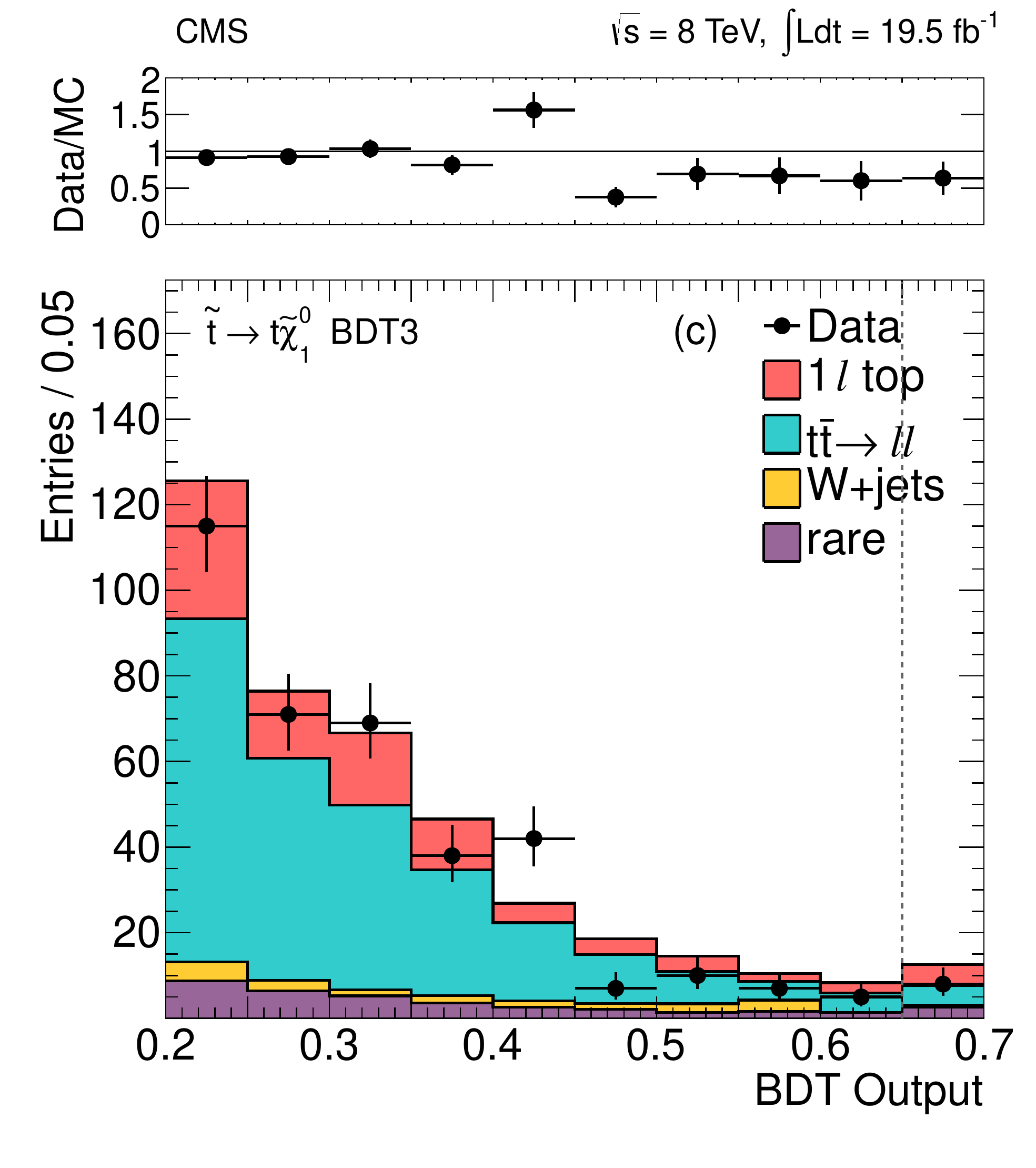}%
    \includegraphics[width=0.49\textwidth]{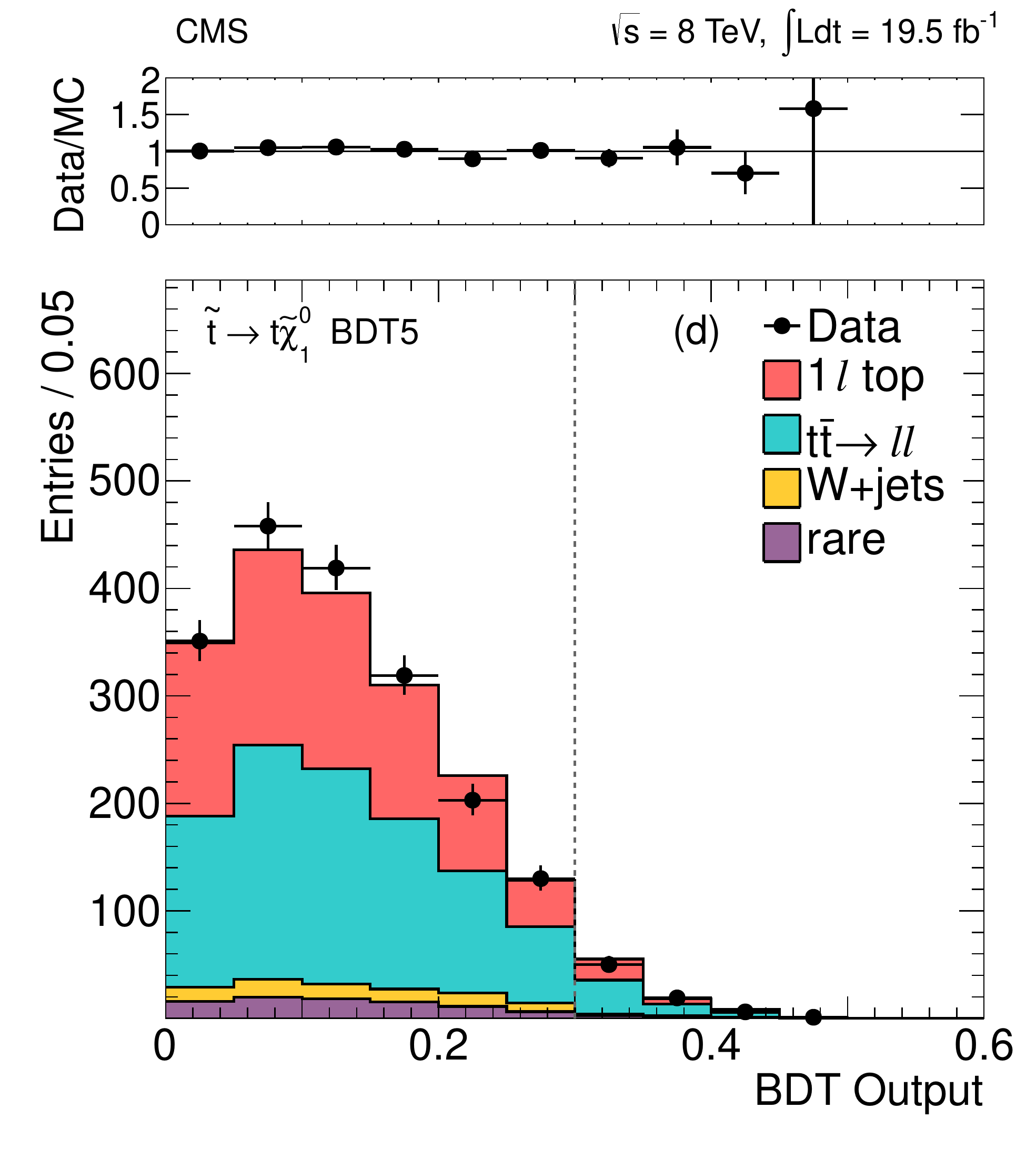} \\
    \caption{
      Comparison of data and MC simulation for the distributions of BDT output and \MT\ corresponding
      to the \Ttt\ scenario in training regions 3 and 5.  The \MT\ distributions
      are shown after the requirement on the BDT output, and the BDT output distributions are shown after the
      $\MT> 120\GeV$ requirement (these requirements are also indicated by vertical dashed lines
      on the respective distributions).
      (a) \MT\ after the cut on the BDT3 output; 
      (b) \MT\ after the cut on the BDT5 output;
      (c) BDT3 output after the \MT\ cut;
      (d) BDT5 output after the \MT\ cut.  In all distributions the last bin contains the overflow.
      \label{fig:plots2}
    }
      \end{center}
\end{figure*}

\begin{figure*}[htb]
  \begin{center}
    \includegraphics[width=0.49\linewidth]{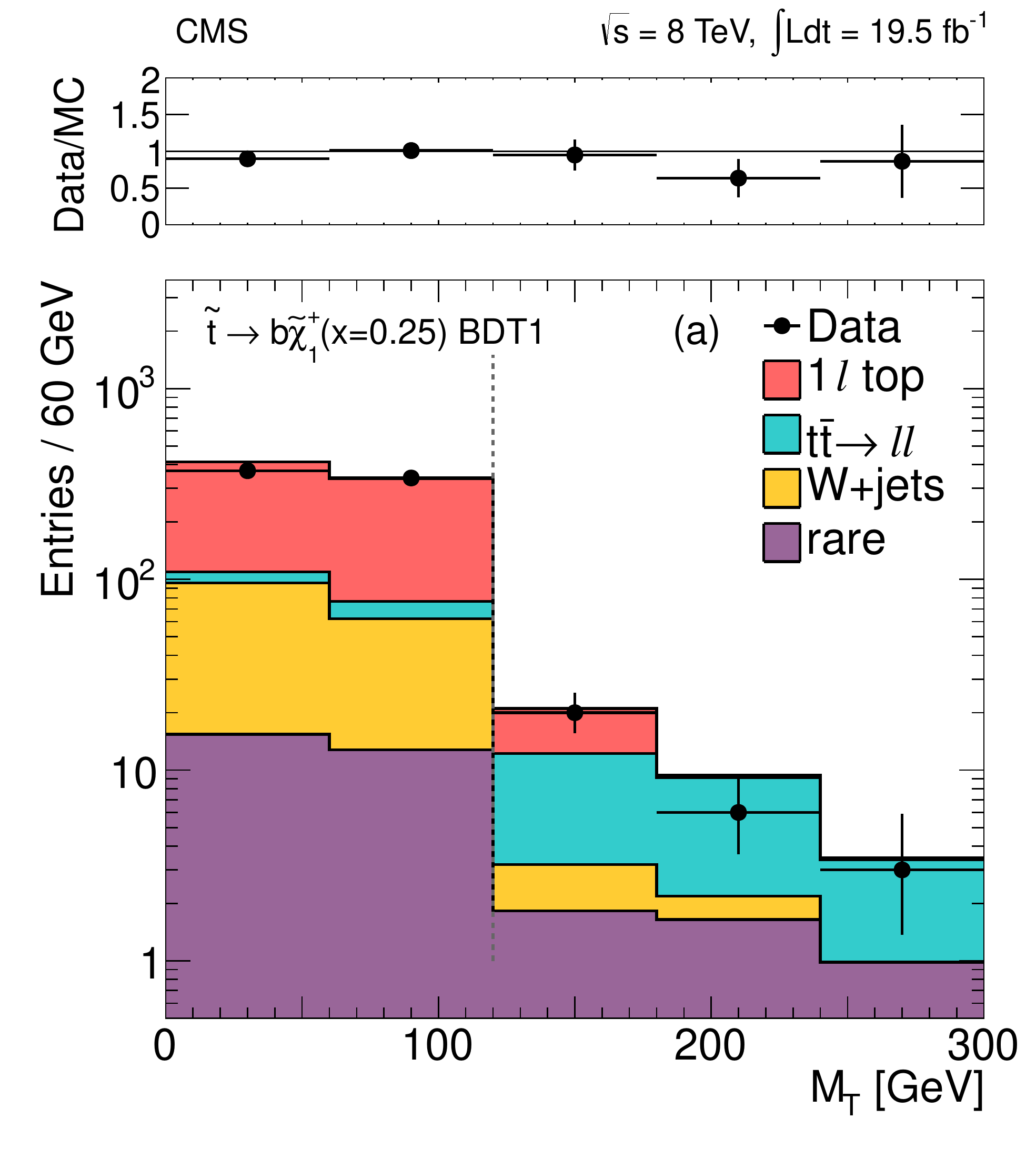}%
    \includegraphics[width=0.49\linewidth]{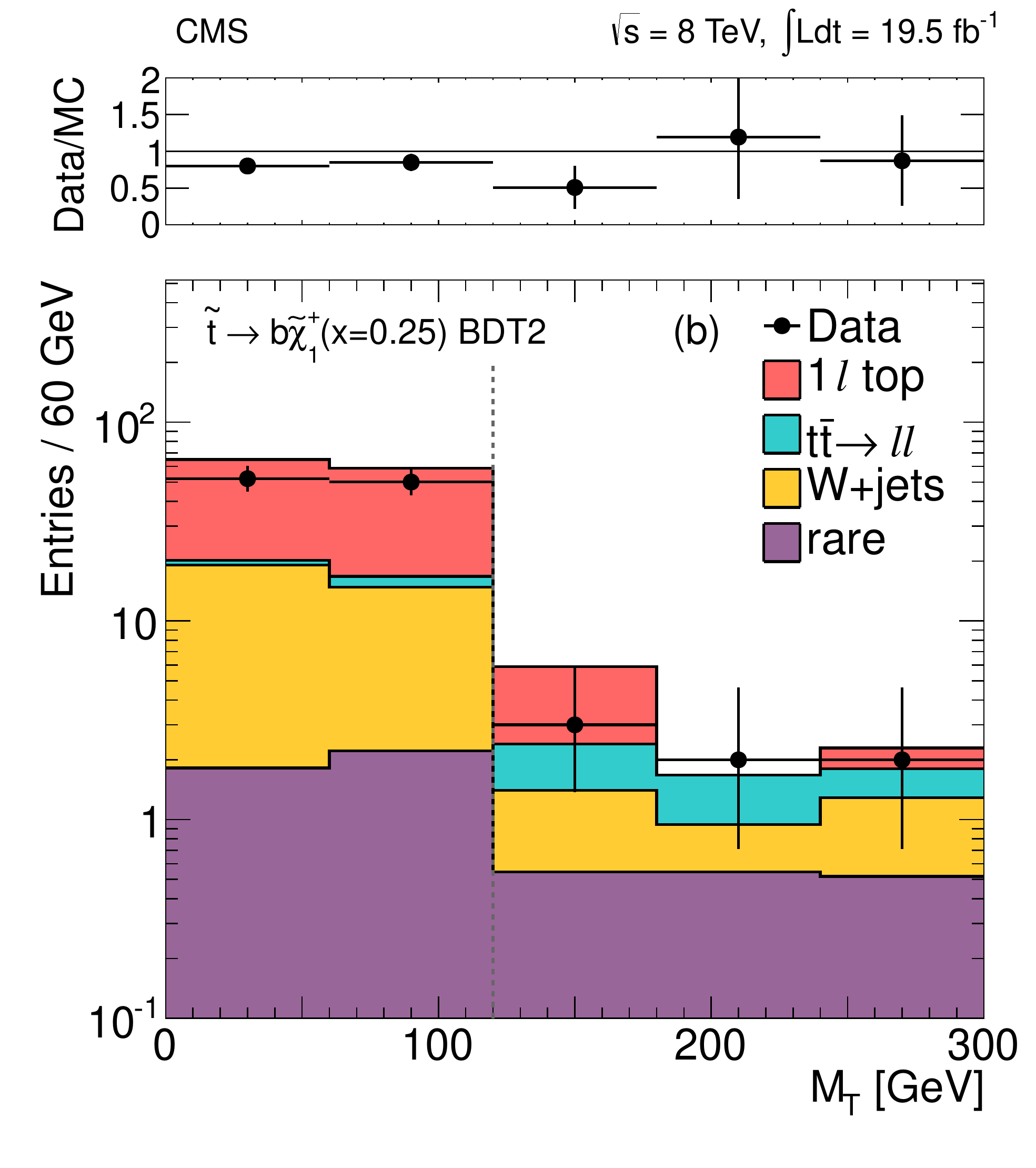} \\
    \includegraphics[width=0.49\linewidth]{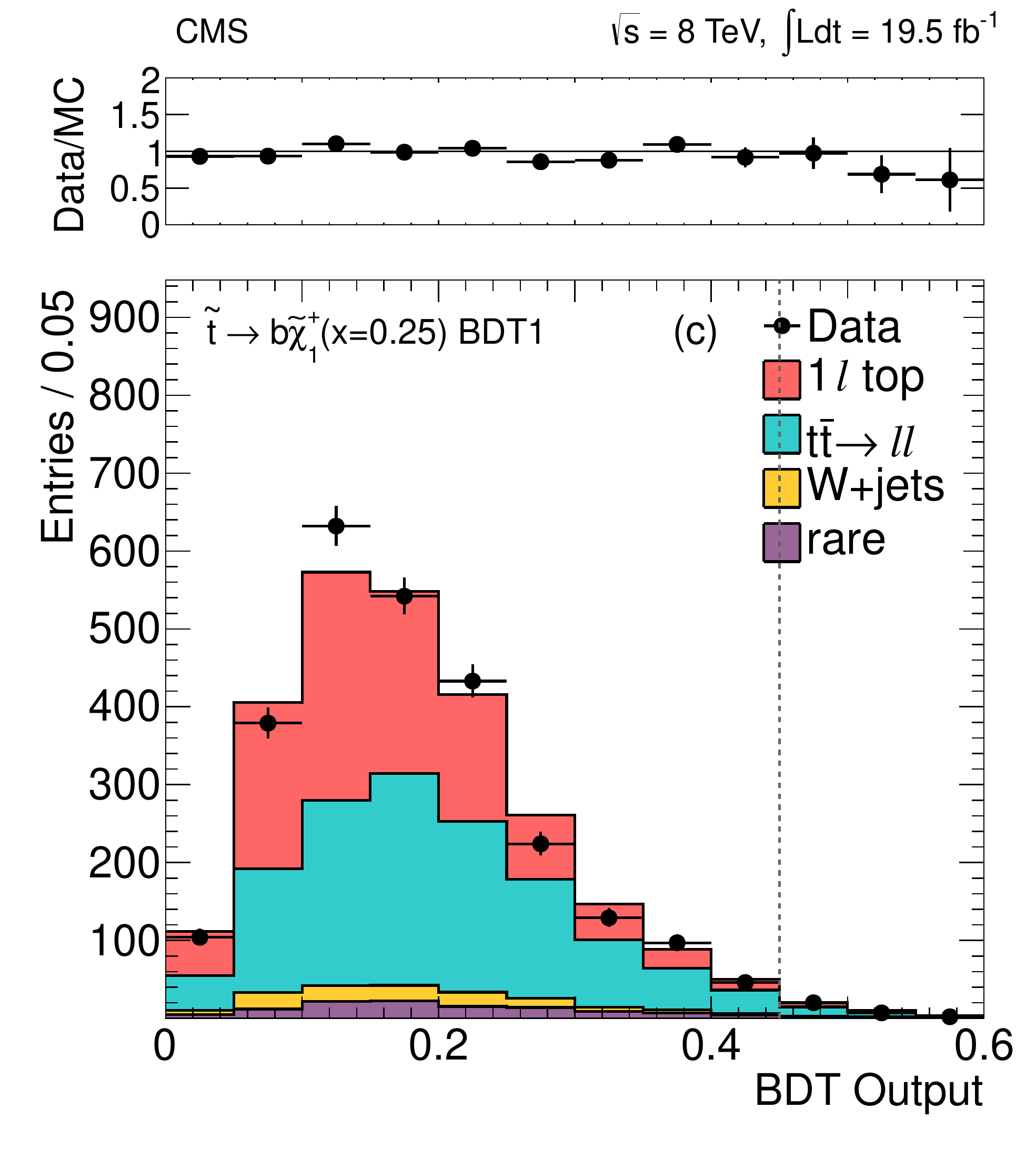}%
    \includegraphics[width=0.49\linewidth]{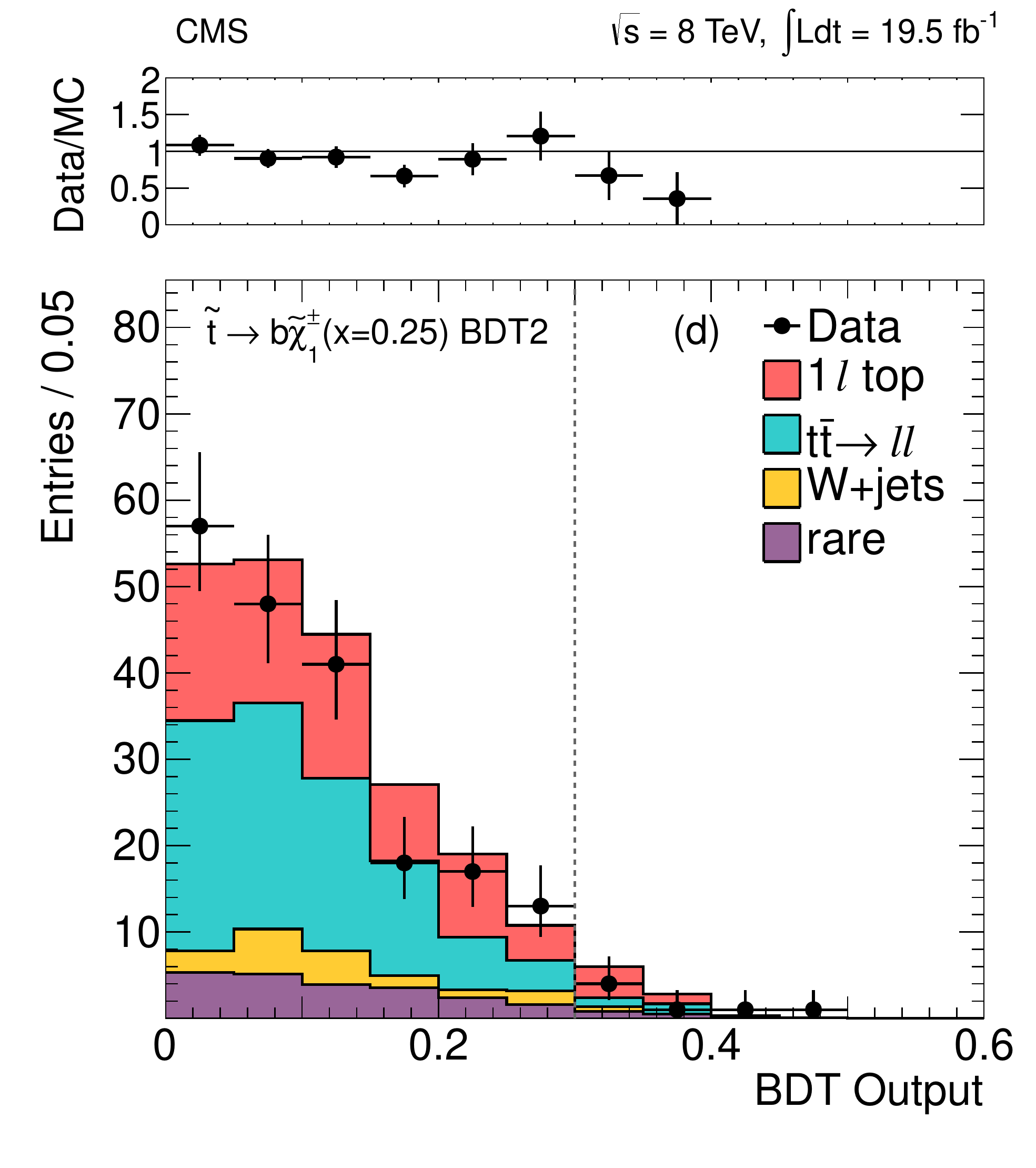} \\
    \caption{
      Comparison of data and MC simulation for the distributions of BDT output and \MT\ corresponding
      to the $x=0.25$ \TbW\ scenario in training regions 1 and 2.  The \MT\ distributions
      are shown after the requirement on the BDT output, and the BDT output distributions are shown after the
      $\MT> 120\GeV$ requirement (these requirements are also indicated by vertical dashed lines
      on the respective distributions).
      (a) \MT\ after the cut on the BDT1 output; 
      (b) \MT\ after the cut on the BDT2 output;
      (c) BDT1 output after the \MT\ cut;
      (d) BDT2 output after the \MT\ cut. In all distributions the last bin contains the overflow.
      \label{fig:plots3}
    }
      \end{center}
\end{figure*}

\begin{figure*}[htb]
  \begin{center}
    \includegraphics[width=0.49\textwidth]{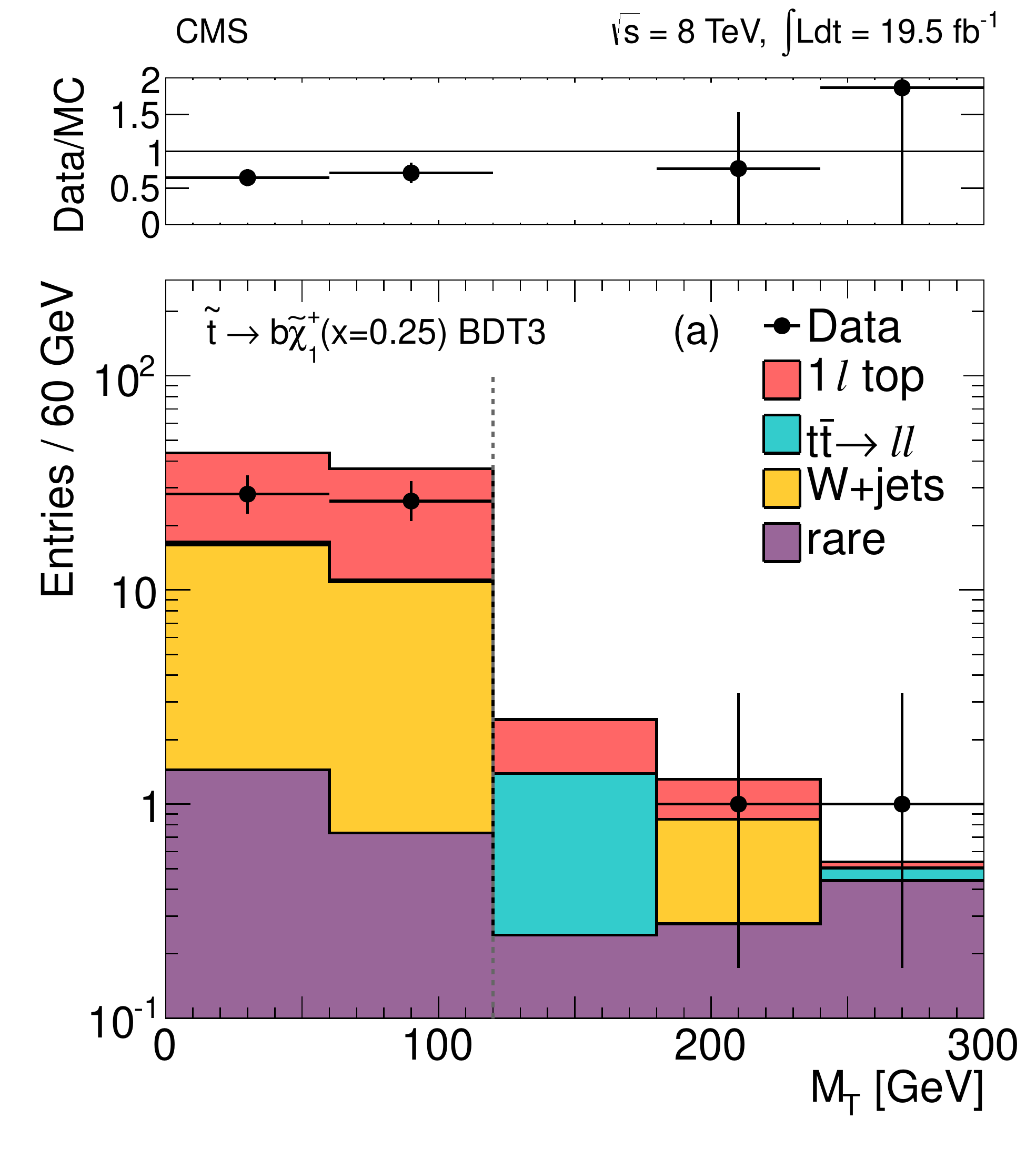}%
    \includegraphics[width=0.49\textwidth]{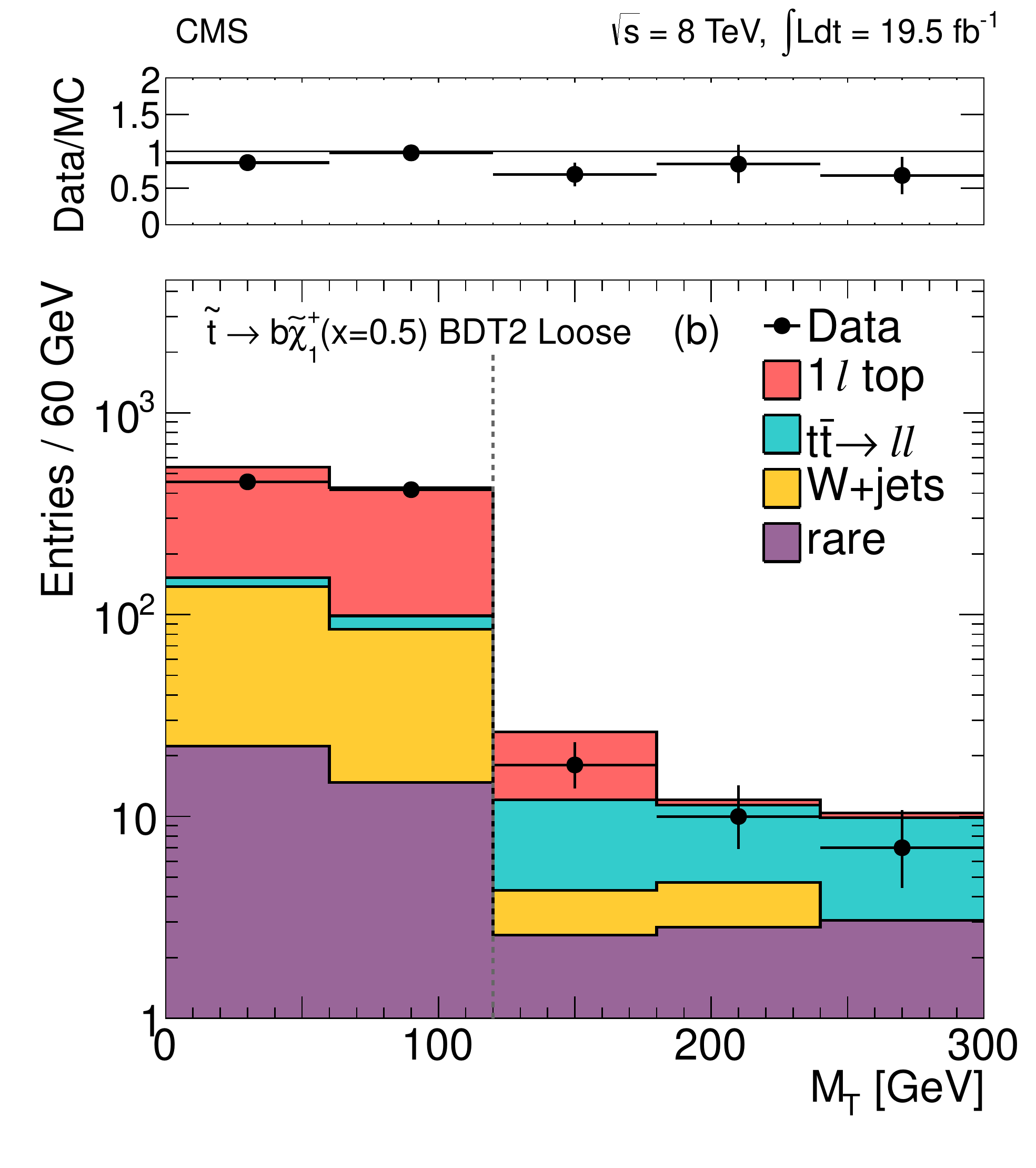} \\
    \includegraphics[width=0.49\textwidth]{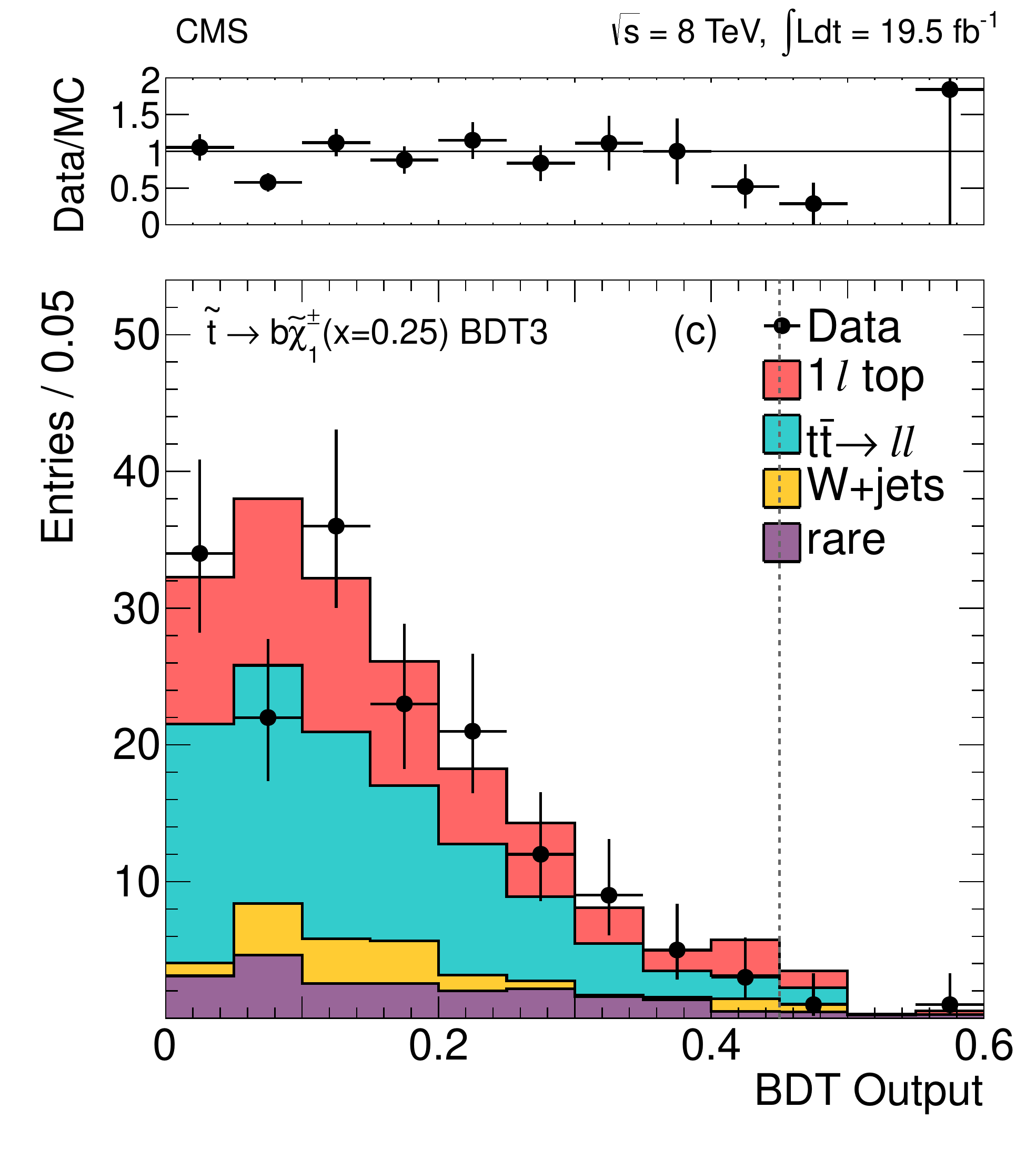}%
    \includegraphics[width=0.49\textwidth]{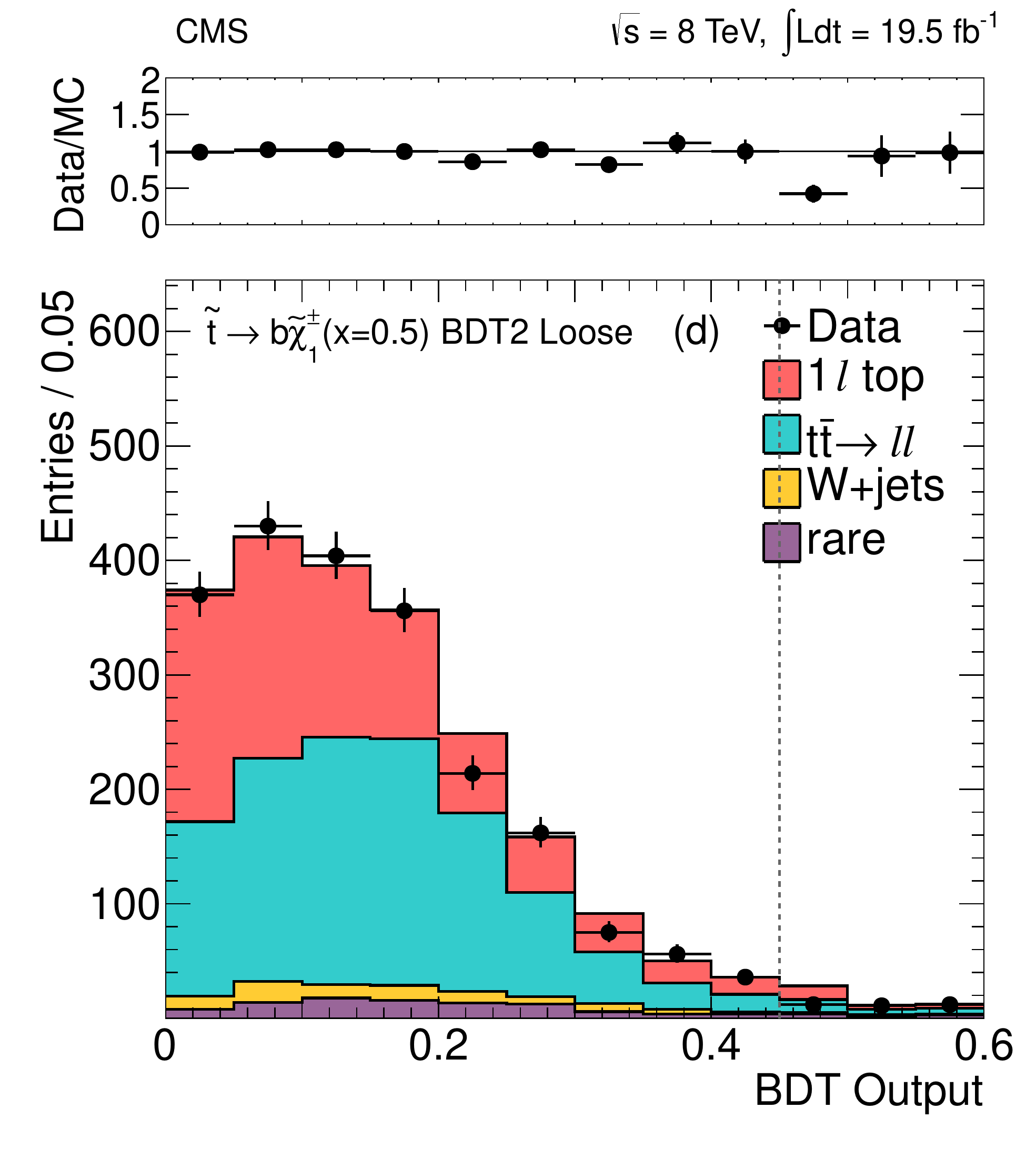} \\
    \caption{
      Comparison of data and MC simulation for the distributions of BDT output and \MT\ corresponding
      to the \TbW\ scenario in training regions 3 (for $x=0.25$) and 2 (for $x=0.5$).  The \MT\ distributions
      are shown after the requirement on the BDT output, and the BDT output distributions are shown after the
      $\MT> 120\GeV$ requirement (these requirements are also indicated by vertical dashed lines
      on the respective distributions).
      (a) \MT\ after the cut on the BDT3 ($x=0.25$) output; 
      (b) \MT\ after the loose cut on the BDT2 ($x=0.5$) output;
      (c) BDT3 ($x=0.25$) output after the \MT\ cut;
      (d) BDT2 ($x=0.5$) output after the \MT\ cut. In all distributions the last bin contains the overflow.
      \label{fig:plots4}
    }
      \end{center}
\end{figure*}

\begin{figure*}[htb]
  \begin{center}
    \includegraphics[width=0.49\textwidth]{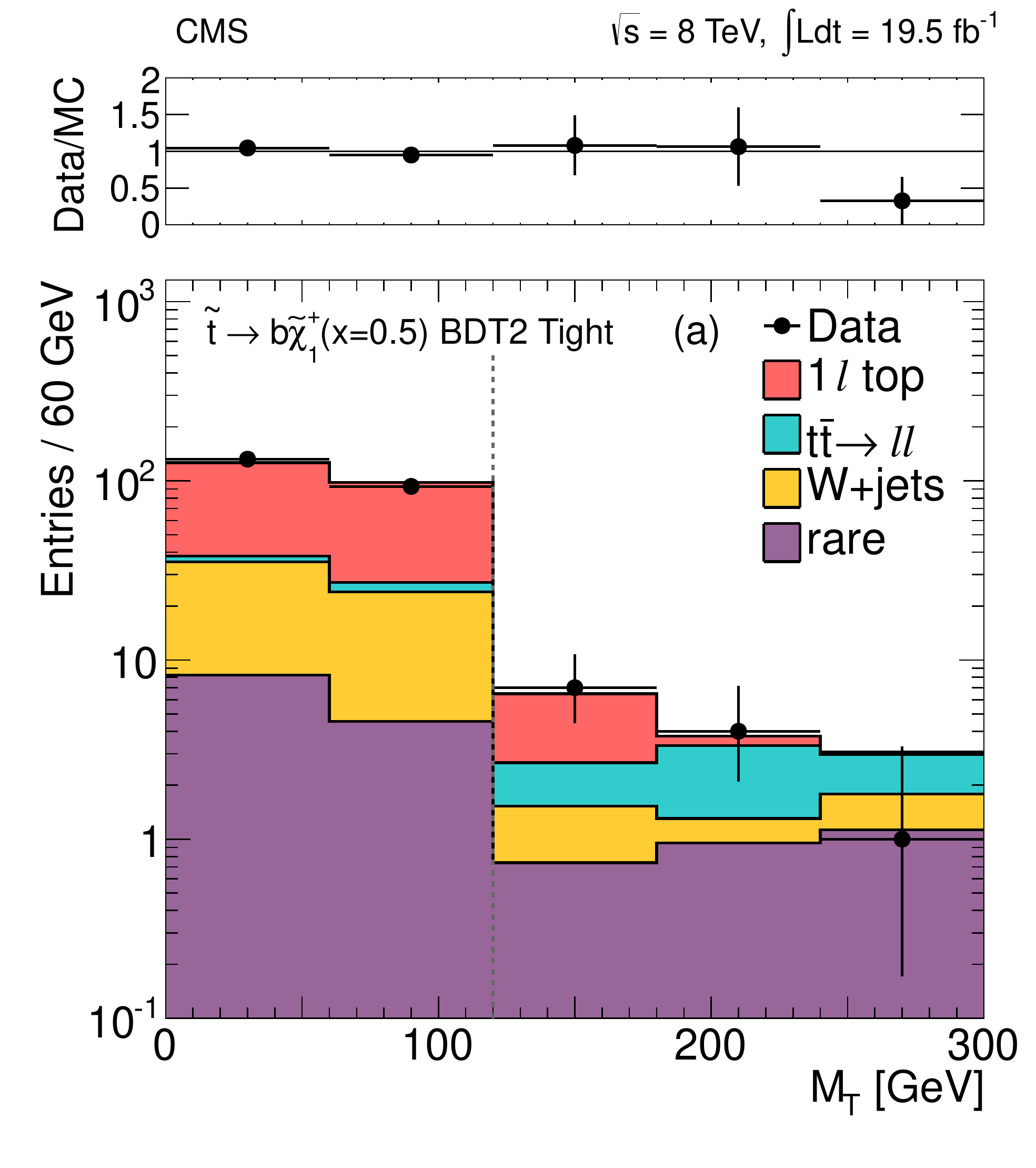}%
    \includegraphics[width=0.49\textwidth]{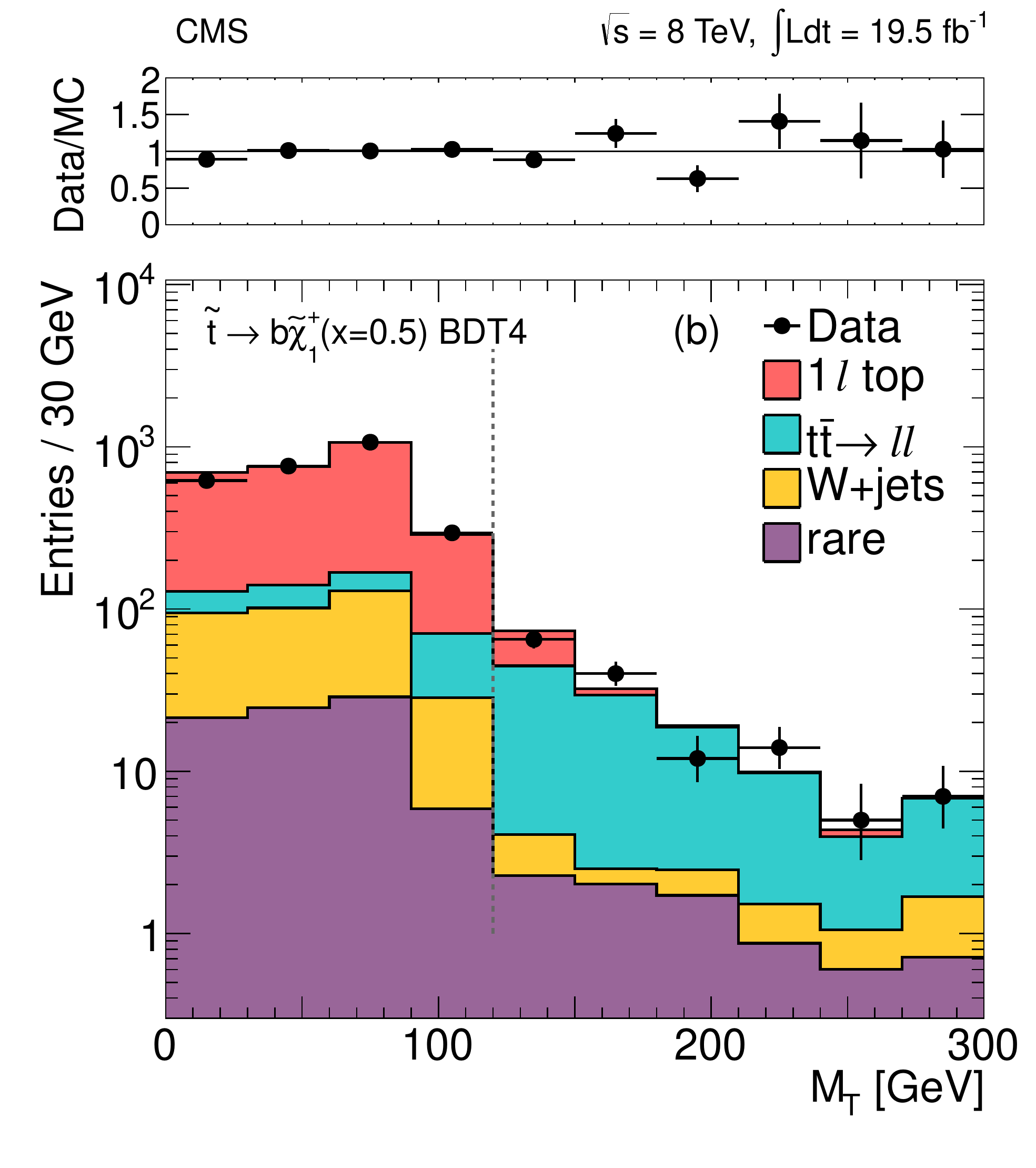} \\
    \includegraphics[width=0.49\textwidth]{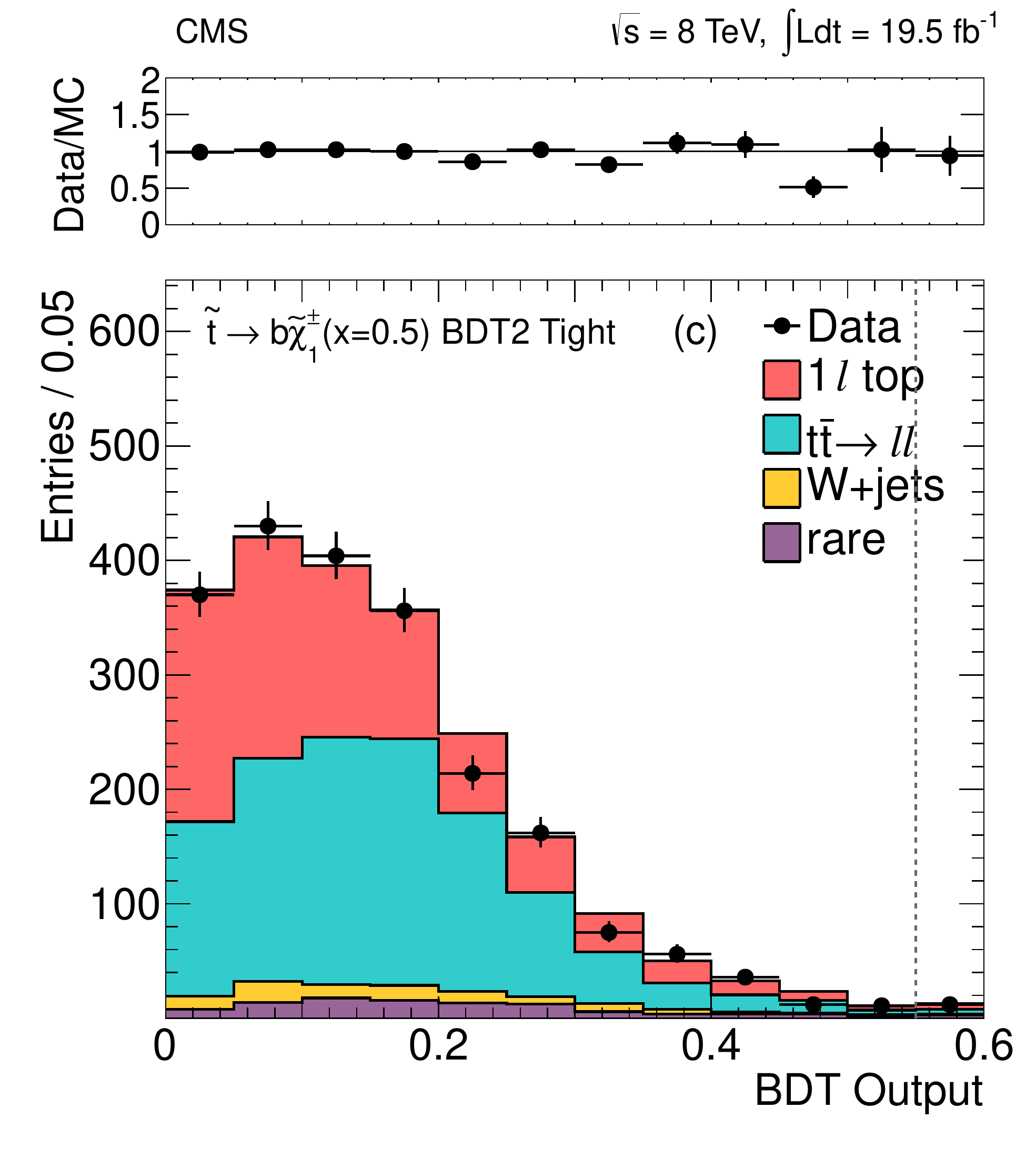}%
    \includegraphics[width=0.49\textwidth]{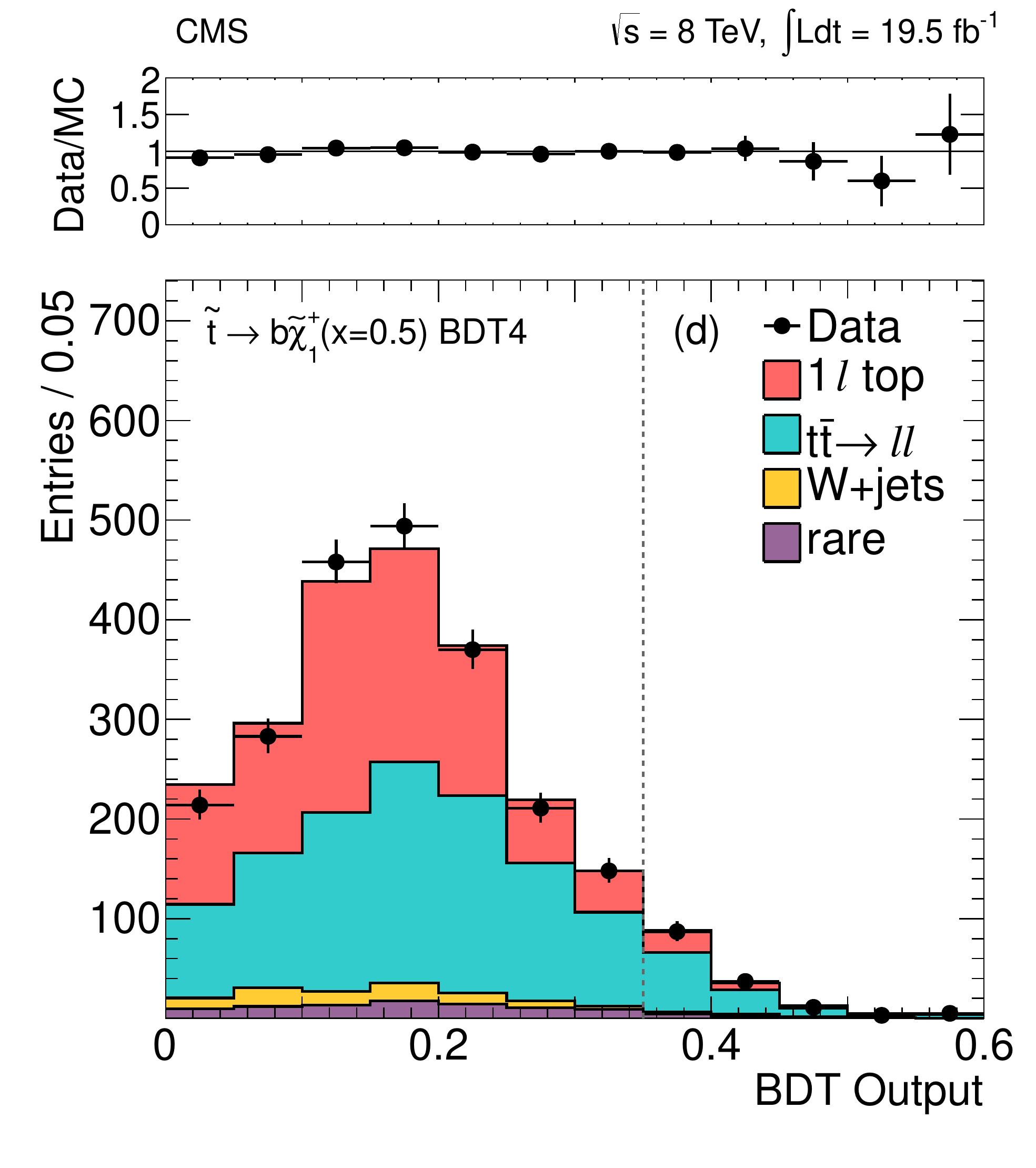} \\
    \caption{
      Comparison of data and MC simulation for the distributions of BDT output and \MT\ corresponding
      to the $x=0.5$ \TbW\ scenario in training regions 2 and 4.  The \MT\ distributions
      are shown after the requirement on the BDT output, and the BDT output distributions are shown after the
      $\MT> 120\GeV$ requirement (these requirements are also indicated by vertical dashed lines
      on the respective distributions).
      (a) \MT\ after the tight cut on the BDT2 output; 
      (b) \MT\ after the cut on the BDT4 output;
      (c) BDT2 output after the \MT\ cut;
      (d) BDT4 output after the \MT\ cut.  In all distributions the last bin contains the overflow.
      \label{fig:plots5}
    }
      \end{center}
\end{figure*}

\begin{figure*}[htb]
  \begin{center}
    \includegraphics[width=0.49\textwidth]{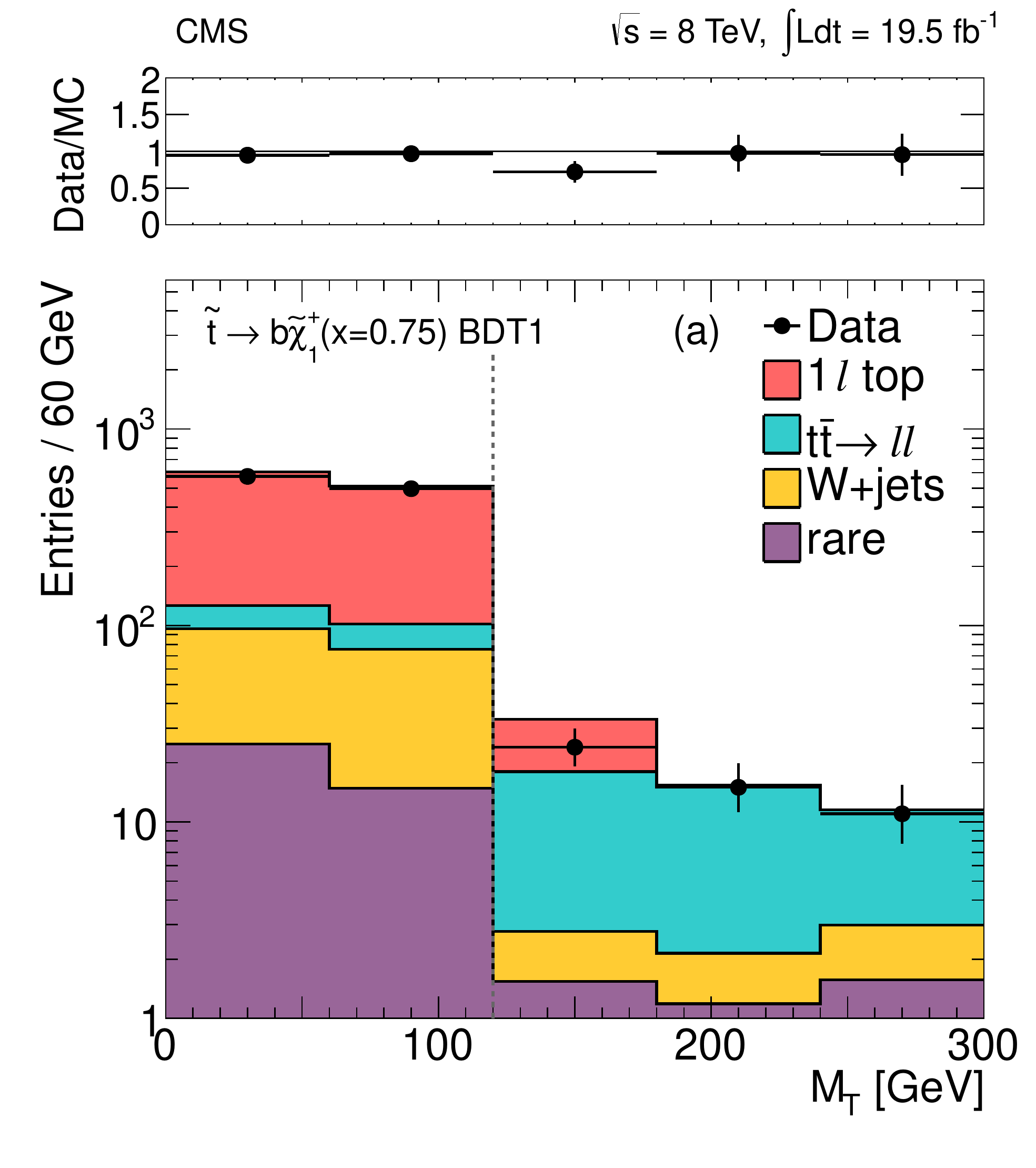}%
    \includegraphics[width=0.49\textwidth]{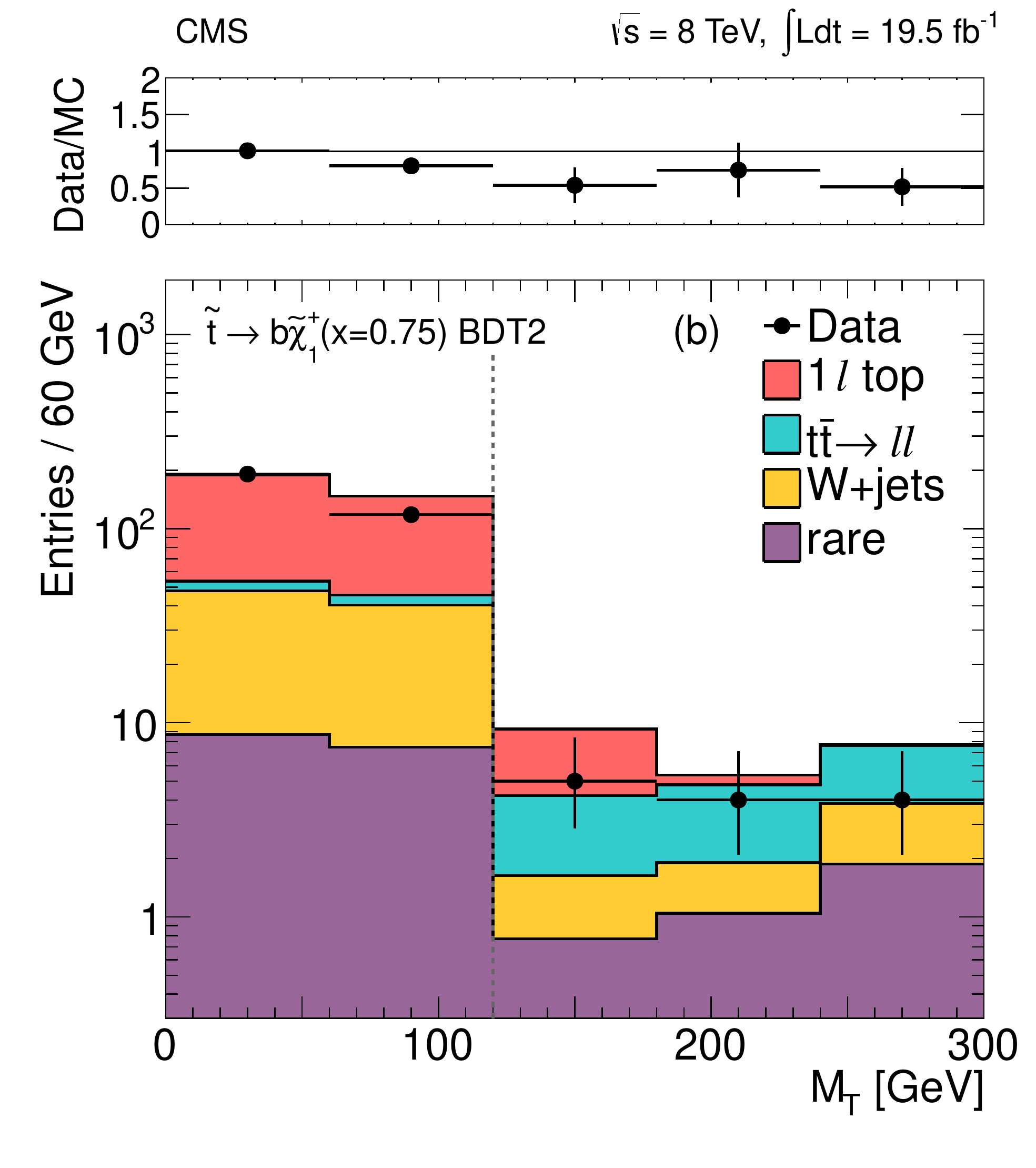} \\
    \includegraphics[width=0.49\textwidth]{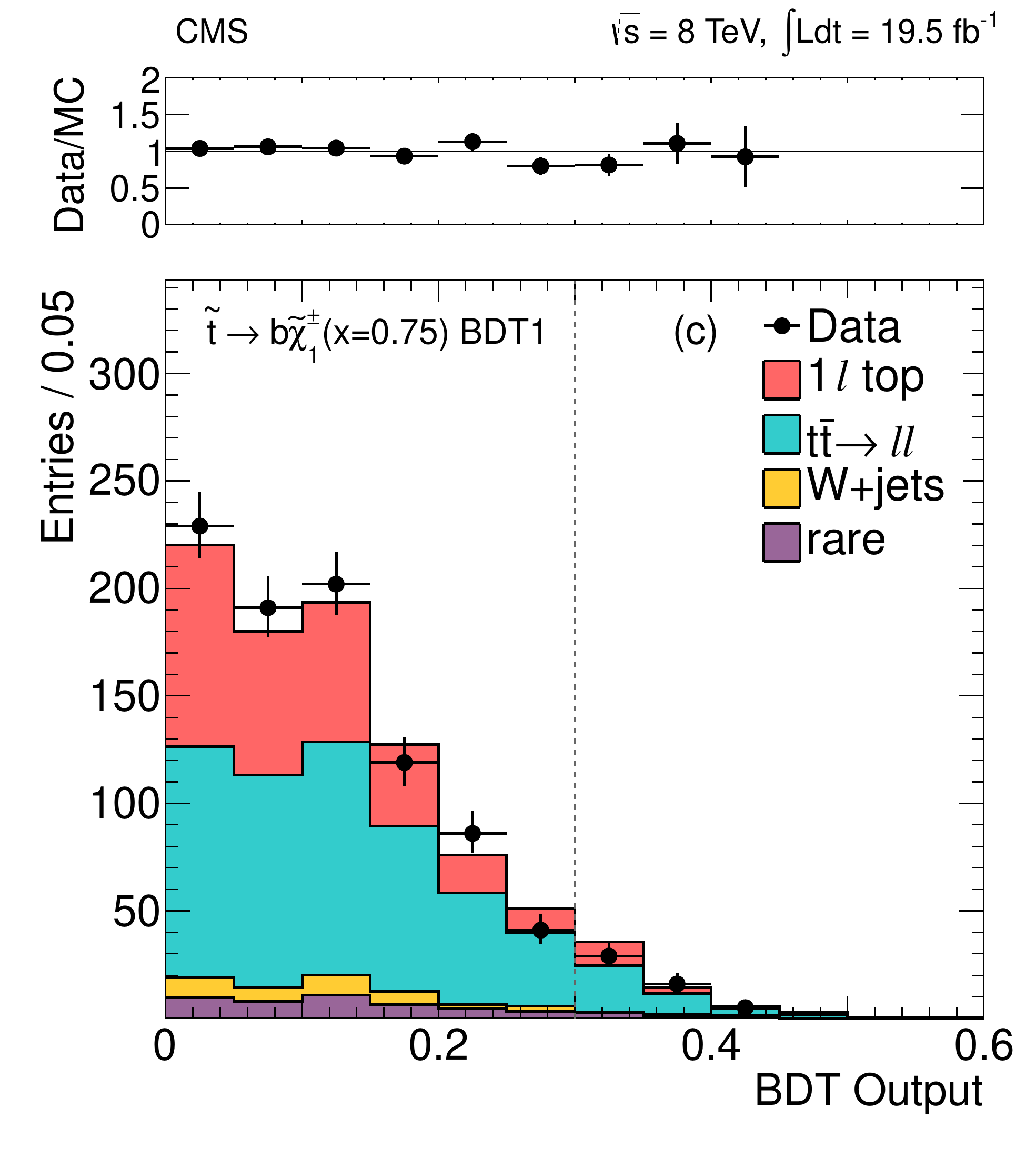}%
    \includegraphics[width=0.49\textwidth]{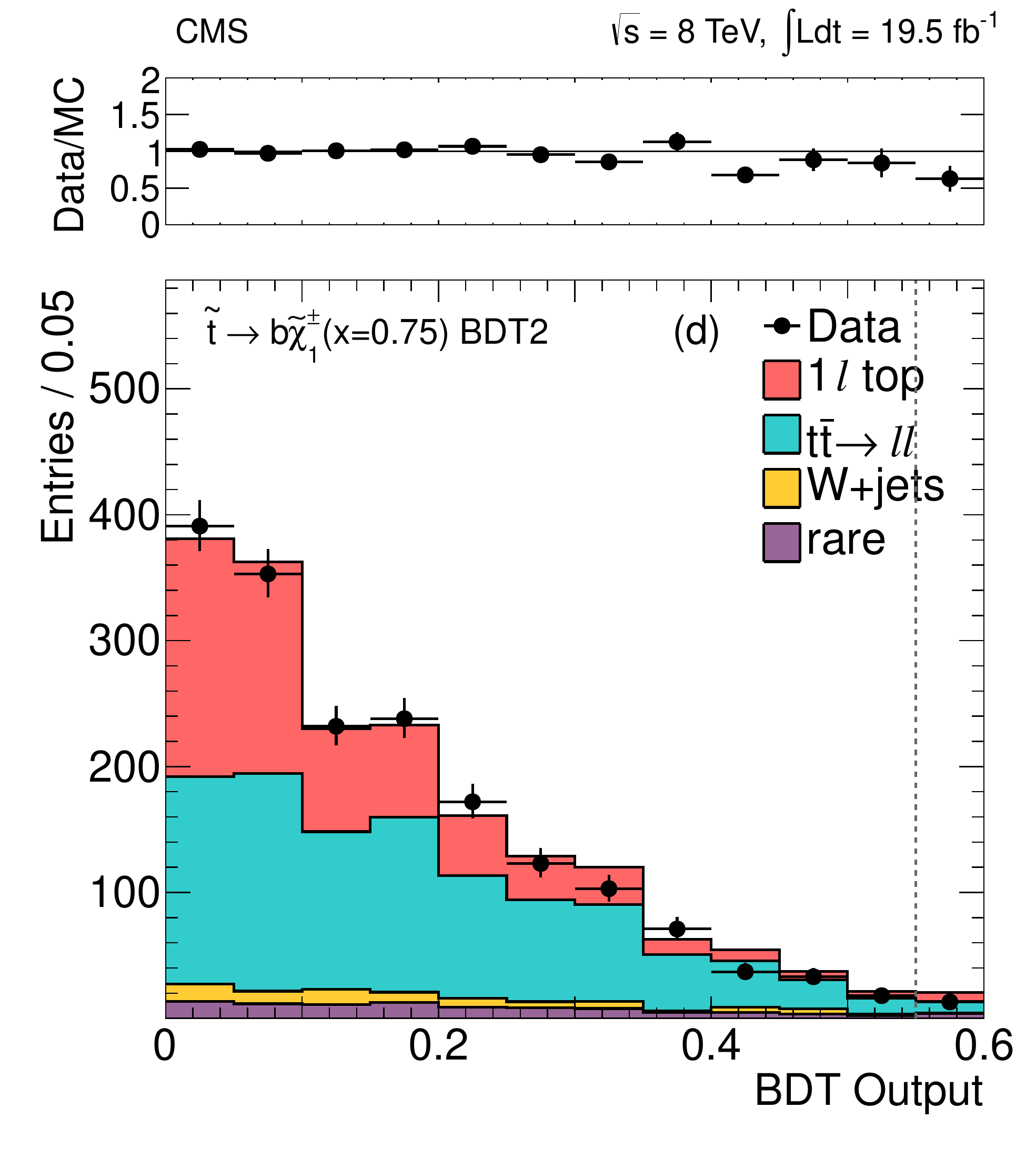} \\
    \caption{
      Comparison of data and MC simulation for the distributions of BDT output and \MT\ corresponding
      to the $x=0.75$ \TbW\ scenario in training regions 1 and 2.  The \MT\ distributions
      are shown after the requirement on the BDT output, and the BDT output distributions are shown after the
      $\MT> 120\GeV$ requirement (these requirements are also indicated by vertical dashed lines
      on the respective distributions).
      (a) \MT\ after the cut on the BDT1 output; 
      (b) \MT\ after the cut on the BDT2 output;
      (c) BDT1 output after the \MT\ cut;
      (d) BDT2 output after the \MT\ cut. In all distributions the last bin contains the overflow.
      \label{fig:plots6}
    }
      \end{center}
\end{figure*}

\begin{figure*}[htb]
  \begin{center}
    \includegraphics[width=0.49\textwidth]{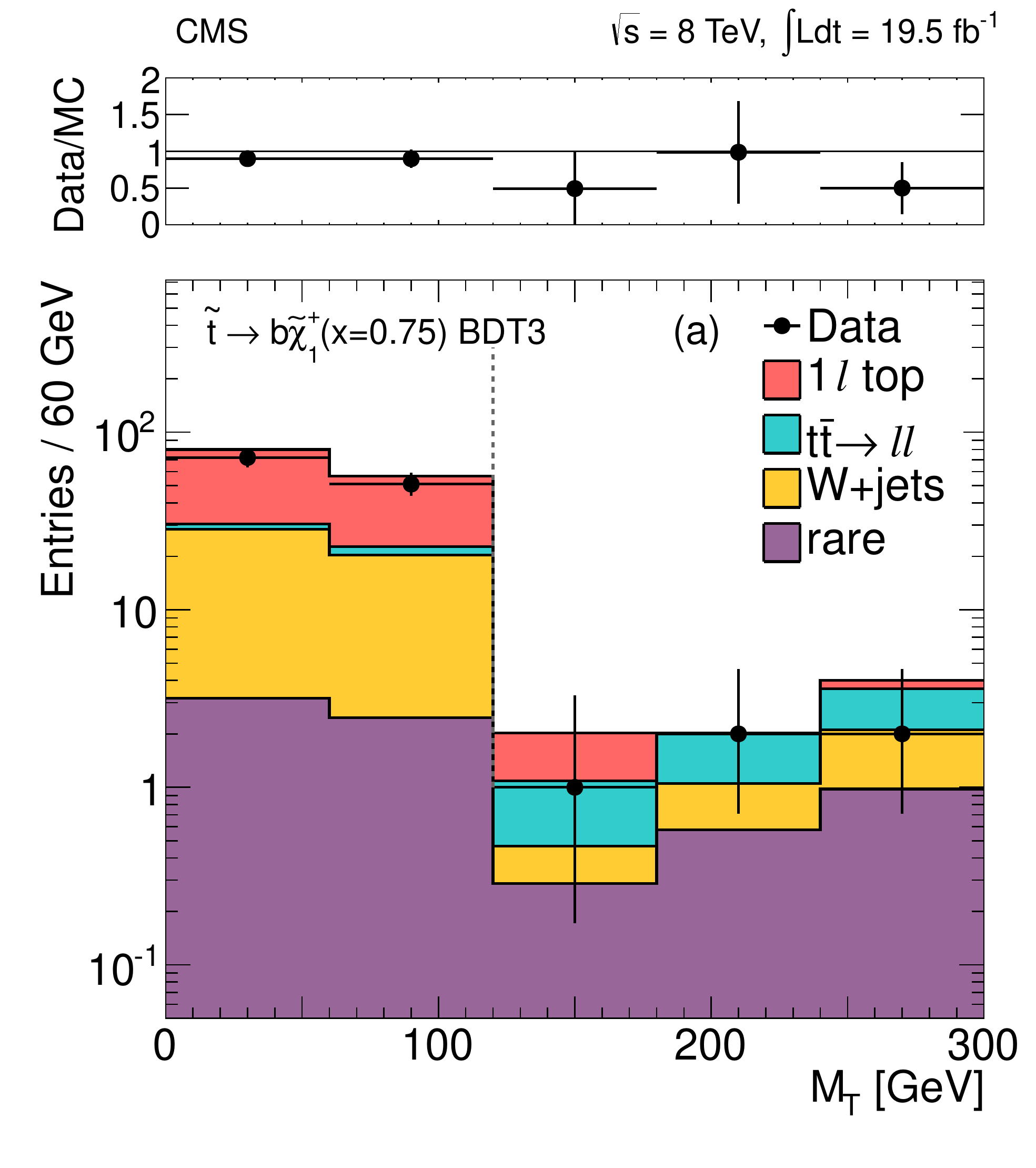}%
    \includegraphics[width=0.49\textwidth]{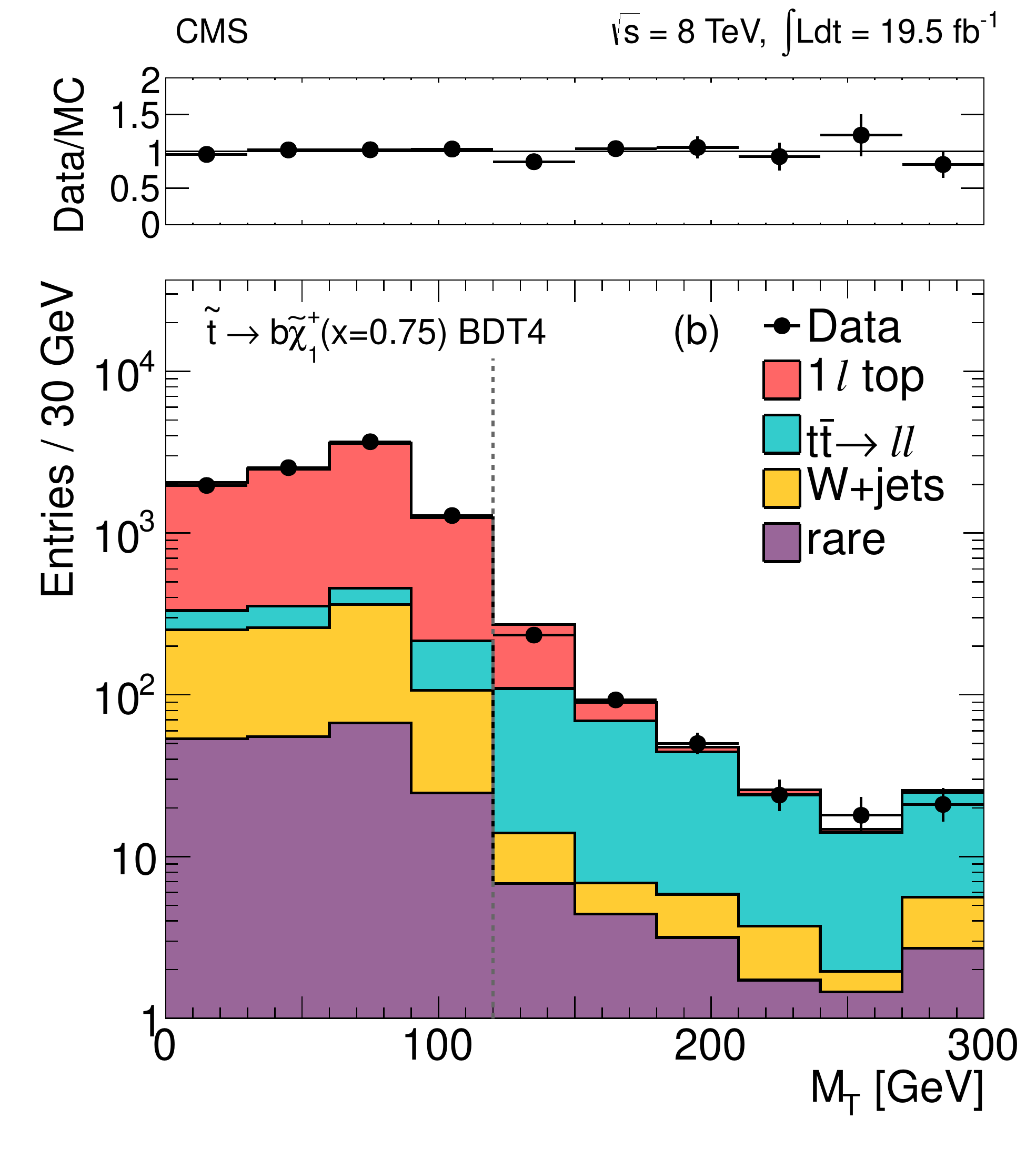} \\
    \includegraphics[width=0.49\textwidth]{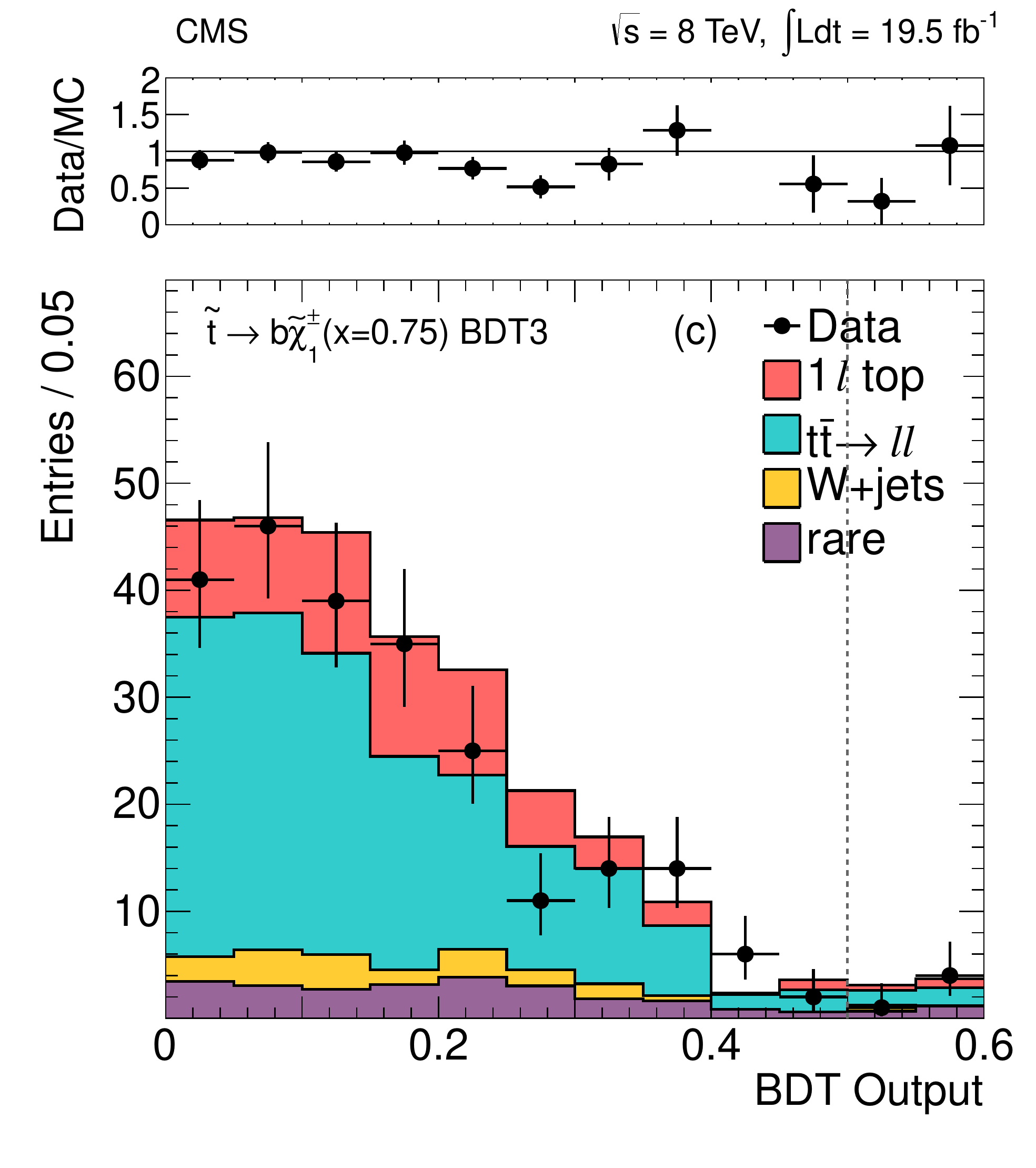}%
    \includegraphics[width=0.49\textwidth]{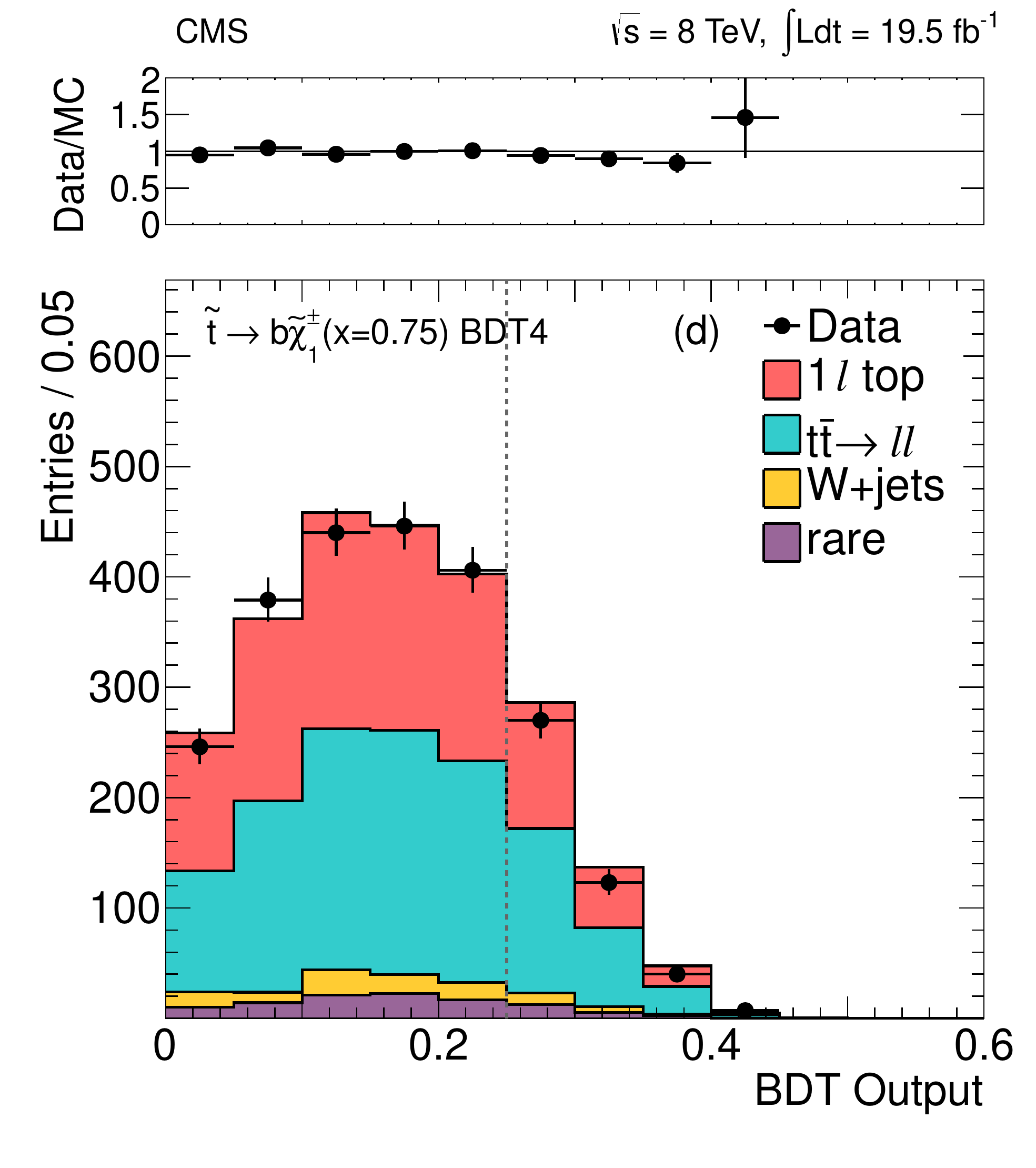} \\
    \caption{
      Comparison of data and MC simulation for the distributions of BDT output and \MT\ corresponding
      to the $x=0.75$ \TbW\ scenario in training regions 3 and 4.  The \MT\ distributions
      are shown after the requirement on the BDT output, and the BDT output distributions are shown after the
      $\MT> 120\GeV$ requirement (these requirements are also indicated by vertical dashed lines
      on the respective distributions).
      (a) \MT\ after the cut on the BDT3 output; 
      (b) \MT\ after the cut on the BDT4 output;
      (c) BDT3 output after the \MT\ cut;
      (d) BDT4 output after the \MT\ cut.  In all distributions the last bin contains the overflow.
      \label{fig:plots7}
    }
      \end{center}
\end{figure*}

In this section, \MT\ and BDT-output distributions
in addition to those shown in Figs.~\ref{fig:mt_t2tt} and~\ref{fig:mt_t2bw}
are presented for the \Ttt\ (Figs.~\ref{fig:plots1}--\ref{fig:plots2})
and \TbW\ (Figs.~\ref{fig:plots3}--\ref{fig:plots7}) BDT signal regions.

\ifthenelse{\boolean{cms@external}}{}{\clearpage}
\subsection{Further information about model interpretations}
\label{app:models}

\begin{figure*}[htbp]
\centering
\includegraphics[width=0.49\textwidth]{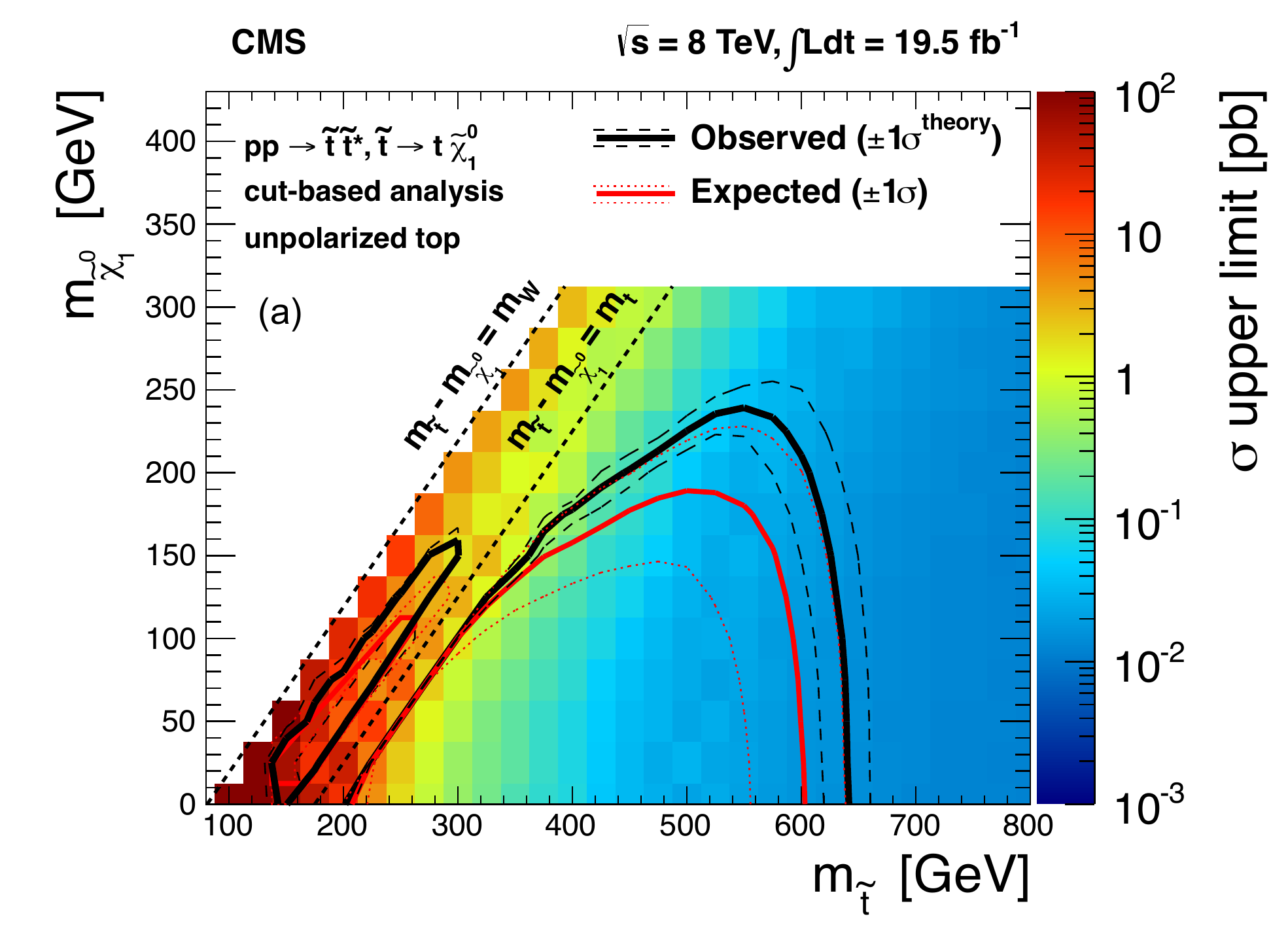}
\includegraphics[width=0.49\textwidth]{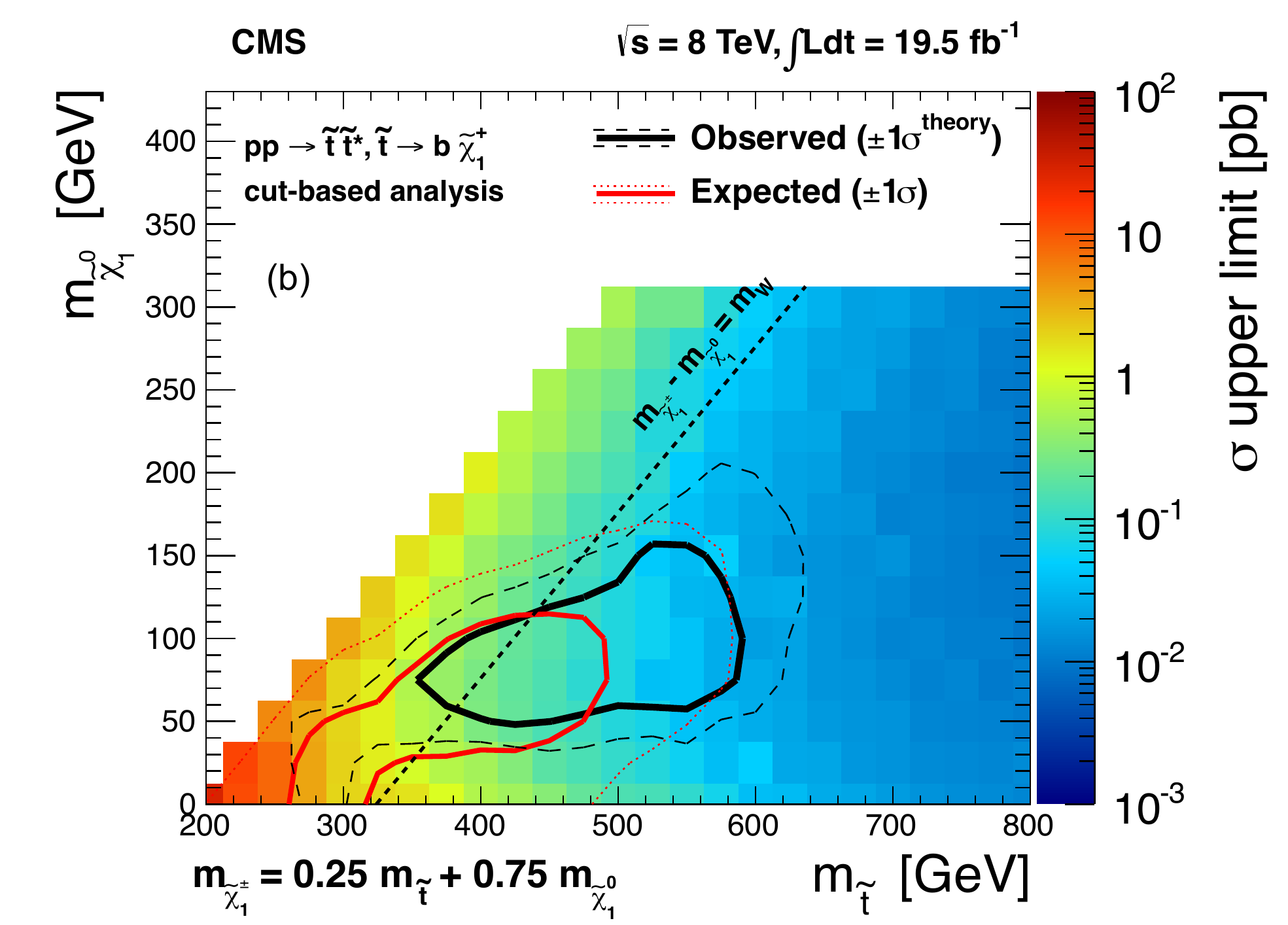}
\includegraphics[width=0.49\textwidth]{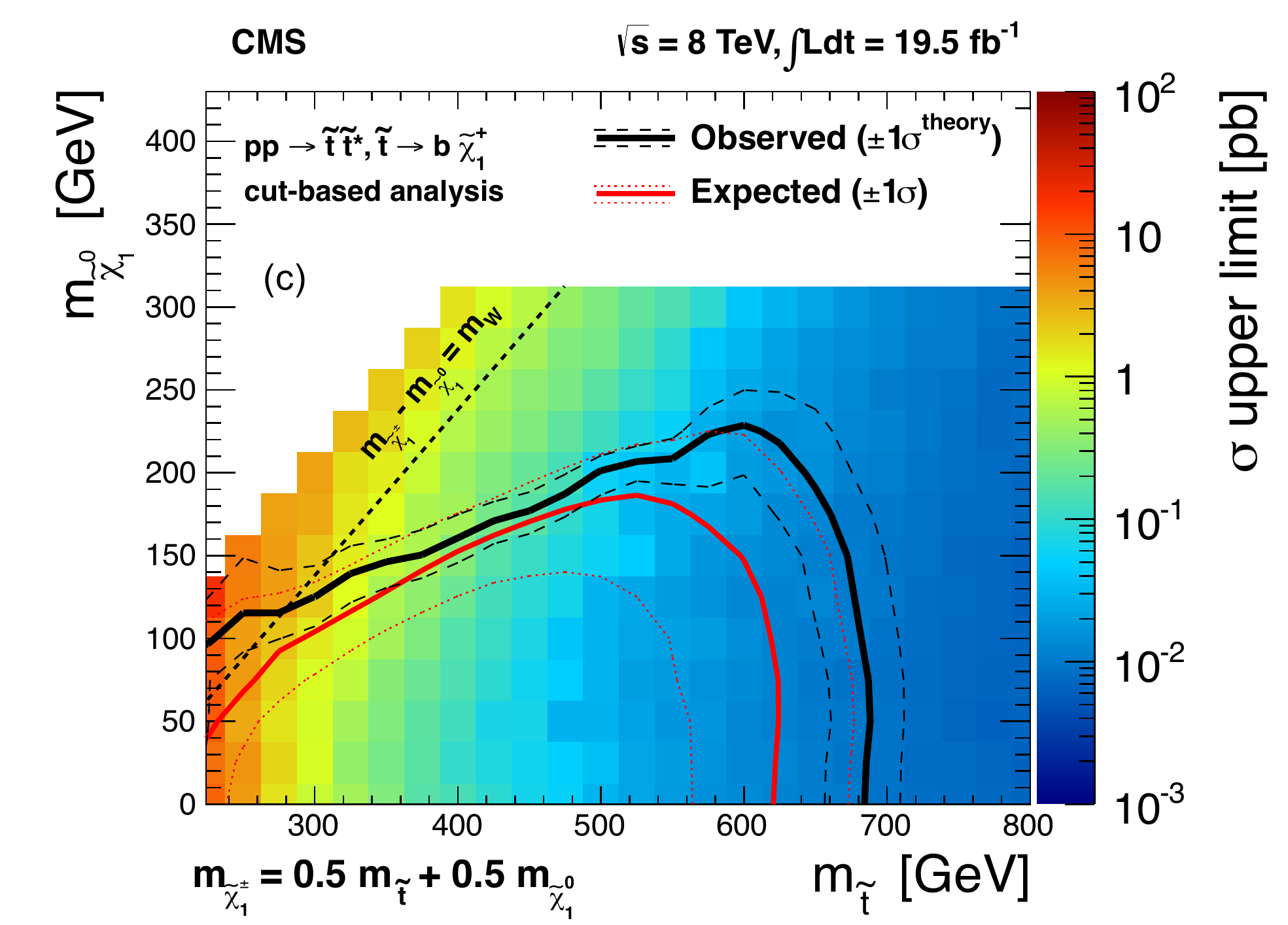}
\includegraphics[width=0.49\textwidth]{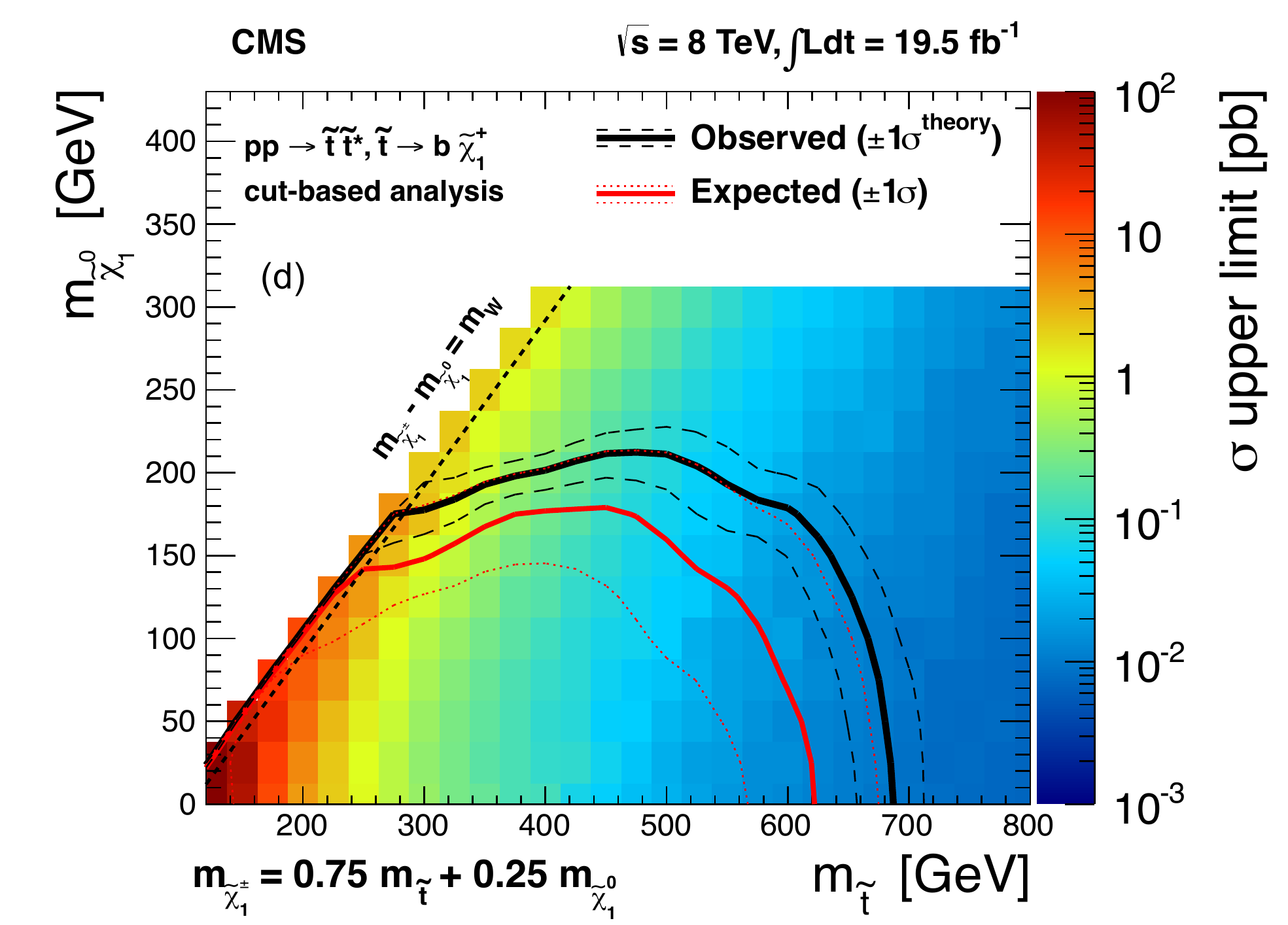}
\caption{
Interpretations based on the results of the cut-based analysis.
(a) \Ttt\ model; 
(b) \TbW\ model with $x=0.25$;
(c) \TbW\ model with $x=0.50$;
(d) \TbW\ model with $x=0.75$;
The color scale indicates the observed cross section upper limit.
The observed (solid black), median expected (solid red), and
${\pm}1\sigma$ expected (dotted red) 95\% CL exclusion contours
are indicated.
The variations in the excluded region due to ${\pm}1\sigma$ uncertainty
of the theoretical prediction of the cross section for top-squark pair
production are also indicated.
\label{fig:cac_interpretations}
}
\end{figure*}

\begin{figure*}[htbp]
\begin{center}
\begin{tabular}{cc}
\includegraphics[width=0.49\textwidth]{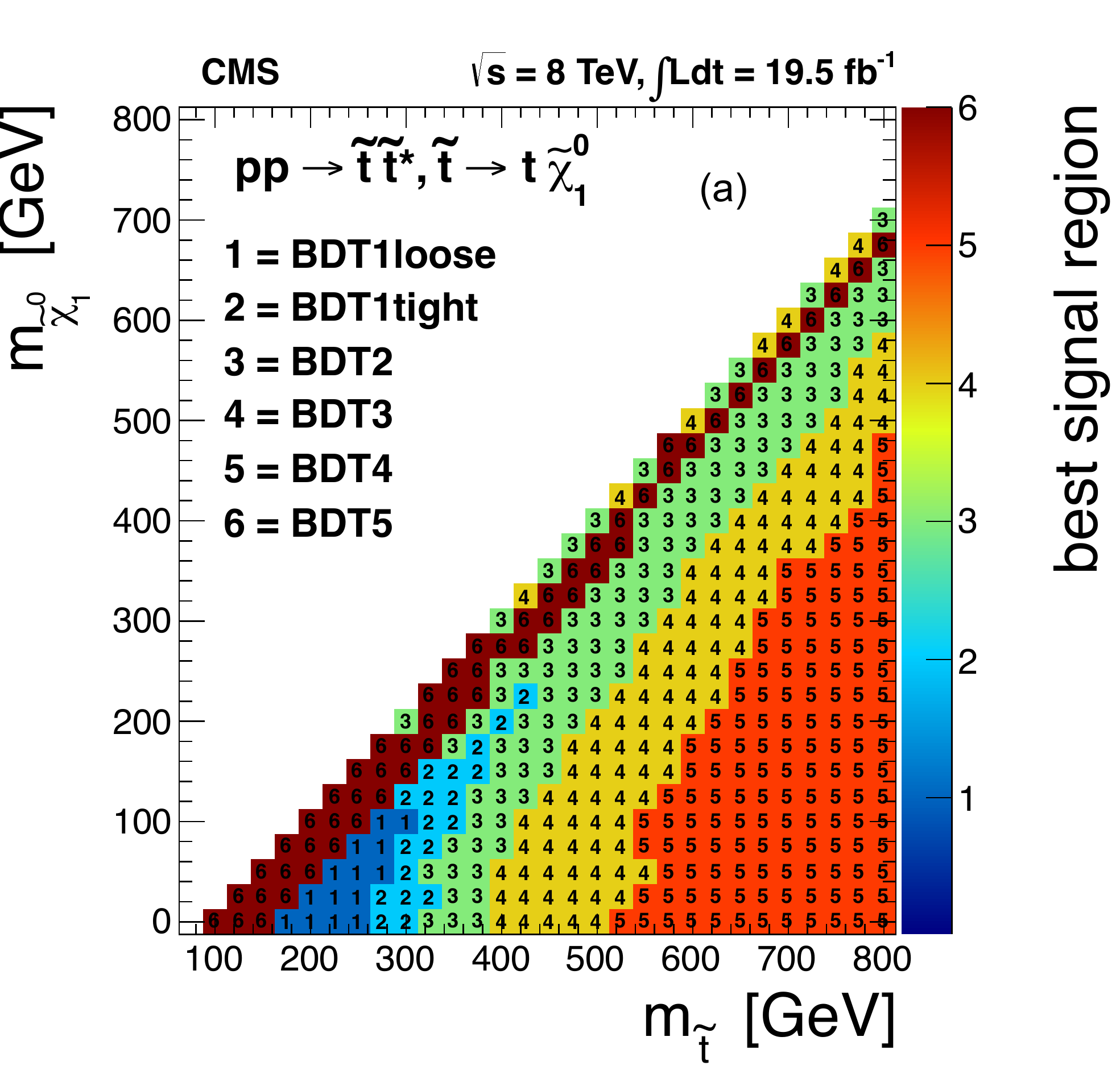} %
\includegraphics[width=0.49\textwidth]{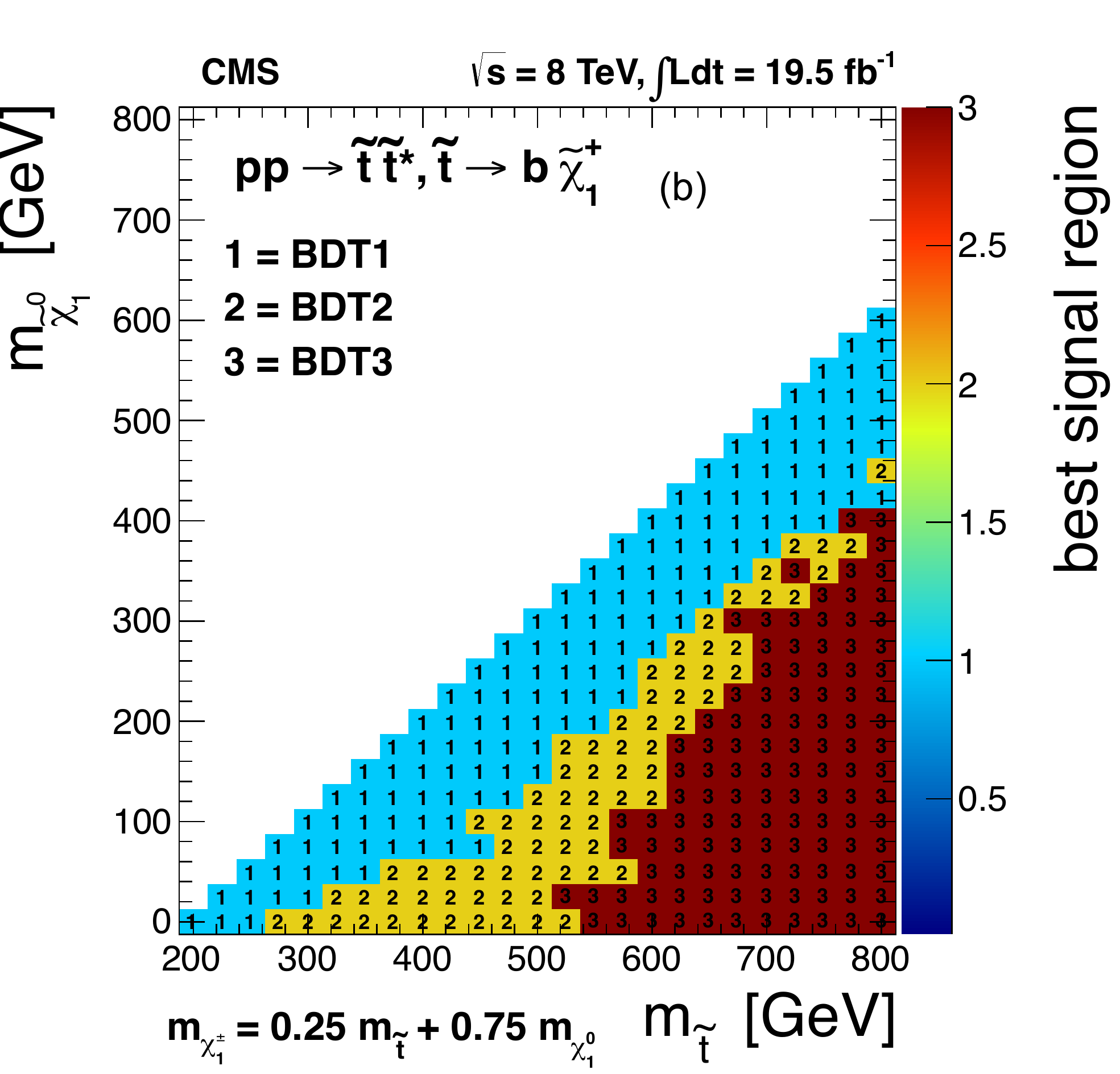} \\
\includegraphics[width=0.49\textwidth]{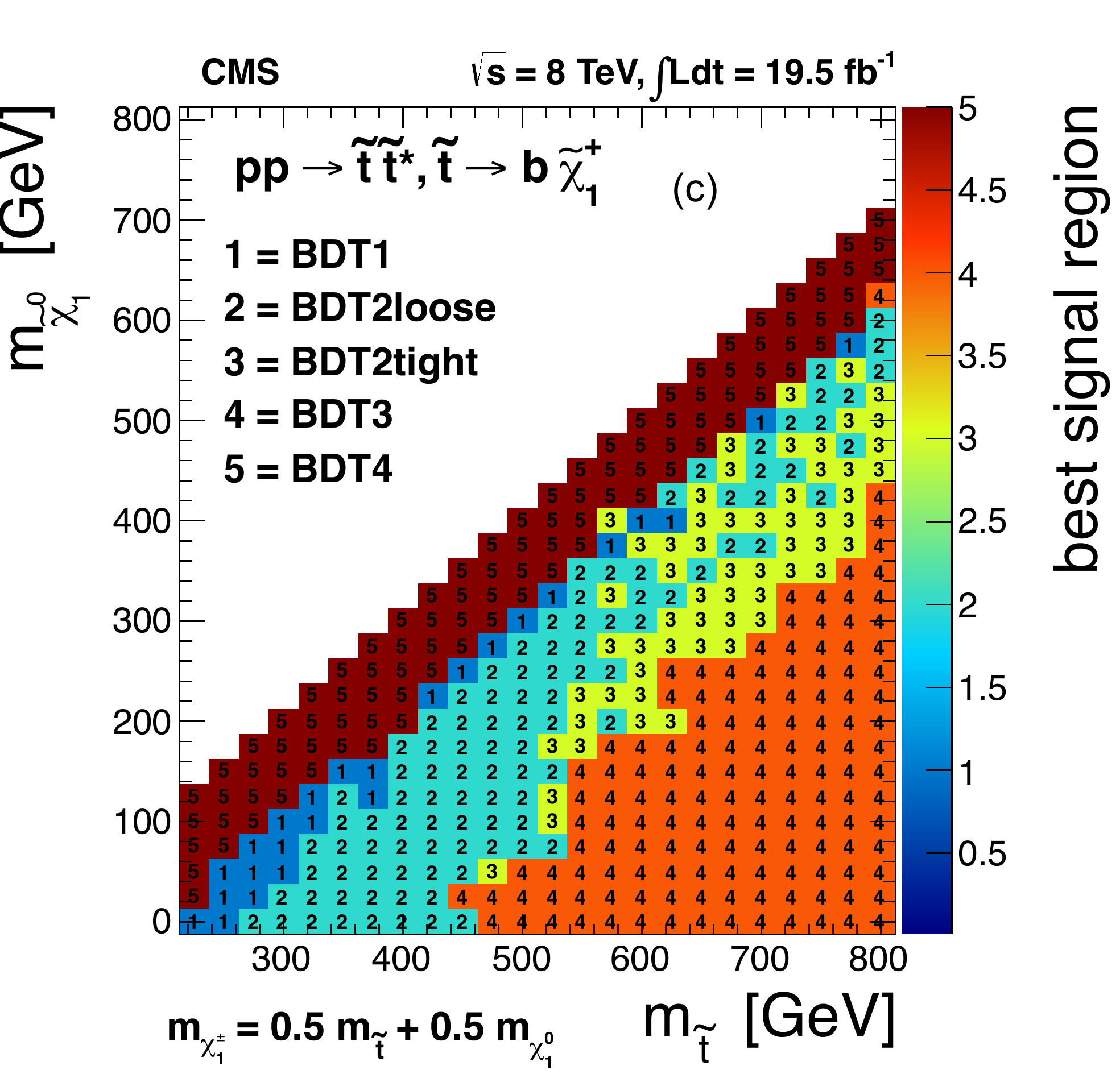} %
\includegraphics[width=0.49\textwidth]{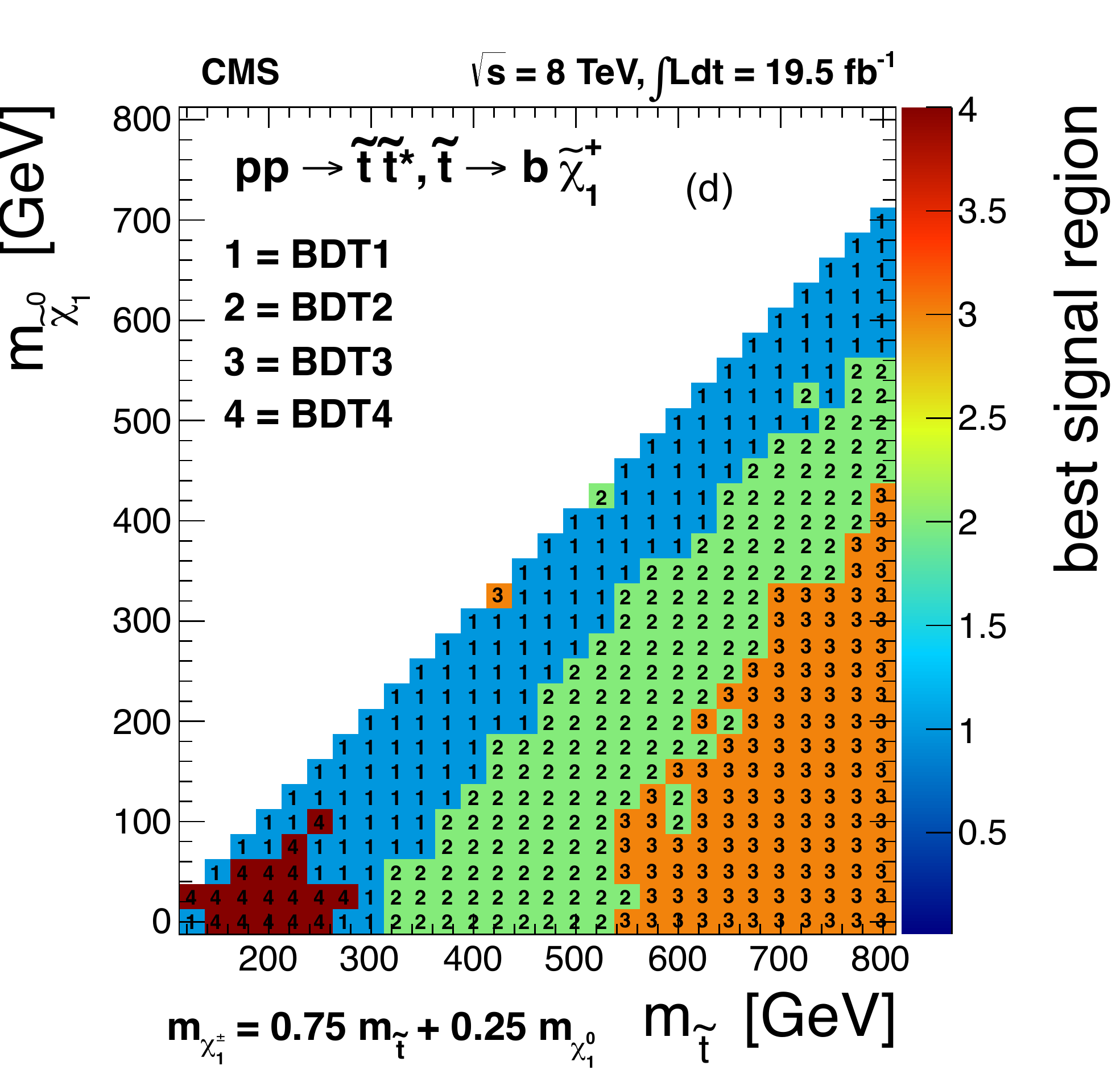}
\end{tabular}
\caption{ The most sensitive signal region in the $m_{\lsp}$ vs. $m_{\PSQt}$ parameter space
in the BDT analysis,
for the (a) \Ttt\ model, and the
\TbW\ model with chargino mass parameter (b) $x=0.25$,
(c) 0.5, and (d) 0.75. The number indicates the BDT training region.
\label{fig:bdt_bestregion}
}
\end{center}
\end{figure*}

\begin{figure*}[htbp]
\begin{center}
\includegraphics[width=0.49\textwidth]{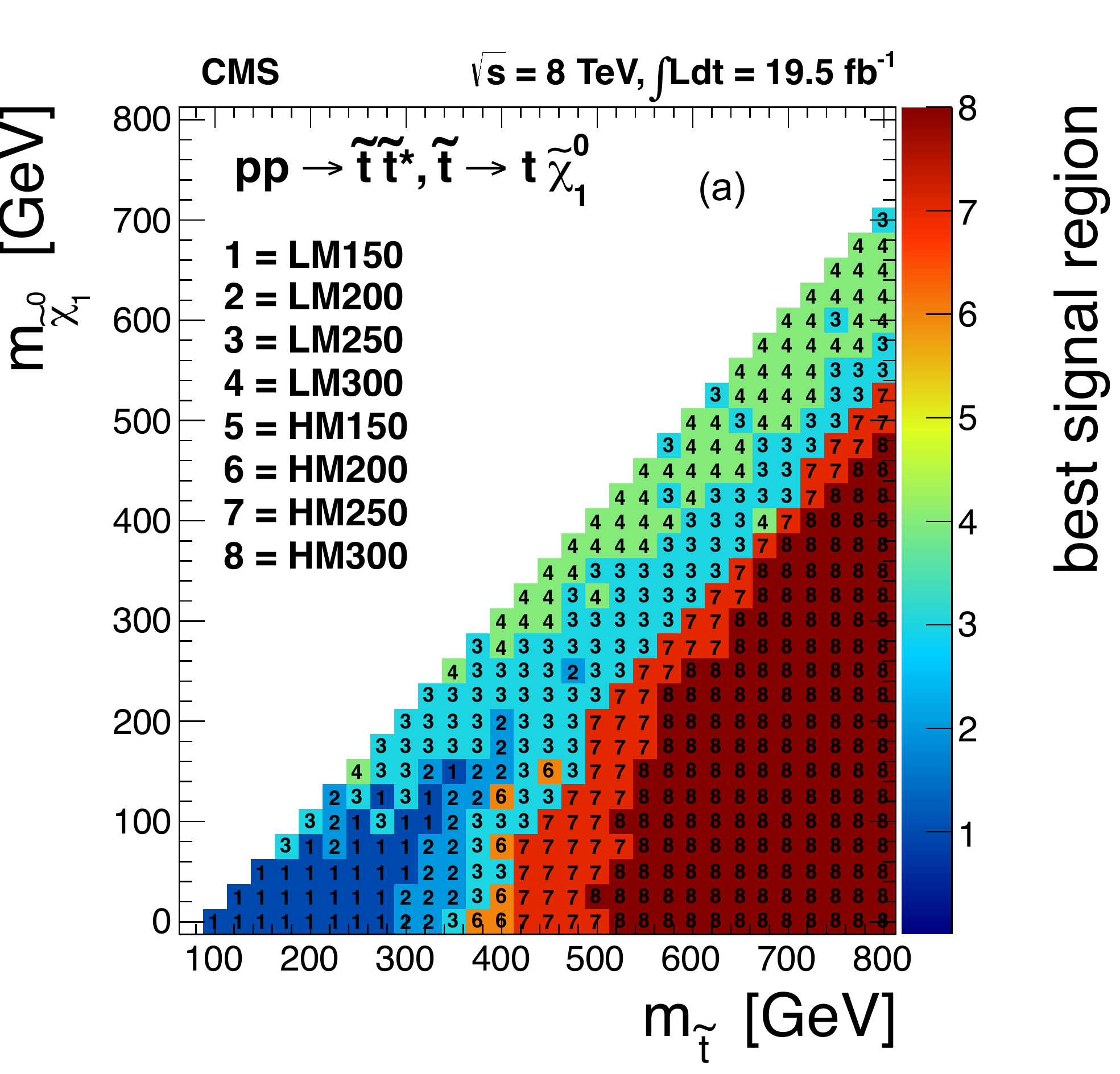}
\includegraphics[width=0.49\textwidth]{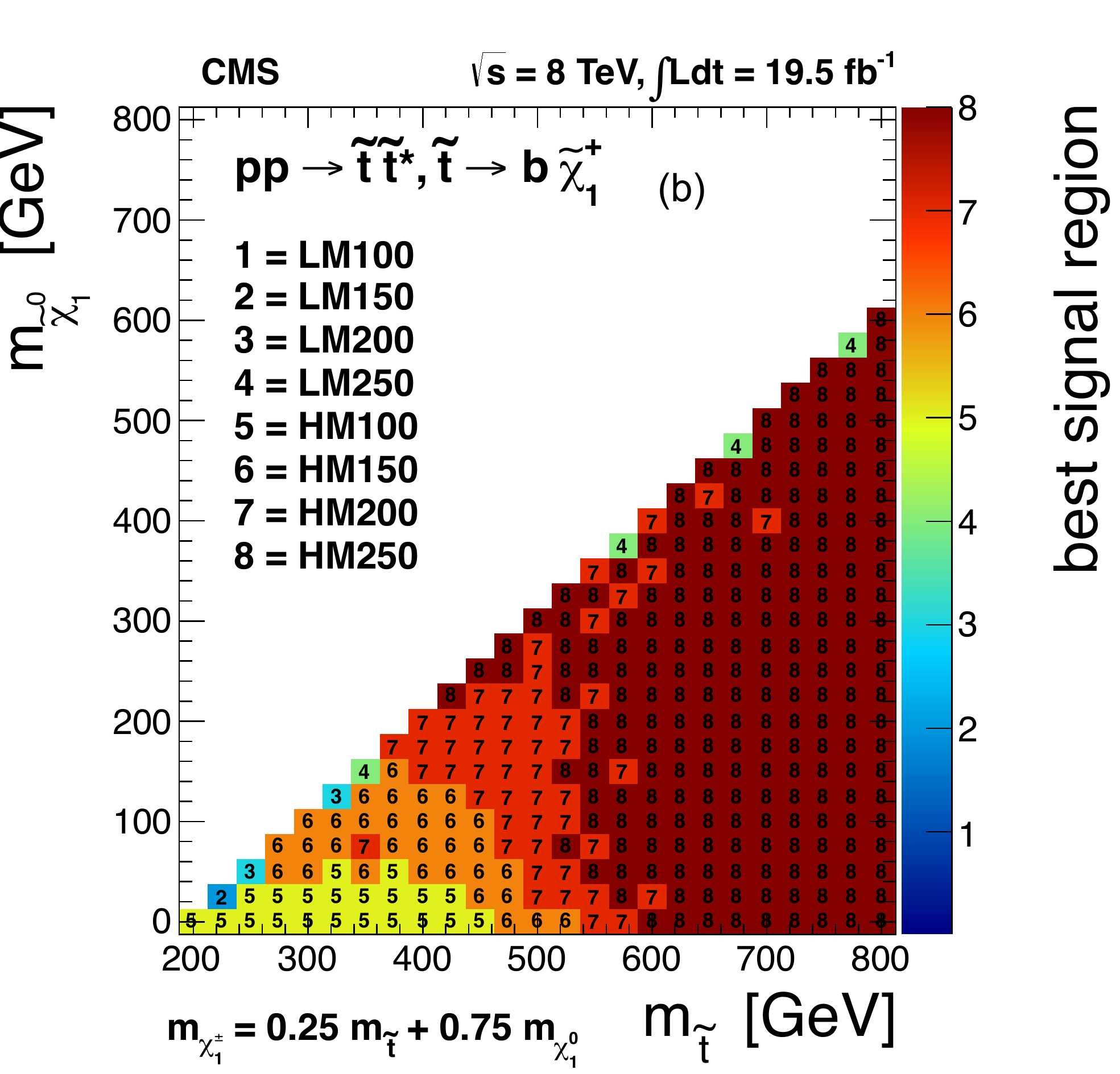}
\includegraphics[width=0.49\textwidth]{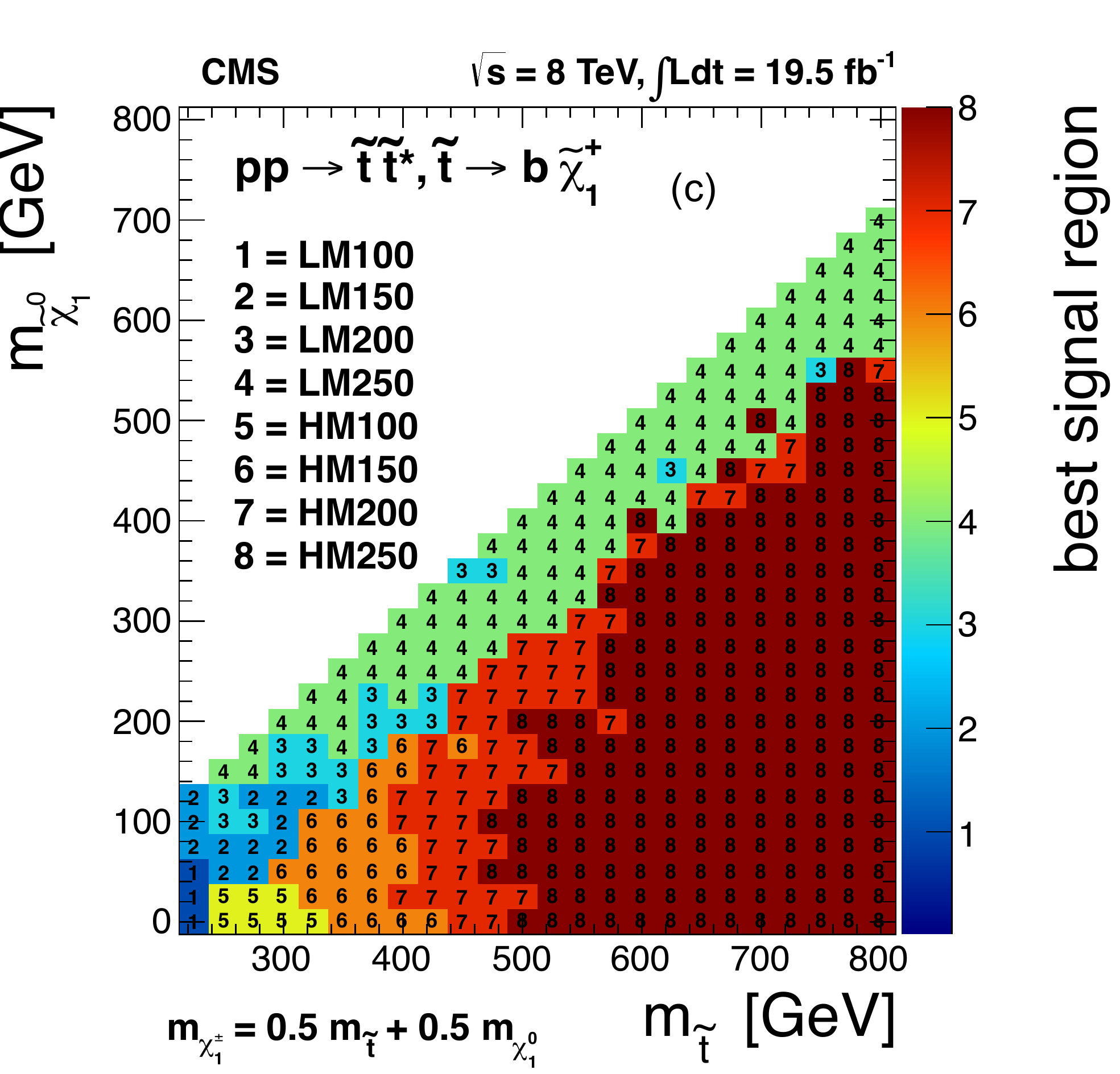}
\includegraphics[width=0.49\textwidth]{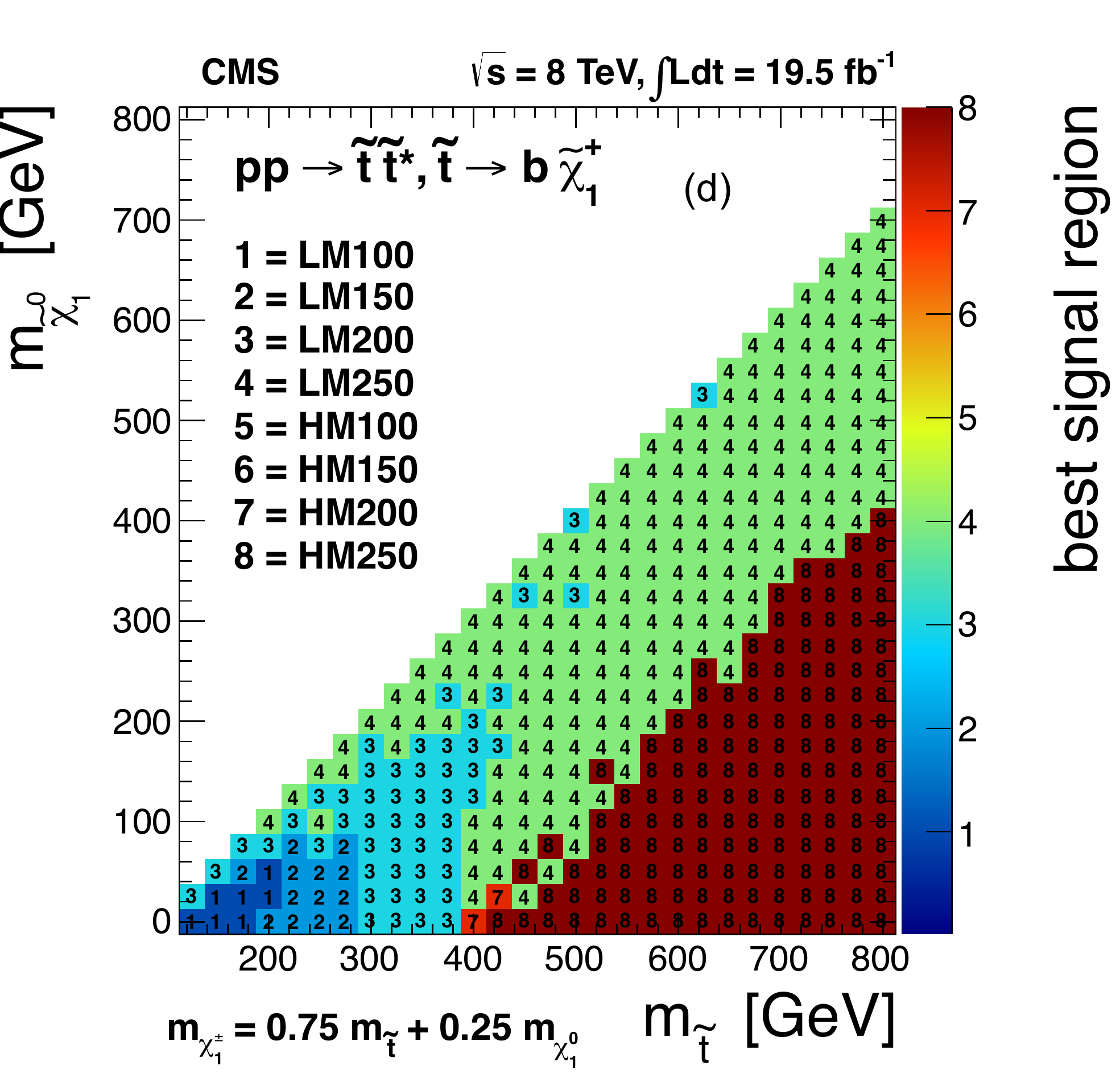}
\caption{ The most sensitive signal region in the $m_{\lsp}$ vs. $m_{\PSQt}$ parameter space
in the cut-based analysis,
for the (a) \Ttt\ model, and the
\TbW\ model with chargino mass parameter (b) $x=0.25$,
(c) 0.5, and (d) 0.75. LM and HM refer to
low $\Delta M$ and high $\Delta M$,
respectively, and the number indicates the \MET\ requirement.
\label{fig:cac_bestregion}
}
\end{center}
\end{figure*}

\begin{figure*}[htbp]
\begin{center}
\includegraphics[width=0.48\textwidth]{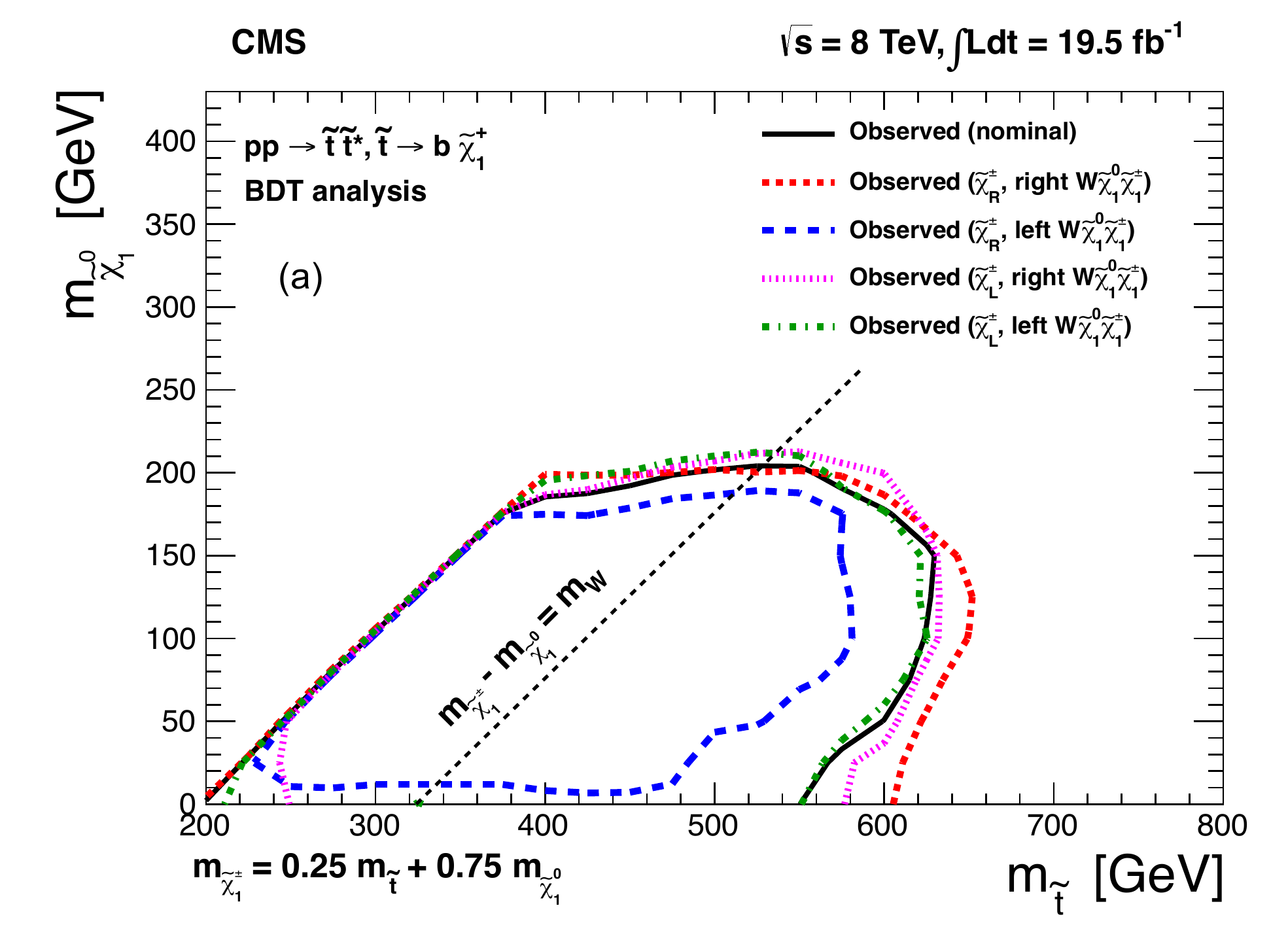}
\includegraphics[width=0.48\textwidth]{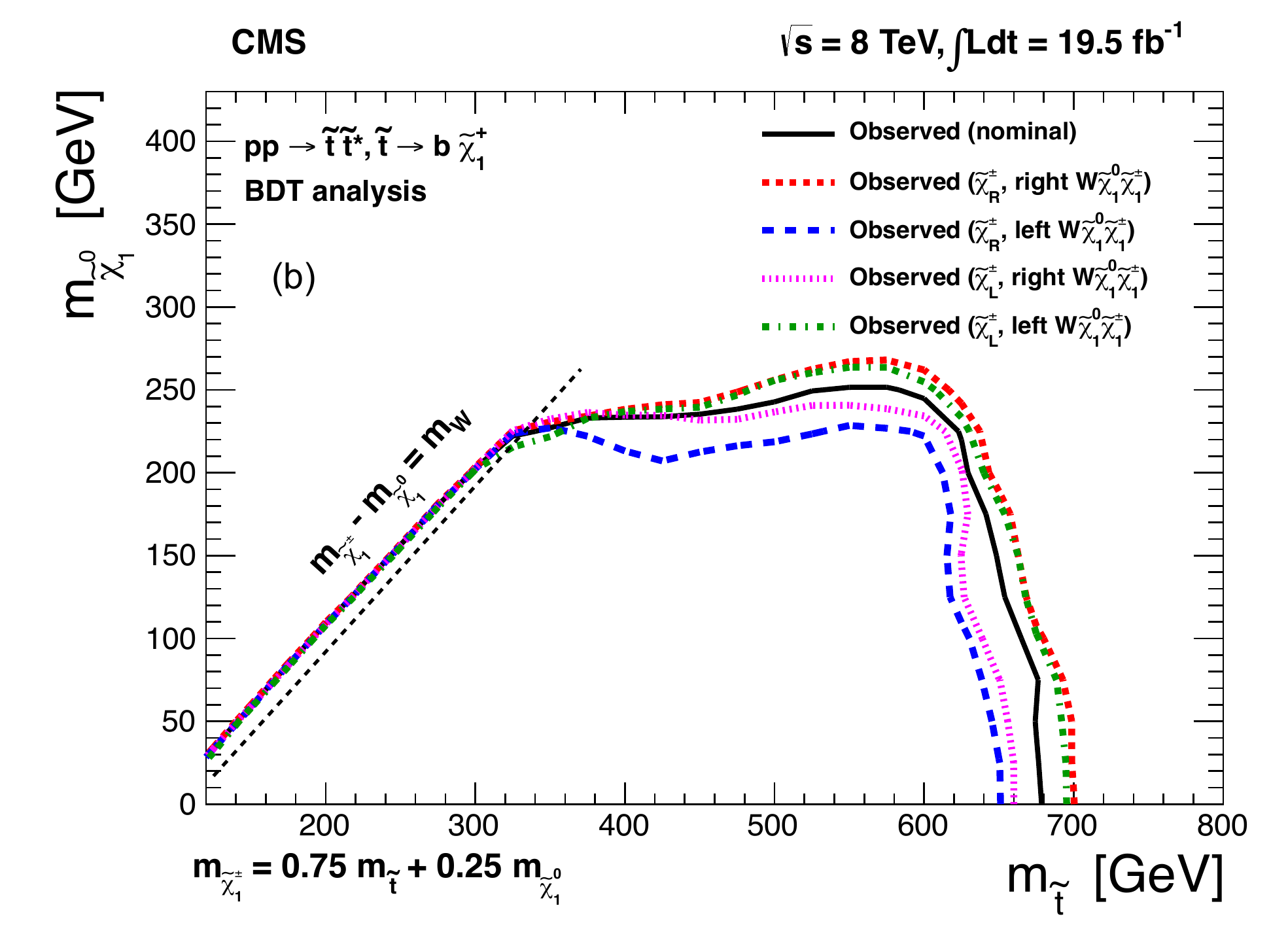}
\caption{
The observed 95\% CL excluded regions for the \TbW\ model with (a)
$x=0.25$ and (b) $0.75$ for the nominal
scenario, right- vs. left-handed charginos ($\tilde{\chi}_{R}^{\pm}$ and $\tilde{\chi}_{R}^{\pm}$, respectively),
and right- vs. left-handed \PW\lsp\chipo\ couplings.
\label{fig:T2bw_polarization}
}
\end{center}
\end{figure*}

The interpretations for the \Ttt\ and \TbW\ scenarios, using the cut-based analysis,
are presented in Fig.~\ref{fig:cac_interpretations}.
Maps of the most sensitive signal regions for the cut-based and BDT
searches are
shown in
Figs.~\ref{fig:bdt_bestregion} and \ref{fig:cac_bestregion}.
The variations in the \TbW\ $x=0.25$ and $0.75$ limits due to assumptions about particle polarizations
are presented in Fig.~\ref{fig:T2bw_polarization}.

\section{Monte Carlo modeling of initial-state radiation}
\label{app:isr}

The experimental acceptance for signal events depends on initial-state radiation (ISR).  As the simulation is not necessarily expected to model ISR well, we validate the simulation by comparing \MADGRAPH MC predictions with data.  The predicted \pt\ spectrum of the system recoiling against the ISR jets is compared with data in \zjets, \ttbar, and $\PW\Z$ final states.  These processes can be measured with good statistical precision in data and cover a variety of masses and initial states.

\begin{figure*}[htbp]
\begin{center}
\includegraphics[width=0.47\textwidth]{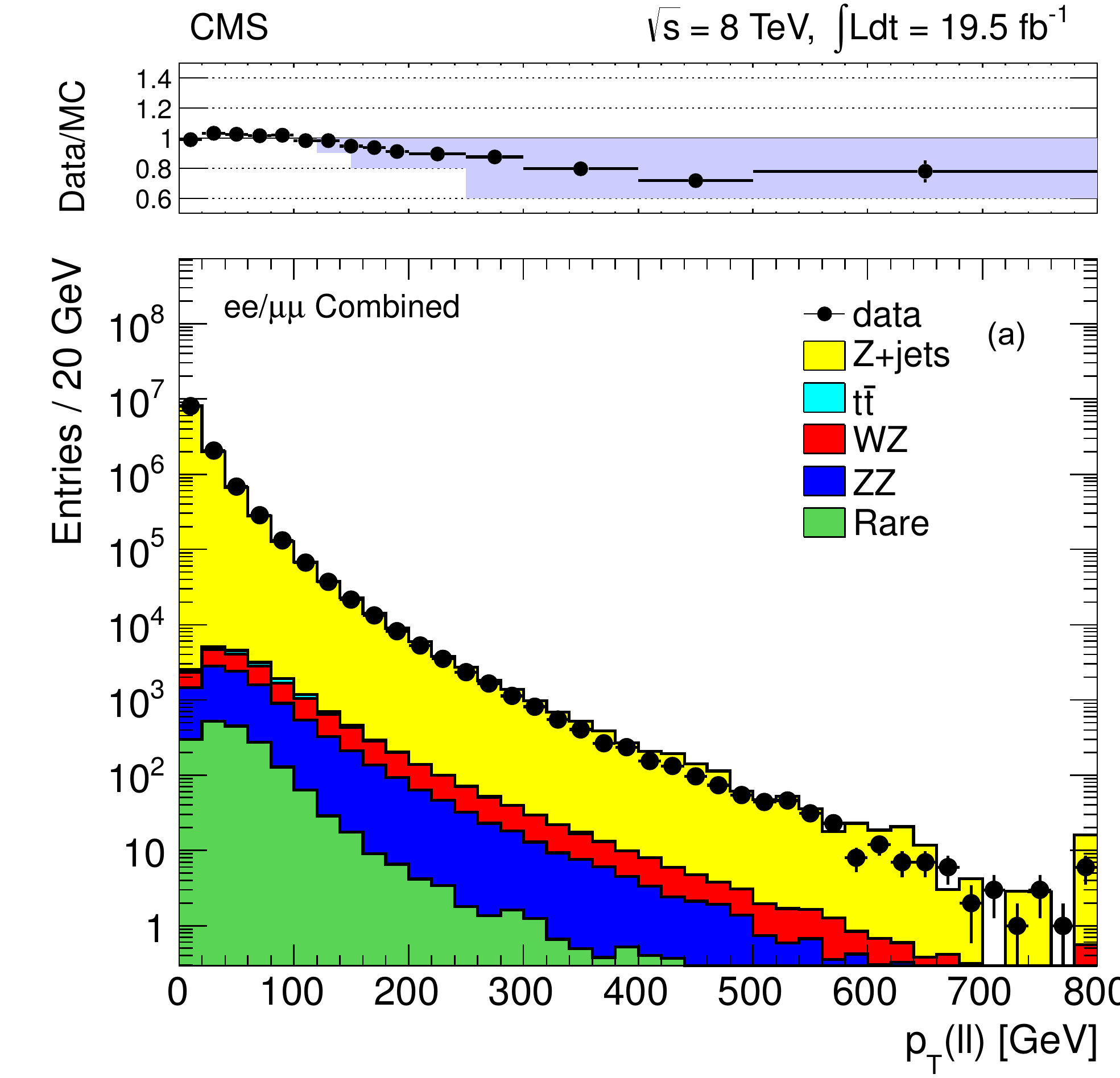}
\includegraphics[width=0.47\textwidth]{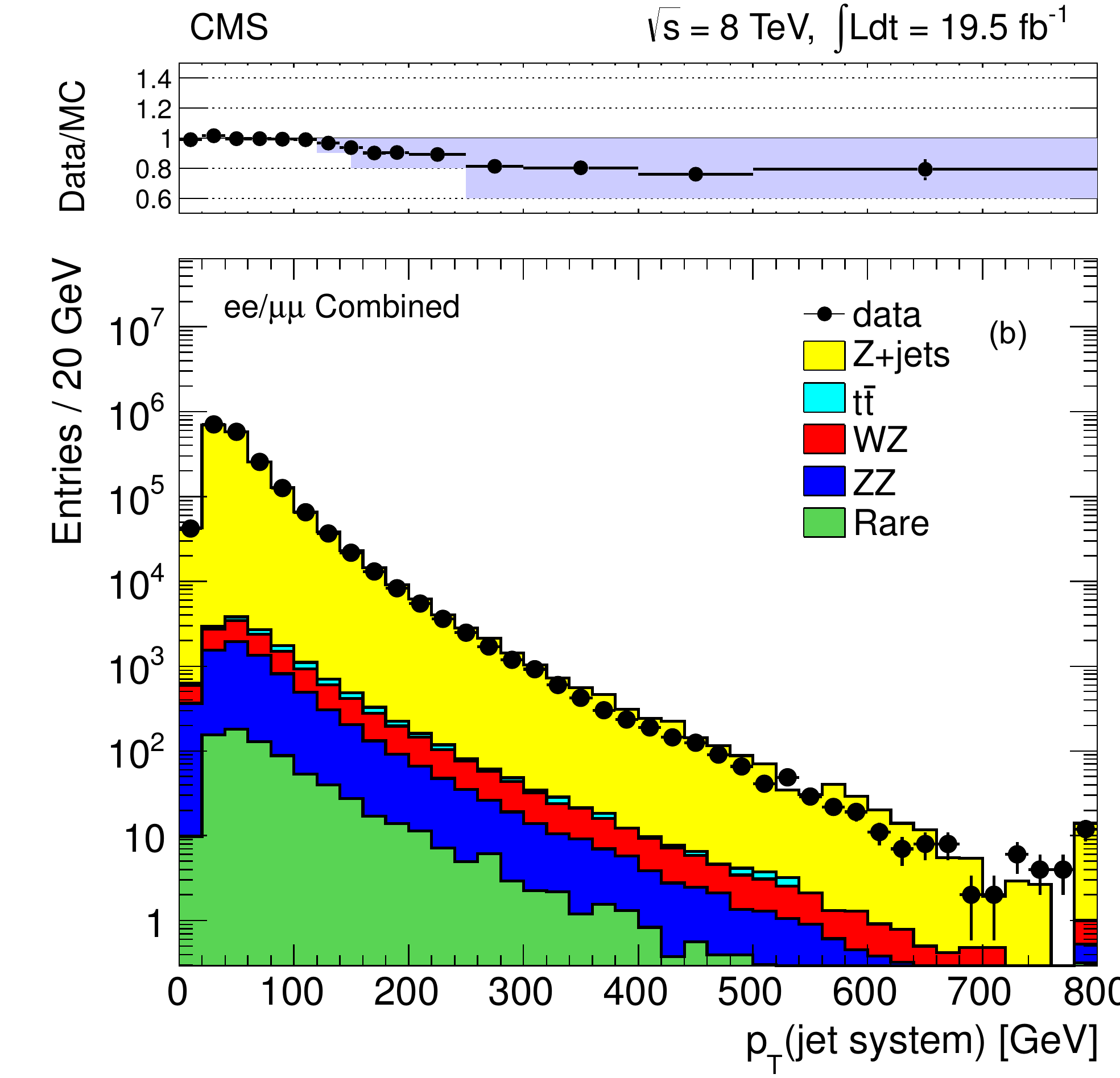}
\caption{
Comparison of data to MC predictions for the (a) dilepton \pt\ and (b)
jet recoil system \pt\ in \zjets\ events.
The MC prediction is normalized to the total data yield.
The data/MC ratio is also shown.  The shaded band is centered on the weight values.
The width of the band indicates the associated systematic uncertainty.
In both distributions the last bin contains the overflow.
\label{fig:zjets_dilpt}
}
\end{center}
\end{figure*}

\zjets\ events are selected by requiring exactly two opposite-sign, same-flavor leptons (ee or $\mu\mu$) with an invariant mass between 81 and 101\GeV.
These events, as well as the \ttbar\ and $\PW\Z$ samples discussed below,
are collected with dilepton triggers.  Events with at least one b-tagged jet or with additional lepton candidates are vetoed to remove contributions from \ttbar\ and diboson ($\PW\Z/\Z\Z$) production, respectively.  In \zjets\ events, the \Z\ boson is expected to be balanced in transverse momentum with the ISR jet system. The \pt\ of the \Z\ boson is thus computed in two ways: as the \pt\ of the dilepton system, and, for events with at least one reconstructed jet, as the \pt\ of the vector sum of the reconstructed jets, termed the ``jet system'' \pt.  The predicted MC spectrum for each quantity is compared with data, as shown in Fig.~\ref{fig:zjets_dilpt}.
The MC prediction is normalized to the total data yield so that the shapes can be readily compared.
This procedure changes the normalization of the simulation by 4\%,
consitent with the luminosty uncertainty.
Agreement is observed at lower \pt, while at higher \pt\ the MC predictions lie
above the data.
The predictions from simulation exceed the data by
about 10\% for $\pt=150$\GeV and 20\% for $\pt=250$\GeV.
Both quantities show the same trend, validating the jet recoil method of measuring this quantity.  The dilepton \pt\ and jet system \pt\ are also checked for events with exactly one, two, or three jets, as well as at least four jets, and in each case the results are consistent with the inclusive results shown in Fig.~\ref{fig:zjets_dilpt}. The impact of the jet energy scale uncertainty, which only affects the jet system \pt, is found to be much smaller than the observed discrepancies.

\begin{figure}[hbt]
\begin{center}
\includegraphics[width=\cmsFigWidth]{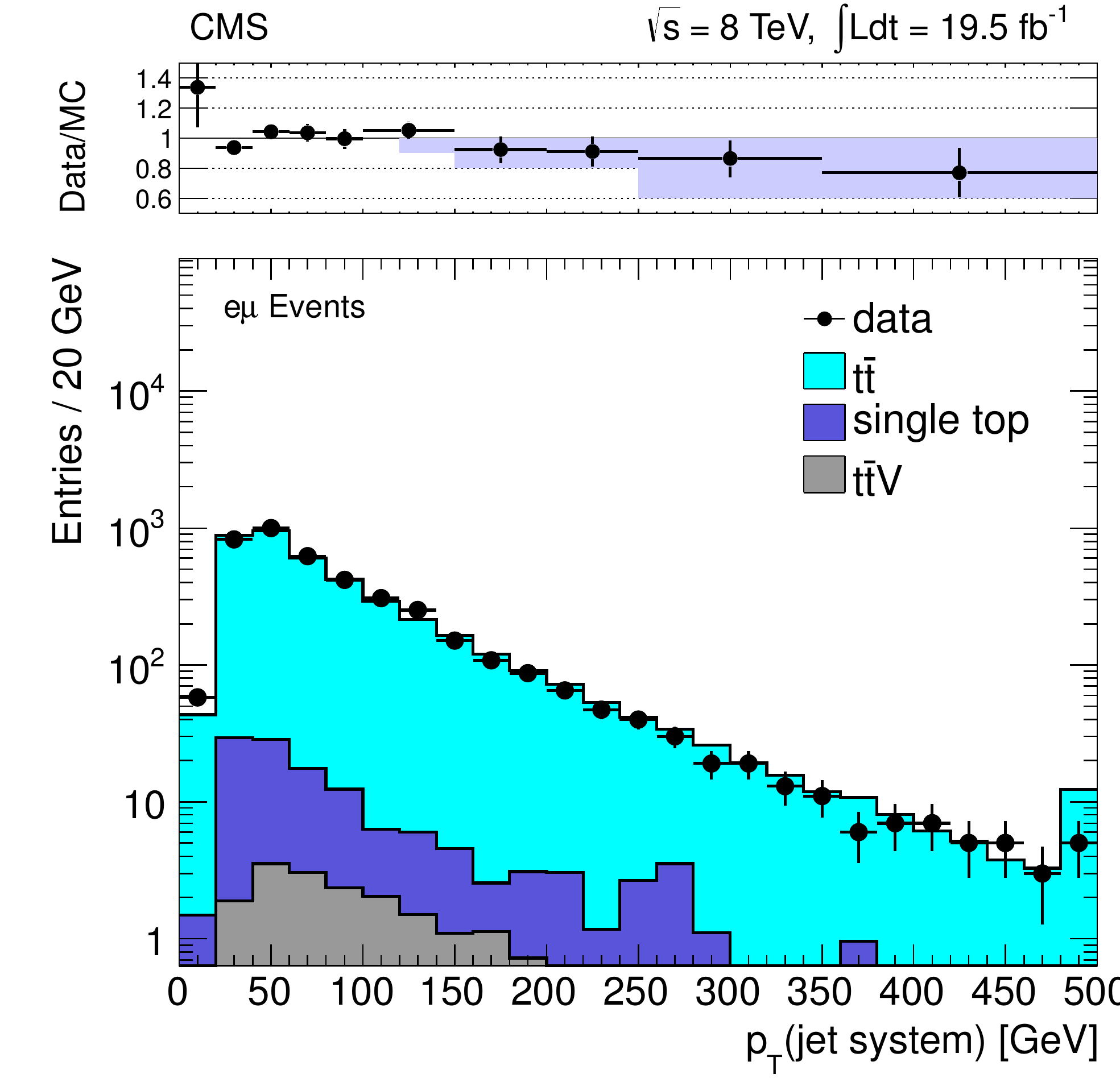}
\caption{
Comparison of data to MC prediction for the jet recoil system \pt\ in \ttbar\ events.
The MC prediction is normalized to the total data yield.
The ratio of data/MC is also shown.
The shaded band shows the weights derived for MC simulation and the variation to assess systematic uncertainties.
The last bin contains the overflow.
\label{fig:ttbar_sumpt}
}
\end{center}
\end{figure}

Dilepton \ttbar\ events are selected by requiring an opposite-sign e$\mu$ pair and exactly two b-tagged jets.  Events containing a third lepton candidate are vetoed.  These requirements select dilepton \ttbar\ events with high purity (about 97\% in simulation) and unambiguously identify all the visible \ttbar\ decay products.  Because of the presence of neutrinos in the \ttbar\ decays, the \pt\ of the \ttbar\ cannot be directly measured but can be inferred from the ISR jet recoil system. Additional jets beyond the two b-tagged jets in these events are thus considered to be ISR jets for the purposes of this study, and the ``jet system'' is formed by the vector sum of ISR jets.  The \pt\ of the jet system defined this way is found in simulation to accurately reproduce the \pt\ of the generated \ttbar\ system.  The predicted jet system \pt\ spectrum is compared with data in Fig.~\ref{fig:ttbar_sumpt}.  Agreement is found at lower \pt. At higher \pt,
the simulation is consistent with the data to within the uncertainties, but it also exhibits a trend to overpredict the data, as in the case of \zjets\ events.  The jet system \pt\ is also checked for events with exactly one, two, or three jets, as well as at least four jets, and in each case the results are consistent with the inclusive results shown in Fig.~\ref{fig:ttbar_sumpt}.
Again, the effect of the jet energy scale uncertainty is examined and found to be small.

Finally, $\PW\Z \to \ell\nu\ell\ell$ events are selected by requiring exactly three leptons, with two opposite-sign same-flavor leptons (ee or $\mu\mu$) consistent with the \Z\ boson mass and a third lepton (e or $\mu$) with $\MT > 50$\GeV.  Events with at least one b-tagged jet are vetoed.  The expected purity of this selection from
simulation is about 83\%, with about 7\% of events coming from $\Z\Z$ production.  As with \ttbar\ events,
the neutrino in the final state prevents a direct measurement of the $\PW\Z$ system \pt, but the jet recoil system can be used and is defined in the
same way as for the \zjets\ sample.
In data, this selection yields on the order of 1000 events, so the statistical uncertainty at high values of jet system \pt\ is large.
As for  the \ttbar\ MC simulated events, the $\PW\Z$ simulation is found to be consistent with the data to within the uncertainties,
but also shows a trend to overpredict the data at large \pt that
is consistent with the level observed for the \zjets\ events.

Given the MC overprediction observed in the high-statistics \zjets\ events, and the consistency of the other final states with this result, weights are derived to correct the MC prediction as a function of the \pt\ of the system recoiling against ISR jets.  These weights are applied to the
\MADGRAPH signal samples used in this analysis, and the full values of the corrections are taken as a systematic uncertainty.
The values of the weights range from 0--20\% depending on the \pt\ of the system recoiling against ISR jets.  The shaded bands shown on the ratio plots in Figs.~\ref{fig:zjets_dilpt}--\ref{fig:ttbar_sumpt} are centered on the weighted MC prediction, with the width of the band showing the associated uncertainty.

\cleardoublepage \section{The CMS Collaboration \label{app:collab}}\begin{sloppypar}\hyphenpenalty=5000\widowpenalty=500\clubpenalty=5000\input{SUS-13-011-authorlist.tex}\end{sloppypar}
\end{document}

%% file: SUS-13-011-authorlist.tex
\textbf{Yerevan Physics Institute,  Yerevan,  Armenia}\\*[0pt]
S.~Chatrchyan, V.~Khachatryan, A.M.~Sirunyan, A.~Tumasyan
\vskip\cmsinstskip
\textbf{Institut f\"{u}r Hochenergiephysik der OeAW,  Wien,  Austria}\\*[0pt]
W.~Adam, T.~Bergauer, M.~Dragicevic, J.~Er\"{o}, C.~Fabjan\cmsAuthorMark{1}, M.~Friedl, R.~Fr\"{u}hwirth\cmsAuthorMark{1}, V.M.~Ghete, N.~H\"{o}rmann, J.~Hrubec, M.~Jeitler\cmsAuthorMark{1}, W.~Kiesenhofer, V.~Kn\"{u}nz, M.~Krammer\cmsAuthorMark{1}, I.~Kr\"{a}tschmer, D.~Liko, I.~Mikulec, D.~Rabady\cmsAuthorMark{2}, B.~Rahbaran, C.~Rohringer, H.~Rohringer, R.~Sch\"{o}fbeck, J.~Strauss, A.~Taurok, W.~Treberer-Treberspurg, W.~Waltenberger, C.-E.~Wulz\cmsAuthorMark{1}
% ...
\vskip\cmsinstskip
% ...
1:~~Also at Vienna University of Technology, Vienna, Austria\\
2:~~Also at CERN, European Organization for Nuclear Research, Geneva, Switzerland\\
% ...


%% file: SUS-13-011_temp.bbl
\providecommand{\href}[2]{#2}\begingroup\raggedright\begin{thebibliography}{10}%
\makeatletter
\providecommand{\hrefCMSnoop }[0]{\@secondoftwo}%
\makeatother
\providecommand{\doi}{\texttt{doi:}\begingroup \urlstyle{tt}\Url}

\bibitem{SUSY1}
\hrefCMSnoop {} {S.~Dimopoulos and S.~Raby, ``{Supercolor}'',} \textit{ Nucl.
  Phys. B} \textbf{ 192} (1981) 353,
\href{http://dx.doi.org/10.1016/0550-3213(81)90430-2}{\doi{10.1016/0550-3213(81)90430-2}}.

\bibitem{SUSY2}
\hrefCMSnoop {} {E.~Witten, ``{Dynamical Breaking of Supersymmetry}'',}
  \textit{ Nucl. Phys. B} \textbf{ 188} (1981) 513,
\href{http://dx.doi.org/10.1016/0550-3213(81)90006-7}{\doi{10.1016/0550-3213(81)90006-7}}.

\bibitem{SUSY3}
\hrefCMSnoop {} {M.~Dine, W.~Fischler, and M.~Srednicki, ``{Supersymmetric
  Technicolor}'',} \textit{ Nucl. Phys. B} \textbf{ 189} (1981) 575,
\href{http://dx.doi.org/10.1016/0550-3213(81)90582-4}{\doi{10.1016/0550-3213(81)90582-4}}.

\bibitem{SUSY4}
\hrefCMSnoop {} {S.~Dimopoulos and H.~Georgi, ``{Softly Broken Supersymmetry
  and SU(5)}'',} \textit{ Nucl. Phys. B} \textbf{ 193} (1981) 150,
\href{http://dx.doi.org/10.1016/0550-3213(81)90522-8}{\doi{10.1016/0550-3213(81)90522-8}}.

\bibitem{SUSY5}
\hrefCMSnoop {} {N.~Sakai, ``{Naturalness in Supersymmetric Guts}'',} \textit{
  Z. Phys. C} \textbf{ 11} (1981) 153,
\href{http://dx.doi.org/10.1007/BF01573998}{\doi{10.1007/BF01573998}}.

\bibitem{SUSY6}
\hrefCMSnoop {} {R.~K. Kaul and P.~Majumdar, ``{Cancellation of quadratically
  divergent mass corrections in globally supersymmetric spontaneously broken
  gauge theories}'',} \textit{ Nucl. Phys. B} \textbf{ 199} (1982) 36,
\href{http://dx.doi.org/10.1016/0550-3213(82)90565-X}{\doi{10.1016/0550-3213(82)90565-X}}.

\bibitem{JINST}
\hrefCMSnoop {} {{ CMS} Collaboration, ``{The CMS experiment at the CERN
  LHC}'',} \textit{ JINST} \textbf{ 3} (2008) S08004,
\href{http://dx.doi.org/10.1088/1748-0221/3/08/S08004}{\doi{10.1088/1748-0221/3/08/S08004}}.

\bibitem{Barbieri:1987fn}
\hrefCMSnoop {} {R.~Barbieri and G.~F. Giudice, ``Upper bounds on
  supersymmetric particle masses'',} \textit{ Nucl. Phys. B} \textbf{ 306}
  (1988) 63,
\href{http://dx.doi.org/10.1016/0550-3213(88)90171-X}{\doi{10.1016/0550-3213(88)90171-X}}.

\bibitem{deCarlos1993320}
\hrefCMSnoop {} {B.~de~Carlos and J.~A. Casas, ``One-loop analysis of the
  electroweak breaking in supersymmetric models and the fine-tuning problem'',}
  \textit{ Phys. Lett. B} \textbf{ 309} (1993) 320,
  \href{http://dx.doi.org/10.1016/0370-2693(93)90940-J}{\doi{10.1016/0370-2693(93)90940-J}},
\href{http://www.arXiv.org/abs/hep-ph/9303291}{\texttt{ arXiv:hep-ph/9303291}}.

\bibitem{Dimopoulos1995573}
\hrefCMSnoop {} {S.~Dimopoulos and G.~F. Giudice, ``Naturalness constraints in
  supersymmetric theories with non-universal soft terms'',} \textit{ Phys.
  Lett. B} \textbf{ 357} (1995) 573,
  \href{http://dx.doi.org/10.1016/0370-2693(95)00961-J}{\doi{10.1016/0370-2693(95)00961-J}},
\href{http://www.arXiv.org/abs/hep-ph/9507282}{\texttt{ arXiv:hep-ph/9507282}}.

\bibitem{Barbieri199676}
\hrefCMSnoop {} {R.~Barbieri, G.~Dvali, and L.~J. Hall, ``Predictions from a
  U(2) flavour symmetry in supersymmetric theories'',} \textit{ Phys. Lett. B}
  \textbf{ 377} (1996) 76,
  \href{http://dx.doi.org/10.1016/0370-2693(96)00318-8}{\doi{10.1016/0370-2693(96)00318-8}},
\href{http://www.arXiv.org/abs/hep-ph/9512388}{\texttt{ arXiv:hep-ph/9512388}}.

\bibitem{Papucci:2011wy}
\hrefCMSnoop {} {M.~Papucci, J.~T. Ruderman, and A.~Weiler, ``Natural {SUSY}
  endures'',} \textit{ JHEP} \textbf{ 09} (2012) 035,
  \href{http://dx.doi.org/10.1007/JHEP09(2012)035}{\doi{10.1007/JHEP09(2012)035}},
\href{http://www.arXiv.org/abs/1110.6926}{\texttt{ arXiv:1110.6926}}.

\bibitem{ATLAS_Higgs}
\hrefCMSnoop {} {{ ATLAS} Collaboration, ``{Observation of a new particle in
  the search for the Standard Model Higgs boson with the ATLAS detector at the
  LHC}'',} \textit{ Phys. Lett. B} \textbf{ 716} (2012) 1,
  \href{http://dx.doi.org/10.1016/j.physletb.2012.08.020}{\doi{10.1016/j.physletb.2012.08.020}},
\href{http://www.arXiv.org/abs/1207.7214}{\texttt{ arXiv:1207.7214}}.

\bibitem{CMS_Higgs}
\hrefCMSnoop {} {{ CMS} Collaboration, ``{Observation of a new boson at a mass
  of 125~GeV with the CMS experiment at the LHC}'',} \textit{ Phys. Lett. B}
  \textbf{ 716} (2012) 30,
  \href{http://dx.doi.org/10.1016/j.physletb.2012.08.021}{\doi{10.1016/j.physletb.2012.08.021}},
\href{http://www.arXiv.org/abs/1207.7235}{\texttt{ arXiv:1207.7235}}.

\bibitem{CMS_HiggsLong}
\hrefCMSnoop {} {{ CMS} Collaboration, ``{Search for a standard-model-like
  Higgs boson with a mass in the range 145 to 1000 GeV at the LHC}'',} \textit{
  Eur. Phys. J. C} \textbf{ 73} (2013) 2469,
  \href{http://dx.doi.org/10.1140/epjc/s10052-013-2469-8}{\doi{10.1140/epjc/s10052-013-2469-8}},
\href{http://www.arXiv.org/abs/1304.0213}{\texttt{ arXiv:1304.0213}}.

\bibitem{ATLAS1}
\hrefCMSnoop {} {{ ATLAS} Collaboration, ``{Search for a supersymmetric partner
  to the top quark in final states with jets and missing transverse momentum at
  $\sqrt{s}$ = 7 TeV with the ATLAS detector}'',} \textit{ Phys. Rev. Lett.}
  \textbf{ 109} (2012) 211802,
  \href{http://dx.doi.org/10.1103/PhysRevLett.109.211802}{\doi{10.1103/PhysRevLett.109.211802}},
\href{http://www.arXiv.org/abs/1208.1447}{\texttt{ arXiv:1208.1447}}.

\bibitem{ATLAS2}
\hrefCMSnoop {} {{ ATLAS} Collaboration, ``{Search for direct top squark pair
  production in final states with one isolated lepton, jets, and missing
  transverse momentum in $\sqrt{s}$ = 7 TeV pp collisions using 4.7 fb$^{-1}$
  of ATLAS data}'',} \textit{ Phys. Rev. Lett.} \textbf{ 109} (2012) 211803,
  \href{http://dx.doi.org/10.1103/PhysRevLett.109.211803}{\doi{10.1103/PhysRevLett.109.211803}},
\href{http://www.arXiv.org/abs/1208.2590}{\texttt{ arXiv:1208.2590}}.

\bibitem{ATLAS3}
\hrefCMSnoop {} {{ ATLAS} Collaboration, ``{Search for light scalar top quark
  pair production in final states with two leptons with the ATLAS detector in
  $\sqrt{s}$ = 7 TeV proton-proton collisions}'',} \textit{ Eur. Phys. J. C}
  \textbf{ 72} (2012) 2237,
  \href{http://dx.doi.org/10.1140/epjc/s10052-012-2237-1}{\doi{10.1140/epjc/s10052-012-2237-1}},
\href{http://www.arXiv.org/abs/1208.4305}{\texttt{ arXiv:1208.4305}}.

\bibitem{ATLAS4}
\hrefCMSnoop {} {{ ATLAS} Collaboration, ``{Search for light top squark pair
  production in final states with leptons and b-jets with the ATLAS detector in
  $\sqrt{s}$ = 7 TeV proton-proton collisions}'',} \textit{ Phys. Lett. B}
  \textbf{ 720} (2013) 13,
  \href{http://dx.doi.org/10.1016/j.physletb.2013.01.049}{\doi{10.1016/j.physletb.2013.01.049}},
\href{http://www.arXiv.org/abs/1209.2102}{\texttt{ arXiv:1209.2102}}.

\bibitem{ATLAS5}
\hrefCMSnoop {} {{ ATLAS} Collaboration, ``{Search for a heavy top-quark
  partner in final states with two leptons with the ATLAS detector at the
  LHC}'',} \textit{ JHEP} \textbf{ 11} (2012) 094,
  \href{http://dx.doi.org/10.1007/JHEP11(2012)094}{\doi{10.1007/JHEP11(2012)094}},
\href{http://www.arXiv.org/abs/1209.4186}{\texttt{ arXiv:1209.4186}}.

\bibitem{CDFstop}
\hrefCMSnoop {} {{ CDF} Collaboration, ``{Search for the supersymmetric partner
  of the top quark in $p\bar{p}$ collisions at $\sqrt{s}$ = 1.96 TeV}'',}
  \textit{ Phys. Rev. D} \textbf{ 82} (2010) 092001,
  \href{http://dx.doi.org/10.1103/PhysRevD.82.092001}{\doi{10.1103/PhysRevD.82.092001}},
\href{http://www.arXiv.org/abs/1009.0266}{\texttt{ arXiv:1009.0266}}.

\bibitem{D0stop}
\hrefCMSnoop {} {{ D\O} Collaboration, ``{Search for pair production of the
  scalar top quark in the electron+muon final state}'',} \textit{ Phys. Lett.
  B} \textbf{ 696} (2011) 321,
  \href{http://dx.doi.org/10.1016/j.physletb.2010.12.052}{\doi{10.1016/j.physletb.2010.12.052}},
\href{http://www.arXiv.org/abs/1009.5950}{\texttt{ arXiv:1009.5950}}.

\bibitem{POWHEG}
\hrefCMSnoop {} {S.~Frixione, P.~Nason, and C.~Oleari, ``{Matching NLO QCD
  computations with parton shower simulations: the POWHEG method}'',} \textit{
  JHEP} \textbf{ 11} (2007) 070,
  \href{http://dx.doi.org/10.1088/1126-6708/2007/11/070}{\doi{10.1088/1126-6708/2007/11/070}},
\href{http://www.arXiv.org/abs/0709.2092}{\texttt{ arXiv:0709.2092}}.

\bibitem{MCatNLO1}
\hrefCMSnoop {} {S.~Frixione and B.~R. Webber, ``{Matching NLO QCD computations
  and parton shower simulations}'',} \textit{ JHEP} \textbf{ 06} (2002) 029,
  \href{http://dx.doi.org/10.1088/1126-6708/2002/06/029}{\doi{10.1088/1126-6708/2002/06/029}},
\href{http://www.arXiv.org/abs/hep-ph/0204244}{\texttt{ arXiv:hep-ph/0204244}}.

\bibitem{MCatNLO2}
\hrefCMSnoop {} {S.~Frixione, P.~Nason, and B.~R. Webber, ``{Matching NLO QCD
  and parton showers in heavy flavor production}'',} \textit{ JHEP} \textbf{
  08} (2003) 007,
  \href{http://dx.doi.org/10.1088/1126-6708/2003/08/007}{\doi{10.1088/1126-6708/2003/08/007}},
\href{http://www.arXiv.org/abs/hep-ph/0305252}{\texttt{ arXiv:hep-ph/0305252}}.

\bibitem{madgraph5}
J.~Alwall\hrefCMSnoop {} { {et~al.}, ``{MadGraph} 5: going beyond'',} \textit{
  JHEP} \textbf{ 06} (2011) 128,
  \href{http://dx.doi.org/10.1007/JHEP06(2011)128}{\doi{10.1007/JHEP06(2011)128}},
\href{http://www.arXiv.org/abs/1106.0522}{\texttt{ arXiv:1106.0522}}.

\bibitem{ct10}
H.-L. Lai\hrefCMSnoop {} { {et~al.}, ``{New parton distributions for collider
  physics}'',} \textit{ Phys. Rev. D} \textbf{ 82} (2010) 074024,
  \href{http://dx.doi.org/10.1103/PhysRevD.82.074024}{\doi{10.1103/PhysRevD.82.074024}},
\href{http://www.arXiv.org/abs/1007.2241}{\texttt{ arXiv:1007.2241}}.

\bibitem{cteq6lm}
J.~Pumplin\hrefCMSnoop {} { {et~al.}, ``{New generation of parton distributions
  with uncertainties from global QCD analysis}'',} \textit{ JHEP} \textbf{ 07}
  (2002) 012,
  \href{http://dx.doi.org/10.1088/1126-6708/2002/07/012}{\doi{10.1088/1126-6708/2002/07/012}},
\href{http://www.arXiv.org/abs/hep-ph/0201195}{\texttt{ arXiv:hep-ph/0201195}}.

\bibitem{xsec_WZ}
\hrefCMSnoop {} {R.~Gavin, Y.~Li, F.~Petriello, and S.~Quackenbush, ``{W
  Physics at the LHC with FEWZ 2.1}'',} \textit{ Comput. Phys. Commun.}
  \textbf{ 184} (2013) 208,
  \href{http://dx.doi.org/10.1016/j.cpc.2012.09.005}{\doi{10.1016/j.cpc.2012.09.005}},
\href{http://www.arXiv.org/abs/1201.5896}{\texttt{ arXiv:1201.5896}}.

\bibitem{xsec_ttbar}
\hrefCMSnoop {} {N.~Kidonakis, ``{Differential and total cross sections for top
  pair and single top production}'',} in \textit{ XX International Workshop on
  Deep-Indelastic Scattering and Related Subjects}, p.~831.
\newblock Bonn, 2012.
\newblock \href{http://www.arXiv.org/abs/1205.3453}{\texttt{ arXiv:1205.3453}}.
\newblock
\href{http://dx.doi.org/10.3204/DESY-PROC-2012-02/251}{\doi{10.3204/DESY-PROC-2012-02/251}}.

\bibitem{xsec_ttbarW}
\hrefCMSnoop {} {J.~M. Campbell and R.~K. Ellis, ``{$\ttbar\PW^{\pm}$
  production and decay at NLO}'',} \textit{ JHEP} \textbf{ 07} (2012) 052,
  \href{http://dx.doi.org/10.1007/JHEP07(2012)052}{\doi{10.1007/JHEP07(2012)052}},
\href{http://www.arXiv.org/abs/1204.5678}{\texttt{ arXiv:1204.5678}}.

\bibitem{xsec_ttbarZ}
\hrefCMSnoop {} {M.~V. Garzelli, A.~Kardos, C.~G. Papadopoulos, and
  Z.~Trocsanyi, ``{$\ttbar\PW^{\pm}$ and $\ttbar{\rm Z}$ hadroproduction at NLO
  accuracy in QCD with Parton Shower and Hadronization Effects}'',} \textit{
  JHEP} \textbf{ 11} (2012) 056,
  \href{http://dx.doi.org/10.1007/JHEP11(2012)056}{\doi{10.1007/JHEP11(2012)056}},
\href{http://www.arXiv.org/abs/1208.2665}{\texttt{ arXiv:1208.2665}}.

\bibitem{xsec_MCFM}
\hrefCMSnoop {} {J.~M. Campbell and R.~K. Ellis, ``{MCFM for the Tevatron and
  the LHC}'',} \textit{ Nucl. Phys. Proc. Suppl.} \textbf{ 205-206} (2010) 10,
  \href{http://dx.doi.org/10.1016/j.nuclphysbps.2010.08.011}{\doi{10.1016/j.nuclphysbps.2010.08.011}},
\href{http://www.arXiv.org/abs/1007.3492}{\texttt{ arXiv:1007.3492}}.

\bibitem{MCatNLO3}
R.~Frederix\hrefCMSnoop {} { {et~al.}, ``{Four-lepton production at hadron
  colliders: aMC@NLO predictions with theoretical uncertainties}'',} \textit{
  JHEP} \textbf{ 02} (2012) 099,
  \href{http://dx.doi.org/10.1007/JHEP02(2012)099}{\doi{10.1007/JHEP02(2012)099}},
\href{http://www.arXiv.org/abs/1110.4738}{\texttt{ arXiv:1110.4738}}.

\bibitem{Pythia}
\hrefCMSnoop {} {T.~Sj{\"o}strand, S.~Mrenna, and P.~Skands, ``{PYTHIA} 6.4
  physics and manual'',} \textit{ JHEP} \textbf{ 05} (2006) 026,
  \href{http://dx.doi.org/10.1088/1126-6708/2006/05/026}{\doi{10.1088/1126-6708/2006/05/026}},
\href{http://www.arXiv.org/abs/hep-ph/0603175}{\texttt{ arXiv:hep-ph/0603175}}.

\bibitem{polarization1}
\hrefCMSnoop {} {M.~Perelstein and A.~Weiler, ``Polarized tops from stop decays
  at the {LHC}'',} \textit{ JHEP} \textbf{ 03} (2009) 141,
  \href{http://dx.doi.org/10.1088/1126-6708/2009/03/141}{\doi{10.1088/1126-6708/2009/03/141}},
\href{http://www.arXiv.org/abs/0811.1024}{\texttt{ arXiv:0811.1024}}.

\bibitem{polarization2}
\hrefCMSnoop {} {I.~Low, ``{Polarized Charginos (and Tops) in Stop Decays}'',}
  (2013).
\href{http://www.arXiv.org/abs/1304.0491}{\texttt{ arXiv:1304.0491}}.

\bibitem{bib-nlo-nll-01}
\hrefCMSnoop {} {W.~Beenakker, R.~H{\"o}pker, M.~Spira, and P.~M. Zerwas,
  ``Squark and gluino production at hadron colliders'',} \textit{ Nucl. Phys.
  B} \textbf{ 492} (1997) 51,
\href{http://dx.doi.org/10.1016/S0550-3213(97)00084-9}{\doi{10.1016/S0550-3213(97)00084-9}}.

\bibitem{bib-nlo-nll-02}
\hrefCMSnoop {} {A.~Kulesza and L.~Motyka, ``{Threshold resummation for
  squark-antisquark and gluino-pair production at the {LHC}}'',} \textit{ Phys.
  Rev. Lett.} \textbf{ 102} (2009) 111802,
\href{http://dx.doi.org/10.1103/PhysRevLett.102.111802}{\doi{10.1103/PhysRevLett.102.111802}}.

\bibitem{bib-nlo-nll-03}
\hrefCMSnoop {} {A.~Kulesza and L.~Motyka, ``{Soft gluon resummation for the
  production of gluino-gluino and squark-antisquark pairs at the {LHC}}'',}
  \textit{ Phys. Rev. D} \textbf{ 80} (2009) 095004,
\href{http://dx.doi.org/10.1103/PhysRevD.80.095004}{\doi{10.1103/PhysRevD.80.095004}}.

\bibitem{bib-nlo-nll-04}
W.~Beenakker\hrefCMSnoop {} { {et~al.}, ``{Soft-gluon resummation for squark
  and gluino hadroproduction}'',} \textit{ JHEP} \textbf{ 12} (2009) 041,
\href{http://dx.doi.org/10.1088/1126-6708/2009/12/041}{\doi{10.1088/1126-6708/2009/12/041}}.

\bibitem{bib-nlo-nll-05}
W.~Beenakker\hrefCMSnoop {} { {et~al.}, ``{Squark and gluino
  hadroproduction}'',} \textit{ Int. J. Mod. Phys. A} \textbf{ 26} (2011) 2637,
\href{http://dx.doi.org/10.1142/S0217751X11053560}{\doi{10.1142/S0217751X11053560}}.

\bibitem{ref:xsec}
M.~Kr{\"a}mer\hrefCMSnoop {} { {et~al.}, ``{Supersymmetry production cross
  sections in pp collisions at $\sqrt{s}$ = 7~TeV}'',} (2012).
\href{http://www.arXiv.org/abs/1206.2892}{\texttt{ arXiv:1206.2892}}.

\bibitem{Abdullin:2011zz}
\hrefCMSnoop {} {{ CMS} Collaboration, ``{The fast simulation of the CMS
  detector at LHC}'',} \textit{ J. Phys. Conf. Ser.} \textbf{ 331} (2011)
  032049,
\href{http://dx.doi.org/10.1088/1742-6596/331/3/032049}{\doi{10.1088/1742-6596/331/3/032049}}.

\bibitem{Geant}
\hrefCMSnoop {} {{ GEANT4} Collaboration, ``{GEANT4} --- a simulation
  toolkit'',} \textit{ Nucl. Instrum. Meth. A} \textbf{ 506} (2003) 250,
  \href{http://dx.doi.org/10.1016/S0168-9002(03)01368-8}{\doi{10.1016/S0168-9002(03)01368-8}}.

\bibitem{EGMPAS}
\href {http://cdsweb.cern.ch/record/1299116} {{ CMS} Collaboration, ``Electron
  Reconstruction and Identification at $\sqrt{s} = 7$ {TeV}'',} CMS Physics
  Analysis Summary CMS-PAS-EGM-10-004, (2010).

\bibitem{MUOART}
\hrefCMSnoop {} {{ CMS} Collaboration, ``{Performance of CMS muon
  reconstruction in $pp$ collision events at $\sqrt{s}=7$ TeV}'',} \textit{
  JINST} \textbf{ 7} (2012) P10002,
  \href{http://dx.doi.org/10.1088/1748-0221/7/10/P10002}{\doi{10.1088/1748-0221/7/10/P10002}},
\href{http://www.arXiv.org/abs/1206.4071}{\texttt{ arXiv:1206.4071}}.

\bibitem{CMS-PAS-PFT-10-002}
\href {http://cdsweb.cern.ch/record/1279341} {{ CMS} Collaboration,
  ``Commissioning of the Particle-Flow Reconstruction in Minimum-Bias and Jet
  Events from {\Pp\Pp} Collisions at 7 {TeV}'',} CMS Physics Analysis Summary
  CMS-PAS-PFT-10-002, (2010).

\bibitem{tau}
\hrefCMSnoop {} {{ CMS} Collaboration, ``{Performance of tau-lepton
  reconstruction and identification in CMS}'',} \textit{ JINST} \textbf{ 7}
  (2012) P01001,
  \href{http://dx.doi.org/10.1088/1748-0221/7/01/P01001}{\doi{10.1088/1748-0221/7/01/P01001}},
\href{http://www.arXiv.org/abs/1109.6034}{\texttt{ arXiv:1109.6034}}.

\bibitem{antikt}
\hrefCMSnoop {} {M.~Cacciari, G.~P. Salam, and G.~Soyez, ``The anti-$k_t$ jet
  clustering algorithm'',} \textit{ JHEP} \textbf{ 04} (2008) 063,
  \href{http://dx.doi.org/10.1088/1126-6708/2008/04/063}{\doi{10.1088/1126-6708/2008/04/063}},
  \href{http://www.arXiv.org/abs/0802.1189}{\texttt{ arXiv:0802.1189}}.

\bibitem{Cacciari:2005hq}
\hrefCMSnoop {} {M.~Cacciari and G.~P. Salam, ``{Dispelling the N$^3$ myth for
  the $k_t$ jet-finder}'',} \textit{ Phys. Lett. B} \textbf{ 641} (2006) 57,
  \href{http://dx.doi.org/10.1016/j.physletb.2006.08.037}{\doi{10.1016/j.physletb.2006.08.037}}.

\bibitem{FastJet}
\hrefCMSnoop {} {M.~Cacciari, G.~P. Salam, and G.~Soyez, ``{FastJet User
  Manual}'',} \textit{ Eur. Phys. J. C} \textbf{ 72} (2012) 1896,
  \href{http://dx.doi.org/10.1140/epjc/s10052-012-1896-2}{\doi{10.1140/epjc/s10052-012-1896-2}},
\href{http://www.arXiv.org/abs/1111.6097}{\texttt{ arXiv:1111.6097}}.

\bibitem{cacciari-2008-659}
\hrefCMSnoop {} {M.~Cacciari and G.~P. Salam, ``Pileup subtraction using jet
  areas'',} \textit{ Phys. Lett. B} \textbf{ 659} (2008) 119,
  \href{http://dx.doi.org/10.1016/j.physletb.2007.09.077}{\doi{10.1016/j.physletb.2007.09.077}}.

\bibitem{ref:btag}
\hrefCMSnoop {} {{ CMS} Collaboration, ``{Identification of b-quark jets with
  the CMS experiment}'',} \textit{ JINST} \textbf{ 8} (2013) P04013,
  \href{http://dx.doi.org/10.1088/1748-0221/8/04/P04013}{\doi{10.1088/1748-0221/8/04/P04013}},
\href{http://www.arXiv.org/abs/1211.4462}{\texttt{ arXiv:1211.4462}}.

\bibitem{mt2w}
\hrefCMSnoop {} {Y.~Bai, H.-C. Cheng, J.~Gallicchio, and J.~Gu, ``Stop the top
  background of the stop search'',} \textit{ JHEP} \textbf{ 07} (2012) 110,
  \href{http://dx.doi.org/10.1007/JHEP07(2012)110}{\doi{10.1007/JHEP07(2012)110}},
\href{http://www.arXiv.org/abs/1203.4813}{\texttt{ arXiv:1203.4813}}.

\bibitem{mt2-1}
\hrefCMSnoop {} {C.~G. Lester and D.~J. Summers, ``Measuring masses of
  semi-invisibly decaying particles pair produced at hadron colliders'',}
  \textit{ Phys. Lett. B} \textbf{ 463} (1999) 99,
  \href{http://dx.doi.org/10.1016/S0370-2693(99)00945-4}{\doi{10.1016/S0370-2693(99)00945-4}},
\href{http://www.arXiv.org/abs/hep-ph/9906349}{\texttt{ arXiv:hep-ph/9906349}}.

\bibitem{mt2-2}
\hrefCMSnoop {} {A.~Barr, C.~Lester, and P.~Stephens, ``{A variable for
  measuring masses at hadron colliders when missing energy is expected;
  $M_{\mathrm T2}$: the truth behind the glamour}'',} \textit{ J. Phys. G}
  \textbf{ 29} (2003) 2343,
  \href{http://dx.doi.org/10.1088/0954-3899/29/10/304}{\doi{10.1088/0954-3899/29/10/304}},
\href{http://www.arXiv.org/abs/hep-ph/0304226}{\texttt{ arXiv:hep-ph/0304226}}.

\bibitem{Burns:2008va}
\hrefCMSnoop {} {M.~Burns, K.~Kong, K.~T. Matchev, and M.~Park, ``{Using
  subsystem $M_{\mathrm{T}2}$ for complete mass determinations in decay chains
  with missing energy at hadron colliders}'',} \textit{ JHEP} \textbf{ 03}
  (2009) 143,
  \href{http://dx.doi.org/10.1088/1126-6708/2009/03/143}{\doi{10.1088/1126-6708/2009/03/143}},
\href{http://www.arXiv.org/abs/0810.5576}{\texttt{ arXiv:0810.5576}}.

\bibitem{JER}
\hrefCMSnoop {} {{ CMS} Collaboration, ``{Determination of jet energy
  calibration and transverse momentum resolution in CMS}'',} \textit{ JINST}
  \textbf{ 6} (2011) P11002,
  \href{http://dx.doi.org/10.1088/1748-0221/6/11/P11002}{\doi{10.1088/1748-0221/6/11/P11002}},
\href{http://www.arXiv.org/abs/1107.4277}{\texttt{ arXiv:1107.4277}}.

\bibitem{Beringer:1900zz}
\hrefCMSnoop {} {{Particle Data Group}, J.~Beringer {et~al.}, ``{Review of
  Particle Physics}'',} \textit{ Phys. Rev. D} \textbf{ 86} (2012) 010001,
  \href{http://dx.doi.org/10.1103/PhysRevD.86.010001}{\doi{10.1103/PhysRevD.86.010001}}.

\bibitem{TMVA}
\href {http://pos.sissa.it/archive/conferences/050/040/ACAT_040.pdf} {H.~Voss,
  A.~H{\"o}cker, J.~Stelzer, and F.~Tegenfeldt, ``{TMVA: Toolkit for
  Multivariate Data Analysis with ROOT}'',} in \textit{ XIth International
  Workshop on Advanced Computing and Analysis Techniques in Physics Research
  (ACAT)}, p.~040.
\newblock 2007.
\newblock
\href{http://www.arXiv.org/abs/physics/0703039}{\texttt{
  arXiv:physics/0703039}}.
\newblock

\bibitem{LUMIPAS}
\href {http://cdsweb.cern.ch/record/1482193} {{ CMS} Collaboration, ``CMS
  luminosity based on pixel cluster counting --- Summer 2012 update'',} CMS
  Physics Analysis Summary CMS-PAS-LUM-12-001, (2012).

\bibitem{Read:2002hq}
\hrefCMSnoop {} {A.~L. Read, ``Presentation of search results: the {$CL_{S}$}
  technique'',} \textit{ J. Phys. G} \textbf{ 28} (2002) 2693,
  \href{http://dx.doi.org/10.1088/0954-3899/28/10/313}{\doi{10.1088/0954-3899/28/10/313}}.

\bibitem{Junk:1999kv}
\hrefCMSnoop {} {T.~Junk, ``{Confidence level computation for combining
  searches with small statistics}'',} \textit{ Nucl. Instrum. Meth. A} \textbf{
  434} (1999) 435,
  \href{http://dx.doi.org/10.1016/S0168-9002(99)00498-2}{\doi{10.1016/S0168-9002(99)00498-2}},
\href{http://www.arXiv.org/abs/hep-ex/9902006}{\texttt{ arXiv:hep-ex/9902006}}.

\bibitem{LHC-HCG}
\href {http://cdsweb.cern.ch/record/1379837} {{ATLAS and CMS Collaborations,
  LHC Higgs Combination Group}, ``Procedure for the {LHC} {H}iggs boson search
  combination in {S}ummer 2011'',} Technical Report ATL-PHYS-PUB 2011-11, CMS
  NOTE 2011/005, (2011).

\bibitem{website}
\href {https://twiki.cern.ch/twiki/bin/view/CMSPublic/PhysicsResultsSUS13011}
  {{ CMS} Collaboration, ``CMS Twiki: PhysicsResultsSUS13011'',} 2013.
\newblock Temporary link for supplementary material. To be moved to a permanent
  location.

\end{thebibliography}\endgroup
